\documentclass[11pt, prd, aps, showpacs, nofootinbib,superscriptaddress]{revtex4-1}
\usepackage{graphicx}
\usepackage{dcolumn}
\usepackage{bm}
\usepackage{amsmath}
\usepackage{comment}
\usepackage[utf8]{inputenc}
\usepackage{physics}
\usepackage{xcolor}
\usepackage{mathtools}
\usepackage{bbm}
\usepackage{cancel} 

\newcommand{\Bern}{Institute for Theoretical Physics, Albert Einstein Center for Fundamental Physics,\\University of Bern, Sidlerstrasse 5, CH-3012 Bern, Switzerland}
\newcommand{\hiskp}{HISKP (Theory), Rheinische Friedrich-Wilhelms-Universit\"at Bonn,\\Nussallee 14-16, 53115 Bonn, Germany}
\newcommand{\hpca}{High Performance Computing and Analytics Lab, Rheinische Friedrich-Wilhelms-Universit\"at Bonn,\\ Friedrich-Hirzebruch-Allee 8, 53115 Bonn, Germany}
\newcommand{\CyprusU}{Department of Physics, University of Cyprus, 20537 Nicosia, Cyprus}
\newcommand{\CyprusI}{Computation-based Science and Technology Research Center, The Cyprus Institute,\\20 Konstantinou Kavafi Street, 2121 Nicosia, Cyprus}

\newcommand{\Parma}{Dipartimento  di  Scienze  Matematiche,  Fisiche  e  Informatiche,  Universit\`a  di  Parma  and  INFN, Gruppo  Collegato  di  Parma,  Parco  Area  delle  Scienze  7/a  (Campus),  43124  Parma,  Italy}
\newcommand{\Romadue}{Dipartimento di Fisica and INFN, Universit\`a di Roma ``Tor Vergata",\\Via della Ricerca Scientifica 1, I-00133 Roma, Italy}
\newcommand{\Romatre}{Dipartimento di Matematica e Fisica, Universit\`a Roma Tre and INFN, Sezione di Roma Tre,\\Via della Vasca Navale 84, I-00146 Rome, Italy}
\newcommand{\RomatreINFN}{Istituto Nazionale di Fisica Nucleare, Sezione di Roma Tre,\\Via della Vasca Navale 84, I-00146 Rome, Italy}
\newcommand{\NIC}{NIC, DESY, Platanenallee 6, D-15738 Zeuthen, Germany}

\newcommand{\be}{\begin{equation}}
\newcommand{\ee}{\end{equation}}
\newcommand{\bea}{\begin{eqnarray}}
\newcommand{\eea}{\end{eqnarray}}

\begin{document}

\title{Lattice calculation of the short and intermediate time-distance hadronic vacuum polarization contributions to the muon magnetic moment using twisted-mass fermions}

\author{C.~Alexandrou}\affiliation{\CyprusU}\affiliation{\CyprusI}
\author{S.~Bacchio}\affiliation{\CyprusI}
\author{P.~Dimopoulos}\affiliation{\Parma}
\author{J.~Finkenrath}\affiliation{\CyprusI}
\author{R.~Frezzotti}\affiliation{\Romadue} 
\author{G.~Gagliardi}\affiliation{\RomatreINFN}
\author{M.~Garofalo}\affiliation{\hiskp}
\author{K.~Hadjiyiannakou}\affiliation{\CyprusU}\affiliation{\CyprusI}
\author{B.~Kostrzewa}\affiliation{\hpca}
\author{K.~Jansen}\affiliation{\NIC}
\author{V.~Lubicz}\affiliation{\Romatre}
\author{M.~Petschlies}\affiliation{\hiskp}
\author{F.~Sanfilippo}\affiliation{\RomatreINFN}
\author{S.~Simula}\affiliation{\RomatreINFN}
\author{C.~Urbach}\affiliation{\hiskp}
\author{U.~Wenger}\affiliation{\Bern}

\begin{abstract}
\vspace{0.05cm}
\centerline{\includegraphics[height=4.7cm]{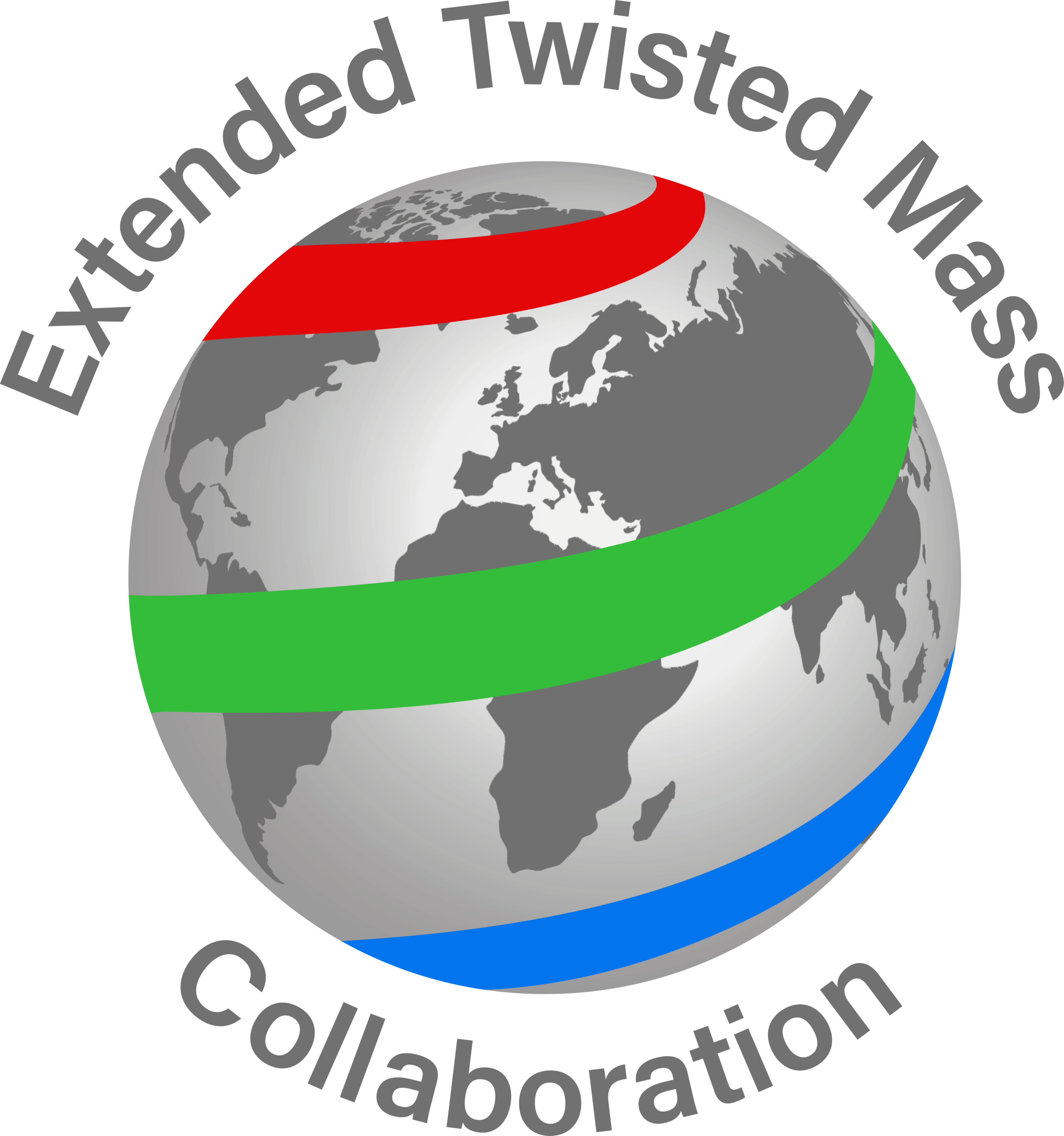}}
\vspace{0.1cm}
We present a lattice determination of the leading-order hadronic vacuum polarization (HVP) contribution to the muon anomalous magnetic moment, $a_{\mu}^{\rm HVP}$, in the so-called short and intermediate time-distance windows, $a_{\mu}^{\rm SD}$ and $a_{\mu}^{\rm W}$, defined by the RBC/UKQCD Collaboration\,\cite{RBC:2018dos}. We employ gauge ensembles produced by the Extended Twisted Mass Collaboration (ETMC) with $N_f = 2 + 1 + 1$ flavors of Wilson-clover twisted-mass quarks with masses of all the dynamical quark flavors tuned close to their physical values. The simulations are carried out at three values of the lattice spacing equal to $\simeq 0.057, 0.068$ and $0.080$ fm with spatial lattice sizes up to $L \simeq 7.6$~fm. For the short distance window we obtain $a_\mu^{\rm SD}({\rm ETMC}) = 69.27\,(34) \cdot 10^{-10}$, which is consistent with the recent dispersive value of $a_\mu^{\rm SD}(e^+ e^-) = 68.4\,(5) \cdot 10^{-10}$\,\cite{Colangelo:2022vok}. 
In the case of the intermediate window we get the value $a_\mu^{\rm W}({\rm ETMC}) = 236.3\,(1.3) \cdot 10^{-10}$, which is consistent with the result $a_\mu^{\rm W}({\rm BMW}) = 236.7\,(1.4) \cdot 10^{-10}$\,\cite{Borsanyi:2020mff} by the BMW collaboration as well as with the recent determination by the CLS/Mainz group of $a_\mu^{\rm W}({\rm CLS}) = 237.30\,(1.46) \cdot 10^{-10}$ \,\cite{Ce:2022kxy}.
However, it is larger than the dispersive result of $a_\mu^{\rm W}(e^+ e^-) = 229.4\,(1.4) \cdot 10^{-10}$\,\cite{Colangelo:2022vok} by approximately $3.6$ standard deviations. The tension increases to approximately $4.5$ standard deviations if we average our ETMC result with those by BMW and CLS/Mainz. Our accurate lattice results in the short and intermediate windows  point to a possible deviation of the $e^+ e^-$ cross section data with respect to Standard Model  predictions in the low and intermediate energy regions, but not in the high energy region. 
\end{abstract}

\maketitle

\section{Introduction}
\label{sec:introduction}

The anomalous magnetic moment of the muon $a_{\mu} \equiv (g-2)/2$, is one of the most  precisely determined quantities in  physics, both experimentally and theoretically. It is a crucial quantity  for which a long-standing tension between the experimental value and the Standard Model (SM) prediction might provide  important evidence for New Physics (NP) beyond the SM. The Fermilab Muon $g-2$ experiment (E989), has recently published the results of the analysis of the Run-1 data collected in 2018\,\cite{Muong-2:2021ojo, Muong-2:2021ovs, Muong-2:2021xzz, Muong-2:2021vma}, finding a remarkable good agreement with the previous E821 measurement at BNL\,\cite{Muong-2:2006rrc}. The current experimental world average\,\cite{Muong-2:2021ojo}
\be
    \label{exp:val}
    a_{\mu}^{exp} = 116\,592\,061 (41) \cdot 10^{-11}~, 
\ee
has a relative uncertainty of $0.35$ ppm. The ongoing data analysis of the second and third runs at Fermilab, will allow to further reduce the uncertainty on the experimental value by a factor of four, and a forthcoming experiment at J-PARC (E34)\,\cite{Abe:2019thb} is expected to reach a similar precision.

From the theoretical side, the dominant source of uncertainty in the determination of $a_{\mu}$ comes from the leading-order Hadronic Vacuum Polarization (HVP) term $a_{\mu}^{\rm HVP}$ of order $\mathcal{O}(\alpha_{em}^{2})$, and to a less extent, from the Hadronic Light-by-Light (HLbL) scattering contributions of order $\mathcal{O}(\alpha_{em}^{3})$. The most precise prediction for the HVP contribution has been obtained till now using a data-driven approach, in which the HVP contribution is reconstructed from the experimental cross section data for electron-positron annihilation into hadrons, using dispersion relations and assuming only SM physics at high energy\,\cite{Davier:2017zfy, Keshavarzi:2018mgv, Colangelo:2018mtw, Hoferichter:2019mqg, Keshavarzi:2019abf, Davier:2019can}. Such dispersive analyses find~\cite{Aoyama:2020ynm} a value of
\be
    \label{eq:HVP_dispersive}
    a_{\mu}^{\rm HVP}({\rm disp.}) = 6\,931 (40) \cdot 10^{-11} ~ , ~
\ee
which corresponds to an overall uncertainty on $a_{\mu}$ of $0.37$ parts per million (ppm). The difference between the experimental result of Eq.~(\ref{exp:val}) and the value $a_\mu^{SM}$, obtained by using Eq.~(\ref{eq:HVP_dispersive}) for the HVP contribution, is 
\be
    \label{eq:anomaly}
    \Delta a_{\mu} = a_{\mu}^{exp} - a_{\mu}^{SM} = 251 (41) (43) \cdot 10^{-11} = 251 (59) \cdot 10^{-11} ~ , ~
\ee
where the first error is from the experiment and the second one from the theory\,\cite{Aoyama:2020ynm}.
The difference given in Eq.\,(\ref{eq:anomaly}) corresponds to a discrepancy of $4.3$ standard deviations ($4.3\,\sigma$).
To match the accuracy of the upcoming experimental results, it is very important to check the result of the dispersive analysis using different methods, and to reduce the theoretical uncertainty. To this end a complementary and powerful approach to compute the HVP term is provided by lattice QCD (LQCD)\,\cite{Aoyama:2020ynm}. In the LQCD formulation, $a_{\mu}^{\rm HVP}$ can be extracted from the zero three-momentum Euclidean correlation function of two electromagnetic (em) currents, $V(t)$, employing the so-called Euclidean time-momentum representation, as described in Section\,\ref{sec:definitions}.

In recent years, impressive progress has been made by the LQCD community enabling the  evaluation of  $a_{\mu}^{\rm HVP}$ with increasing precision,  reaching the goal of a few permille accuracy. A breakthrough concerning the precision achieved came from the recent lattice calculation performed by the BMW Collaboration, that found a values of $a_\mu^{\rm HVP}({\rm BMW}) = 7\,075 (55) \cdot 10^{-11}$\,\cite{Borsanyi:2020mff}, corresponding to a relative uncertainty of $0.8\%$. The result of the BMW Collaboration differs from the dispersive one of Eq.~(\ref{eq:HVP_dispersive}) at the level of $2.1\sigma$ and reduces the difference given by Eq.\,(\ref{eq:anomaly}) to $1.5 \sigma$.  Independent LQCD determinations of the HVP term  with a few permille accuracy are needed in order to confirm the BMW result. This requires a joint effort from the lattice  QCD community because of the large degree of complexity inherent to such calculations and of the delicate task of controlling all sources of systematic errors in order to achieve the targeted precision.

In this respect, the so-called short and intermediate time-distance windows, introduced by the RBC-UKQCD Collaboration\,\cite{RBC:2018dos}, are very important benchmark quantities. They allow for comparisons not only among determinations from lattice methods, which are {\it ab initio} SM predictions, but also with the results obtained by the dispersive approach using the experimental data from $e^+ e^-\rightarrow$ hadrons.
By modifying the integration kernel using suitably defined smooth step-functions that are tailored to exponentially suppress contributions from given time regions, it is possible to decompose the full HVP as the sum of three terms
\be
    \label{eq:amuw}
    a_\mu^{\rm HVP} \equiv a_\mu^{\rm SD} + a_\mu^{\rm W} + a_\mu^{\rm LD} ~ , ~  
\ee
which probe separately short- ($a_\mu^{\rm SD}$), intermediate- ($a_\mu^{\rm W}$) and long-distance physics ($a_\mu^{\rm LD}$), respectively. 
In the long-distance window, since the tail of the correlator $V(t)$ is dominated by light two-pion states,  one typically observes large statistical noise and large finite size effects (FSEs), the treatment of which requires refined techniques. 
On the contrary, in the short- and intermediate-distance  windows, FSEs are moderate and, moreover, the lattice data are more precise, allowing a cleaner comparison across independent lattice calculations. 
The short-distance contribution $a_\mu^{\rm SD}$ suffers from large discretization artifacts, associated to the behaviour of the correlator $V(t)$ at small time distances.
Even though this may present a significant challenge, the comparison among the results obtained with different lattice regularizations provides the opportunity to test the robustness of the continuum limit extrapolation.

In this work, we present the results of the Extended Twisted Mass Collaboration (ETMC) on the short-distance and intermediate window contributions related to the isospin-symmetric up and down ($\ell$), strange ($s$) and charm ($c$) quark connected contributions, as well as the quark disconnected contributions (${\rm disc.}$). The analysis is performed using gauge configurations generated by ETMC with $N_f = 2 + 1 + 1$ flavors of Wilson Clover twisted-mass sea quarks with masses tuned very close~\footnote{The sea $s$ and $c$ quark mass values are fixed by imposing at a few percent accuracy level the physical conditions detailed in Ref.\,\cite{Alexandrou:2018egz}, while the sea light-quark mass values lie for all lattice spacings within 5-10\% from the ``physical" value defined by the $M_\pi^{isoQCD} = 135$~MeV, and a check and/or correction for the effect of the corresponding mismatch on the observables of interest is carefully carried out in our analyses. For more details see Appendix~\ref{sec:simulations}.}
to their physical values~\cite{Alexandrou:2018egz, ExtendedTwistedMass:2020tvp, ExtendedTwistedMass:2021qui, Finkenrath:2022eon}. We will refer to these ensembles as physical point ensembles. These ensembles correspond to three values of the lattice spacing, namely $a \simeq 0.057, 0.068, 0.080$~fm, determined in the meson sector (see  Appendix\,\ref{sec:spacing}), and spatial lattice sizes  ranging from $L \simeq 5.1$~fm to $L \simeq 7.6$~fm. 

 Using such physical point gauge ensembles 
 better controls the systematic error arising from the chiral extrapolation that would be required had one used ensembles simulated with heavier than 140 MeV pions. After the continuum and infinite volume extrapolations, we obtain for the short-distance window 
\bea
    a_\mu^{\rm SD}(\ell) & = & 48.24 (20) \cdot 10^{-10} ~ , ~ \nonumber \\
    \label{eq:amuSD_results}
    a_\mu^{\rm SD}(s) & = & 9.074 (64) \cdot 10^{-10} ~ , ~ \\ 
    a_\mu^{\rm SD}(c) & = & 11.61 (27) \cdot 10^{-10} ~ , ~ \nonumber \\
    a_\mu^{\rm SD}(\textrm{disc.}) & = & -0.006(5) \cdot 10^{-10} ~ , ~ \nonumber
\eea
where the first three results refer to the quark connected contributions to $a_\mu^{\rm SD}$ from light, strange and charm quarks and the latter is the sum of all quark disconnected (flavour diagonal and off-diagonal) contributions~\footnote{The separation of quark connected and disconnected contributions to a given correlator can be expressed in terms of local correlators by formally introducing, when needed, a suitable number of extra  valence flavours (having the same masses as the physical quarks) and the corresponding ghosts. In this work, the different contributions to $a_\mu^{\rm SD,W}$ can be separately extracted from local current-current correlators computed within the renormalizable mixed action lattice setup described in detail in Appendix~\ref{sec:simulations}.}.
Adding to Eq.\,(\ref{eq:amuSD_results}) also the contribution $a_\mu^{\rm SD}(b) = 0.32 \cdot 10^{-10}$ coming from the bottom quark (see also the lattice results of Ref.\,\cite{Hatton:2021dvg}) and the QED correction $a_\mu^{\rm SD}({\rm QED}) = 3 \cdot 10^{-12}$, both estimated in perturbative QCD and QED using the ``rhad" software package\,\cite{Harlander:2002ur}, we get
\be
    \label{eq:amu_SD_final}
    a_\mu^{\rm SD}({\rm ETMC}) = 69.27 (34) \cdot 10^{-10} ~ . ~
\ee
In the case of the intermediate-distance window we obtain
\bea
    \label{eq:amu_W_contributions}
    a_\mu^{\rm W}(\ell) & = & 206.5 (1.3) \cdot 10^{-10} ~ , ~ \nonumber \\
    \label{eq:amuW_results}
    a_\mu^{\rm W}(s) & = & 27.28 (20) \cdot 10^{-10} ~ , ~  \\ 
    a_\mu^{\rm W}(c) & = & 2.90 (12) \cdot 10^{-10} ~ , ~ \nonumber \\ 
    a_\mu^{\rm W}(\textrm{disc.}) & = & -0.78 (21) \cdot 10^{-10}  ~ . ~ \nonumber
\eea
 We note that in this work we do not compute the isospin-breaking (IB) contribution $a_\mu^{\rm W}(IB)$. Taking for the latter the BMW value of Ref.\,\cite{Borsanyi:2020mff}, namely $a_\mu^{\rm W}(IB) = 0.43 (4) \cdot 10^{-10}$, and summing up with the contributions\, of Eq.~(\ref{eq:amu_W_contributions}), we get
\be
    \label{eq:amu_W}
    a_\mu^{\rm W}({\rm ETMC}) = 236.3 (1.3) \cdot 10^{-10} ~ , ~
\ee
which is consistent both with the BMW result $a_{\mu}^{\rm W}({\rm BMW}) = 236.7(1.4) \cdot 10^{-10}$\,\cite{Borsanyi:2020mff} and the recent CLS/Mainz one $a_\mu^{\rm W}({\rm CLS}) = 237.30\,(1.46) \cdot 10^{-10}$ \,\cite{Ce:2022kxy} to better than $1\sigma$ level.
The nice consistency observed among three accurate determinations of $a_\mu^{\rm W}$ represents a remarkable success for LQCD computations.

We can compare our lattice results with those obtained with dispersive methods using the experimental $e^+ e^- \to $ hadrons data. 
The dispersive results obtained in Refs.\,\cite{Keshavarzi:2019abf,KNT} are
\bea
    \label{eq:amu_SD_disp_KNT}
    a_\mu^{\rm SD}(e^+ e^-) & = & 68.44 (48) \cdot 10^{-10} ~ , ~ \\
    \label{eq:amu_W_disp_KNT}
    a_\mu^{\rm W}(e^+ e^-) & = & 229.51 (87) \cdot 10^{-10} ~ , ~
\eea
while the analyses of Refs.\,\cite{Keshavarzi:2018mgv, Colangelo:2018mtw, Hoferichter:2019mqg, Keshavarzi:2019abf} and the merging procedure of Ref.\,\cite{Aoyama:2020ynm}, which takes into account tensions in the $e^+ e^-$ database in a more conservative way, yield\,\cite{Colangelo:2022vok}
\bea
    \label{eq:amu_SD_disp}
    a_\mu^{\rm SD}(e^+ e^-) & = & 68.4 (5) \cdot 10^{-10} ~ , ~ \\
    \label{eq:amu_W_disp}
    a_\mu^{\rm W}(e^+ e^-) & = & 229.4 (1.4) \cdot 10^{-10} ~ . ~ 
\eea
Our result given in Eq.(\ref{eq:amu_SD_final}) agrees with the dispersive one of Eq.~(\ref{eq:amu_SD_disp}) within $\simeq 1.4 \sigma$ in the short-distance window, while there is a $\simeq 3.6 \sigma$ tension between our result of Eq.~(\ref{eq:amu_W}) and the corresponding dispersive one of Eq.~(\ref{eq:amu_W_disp}) in the intermediate window. 
The tension increases to $\simeq 4.2 \sigma$ if we average our result of  Eq.~(\ref{eq:amu_W}) with the BMW one, obtaining $a_\mu^{\rm W} = 236.49 (95) \cdot 10^{-10}$. Taking into account also the recent CLS/Mainz result we get an average of three lattice computations equal to $a_\mu^{\rm W} = 236.73 (80) \cdot 10^{-10}$, which turns out to be in tension with the dispersive result of Eq.\,(\ref{eq:amu_W_disp_KNT}) by $\simeq 6.1 \sigma$ and the more conservative result\,of Eq.~(\ref{eq:amu_W_disp}) by $\simeq 4.5 \sigma$.

 The impact of this work is twofold: Firstly, concerning the intermediate-distance window, we confirm the two recent and most accurate LQCD results by BMW and CLS/Mainz, the consequence of which is to  increase the discrepancy with the corresponding prediction based on $e^+ e^-$ cross section data to the remarkable significance level of $\simeq 4.5$ standard deviations. Secondly, we  accurately compute, for the first time, the contribution from the short-distance window, showing that there is no significant tension with the corresponding dispersive result. This clearly indicates that any deviation between the QCD+QED theory predictions, as used in the low-energy SM framework used in  lattice calculations, and the $e^+ e^-$ cross section experiments is unlikely to occur at high values of the center-of-mass energy, which corresponds to small values of the Euclidean time distance. Instead, a significant deviation of QCD+QED predictions from $e^+ e^-$ cross section data may occur in the low and/or intermediate energy regions. Such a possibility has been also discussed in recent works\,\cite{Keshavarzi:2020bfy, Crivellin:2020zul, Malaescu:2020zuc, DiLuzio:2021uty, Ce:2022eix} exploiting the constraints from SM electroweak precision tests and low energy observables.

The paper is organized as follows: In Section~\ref{sec:definitions}, we provide the relevant notations and the definitions of the time windows. 
In Section~\ref{sec:connected} we present our determinations of the light-, strange- and charm-quark connected contributions to the vector correlator and we describe some basic steps of our strategy to reach the physical point for all the time windows. The subsections~\ref{sec:amuSD} and \ref{sec:amuW} contain, respectively, the results of our detailed analysis of the continuum limit for the short- and intermediate-distance windows for all the quark flavors. In Section~\ref{sec:disconnected}, we evaluate the sum of all quark disconnected flavour diagonal and off-diagonal contributions. Section~\ref{sec:comparison} is devoted to comparing with other available LQCD calculations as well as with the most recent dispersive results available for the HVP time-window observables according to Refs.\,\cite{Colangelo:2022vok,KNT}. 
The outcome of the comparison with dispersive predictions and its phenomenological implications are briefly discussed.
Our conclusions are summarized in Section\,\ref{sec:conclusions}.

Further in depth technical information is given in the Appendices as follows:
In Appendix\,\ref{sec:simulations}, we briefly describe our mixed-action setup and give details about the lattice simulations, including an improved determination of the lattice spacing with respect to the one carried out in Ref.\,\cite{ExtendedTwistedMass:2021qui}.
In Appendix\,\ref{sec:renormalization}, we evaluate the scale-invariant renormalization constants (RCs) of the vector and axial-vector local quark currents, $Z_V$ and $Z_A$, employing a hadronic method based on Ward Identities (WIs). Combined with a high statistics determination of the relevant correlators, we achieve a high precision determination of $Z_V$ and $Z_A$, as needed to guarantee a final accuracy of $\simeq 0.5 \%$ for the short and intermediate time-distance windows.
In Appendix\,\ref{sec:masses}, we briefly describe our strategy to reach the physical values of the strange- and charm-quark masses, $m_s^{phys}$ and $m_c^{phys}$, using various hadronic inputs.
In Appendix\,\ref{sec:appD}, we show that for our lattice setup the flavor-singlet local vector current renormalizes with the same renormalization constants (RCs) as the non-singlet vector current at all orders in the strong coupling. In Appendix\,\ref{sec:appE}, we collect the relevant analytic formulae for the evaluation of the leading lattice spacing artifacts at short time-distances in the free theory, i.e.~at order ${\cal{O}}(\alpha_s^0)$. In Appendix\,\ref{sec:appF}, we provide some details of our parameterization of finite size effects (FSEs) in the time windows.


\section{Time-momentum representation}
\label{sec:definitions}

Following our previous works\,\cite{Giusti:2017jof,Giusti:2018mdh,Giusti:2019xct}, we adopt the time momentum representation\,\cite{Bernecker:2011gh} and evaluate the HVP contribution to the muon anomalous magnetic moment $a_{\mu}^{\rm HVP}$ as
\be
    \label{eq:amu_HVP}
    a_{\mu}^{\rm{HVP}} = 2 \alpha_{em}^2 \int_0^\infty ~ dt \, t^2 \, K(m_\mu t) \,V(t) ~ , ~  
\ee
where $t$ is the Euclidean time and the kernel function $K(m_{\mu} t)$ is defined as\footnote{The leptonic kernel $K(z)$ is proportional to $z^2$ at small values of $z$ and it goes to $1$ for $z \to \infty$.}
\be
    \label{eq:kernel}
    K(z) = 2 \int_0^1 dy ( 1- y) \left[ 1 - j_0^2 \left(\frac{z}{2}\frac{y}{\sqrt{1 - y}} \right) \right]~,\qquad j_{0}(y) = \frac{\sin{(y)}}{y} ~ . ~
\ee
The Euclidean vector correlator $V(t)$ is defined as
 \be
     \label{eq:VV}
     V(t) \equiv \frac{1}{3} \sum_{i=1,2,3} \int d^3{x} ~ \langle J_i(\vec{x}, t) J_i^\dagger(0) \rangle 
 \ee
with $J_\mu(x)$ being the electromagnetic (em) current operator
 \be
      \label{eq:Jmu}
     J_\mu(x) \equiv \sum_{f = u, d, s, c, ...} q_{em,f} ~ \overline{\psi}_f(x) \gamma_\mu \psi_f(x) ~ 
 \ee
 and $q_{em,f}$ the electric charge for the quark flavor $f$ (in units of the absolute value of the electron charge). 
The vector correlator $V(t)$ can be evaluated on a lattice with spatial volume $V = L^3$ and time extent $T$ at discretized values of the time distance $t / a$, ranging from $0$ to $T / a$.

\subsection{The RBC/UKQCD windows in the time-momentum representation}
\label{sec:windows_t}

\begin{figure}[htb!]
\begin{center}
\includegraphics[scale=0.60]{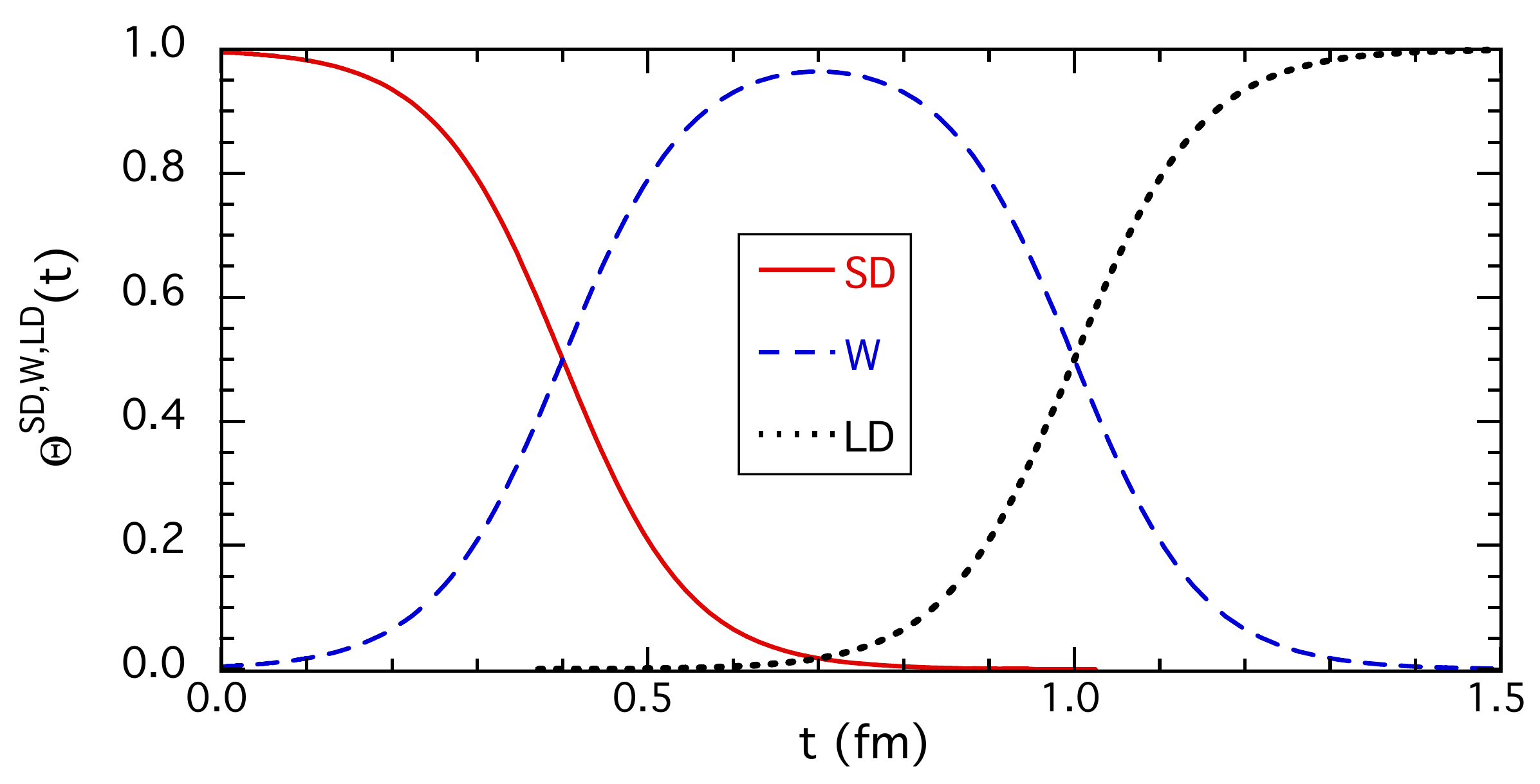}
\vspace{-0.5cm}
\caption{\it \small The time-modulating function $\Theta^w(t)$ for $w = \{\rm SD, W, LD \}$, defined in Eqs.\,(\ref{eq:Mt_SD})-(\ref{eq:Mt_LD}), versus the time distance $t$ for the values of the parameters $t_0, t_1$ and $\Delta$ given in Eq.\,(\ref{eq:parms}).}
\label{fig:windows_t}
\end{center}
\end{figure}

Following the analysis of the RBC/UKQCD Collaboration\,\cite{RBC:2018dos}, each of the three terms appearing in Eq.~(\ref{eq:amuw}) can be obtained from  Eq.\,(\ref{eq:amu_HVP})  with integration kernel $K(m_\mu t)$ multiplied by suitably smoothed Heaviside step-functions, namely
\be
    \label{eq:amu_w}
    a_\mu^w = 2 \alpha_{em}^2 \int_0^\infty ~ dt \, t^2 \, K(m_\mu t) \, \Theta^w(t)\,V(t) ~ \qquad w = \{\rm SD, W, LD \} ~ , ~ 
\ee
where the time-modulating function $\Theta^w(t)$ is given by
\bea
      \label{eq:Mt_SD}
      \Theta^{\rm SD}(t) & \equiv & 1 -  \frac{1}{1 + e^{- 2 (t - t_0) / \Delta}} ~ , ~ \\[2mm]
      \label{eq:Mt_W}
      \Theta^{\rm W}(t) & \equiv & \frac{1}{1 + e^{- 2 (t - t_0) / \Delta}} -  \frac{1}{1 + e^{- 2 (t - t_1) / \Delta}} ~ , ~ \\[2mm]      
      \label{eq:Mt_LD}
      \Theta^{\rm LD}(t) & \equiv & \frac{1}{1 + e^{- 2 (t - t_1) / \Delta}} ~  
\eea
with the parameters $t_0, t_1, \Delta$ chosen\,\cite{RBC:2018dos} to be equal to
\be
    \label{eq:parms}
    t_0 = 0.4 ~ \rm{fm} ~ , ~ \qquad t_1 = 1 ~ {\rm fm} ~ , ~ \qquad \Delta = 0.15~{\rm fm} ~ . ~    
\ee
The resulting time-modulating functions $\Theta^{\rm SD, W, LD}(t)$ are shown in Fig.\,\ref{fig:windows_t}.

In this work, we focus on the determination of the first two terms, i.e.~$w = \{\rm SD, W \}$ corresponding to the short- and intermediate-distance window contributions, postponing the analysis of the more demanding long-distance (LD) term to a future work.

The fermionic Wick contractions appearing in the right hand side (r.h.s.)~of Eq.~(\ref{eq:VV}) give rise to two distinct topologies of Feynman diagrams, namely to the quark-connected and quark-disconnected contributions.
Connected contributions are flavor diagonal, while the disconnected ones have both diagonal and off-diagonal flavor components. 
In what follows we decompose $a_\mu^w$ into the following contributions
\be
    \label{eq:amuw_decomposition}
    a_\mu^w = a_\mu^w(\ell) + a_\mu^w(s) + a_\mu^w(c) + a_\mu^w({\rm disc.}) + \ldots ~ , ~   
\ee
where the first three terms correspond to the quark-connected contributions of mass degenerate up and down ($\ell$) quarks, and a strange ($s$) and a charm ($c$) quark, respectively, while the fourth term represents all quark-disconnected contributions. In Eq.\,(\ref{eq:amuw_decomposition}) the ellipses corresponds to subleading terms that we do not address directly in this work, namely the isospin breaking (IB) effects of order $\mathcal{O}(\alpha_{em}^{3})$ and $\mathcal{O}(\alpha_{em}^{2}(m_d - m_u))$, as well as the contribution of quarks heavier than the charm. Moreover, for the disconnected term $a_\mu^{w}({\rm disc.})$ we evaluate both the flavor-diagonal and the off-diagonal light, strange and charm-quark contributions.

\subsection{The RBC/UKQCD windows in energy}
\label{sec:windows_E}

Let us here make contact with the dispersive approach used in Refs.\,\cite{Keshavarzi:2019abf,KNT}. Using the once-subtracted dispersion relations, the vector correlator $V(t)$ can be written as (see, e.g., Ref.\,\cite{Giusti:2017jof}), 
\be
     V(t) =  \frac{1}{12\pi^2} \int_{E_{thr}}^{\infty} dE E^2 R^{had}(E) e^{-E t}\,,
\ee
where $R^{\rm had}(E)$ is related to the one photon $e^+ e^-$ annihilation cross section into hadrons, $\sigma^{\rm had}(E)$, by
 \be
      \sigma^{had}(E) = \frac{4 \pi \alpha_{em}^2}{3 E^2} R^{had}(E) 
      \label{eq:Rhad}
 \ee
with $E$ being the $e^+ e^-$ center-of-mass energy and $E_{thr} = M_{\pi^0}$ in QCD+QED.
In terms of $R^{had}(E)$ the HVP term $a_\mu^{\rm HVP}$ is given by
 \be
      \label{eq:amu_HVP_E}
      a_\mu^{\rm HVP} = \frac{2 \alpha_{em}^2 m_\mu^2}{9 \pi^2} \, \int_{E_{thr}}^{\infty} dE \frac{1}{E^3} ~ \widetilde{K}\left( \frac{E}{m_\mu} \right) \, R^{had}(E) ~ ,
 \ee
where the leptonic kernel $\widetilde{K}(x)$ is defined as\footnote{The leptonic kernel $\widetilde{K}(x)$ is proportional to $x^2$ at small values of $x$ and it goes to $1$ for $x \to \infty$. At the two-pion threshold one has $\widetilde{K}(2 M_\pi / m_\mu) \simeq 0.63$.}
 \be
      \label{eq:Kx}
      \widetilde{K}(x) = \frac{3}{4} x^5 \int_0^\infty dz ~ z^2 ~ e^{-x \, z} ~ K(z) = 3 x^2 \int_0^1 dy \, (1 - y) \frac{y^2}{y^2 + (1 - y) x^2} ~ . ~
  \ee

\begin{figure}[htb!]
\begin{center}
\includegraphics[scale=0.60]{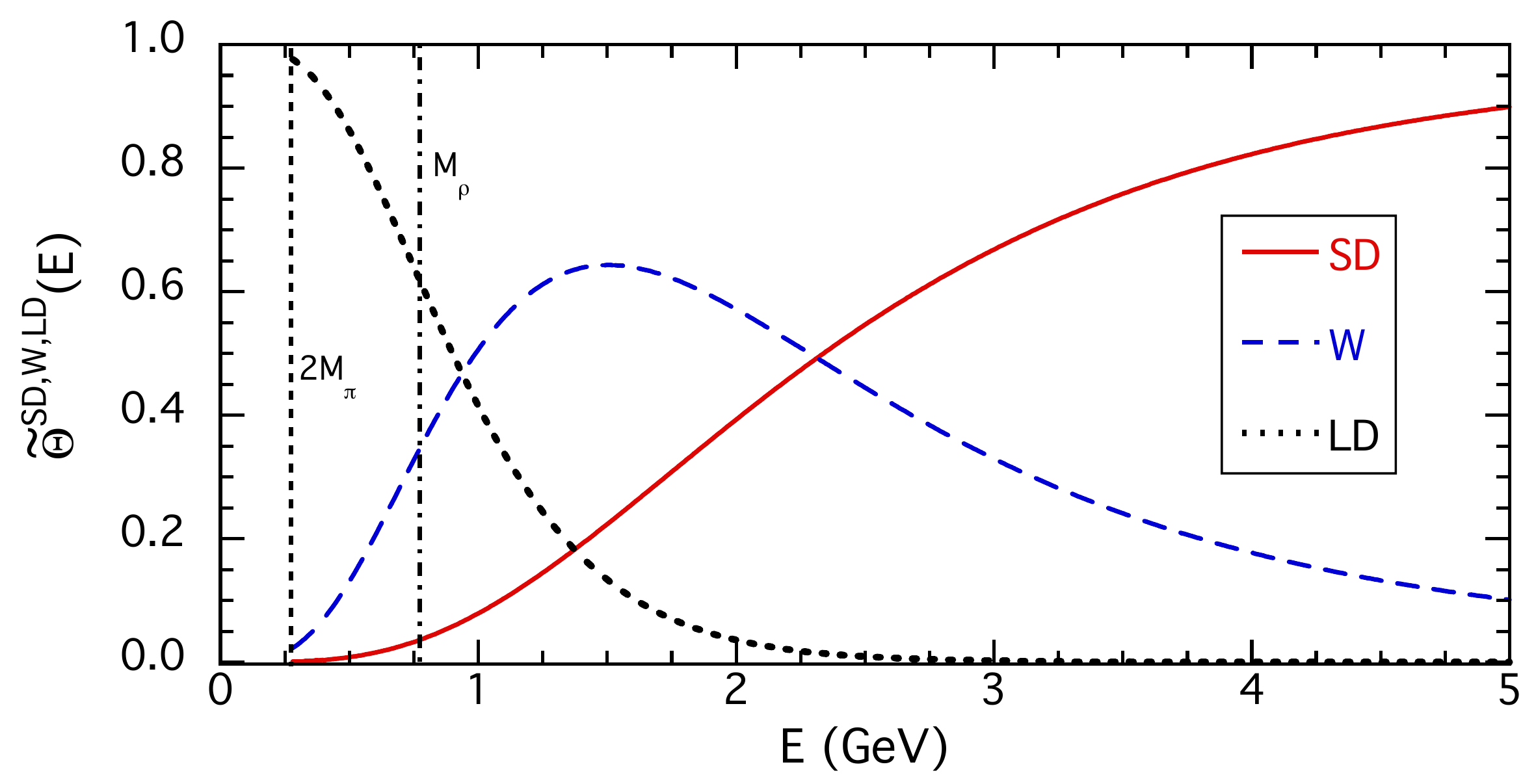}
\vspace{-0.5cm}
\caption{\it \small The energy-modulating function $\widetilde{\Theta}^w(E)$ for $w = \{ \rm SD, W, LD \}$, defined in Eq.\,(\ref{eq:ME_w}), versus the energy $E$. The vertical lines represent the location of the two-pion threshold (short-dashed) and of the $\rho$-meson resonance (dot-dashed).}
\label{fig:windows_E}
\end{center}
\end{figure}

Consequently, the time-window contribution\,(\ref{eq:amu_w}) can be written as \be
      \label{eq:amu_w_E}
      a_\mu^w = \frac{2 \alpha_{em}^2 m_\mu^2}{9 \pi^2} \, \int_{E_{thr}}^{\infty} dE \frac{1}{E^3} ~ \widetilde{K}\left( \frac{E}{m_\mu} \right) \, \widetilde{\Theta}^w(E) \, R^{had}(E) ~ , ~
\ee
where the energy-modulating function $\widetilde{\Theta}^w(E)$ is given by
\be
   \label{eq:ME_w}
   \widetilde{\Theta}^w(E)= \frac{\int_0^\infty dt ~ t^2 ~ e^{- E \, t} ~ K(m_\mu t) ~ \Theta^w(t)}{\int_0^\infty dt ~ t^2 ~ e^{- E \, t} ~ K(m_\mu t)}
\ee
and shown in Fig.\,\ref{fig:windows_E} for $w = \{ \rm SD, W, LD \}$ (cf.~also Ref.\,\cite{Colangelo:2022vok}).


\section{ The connected contributions to $a_{\mu}^{\rm SD}$ and $a_{\mu}^{\rm W}$}
\label{sec:connected}

In this work, we analyse four gauge ensembles recently produced by ETMC in isospin-symmetric QCD (isoQCD) with $N_f = 2 + 1 + 1$ flavors of Wilson-clover twisted-mass quarks as described in Refs.\,\cite{Alexandrou:2018egz, ExtendedTwistedMass:2020tvp, ExtendedTwistedMass:2021qui, Finkenrath:2022eon}.
The parameters of the ensembles are given in Table\,\ref{tab:simudetails} of Appendix\,\ref{sec:simulations}, where our lattice setup and technical details are thoroughly discussed.

In this Section, we present the evaluation of the light-, strange- and charm-quark connected vector correlators (see Eq.\,(\ref{eq:Vregdetail}))
\bea 
   V_\ell^{reg}(t) \; & \equiv & \; \frac{1}{3} a^3\sum_{\bf x} \sum_{i=1,3} \frac{4+1}{9} \left\langle J_{i, reg}^{\ell\ell'}(x) [J_{i, reg}^{\ell\ell'}]^\dagger(0) \right\rangle^{\!(C)} ~ , ~ \\[2mm]  
   V_{s}^{reg}(t)  \; & \equiv & \; \frac{1}{3} a^3\sum_{\bf x} \sum_{i=1,3} \frac{1}{9} \left\langle J_{i, reg}^{ss'}(x) [J_{i, reg}^{ss'}]^\dagger(0) \right\rangle^{\!(C)} ~ , ~ \\[2mm]  
   V_{c}^{reg}(t) \; & \equiv & \; \frac{1}{3} a^3\sum_{\bf x} \sum_{i=1,3} \frac{4}{9} \left\langle J_{i, reg}^{cc'}(x) [J_{i, reg}^{cc'}]^\dagger(0) \right\rangle^{\!(C)}  ~ , ~  
\eea
where $reg \in \{ {\rm tm,OS} \, \}$ specifies the two types of ultra-violet (UV) regularization employed for the renormalized local vector currents $J_{\mu, tm}$ and $J_{\mu, OS}$ defined in Eq.\,(\ref{eq:MAVcurr}) for each quark flavor. The suffices $\ell\ell'$, $s s'$ and $c c'$  on the currents denote that the quark and antiquark fields in each current correspond to different valence replica, thereby giving rise to connected ($C$) Wick contractions only.

Our high precision determination of the two scale-invariant RCs $Z_{V}$ and $Z_{A}$, needed to renormalize the local vector currents in the $tm$ and $OS$ regularizations, is described in Appendix\,\ref{sec:renormalization}.

\begin{figure}[htb!]
\begin{center}
\includegraphics[scale=0.80]{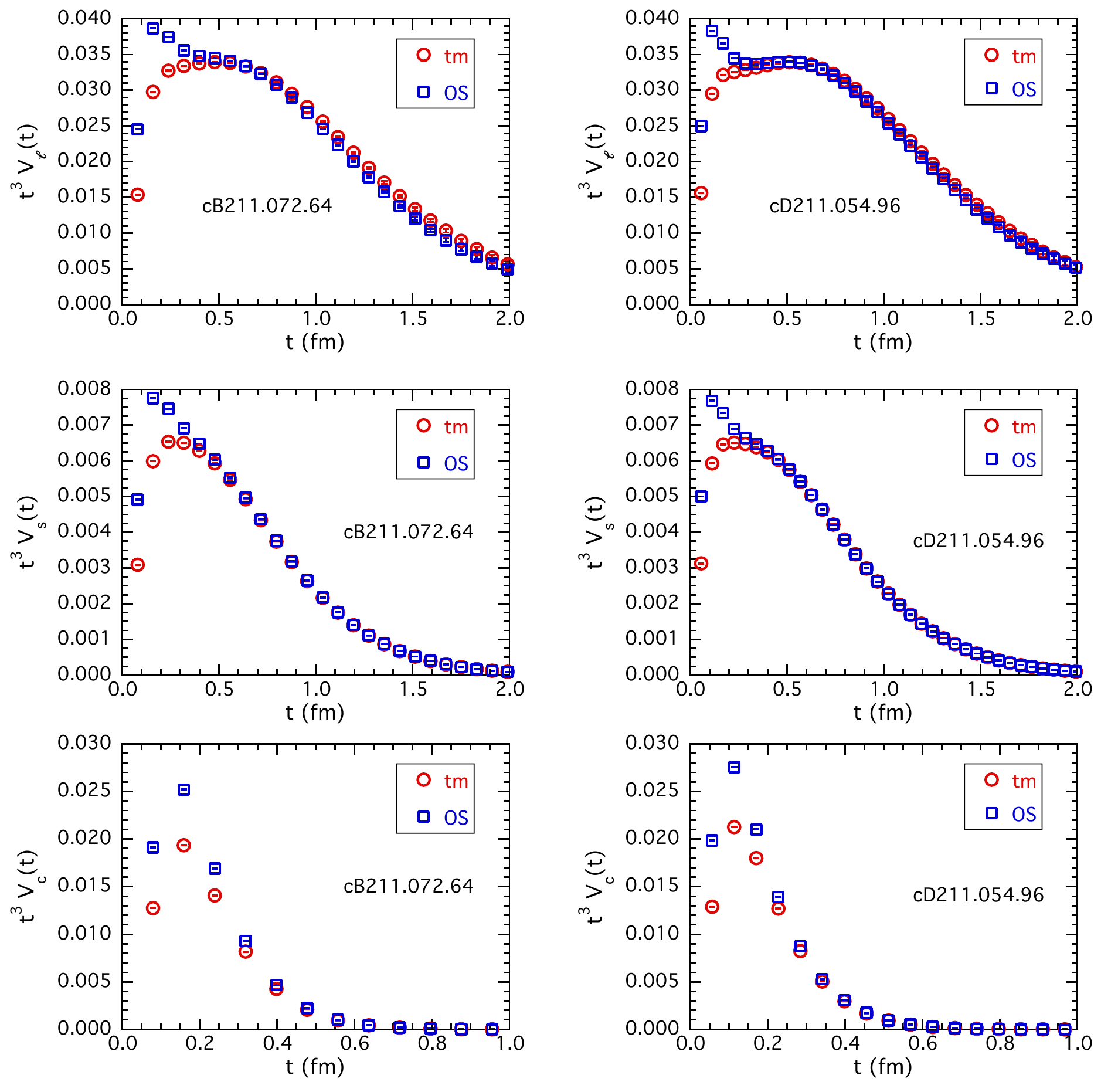}
\vspace{-0.5cm}
\caption{\it \small The connected vector correlators $t^3 V_\ell(t)$ (top), $t^3 V_{s}(t)$ (middle) and $t^3 V_{c}(t)$ (bottom) evaluated for the ETMC ensembles cB211.072.64 (left) and cD211.054.96 (right) using the two UV regularizations denoted by ``tm" for twisted mass quarks (red circles) and``OS" for Osterwalder-Seiler valence quarks (blue squares) versus the time distance $t$. For both ensembles, the bare strange- and charm-quark masses are respectively equal to $\mu_s = \mu_s^L$ and $\mu_c = \mu_c^L$ (see Table\,\ref{tab:simulated_ms} and \ref{tab:simulated_mc} of Appendix\,\ref{sec:masses}.)}
\label{fig:VV_connected}
\end{center}
\end{figure}

 Results for the  correlators $V_f^{\rm tm}(t)$ and $V_f^{\rm OS}(t)$ for $f = \{ \ell, s, c \}$ evaluated on the ETMC ensembles cB211.072.64 and cD211.054.96 are shown in Fig.\,\ref{fig:VV_connected}. For all ensembles, regularizations and quark flavors, the connected vector correlator are precise at the level of percent or better up to time distances of $\simeq 1.5$ fm. Such a range covers the whole time region relevant for the determination of the short- and intermediate-distance window contributions (see Fig.\,\ref{fig:windows_t}). 
We note that for each quark flavor the correlators $V_f^{\rm tm}(t)$ and $V_f^{\rm OS}(t)$ should differ only by discretization effects  of order ${\cal{O}}(a^2)$. From Fig.\,\ref{fig:VV_connected}, it can clearly be seen that at very small time distances, $t \lesssim 0.2 - 0.3$ fm the discretization artifacts are large, while they are small for $t \gtrsim 0.3$ fm.

For each of the four ensembles of Table~\ref{tab:simudetails} in Appendix\,\ref{sec:simulations}, the light-quark correlators $V_\ell^{\rm tm}(t)$ and $V_\ell^{\rm OS}(t)$ are computed using $N_{\rm source}=10^3$ stochastic spatial sources per gauge configuration. The sources are randomly distributed in the time-slice, diagonal in spin and diluted in the color variable. The statistical errors are found to scale as $1/\sqrt{N_{source}}$ up to time distances of $\simeq 1.5$ fm. 

The strange-quark connected vector correlators $V_s^{\rm tm}(t)$ and $V_s^{\rm OS}(t)$ are also computed using the four ensembles of Table~\ref{tab:simudetails} of Appendix\,\ref{sec:simulations}. Depending on the ensemble considered, up to $64$ spatial stochastic sources are used for the inversions of the Dirac operator. As described in the Appendix\,\ref{sec:masses}, for each ensemble we perform simulations at two values of the valence bare strange-quark mass, $a\mu_s$, in order to interpolate the results for $a_\mu^{\rm SD}(s)$ and $a_\mu^{\rm W}(s)$ to the physical strange-quark mass $m_s^{phys}$.

Unlike the light and strange sectors, the charm-quark connected vector correlators $V_c^{\rm tm}(t)$ and $V_c^{\rm OS}(t)$ are computed using the six ETMC ensembles listed in Table\,\ref{tab:simulated_mc} of Appendix\,\ref{sec:masses}. Namely, beyond the three physical point ensembles cB211.072.64, cC211.060.80 and cD211.054.96, we include in this analysis three ensembles at a coarser lattice spacing, namely cA211.53.24, cA211.40.24 and cA211.30.32\,\cite{ExtendedTwistedMass:2021gbo, ExtendedTwistedMass:2021qui}, aiming at a better control of discretization effects in the charm sector\footnote{The pion mass and the value of $M_\pi L$ for the cA211.53.24, cA211.40.24 and cA211.30.32 ensembles are equal to $M_\pi \simeq 365, 302, 261$ MeV and $M_\pi L \simeq 4.0, 3.5, 4.0$,  respectively. FSEs are expected to be negligible for the vector correlator $V_c(t)$ and thus, we do not include the cB211.072.96 ensemble in the charm analysis.
}. 
The improved value of the lattice spacing $a$ for the A ensembles is given in Table\,\ref{tab:spacing} of Appendix\,\ref{sec:simulations}. Depending on the ensemble considered, up to $24$ spatial stochastic sources are used for the inversions of the Dirac operator. For each ensemble, we perform simulations at three values of the valence bare charm-quark mass, $a\mu_c$, in order to interpolate the results for $a_\mu^{\rm SD}(c)$ and $a_\mu^{\rm W}(c)$ to the physical charm-quark mass $m_c^{phys}$ as determined in Appendix\,\ref{sec:masses}.

We interpolate/extrapolate our data for $a_\mu^{\rm w}(s)$ and $a_\mu^{\rm w}(c)$ to the physical strange- and charm-quark masses $m_s^{phys}$ and $m_c^{phys}$ using a linear Ansatz
\be
    a_\mu^{\rm w}(f, m_f) = a_\mu^{\rm w}(f, m_f^{phys}) \cdot \left[ 1 + A_f^w \left( m_f - m_f^{phys} \right) \right] ~,    
\ee
where $f = \{s, c \}$, and $a_\mu^{\rm w}(f, m_f^{phys})$ and $A_f^w$ are fitting parameters. 
In what follows, we consider two different branches of analysis, in which $a_\mu^{\rm w}(s, m_s^{phys})$ ($a_\mu^{\rm w}(c, m_c^{phys})$) is determined using the values of $m_s^{phys}$ ($m_c^{phys}$) obtained using either the $\eta_s$ ($\eta_c$) or the $\phi$ ($J/\Psi$) meson masses.
Then, we perform a separate continuum limit extrapolation for both determinations. Any discrepancy between the continuum extrapolated values obtained using the two hadronic inputs will be added as a systematic error in the final error budget. 

In order to avoid the use of fitting procedures to take into account the slight mistuning of the  pion mass in the simulations as compared to its physical value ($M_\pi^{phys} = M_\pi^{isoQCD} = 135.0\,(2)$ MeV\,\cite{ExtendedTwistedMass:2021qui}), as well as the possible impact of FSEs, we implement in our analysis of all the windows the following three steps:

\begin{itemize}

\item Interpolation to the physical value of the pion mass for each gauge ensemble through explicit simulations at a slightly different value of the light-quark valence bare mass $a \mu_\ell$. By using the same gauge configurations and stochastic sources we get a statistically good determination of the mistuning of the light-quark valence mass, which turns out to be at the level of approximately one to two standard deviations. A further smaller correction due to the slight mistuning of the light-quark sea mass is evaluated adopting the RM123 expansion method\,\cite{deDivitiis:2011eh, deDivitiis:2013xla, Giusti:2019xct}.

\item Usage of a common reference lattice size $L_{ref} = 5.46$ fm through a smooth interpolation of the results for the ensembles cB211.072.64 and cB211.072.96 using a linear fit in the variable $e^{-M_\pi L}$. For the other two ensembles cC211.060.80 and cD211.054.96, when using the improved determination of the lattice spacing (see Table\,\ref{tab:spacing} of Appendix\,\ref{sec:simulations}), the lattice size $L$ is at the correct reference value (see Table~\ref{tab:simudetails} of Appendix\,\ref{sec:simulations}). By comparing the window results for the ensembles cB211.072.64 and  cB211.072.96, we observe as a general trend that, at the lattice spacing of $a\simeq 0.08~{\rm fm}$, FSEs are small in the ``tm" case and practically absent in the ``OS" case. Moreover, once the data are interpolated at the reference spatial lattice size $L_{ref} = 5.46$ fm, the infinite volume limit is  obtained within a fraction of the uncertainties, for all windows except for the case of the light-quark contribution to the intermediate window, $a_\mu^W(\ell)$.

\item For $a_{\mu}^{\rm W}(\ell)$, after taking the continuum limit, we apply a final correction $\Delta a_\mu^{\rm W}(\ell; L_{ref})$ to obtain the infinite volume result (see Eq.\,(\ref{eq:FSE}) of Section\,\ref{sec:amuW}). The correction is evaluated assuming dominance of the FSEs related to intermediate two-pion states in the correlator $V_\ell(t)$, as already observed in Ref.\,\cite{Giusti:2018mdh}. Its explicit expression is given by Eq.\,(\ref{eq:Delta_amuw}) with $w = W$ in Appendix\,\ref{sec:appF}. It contains no free parameters and we will refer to it as the Meyer-Lellouch-L\"uscher-Gounaris-Sakurai (MLLGS) model\,\cite{Luscher:1985dn, Luscher:1986pf, Luscher:1990ux, Luscher:1991cf, Lellouch:2000pv, Meyer:2011um, Francis:2013fzp, Gounaris:1968mw} for FSEs.

\end{itemize}

In what follows, we analyze our lattice data of $a_\mu^w(f)$ for $w = \{SD, W \}$ and $f = \{\ell, s, c \}$ already interpolated at the physical pion mass $M_\pi^{phys} = M_\pi^{isoQCD} = 135.0\,(2)$ MeV and at the reference lattice size $L_{ref} = 5.46$ fm.


\subsection{The short-distance window contributions  $a_{\mu}^{\rm SD}(\ell)$, $a_{\mu}^{\rm SD}(s)$ and $a_{\mu}^{\rm SD}(c)$}
\label{sec:amuSD}

The connected contribution $a_\mu^{\rm SD}(f)$ to the short-distance window is given by
\be
    \label{eq:amuSD}
    a_\mu^{\rm SD}(f) = 2 \alpha_{em}^2 \int_0^\infty ~ dt \, t^2 \, K(m_\mu t) \, \Theta^{\rm SD}(t)\,V_{f}(t) ~, 
\ee
where $f = \{\ell, s, c \}$ and $\Theta^{\rm SD}(t)$ is given by Eq.\,(\ref{eq:Mt_SD}).
In what follows,  window quantities, like $a_\mu^{\rm SD}(f)$, are obtained on each gauge ensemble by replacing the time integral with a discrete sum over time-slices from $t = a$ up to $t = T/2$.

Even if the lattice data for the vector correlators are ${\cal{O}}(a)$-improved thanks to our maximally twisted lattice setup (see Appendix\,\ref{sec:simulations}), care should be taken when considering the continuum limit
of the short-distance window contributions $a_{\mu}^{\rm SD}(\ell)$, $a_{\mu}^{\rm SD}(s)$ and $a_{\mu}^{\rm SD}(c)$. We illustrate this point in the case of the light-quark contribution $a_{\mu}^{\rm SD}(\ell)$, but similar conclusions hold as well also in the case of $a_{\mu}^{\rm SD}(s)$ and $a_{\mu}^{\rm SD}(c)$.

\begin{figure}[htb!]
\begin{center}
\includegraphics[scale=0.60]{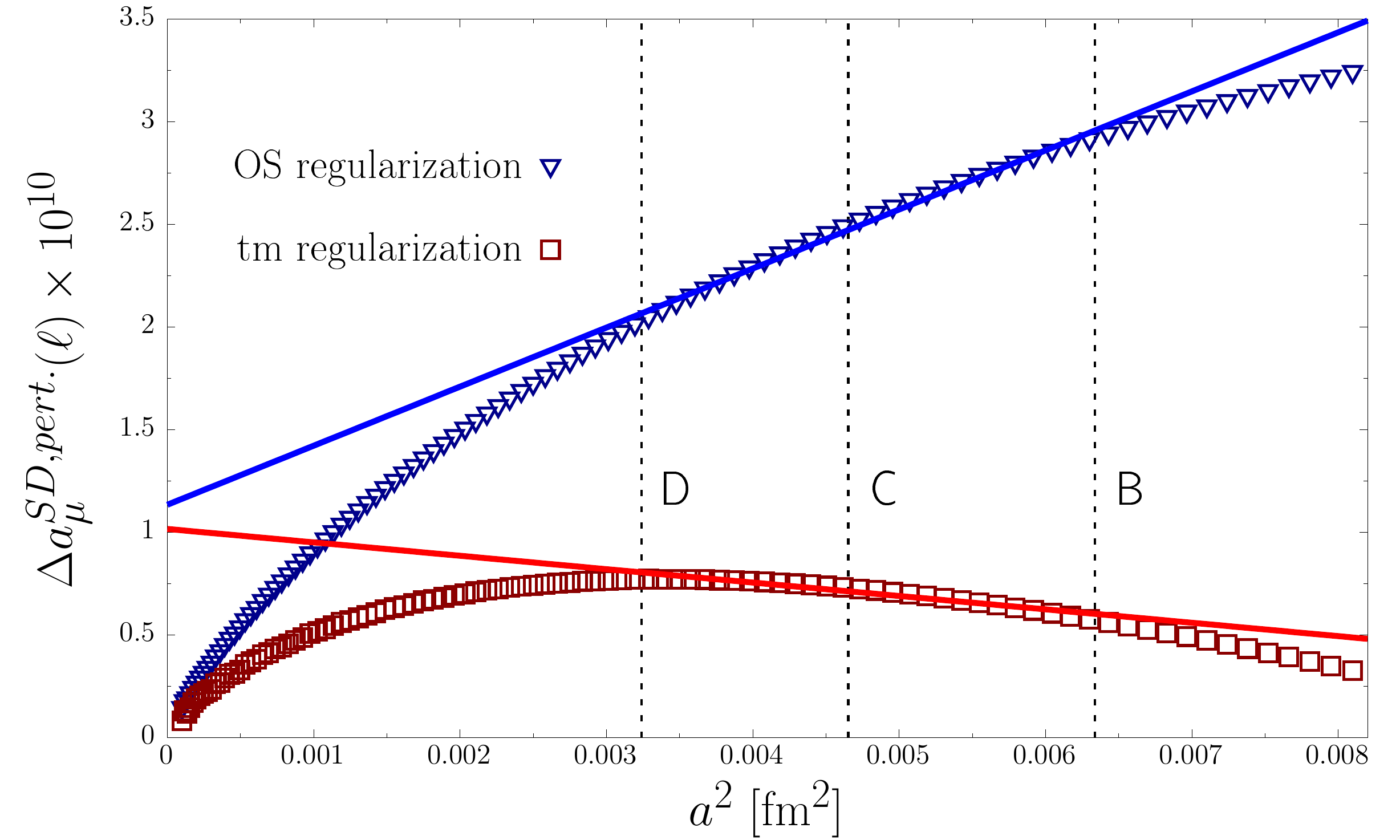}
\vspace{-0.5cm}
\caption{\it \small The discretization effects on $a_\mu^{\rm SD}(\ell)$ evaluated numerically in the free theory and in the massless limit, i.e.~the quantity $\Delta a_\mu^{\rm SD, pert}(\ell)$ given by Eq.\,(\ref{eq:Delta_amu_SD_pert}). The vertical dotted lines indicate the simulated values of the squared lattice spacing. The red squares and the blue triangles correspond respectively to the ``tm" and ``OS" regularizations. The solid lines represent a {\it naive} $a^2$-scaling performed using only the results corresponding to the range of simulated values of the lattice spacing.}
\label{fig:scaling_pert_SD}
\end{center}
\end{figure}

As discussed in Refs.~\cite{DellaMorte:2008xb,Ce:2021xgd}, power counting suggests that for short time distances, i.e.~$t \ll \Lambda_{QCD}^{-1}$, the lattice spacing artifacts in $V_\ell(t)$ can be described by an expansion of the type
\be
    \label{eq:Vud_a2}
    V_\ell(t) = V_\ell^{cont}(t)\cdot\left[ 1 + b_2 \frac{a^2}{t^2} + \sum_{n = 2}^\infty b_{2n}\frac{a^{2n}}{t^{2n}} \right],   
\ee
where $b_{2n}$ ($n = 1, 2, \ldots$) are constants up to logarithmic corrections\,\cite{Husung:2019ytz} and $V_\ell^{cont}(t)$ is the renormalized light-quark correlator in the continuum limit.
The lattice spacing artifacts appearing in $V_\ell(t)$ induce discretization effects on the short-distance light-quark contribution $a_\mu^{\rm SD}(\ell)$, as follows 
\be
    \label{eq:amu_SD}
    a_\mu^{\rm SD}(\ell) = 2 \alpha_{em}^2 \int_a^\infty ~ dt \, t^2 \, K(m_\mu t) \, \Theta^{\rm SD}(t)\,V_\ell^{cont}(t)\cdot\left[ 1 + b_2 \frac{a^2}{t^2} + \sum_{n = 2}^\infty b_{2n}\frac{a^{2n}}{t^{2n}} \right] ~ . ~ 
\ee
Taking into account that, at small values of $t$, the leptonic kernel $K(m_\mu t)$ is proportional only to $m_\mu^2 t^2$ and $V_\ell^{cont}(t)$ is dominated by its perturbative value, which at order ${\cal{O}}(\alpha_s^0)$ reads as (see, e.g., Ref.\,\cite{Giusti:2017jof})
\be
   \label{eq:Vud_pert}
   V_\ell^{cont}(t) ~ _{\overrightarrow{t \ll \Lambda_{QCD}^{-1}}} ~ \frac{5}{18 \pi^2 t^3} ~ , ~
\ee
the discretization effects on $a_\mu^{\rm SD}(\ell)$ are of order ${\cal{O}}(a^2)$ for $n \geq 2$ and of order $a^2 \, \mbox{log}(a)$ for $n = 1$. Beyond the leading order ${\cal{O}}(\alpha_s^0)$, perturbative corrections can induce further discretization effects of order $a^2 \, \mbox{log}^p(a)$ with $p \leq 0$\,\cite{Husung:2019ytz}.
The crucial point here is that discretization effects of the type $a^2 \, \mbox{log}(a)$, containing a positive power of the logarithm, are dangerous. They slow down the convergence with respect to a pure $a^2$-scaling and may not be visible unless simulations at very small lattice spacing are performed. This behaviour is illustrated in Fig.\,\ref{fig:scaling_pert_SD}, where the lattice artifacts $\Delta a_\mu^{\rm SD, pert}(\ell)$, given by 
\be
    \label{eq:Delta_amu_SD_pert}
    \Delta a_\mu^{\rm SD, pert}(\ell) = 2 \alpha_{em}^2 \int_a^\infty ~ dt \, t^2 \, K(m_\mu t) \, \Theta^{\rm SD}(t)\,\frac{5}{18 \pi^2 t^3} \cdot\left[ b_2 \frac{a^2}{t^2} + \sum_{n = 2}^\infty b_{2n}\frac{a^{2n}}{t^{2n}} \right] ~ , ~ 
\ee
are shown for both the ``tm" and ``OS" regularizations. These are evaluated numerically in the free theory and in the massless limit.
It can be  clearly  seen that a {\it naive} fit to  an $a^2$-scaling of the results within the range of the available values of the lattice spacing, indicated by the vertical dotted lines in Fig.\,\ref{fig:scaling_pert_SD}, would lead to an incorrect, non-vanishing continuum limit equal to $\simeq 1 \cdot 10^{-10}$. 

The curvature visible in Fig.\,\ref{fig:scaling_pert_SD}, which yields the correct vanishing continuum limit for $\Delta a_\mu^{\rm SD, pert}(\ell)$, is generated by the term proportional to $b_2$ in Eq.\,(\ref{eq:Delta_amu_SD_pert}).
We  calculate analytically the relative ${\cal{O}}(a^2/t^2)$ artifacts affecting $V_\ell(t)$ at order ${\cal{O}}(\alpha_s^0)$ in lattice perturbation theory with $N_f = 2$ massless twisted-mass fermions. The outcome of such an analysis is that $b_2 = 1$ for both the ``tm" and ``OS" local vector currents, see Appendix\,\ref{sec:appE} for details.

In Fig.~\ref{fig:SD_ell_corr}, we show our determinations of $a_\mu^{\rm SD}(\ell)$ for the ``tm" and ``OS" currents  for all the four ETMC ensembles of Table\,\ref{tab:simudetails} both before (green markers) and after (blue markers) applying the analytic perturbative subtraction of the $a^2 / t^5$ discretization effects in $V_\ell(t)$.
\begin{figure}[htb!]
\begin{center}
\includegraphics[scale=0.60]{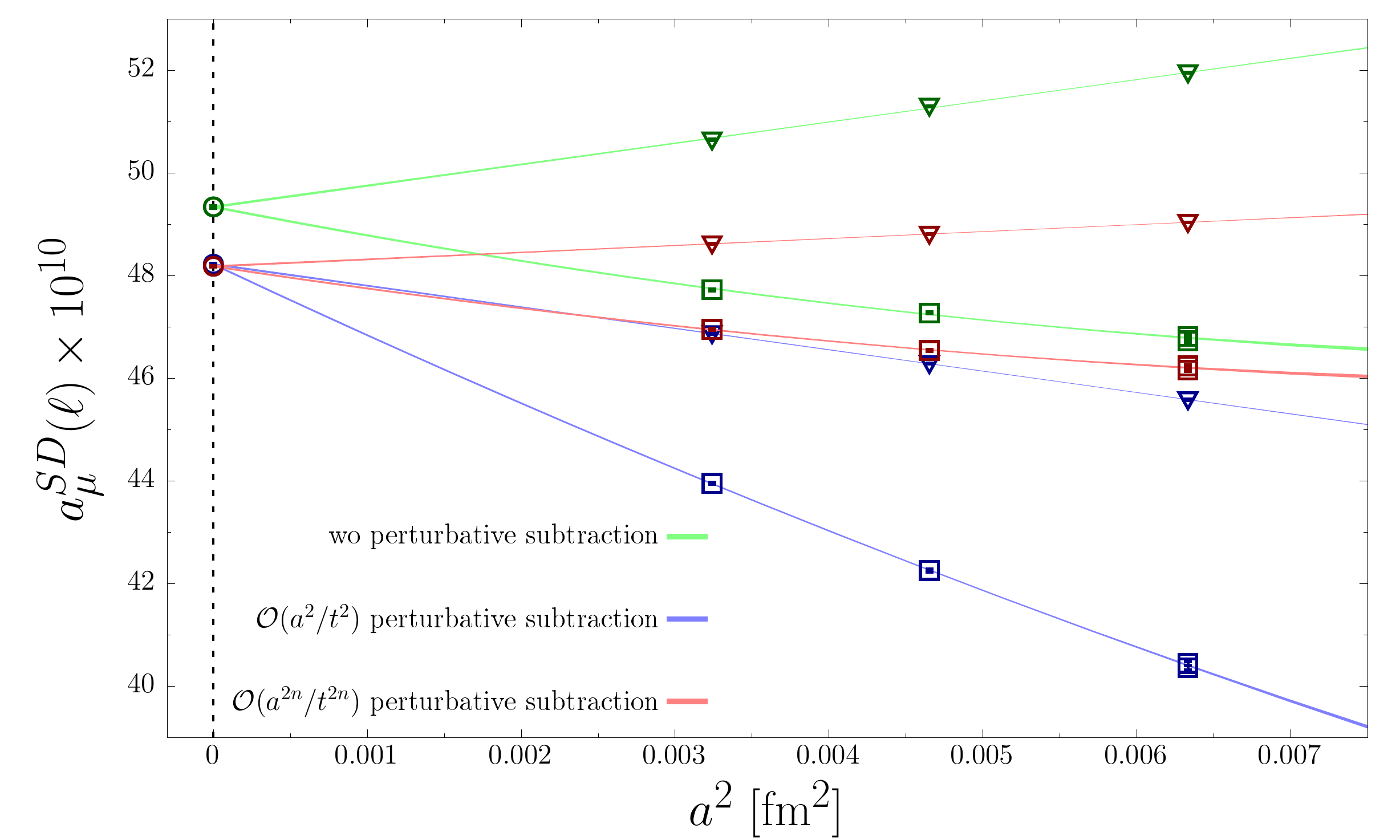}
\vspace{-0.5cm}
\caption{\it \small The light-quark connected contribution to the short-distance window $a_\mu^{\rm SD}(\ell)$, given in Eq.\,(\ref{eq:amu_SD}), versus the squared lattice spacing $a^2$ in physical units using both the ``tm" (squares) and ``OS" (triangles) local currents\,(\ref{eq:MAVcurr}). The green markers correspond to the lattice data for the four ETMC ensembles of Table\,\ref{tab:simudetails}. The blue markers include the subtraction of the analytic perturbative $a^2 / t^5$ discretization effects in $V_\ell(t)$. The red markers represent the lattice data after the subtraction of the lattice artifacts in $V_\ell(t)$ evaluated numerically in the free theory and in the massless limit at all orders in $a^2$. The solid lines are the results of the simple combined fits given in Eq.\,(\ref{eq:naive_fits}) with $D_2^{\rm OS}(\ell) = 0$.}
\label{fig:SD_ell_corr}
\end{center}
\end{figure}
We show also the results of a {\it combined} polynomial fit in powers of $a^2$ of the general type
 \be
     \label{eq:naive_fits}
     a_\mu^{\rm SD}(f) = a_\mu^{\rm SD, cont}(f) \cdot \left[ 1 + D_1^{reg}(f) a^2 + D_2^{reg}(f) a^4 \right] ~ , ~
 \ee
where $reg = \{ {\rm tm}, {\rm OS} \}$ and $a_\mu^{\rm SD, cont}(f)$ is the same value at $a^2 = 0$ for the two regularizations. Since the variation of the logarithmic term $\mbox{log}(a)$ is too mild in the range of the available values of the lattice spacing, we still observe an approximate ${\cal{O}}(a^2)$ scaling in both the unsubtracted and subtracted lattice data. However, as already discussed in connection to Fig.\,\ref{fig:scaling_pert_SD}, the continuum extrapolation for the unsubtracted data misses the correct value by approximately $2 \%$, which is well above our statistical uncertainty and larger than any other source of systematic error.

The dangerous $a^2 / t^5$ discretization effects in $V_\ell(t)$ turn out to be equal for both the ``tm" and ``OS" regularizations, which, however, exhibit quite different lattice artifacts at short time distances at all orders in $\alpha_s$, as can be seen in Fig.\,\ref{fig:VV_connected}. Thus, the question is whether the subtraction of the discretization effects in $V_\ell(t)$ evaluated numerically in the free theory and in the massless limit at all orders in $a^2$ is beneficial. To answer this question, we show in Fig.\,\ref{fig:SD_ell_corr} by the red markers the lattice data after the subtraction of all the lattice artifacts at order ${\cal{O}}(\alpha_s^0)$. The subtracted data exhibit indeed much smaller discretization effects in both regularizations and this fact makes more robust the extrapolation to the continuum limit.

\begin{figure}[htb!]
\begin{center}
\includegraphics[scale=0.80]{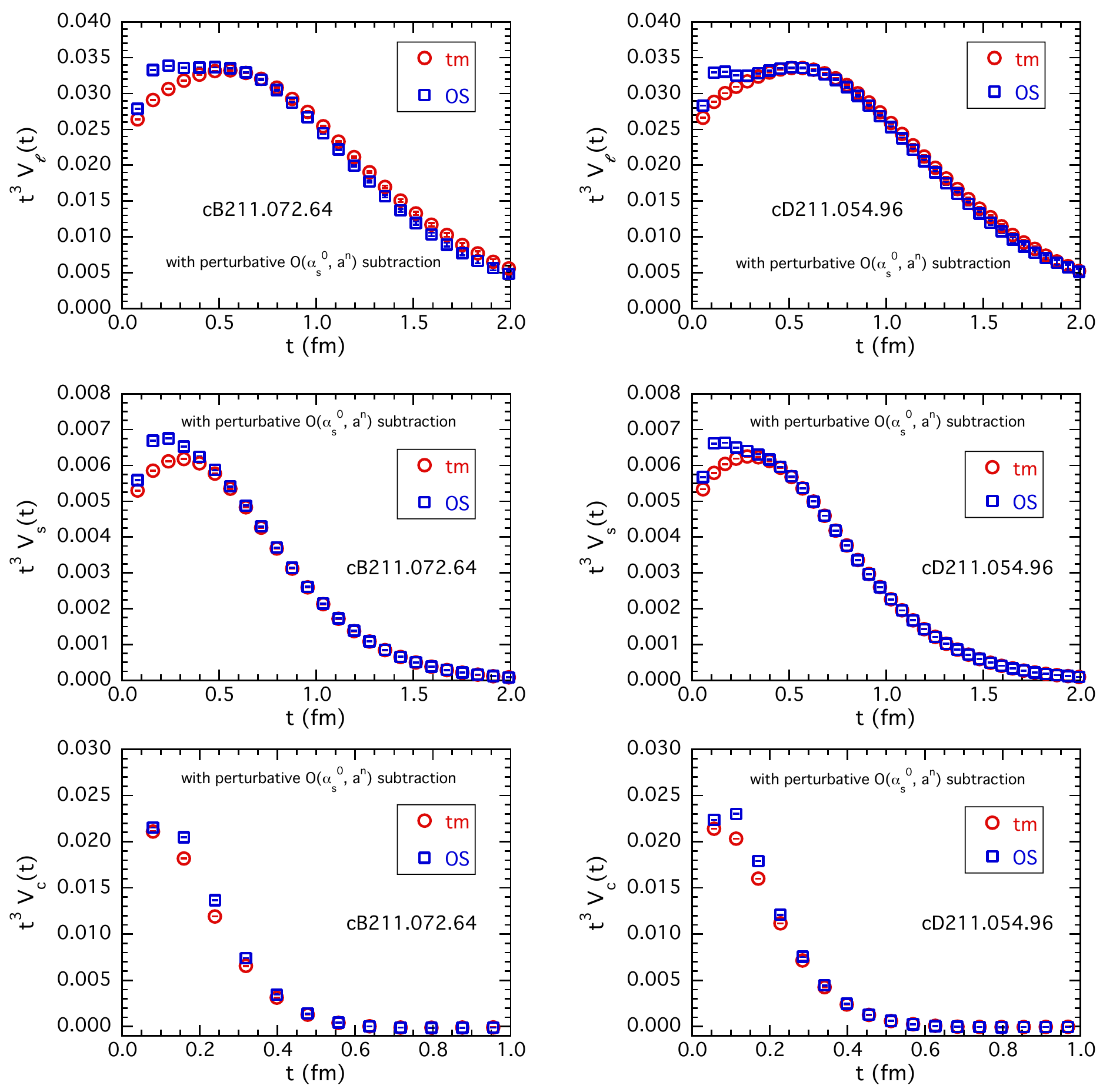}
\vspace{-0.5cm}
\caption{\it \small The same as in Fig.\,\ref{fig:VV_connected}, but after subtracting the perturbative lattice artifacts evaluated numerically in the free-theory, i.e.\,at order ${\cal{O}}(\alpha_s^0)$, at finite values of the bare quark masses and of the lattice spacing.}
\label{fig:VV_connected_sub}
\end{center}
\end{figure}

The strange- and charm-quark contributions to the short-distance window, $a_\mu^{\rm SD}(s)$ and $a_\mu^{\rm SD}(c)$, display the same cutoff dependence as the light-quark one, $a_\mu^{\rm SD}(\ell)$, due to the dangerous massless $a^2\log{a}$ artifacts, which are ``dynamically" generated in the time integral by the region of small time distances of the order $t\sim \mathcal{O}(a)$. In complete analogy with the case of the light-quark contribution, we remove the leading $a^2\log(a)$ cut-off effects from our lattice data, by subtracting from the renormalized strange- and charm-quark vector correlators $V_s(t)$ and $V_c(t)$ the lattice artifacts of the perturbative one, evaluated numerically at order ${\cal{O}}(\alpha_s^0)$ and at finite values of the bare quark masses. The impact of the above subtraction on the vector correlators $V_f^{\rm tm}(t)$ and $V_f^{\rm OS}(t)$ is illustrated in the plots of Fig.\,\ref{fig:VV_connected_sub}, which can be compared with analogous plots of Fig.\,\ref{fig:VV_connected} for the unsubtracted data.

The lattice data for $a_\mu^{\rm SD}(\ell)$ shown in Fig.\,\ref{fig:SD_ell_corr} exhibit a very high statistical precision of the order of $0.05 \%$. Instead, the accuracy we reached for the lattice spacing is only of the order of $0.2 \%$ (see Table\,\ref{tab:spacing}). The reason is that the short-distance window is largely insensitive to the scale setting and therefore to its uncertainty. Indeed, we notice that in the continuum limit and at short time distances the correlator $V_\ell(t)$ is dominated by its perturbative massless term\,(\ref{eq:Vud_pert}). 
Thus, after replacing the modulating function $\Theta^{\rm SD}(t)$ with a Heaviside step function $\theta(t_0 - t)$, the physical value of $a_\mu^{\rm SD}(\ell)$ is almost saturated by the perturbative term $(5 \alpha_{em}^2 / 9 \pi^2) \int_0^1 dx \, K(m_\mu t_0 x) / x $, which does not depend upon the scale setting. For a more quantitative discussion see Appendix\,\ref{sec:spacing}.

\begin{figure}[htb!]
\begin{center}
\includegraphics[scale=0.375]{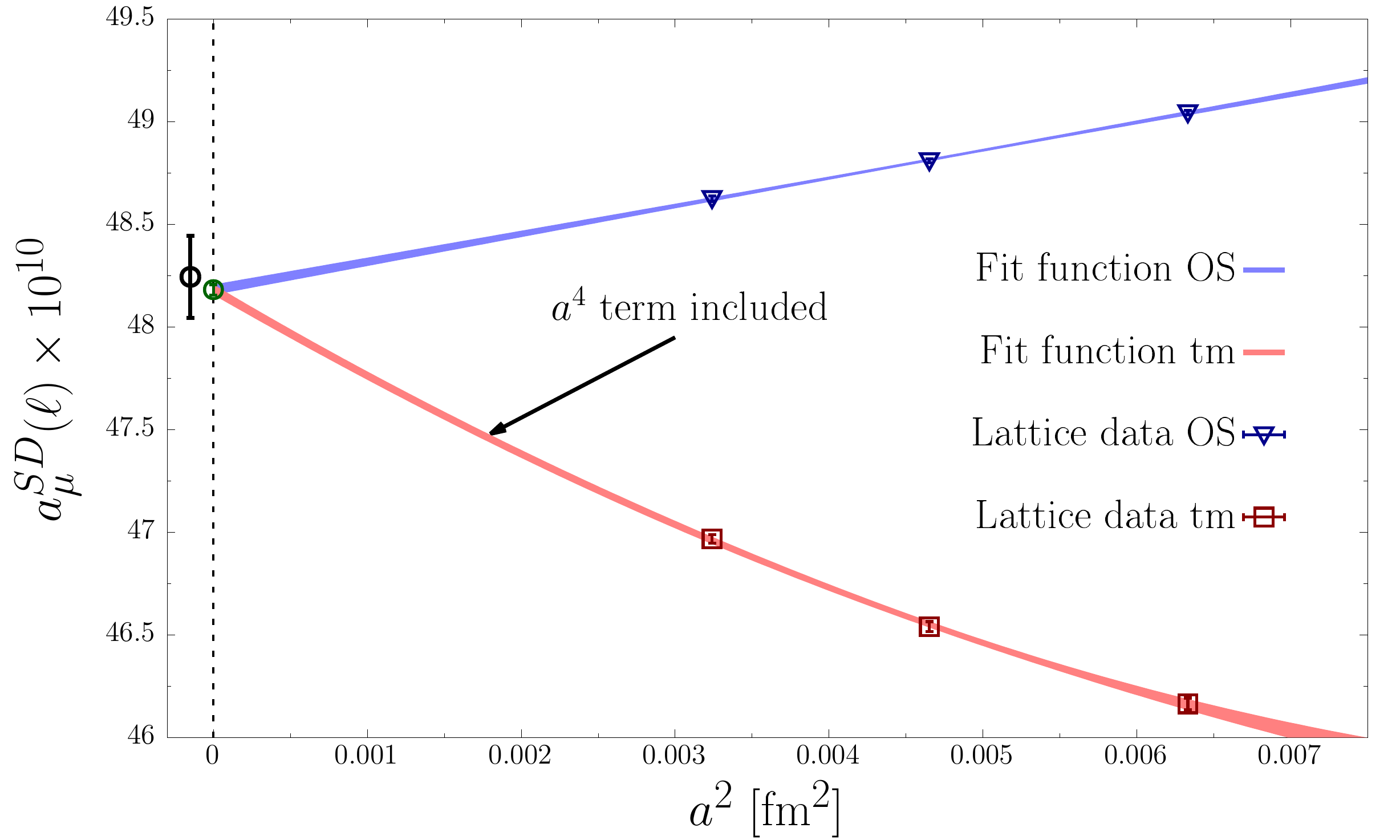}\\
\includegraphics[scale=0.375]{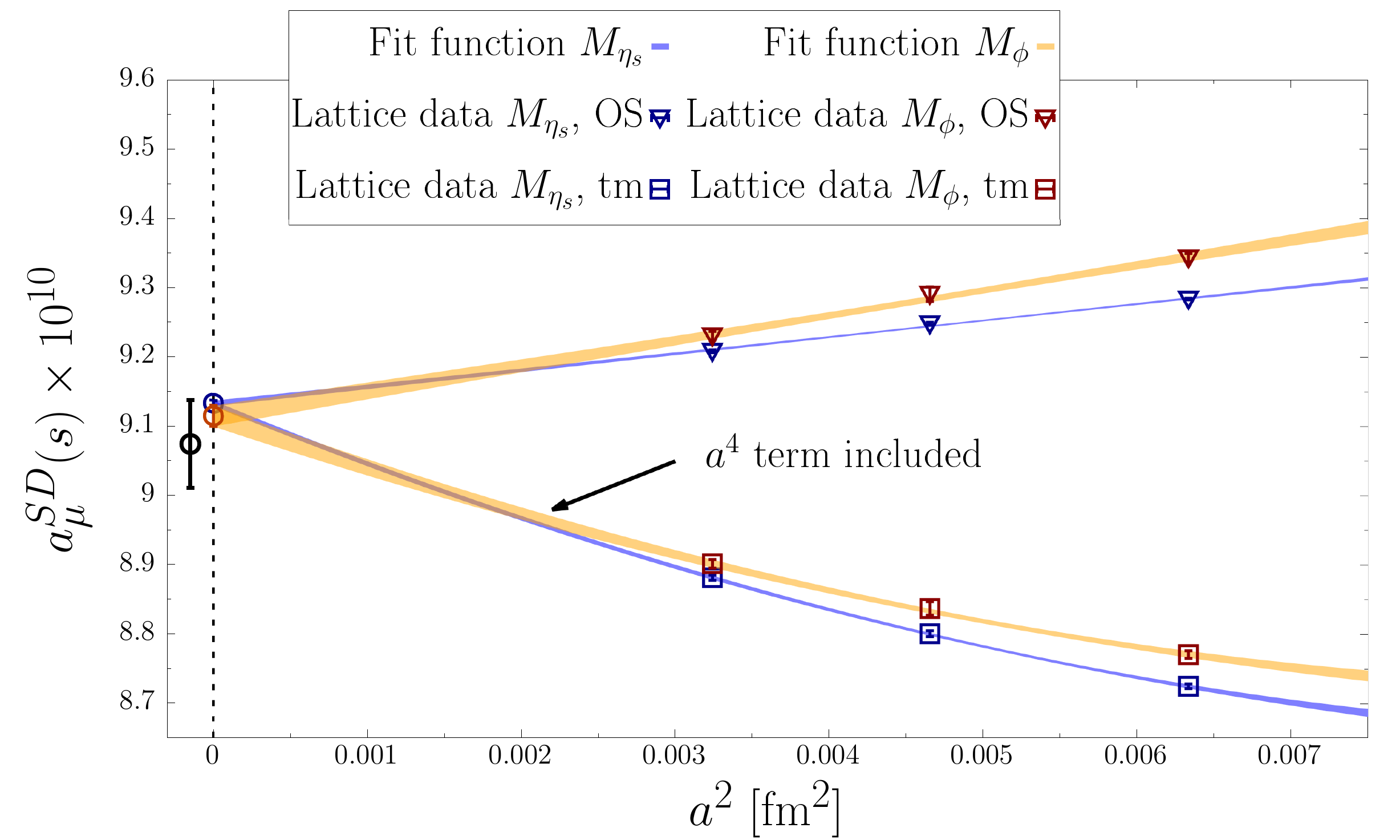}\\
\includegraphics[scale=0.375]{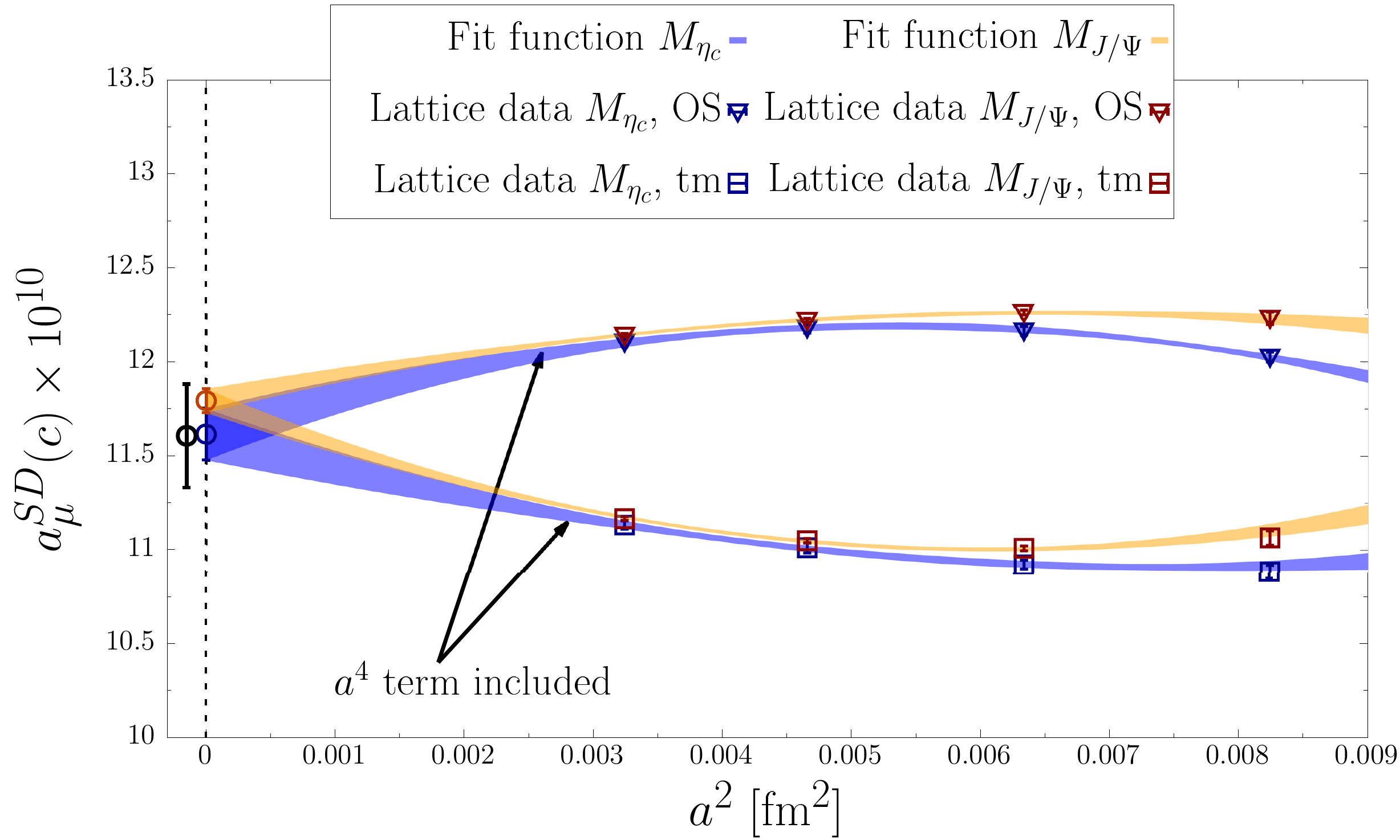}
\vspace{-0.5cm}
\caption{\it \small The light-quark (top), strange-quark (middle) and charm-quark (bottom) connected contributions to the short-distance window $a_\mu^{\rm SD}$ versus the squared lattice spacing $a^2$ in physical units using both the ``tm" (triangles) and ``OS" (squares) regularizations after subtraction of the perturbative lattice artifacts at order ${\cal{O}}(\alpha_s^0)$. In the middle (bottom) panel the blue and red points correspond to the lattice data obtained using the masses of the $\eta_s$ ($\eta_c$) and $\phi$ ($J/\Psi$) mesons to obtain the physical strange (charm) quark mass. The solid lines correspond to the results of the combined fitting procedure given in Eq.\,(\ref{eq:naive_fits}) with $D_2^{\rm OS}(\ell) = D_2^{\rm OS}(s) = 0$ and $D_2^{\rm OS}(c) \neq 0$. The extrapolated values in the continuum limit are shown at $a^2 = 0$ together with our final results given by Eqs.\,(\ref{eq:amuSD_ell_final})-(\ref{eq:amuSD_charm_final}).}
\label{fig:SD_cont_lim}
\end{center}
\end{figure}

The values of $a_\mu^{\rm SD}(\ell)$, $a_\mu^{\rm SD}(s)$ and $a_\mu^{\rm SD}(c)$ obtained after subtraction of the perturbative lattice artifacts at order ${\cal{O}}(\alpha_s^0)$ are shown in Fig.\,\ref{fig:SD_cont_lim} for both the ``tm" and ``OS" regularizations. In the case of $a_\mu^{\rm SD}(s)$ ($a_\mu^{\rm SD}(c)$) our data correspond to the two branches of the analysis in which we set the physical strange (charm) quark mass using either the mass of the $\eta_s$ ($\eta_c$) pseudoscalar meson or that of the $\phi$ ($J/\Psi$) vector meson. 

Discretization effects on $a_\mu^{\rm SD}(\ell)$ (see top panel of Fig.\,\ref{fig:SD_cont_lim}) are consistent with $a^2$-scaling within tiny errors in the ``OS" regularization, while higher order corrections are clearly present in the ``tm" case. The result of the combined fit based on the Ansatz of Eq.\,(\ref{eq:naive_fits}), in a representative case where the fit parameter $D_2^{OS}(\ell)$ is set to zero, is shown by the solid lines.
The extrapolated value of $a_\mu^{\rm SD}(\ell)$ in the continuum limit has a remarkable statistical error of less than $0.1 \%$.

The statistical errors of our lattice data for $a_\mu^{\rm SD}(s)$ (see middle panel of Fig.\,\ref{fig:SD_cont_lim}) are typically of order $\mathcal{O}(0.1 \%)$, with the data obtained using $M_{\eta_s}$ as hadronic input displaying an accuracy of $\simeq 0.05 \%$. 
The solid lines correspond to the results of the combined fit in Eq.\,(\ref{eq:naive_fits}) for $f = s$, in a representative case where the fit parameter $D_2^{OS}(s)$ is set to zero.
 
As for $a_\mu^{\rm SD}(c)$ (see bottom panel of Fig.\,\ref{fig:SD_cont_lim}), the statistical errors of our data are typically of order $\mathcal{O}(0.1 \%)$ for both choices of the reference hadron mass. Discretization effects appear to be at the level of ${\cal{O}}(5 \%)$ with opposite signs between the ``tm" and ``OS" regularizations. The size of discretization effects is limited thanks to the subtraction of the perturbative cutoff effects at order ${\cal{O}}(\alpha_s^0)$ evaluated at the charm-quark mass.
The solid lines correspond to the results of the combined fit in Eq.\,(\ref{eq:naive_fits}) for $f = c$, in a representative case where the quartic $a^{4}$ terms are included for both regularizations.

In order to estimate the systematic uncertainty related to the continuum limit we consider combined fits adopting for all the windows the following generic Ansatz
\be
   \label{eq:amuw_fit}
   a_\mu^{\rm w}(f) = a_\mu^{\rm w, cont}(f) \cdot \left[ 1 + D_1^{reg}(f) \, a^2 + D_{1L}^{reg}(f) \, \frac{a^2}{[\mbox{log}(a^2 \Lambda_0^2)]^{n^{reg}(f)}} + D_2^{reg}(f) \, a^4 \right] ~ , ~
\ee
where $w = \{SD, W \}$ and $f = \{ \ell, s, c \}$, while $D_{1, 1L, 2}^{reg}(f)$ and $a_\mu^{\rm w, cont}(f)$ are free parameters to be fitted to the data. Because of the limited number of data points the case in which all the free parameters are simultaneously non-zero is not considered. We remind that in our combined fits the parameter $a_\mu^{\rm w, cont}(f)$ does not depend upon the regularization $reg = \{ {\rm tm}, {\rm OS} \}$. 

In Eq.\,(\ref{eq:amuw_fit}) we have included possible logarithmic terms of the form $a^2 / \mbox{log}(a^2\Lambda_0^2)]^{n^{reg}(f)}$, where the power $n^{reg}(f)$ represents an effective anomalous dimension for perturbative corrections beyond the leading order ${\cal{O}}(\alpha_s^0)$\,\cite{Husung:2019ytz}. In what follows we will consider the representative cases $n^{reg}(f) = 1, 2, 3$ when $D_{1L}^{reg}(f) \neq 0$. The energy scale $\Lambda_0$ is taken to assume two different values, namely $\Lambda_0 = 1 / w_0 \simeq 1.14$ GeV and $\Lambda_0 = 1 / (3 w_0) \simeq 380$ MeV, where $w_0$ is the gradient-flow scale found to be equal to $w_0 = 0.17383\,(63)$ fm in Ref.\,\cite{ExtendedTwistedMass:2021qui}. In addition to the aforementioned fits, we also performed extrapolations to the continuum limit by leaving out data at the coarsest lattice spacing, as well as separate linear extrapolations for the two regularizations.

In order to reach the continuum limit we have considered also an alternative strategy, based on considering the difference and the ratio of $a_\mu^{\rm w}(f)$ in the two regularizations, namely
\bea
    D^w(f) & \equiv & a_\mu^{\rm w}(f)|_{tm} - a_\mu^{\rm w}(f)|_{OS} ~ , ~ \\[2mm]
    R^w(f) & \equiv & a_\mu^{\rm w}(f)|_{tm} ~ / ~ a_\mu^{\rm w}(f)|_{OS} ~ . ~
\eea
Since the continnum limit of the difference $D^w(f)$ should exactly vanish, while the one of the ratio $R^w(f)$ should be equal to unity, we consider the following fitting functions
\bea
    \label{eq:diff}
    D^w(f) & = & D_1 a^2 + D_{1L} \frac{a^2}{[\mbox{log}(a^2 \Lambda_0^2)]^{n(f)}} + D_2 a^4 ~ , ~ \\[2mm]
    \label{eq:ratio}
    R^w(f) & = &  1 + R_1 a^2 + R_{1L} \frac{a^2}{[\mbox{log}(a^2 \Lambda_0^2)]^{n(f)}} + R_2 a^4 ~ , ~
\eea
where we have assumed that $n^{tm}(f) = n^{OS}(f) = n(f)$. The continuum value $a_\mu^{\rm w, cont}(f)$ is given by
\bea
   a_\mu^{\rm w, cont}(f) & = & \frac{D_1}{R_1} \qquad \mbox{if ~} D_1 \neq 0 \mbox{~and~} R_1 \neq 0 ~ \\[2mm]
                            & = & \frac{D_{1L}}{R_{1L}} \qquad \mbox{otherwise} ~ . ~
\eea
As for Eq.\,(\ref{eq:amuw_fit}), the fitting procedure in which all the free parameters appearing in Eqs.\,(\ref{eq:diff})-(\ref{eq:ratio}) are varied is not considered.

Using Eq.\,(\ref{eq:amuw_fit}) and Eqs.\,(\ref{eq:diff})-(\ref{eq:ratio}) we have carried out hundreds of combined fits of our lattice data for the two regularizations ``tm" and ``OS". In the fitting procedure we have minimized the $\chi^2$-variable constructed taking into account the correlations between the ``tm" and ``OS" correlators corresponding to the same gauge ensemble. We have evaluated the correlation matrix using a jackknife sampling procedure and found that its entries are smaller than $0.5$ for the light-quark contribution, and typically larger (reaching up to $\approx 0.99$) for the heavier flavours.

In order to average the different analyses of the same lattice data, we make use of the procedure developed in Ref.\,\cite{EuropeanTwistedMass:2014osg}: starting from $N$ computations with mean values $x_k$ and uncertainties $\sigma_k$ ($k=1,\cdots,N$), based on the same set of input data, their average $x$ and uncertainty $\sigma_x$ are given by
\be
    \label{eq:averaging}
    x = \sum_{k=1}^N \omega_k ~ x_k ~ , ~ \qquad
    \sigma_x^2 = \sum_{k=1}^N \omega_k ~ \sigma_k^2 + \sum_{k=1}^N \omega_k ~ (x_k - x)^2 ~ , ~
\ee
where $\omega_k$ represents the weight associated with the $k$-th determination. 

We have excluded from the average all fits having $d.o.f. = 1$ in order to avoid overfitting. Then, we have considered two different choices for the remaining weights $\omega_k$. The first one is based on the Akaike Information Criterion (AIC)\,\cite{Akaike}, namely 
\be
    \label{eq:AIC}
    \omega_k \propto e^{- (\chi_k^2 + 2 N_{parms} - N_{data}) / 2} ~ , ~
\ee
where $\chi_k^2$ is the value of the $\chi^2$-variable for the $k$-th computation, $N_{parms}$ is the number of free parameters and $N_{data}$ the number of data points\footnote{We have verified that the use of the slightly different definition proposed in Ref.\,\cite{Neil:2022joj}, namely $\omega_k \propto \mbox{exp}[- (\chi^2 + 2 N_{parms} - 2 N_{data}) / 2]$ leads to very similar averages and errors as compared with those corresponding to the use of Eq.\,(\ref{eq:AIC}).}.
Since in our fits the number of d.o.f. is limited, we adopt also a second choice for $\omega_k$ given by a step function 
\be
   \label{eq:stepF}
   \omega_k \propto \Theta\left[ 1 + 2 \sqrt{ \frac{2}{d.o.f.}} - \frac{\chi_k^2}{d.o.f.} \right] ~ , ~ 
\ee
where $1$ is the mean value and $\sqrt{2 / d.o.f.}$ is the standard deviation of the $\chi^2/d.o.f.$ distribution.
The results obtained with the above two choices of $\omega_k$ are reassuringly very similar and their small difference is added as a systematic error in the final error budget.
At the physical point we get
\bea
    \label{eq:amuSD_ell_final}
    a_\mu^{\rm SD}(\ell) & = & 48.24 ~ (3)_{stat} ~ (20)_{syst} \cdot 10^{-10} = 48.24 ~ (20) \cdot 10^{-10} ~ , ~ \\[2mm]
    \label{eq:amuSD_strange_final}
    a_\mu^{\rm SD}(s) & = & 9.074 ~ (14)_{stat} ~ (62)_{syst} \cdot 10^{-10} = 9.074 ~ (64) \cdot 10^{-10} ~ . ~ \\[2mm]
    \label{eq:amuSD_charm_final}    
    a_\mu^{\rm SD}(c) & = & 11.61 ~ (9)_{stat} ~ (25)_{syst} \cdot 10^{-10} = 11.61 ~ (27) \cdot 10^{-10} ~ . ~ 
\eea
where 
\begin{itemize}
    \item $()_{stat}$ includes the statistical uncertainty of the Monte Carlo samplings and the one due to the fitting procedure;
    \item $()_{syst}$ represents the systematic error coming from discretization effects, evaluated according to Eq.\,(\ref{eq:averaging}) from the results of the fits based on the Ansatz in Eq.\,(\ref{eq:amuw_fit}) and Eqs.\,(\ref{eq:diff})-(\ref{eq:ratio}).
\end{itemize}
The final error corresponds to the statistical and systematic errors added in quadrature. 

\begin{figure}[htb!]
\begin{center}
\includegraphics[scale=0.32]{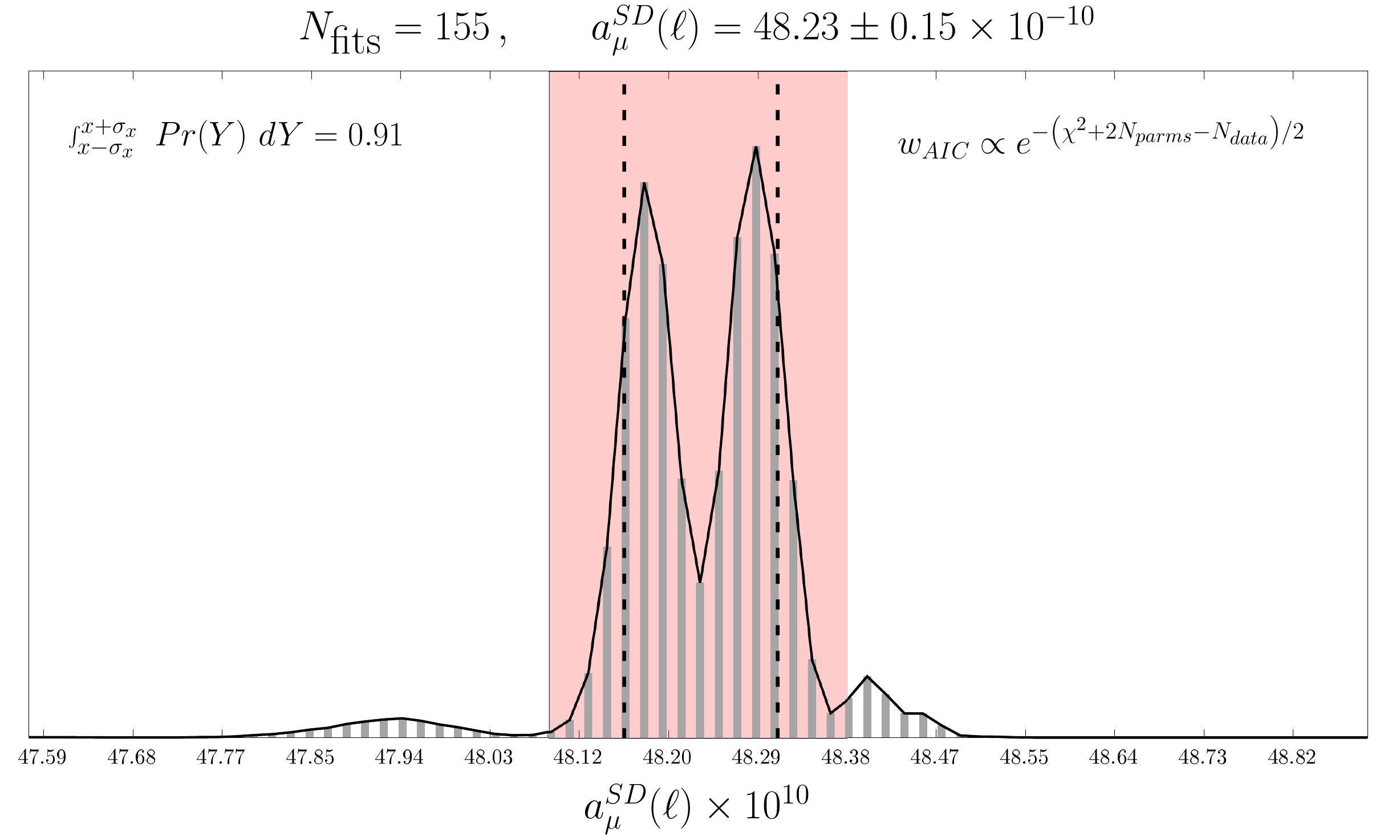} ~ 
\includegraphics[scale=0.32]{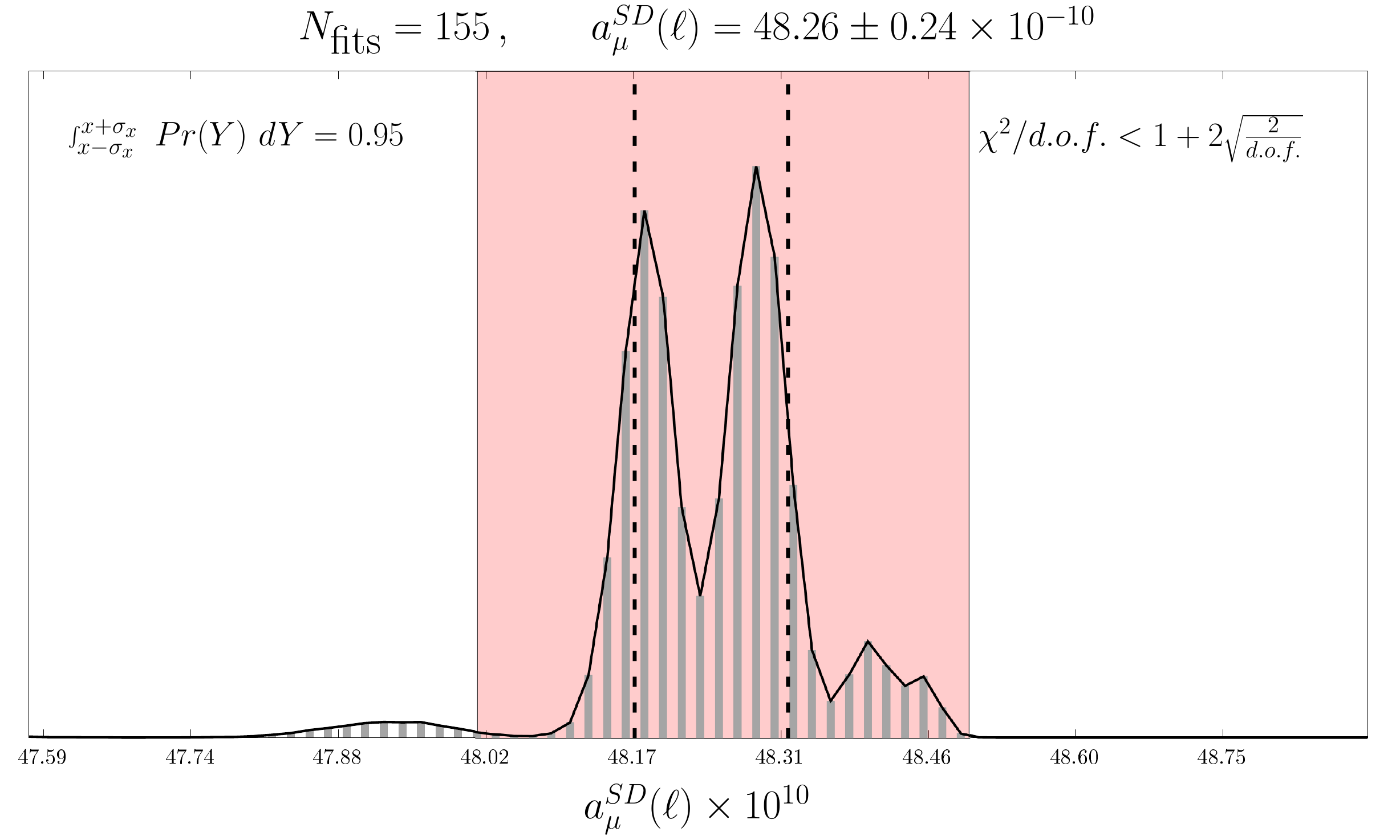}\\
\includegraphics[scale=0.32]{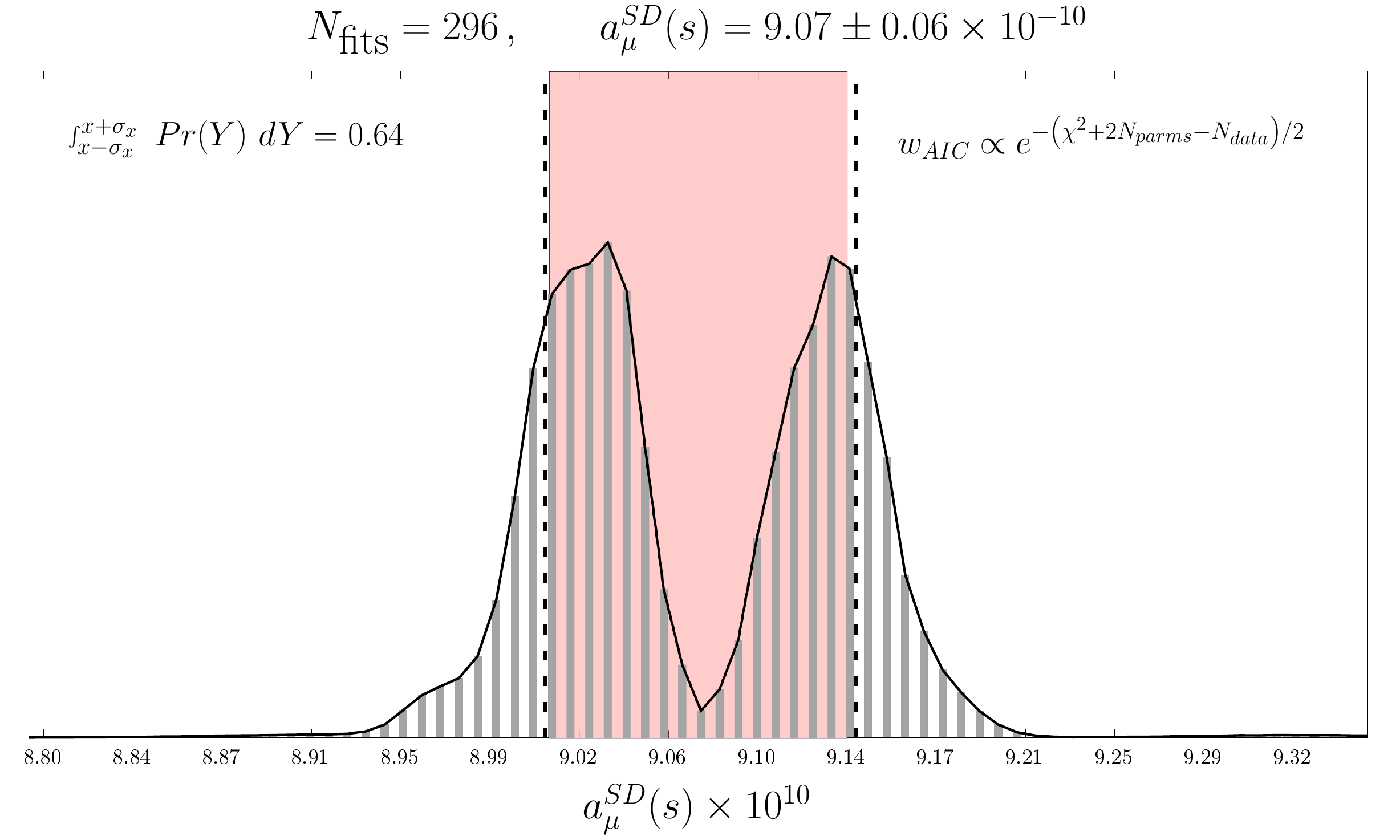} ~
\includegraphics[scale=0.32]{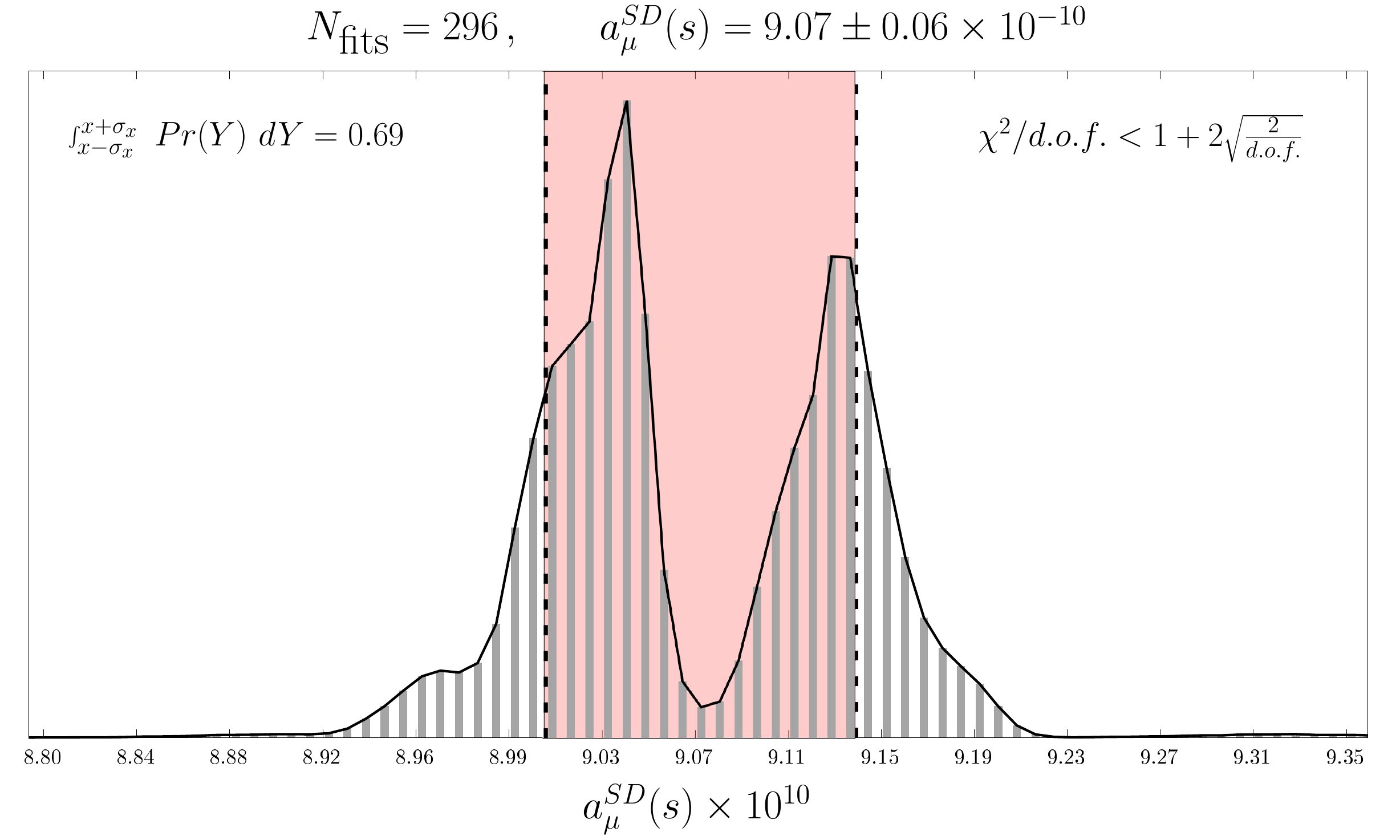}\\
\includegraphics[scale=0.32]{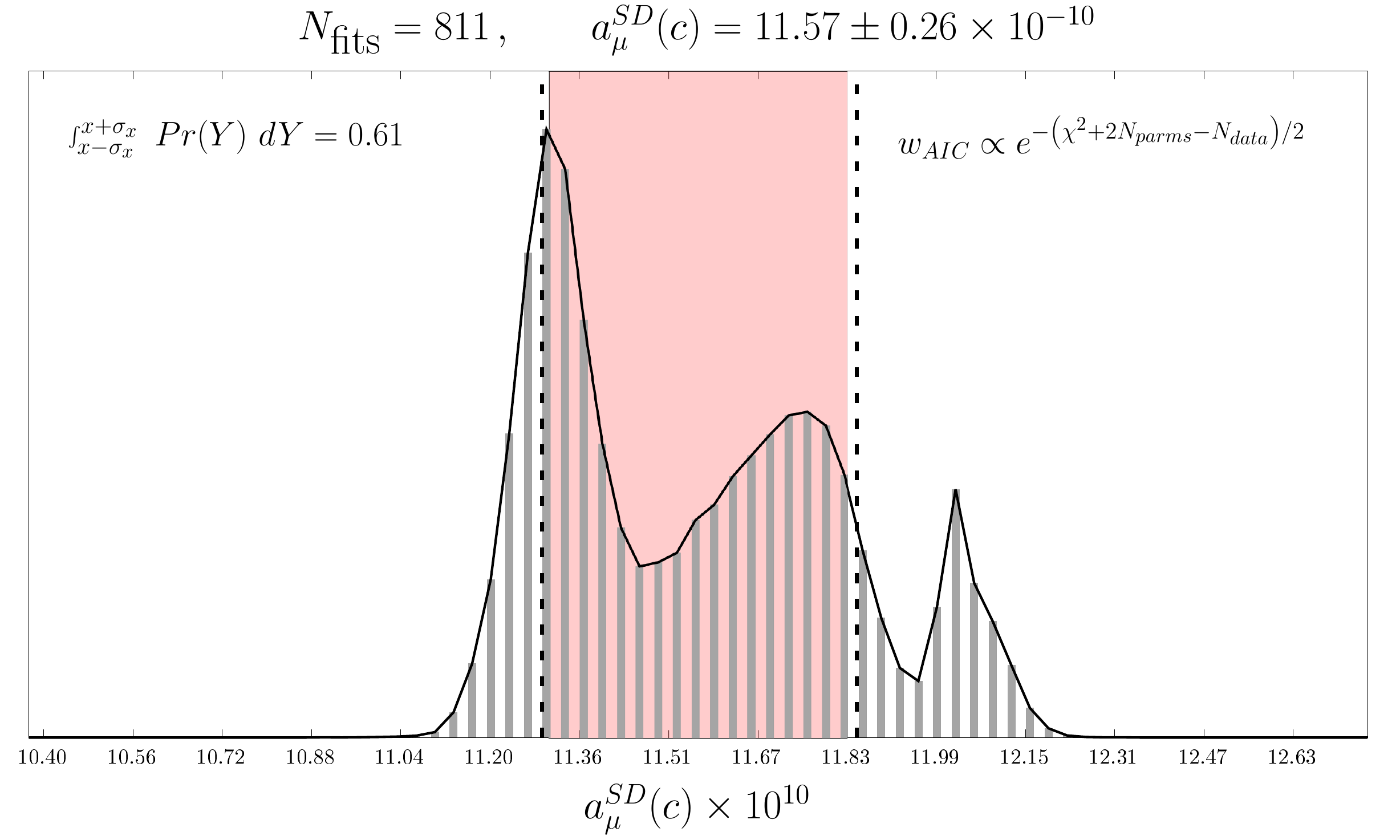} ~ 
\includegraphics[scale=0.32]{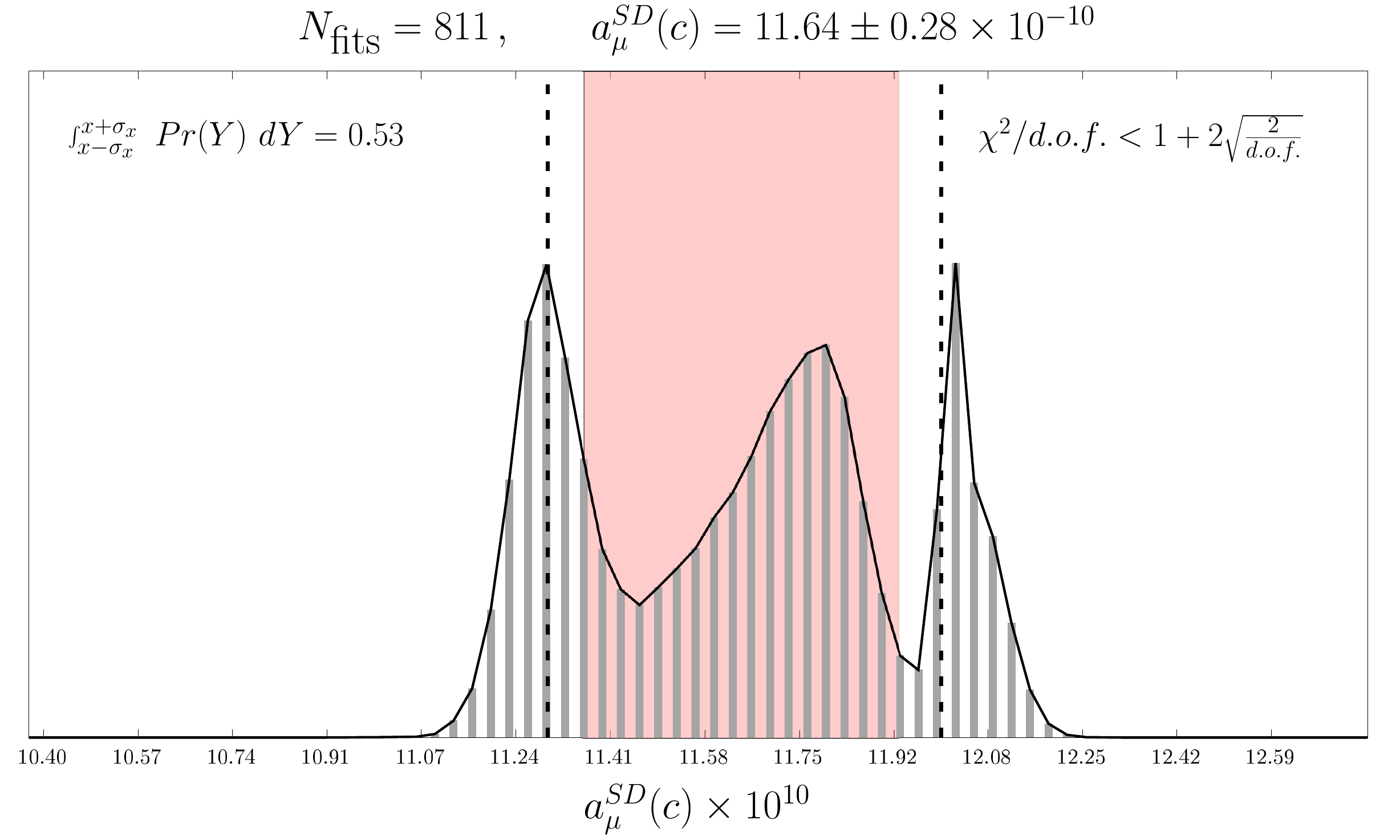}
\vspace{-0.5cm}
\caption{\it \small Histograms of the results at the physical point obtained by our fitting procedures based on Eq.\,(\ref{eq:amuw_fit}) and Eqs.\,(\ref{eq:diff})-(\ref{eq:ratio}) applied to our lattice data of $a_\mu^{\rm SD}(\ell)$ (top panels), $a_\mu^{\rm SD}(s)$ (middle panels) and $a_\mu^{\rm SD}(c)$ (bottom panels) adopting for the weights $\omega_k$ either the AIC (left panels) or the step function (right panels), described respectively by Eqs.\,(\ref{eq:AIC}) and (\ref{eq:stepF}). The red bands correspond to our final results\,(\ref{eq:amuSD_ell_final})-(\ref{eq:amuSD_charm_final}). In each panel we show the number of fits, the average ($x$) and the error ($\sigma_x$) evaluated according to Eq.\,(\ref{eq:averaging}) and the cumulative probability corresponding to the interval $[x - \sigma_x, x + \sigma_x]$. The vertical short-dashed lines correspond to the 16-th and 84-th percentiles of the p.d.f.~$Pr(Y)$.}
\label{fig:amuSD}
\end{center}
\end{figure}

In Fig.\,\ref{fig:amuSD} we show the histograms of the results at the physical point obtained by our fitting procedures based on Eq.\,(\ref{eq:amuw_fit}) and Eqs.\,(\ref{eq:diff})-(\ref{eq:ratio}) applied to our lattice data of $a_\mu^{\rm SD}(\ell)$, $a_\mu^{\rm SD}(s)$ and $a_\mu^{\rm SD}(c)$ for the two choices\,(\ref{eq:AIC}) and (\ref{eq:stepF}) for the weights $\omega_k$ appearing in Eq.\,(\ref{eq:averaging}).
The distributions exhibit multiple peaks. This feature is related to the fact that the statistical uncertainties are significantly smaller than the systematic ones. We stress that such a situation is ideal for the application of the averaging procedure given by Eq.\,(\ref{eq:averaging}).

Before closing the subsection, we show the results of a crosscheck we performed to exclude that possible residual cut-off effects of the type $a^{2}/[\log(a^{2} \Lambda_0^2)]^{n^{reg}}$ with $n^{reg} < 0$ may spoil our continuum limit extrapolation of $a_\mu^{\rm SD}$. To this end we have carried out a slightly different analysis of $a_{\mu}^{\rm SD}(\ell)$, in which we consider a truncated version of Eq.~(\ref{eq:amu_SD}), where the lower bound of integration is fixed to a non-zero $t_{min}$, i.e. 
\be
\label{eq:amu_SD_truncated}
a_\mu^{\rm SD}(\ell, t_{min}) \equiv 2 \alpha_{em}^2 \int_{t_{min}}^\infty ~ dt \, t^2 \, K(m_\mu t) \, \Theta^{\rm D}(t)\,V_\ell(t) ~ , 
\ee
where $t_{min}$ is kept fixed in physical units for all ensembles. Clearly, one has $a_{\mu}^{\rm SD}(\ell, t_{min} \to 0) = a_{\mu}^{\rm SD}(\ell)$. The idea is to perform first the continuum extrapolation at fixed $t_{min}$ and then to look at the behaviour of $a_{\mu}^{\rm SD}(\ell, t_{min})$ as $t_{min}$ is decreased towards zero. In $a_{\mu}^{\rm SD}(\ell, t_{min})$, the logarithmic $a^{2}/[\log(a^{2}/w_0^{2})]^{n^{reg}}$ cut-off effects generated in $a_{\mu}^{\rm SD}(\ell)$ by the integration at short times, become simple $a^{2}$-like lattice artifacts with potentially large $1/[\log(t_{min}^{2}/w_0^{2})]^{n^{reg}}$ coefficients, which can be then safely extrapolated to zero. We use values of $t_{min}$ in the range $[ 0.08, 0.15]~$fm, which correspond to $t_{min} > a$ for all the ensembles of Table~\ref{tab:simudetails}. 

\begin{figure}[htb!]
\begin{center}
\includegraphics[scale=0.60]{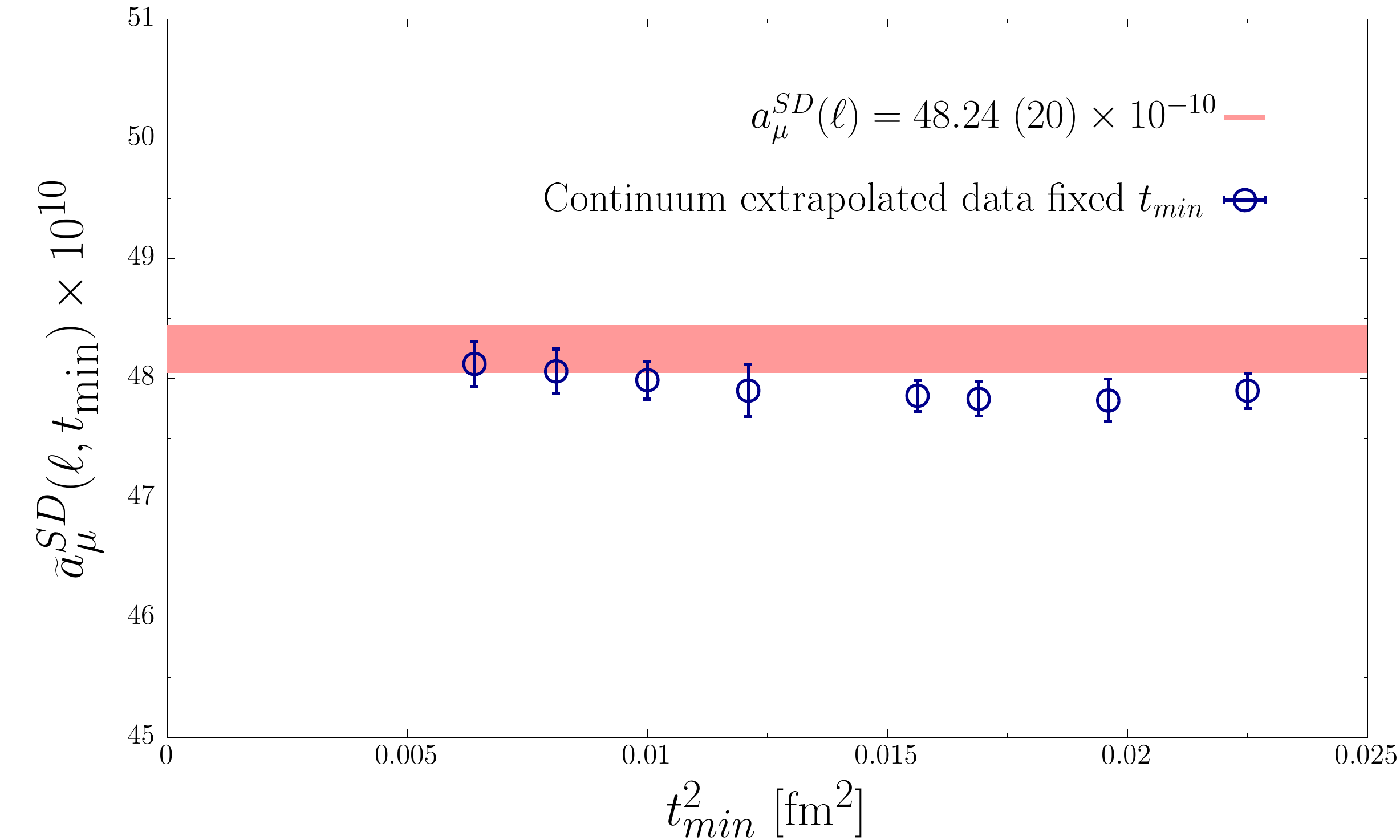}
\vspace{-0.25cm}
\caption{\it \small Results for the modified short-distance window $\widetilde{a}_{\mu}^{\rm SD}(\ell,t_{min})$ (see  Eq.\,(\ref{eq:amu_SD_truncated_modified})), obtained for various values of $t_{min}$ after extrapolation to the continuum limit using the combined fitting procedures based on Eq.~(\ref{eq:amuw_fit}) and Eqs.\,(\ref{eq:diff})-(\ref{eq:ratio}). The red band corresponds to our final result given by Eq.\,(\ref{eq:amuSD_ell_final}).}
\label{fig:SD_cont_lim_tmins}
\end{center}
\end{figure}

It is useful to consider the following quantity
\be
\label{eq:amu_SD_truncated_modified}
\widetilde{a}_{\mu}^{\rm SD}(\ell,t_{min}) = a_{\mu}^{\rm SD}(\ell, t_{min}) + \Delta a_{\mu}^{\rm SD, pert}(\ell, t_{min})~,
\ee
where 
\be
\Delta a_{\mu}^{\rm SD, pert}(\ell, t_{min}) \equiv  2 \alpha_{em}^2 \int_{0}^{t_{min}} ~ dt \, t^2 \, K(m_\mu t) \, \Theta^{\rm SD}(t)\,V_\ell^{cont}(t)    
\ee
and $V_\ell^{cont}(t)$ is the light-quark correlator in the continuum limit, obtained using the ``rhad" software package\,\cite{Harlander:2002ur} at order $\mathcal{O}(\alpha_{s}^{4})$. 
The difference $a_{\mu}^{\rm SD}(\ell) - \widetilde{a}_{\mu}^{\rm SD}(\ell, t_{\min})$ is thus expected to be of order ${\cal{O}}(\alpha_s^5(1 / t_{min}) ~ t_{min}^2)$.
In Fig.~\ref{fig:SD_cont_lim_tmins} we show our determinations of $\widetilde{a}_{\mu}^{\rm SD}(\ell, t_{min})$ after extrapolation of $a_{\mu}^{\rm SD}(\ell, t_{min})$ to the continuum limit using the combined fit procedures based on Eq.~(\ref{eq:amuw_fit}) and Eqs.\,(\ref{eq:diff})-(\ref{eq:ratio}). The data exhibit a nice flat behavior in $t_{min}^2$ with a very small residual slope due to effects at order ${\cal{O}}(\alpha_s^5(1 / t_{min}) ~ t_{min}^2)$. It is reassuring that the data for $\widetilde{a}_{\mu}^{\rm SD}(\ell, t_{min})$ are consistent for $t_{min} \lesssim 0.1$ fm with our final short-distance result of Eq.\,(\ref{eq:amuSD_ell_final}) within one standard deviation.


\subsection{The intermediate windows $a_{\mu}^{\rm W}(\ell)$, $a_{\mu}^{\rm W}(s)$ and $a_{\mu}^{\rm W}(c)$}
\label{sec:amuW}

\begin{figure}[htb!]
\begin{center}
\includegraphics[scale=0.375]{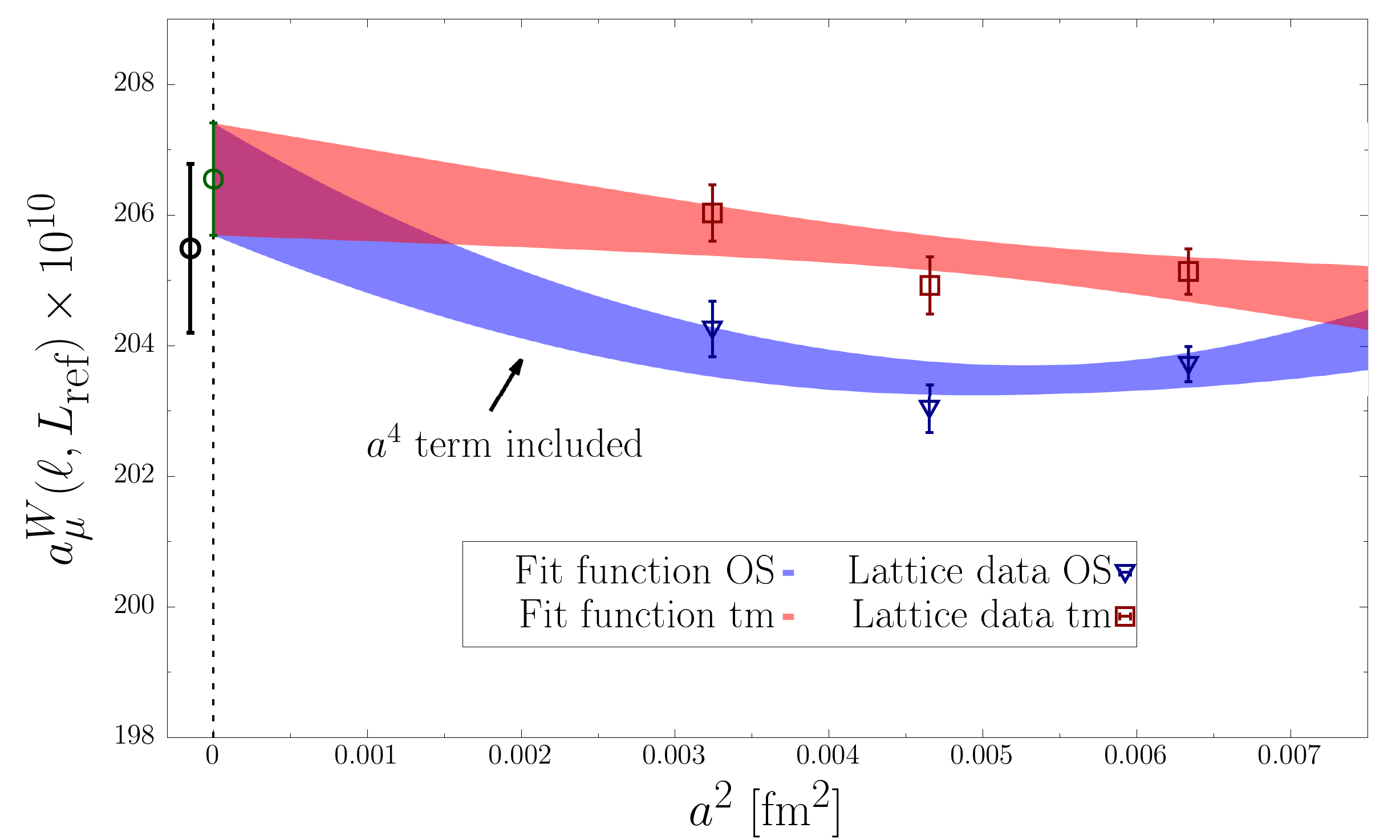}\\
\includegraphics[scale=0.375]{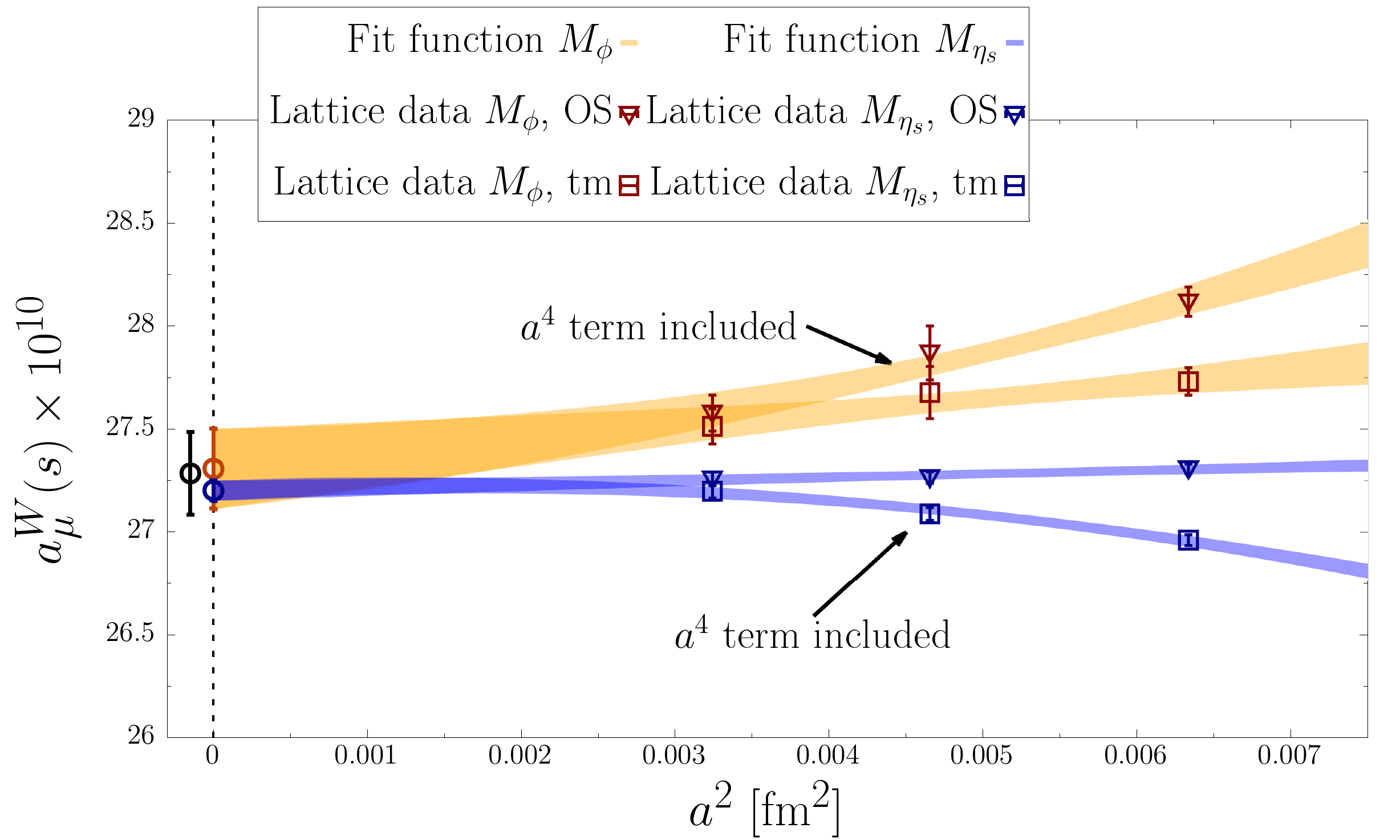}\\
\includegraphics[scale=0.375]{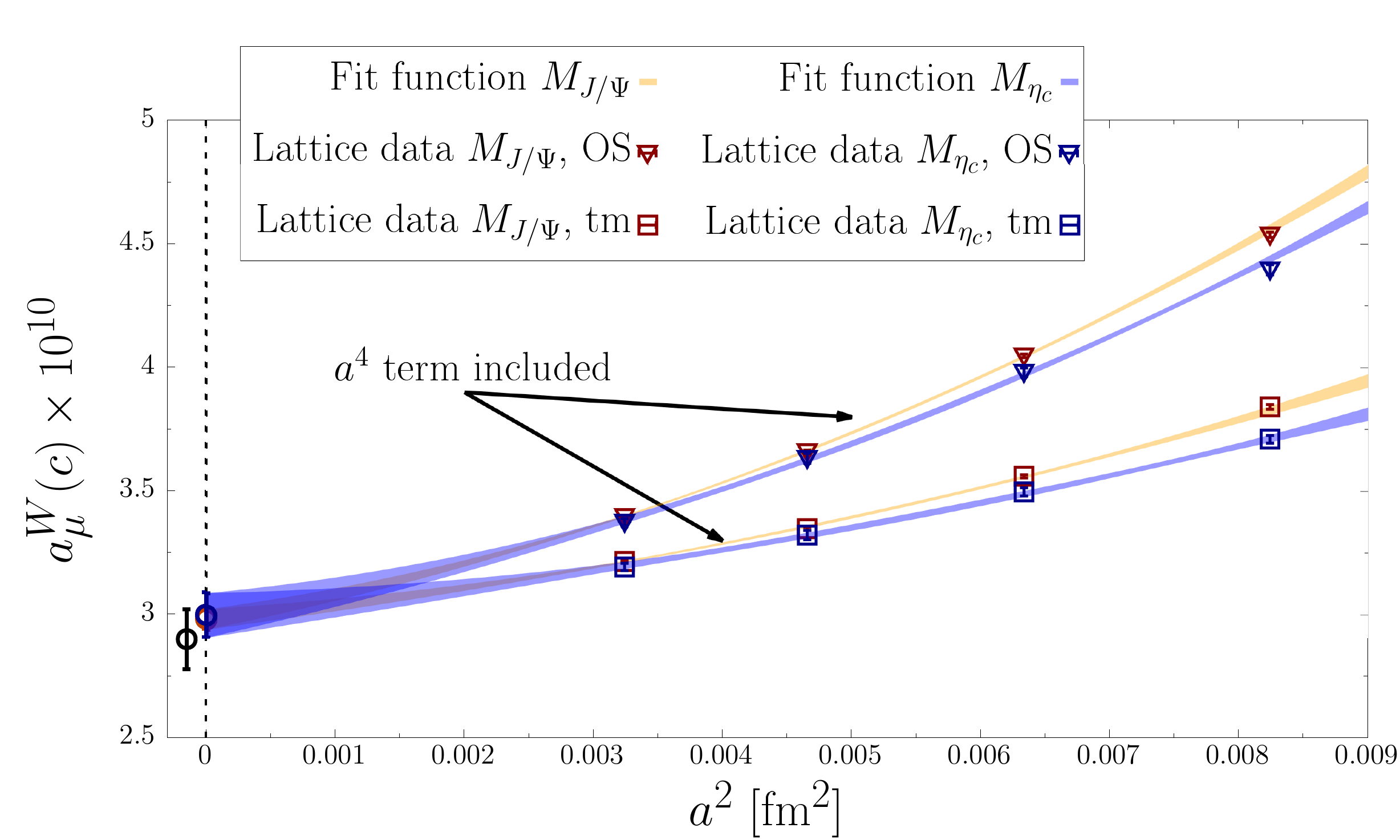}
\vspace{-0.5cm}
\caption{\it \small The light-quark (top), strange-quark (middle) and charm-quark (bottom) connected contributions to the intermediate window $a_\mu^{\rm W}$ versus the squared lattice spacing $a^2$ in physical units using both the ``tm" (triangles) and ``OS" (squares) regularizations. In the middle (bottom) panel the blue and red points correspond to the lattice data obtained using the masses of the $\eta_s$ ($\eta_c$) and $\phi$ ($J/\Psi$) mesons to obtain the physical strange (charm) quark mass. The solid lines correspond to representative examples of continuum extrapolation obtained using the Ansatz in Eq.\,(\ref{eq:amuw_fit}) with $D_{1L}^{\rm tm}(f) = D_{1L}^{\rm OS}(f) = 0$ (polynomial fits). The extrapolated values in the continuum limit are shown at $a^2 = 0$ together with our final results given by Eqs.\,(\ref{eq:amuW_ell_Lref})-(\ref{eq:amuW_charm_final}).}
\label{fig:W_cont_lim}
\end{center}
\end{figure}

The connected contribution $a_\mu^{\rm W}(f)$ to the intermediate window is given by
\be
    \label{eq:amuW}
    a_\mu^{\rm W}(f) = 2 \alpha_{em}^2 \int_0^\infty ~ dt \, t^2 \, K(m_\mu t) \, \Theta^{\rm W}(t)\,V_f(t) ~ 
\ee
where $f = \{\ell, s, c \}$ and $\Theta^{\rm W}(t)$ is given by Eq.\,(\ref{eq:Mt_W}).
Our results corresponding to the ``tm" and ``OS" regularizations, at the physical pion mass $M_\pi^{phys} = M_\pi^{isoQCD} = 135.0\,(2)$ MeV and at the reference lattice size $L_{ref} = 5.46$ fm, are shown in Fig.\,\ref{fig:W_cont_lim} together with a representative example of continuum extrapolation.
We note that, in contrast with the short-distance window, there are no discretization effects of the type $a^2 \, \mbox{log}(a)$, thanks to the exponential suppression of the modulating function $\Theta^{\rm W}(t)$ at small values of $t \approx a$ (see Fig.\,\ref{fig:windows_t}).
Therefore, we do not carry out any subtraction of the tree-level perturbative lattice artifacts.

The statistical precision of the lattice data for $a_\mu^W(\ell, L_{ref})$ is of the order $\mathcal{O}(0.2 \%)$.
Also the results for $a_\mu^W(s)$ obtained using $M_{\eta_s}$ have a very good precision of order $\mathcal{O}(0.2 \%)$, while the ones obtained using $M_\phi$ have typically larger errors by a factor of $\simeq 3$. This originates from the fact that the plateaux of the $\phi$-meson effective mass are substantially noisier than the ones of the pseudoscalar $\eta_s$ meson (see Fig.~\ref{fig:eff_mass_phi_etas}).
Finally, the results for $a_\mu^W(c)$ exhibit a very good precision of order $\mathcal{O}(0.5 \%)$ when we use $M_{\eta_c}$ and of order $\mathcal{O}(0.2 \%)$ when we use $M_{J/\Psi}$.  

\begin{figure}[htb!]
\begin{center}
\includegraphics[scale=0.32]{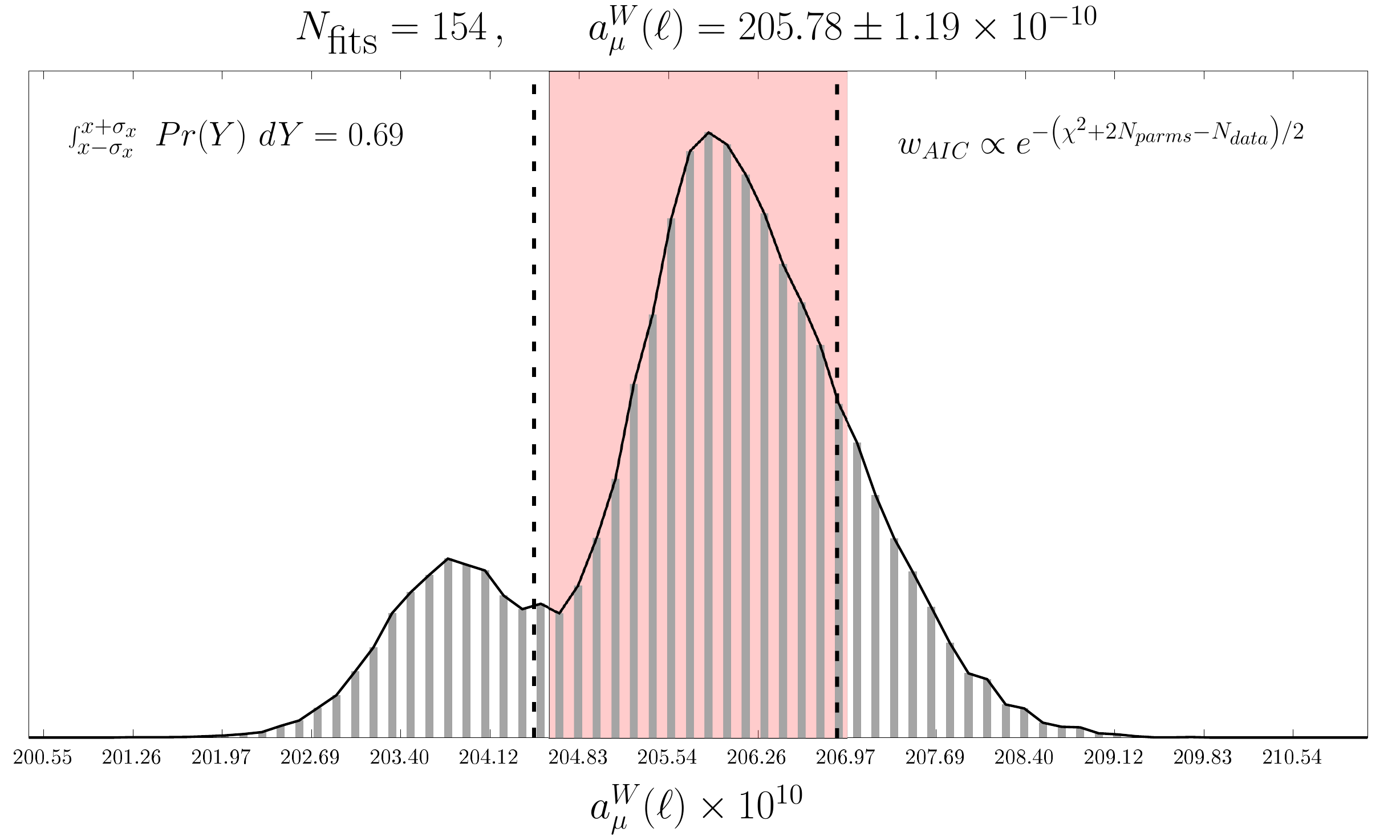} ~ 
\includegraphics[scale=0.32]{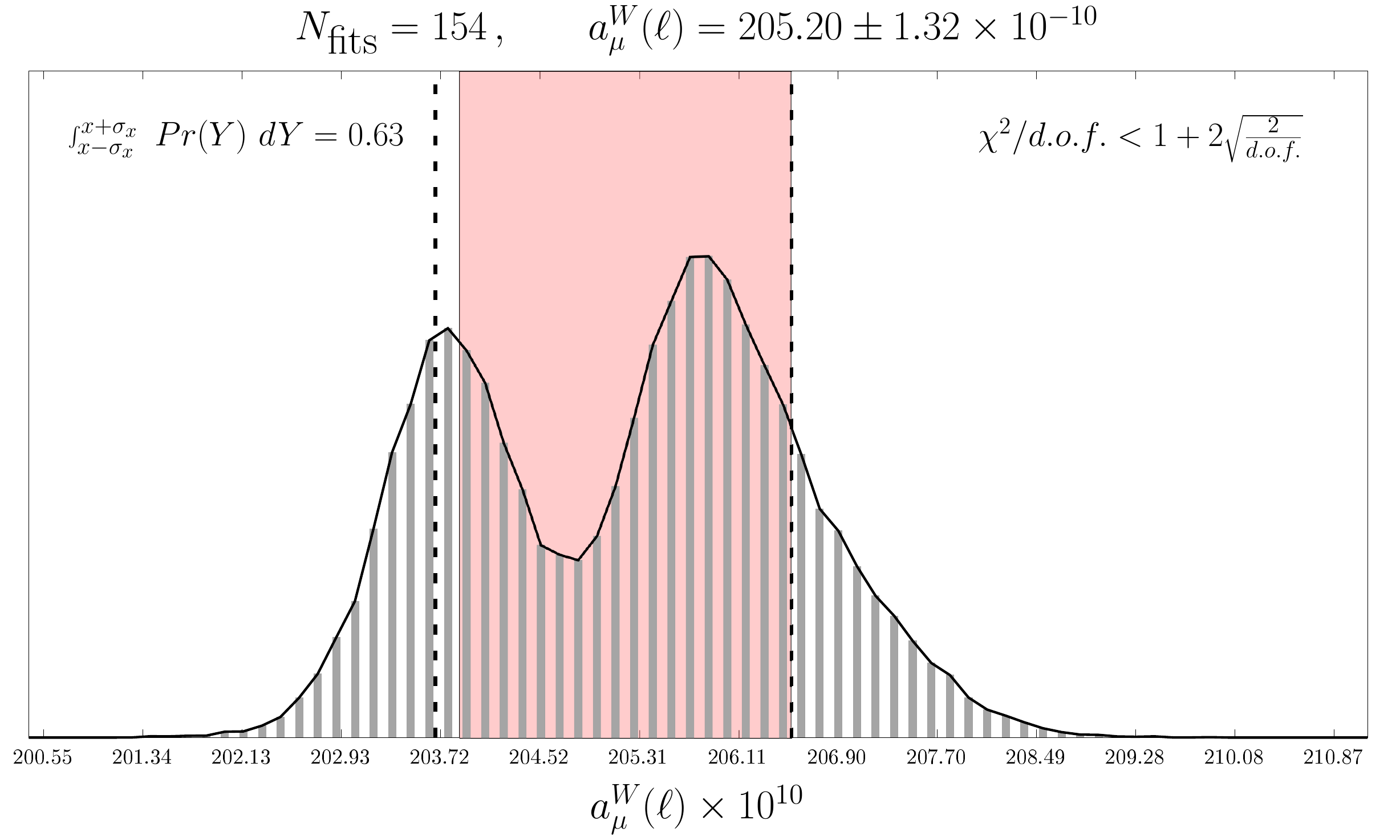}\\
\includegraphics[scale=0.32]{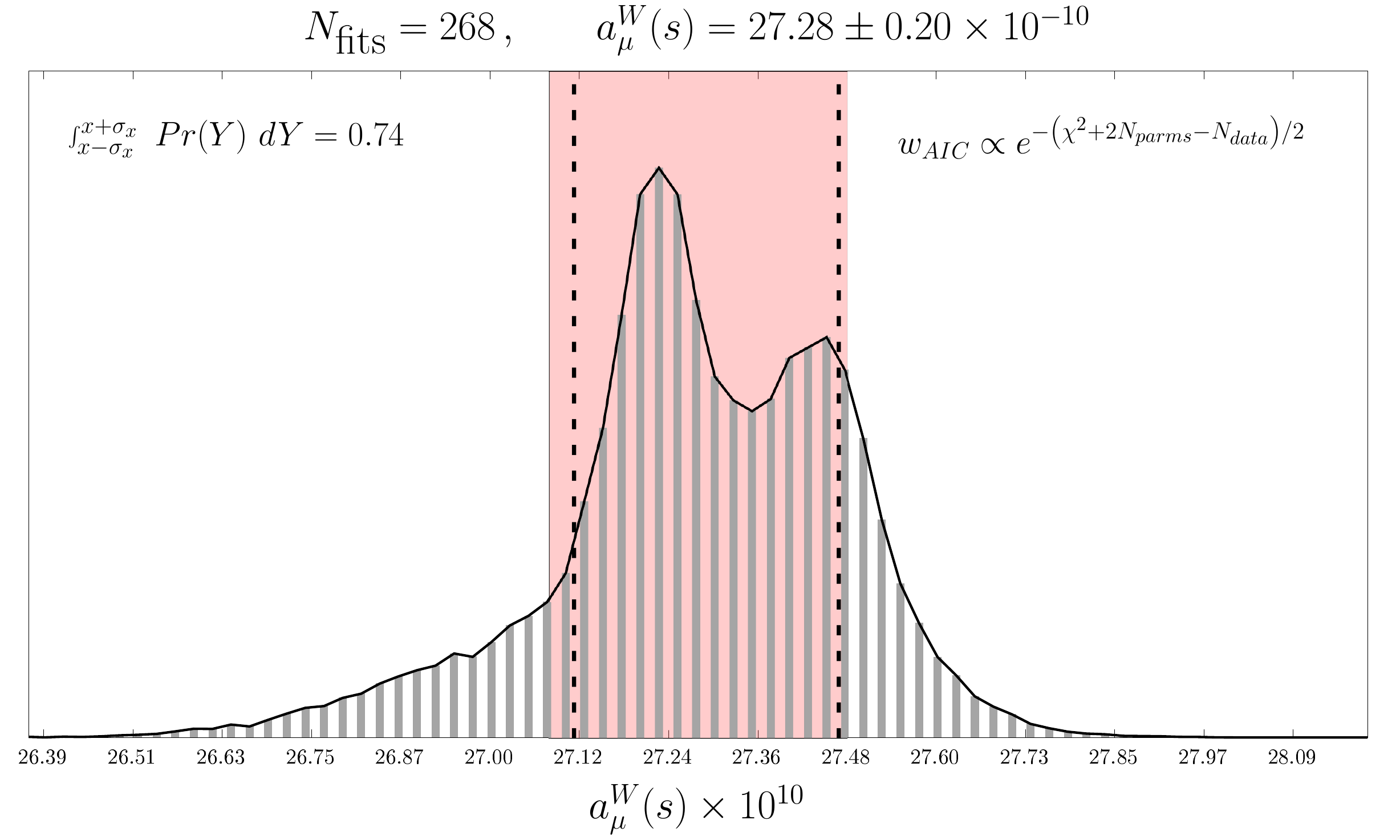} ~
\includegraphics[scale=0.32]{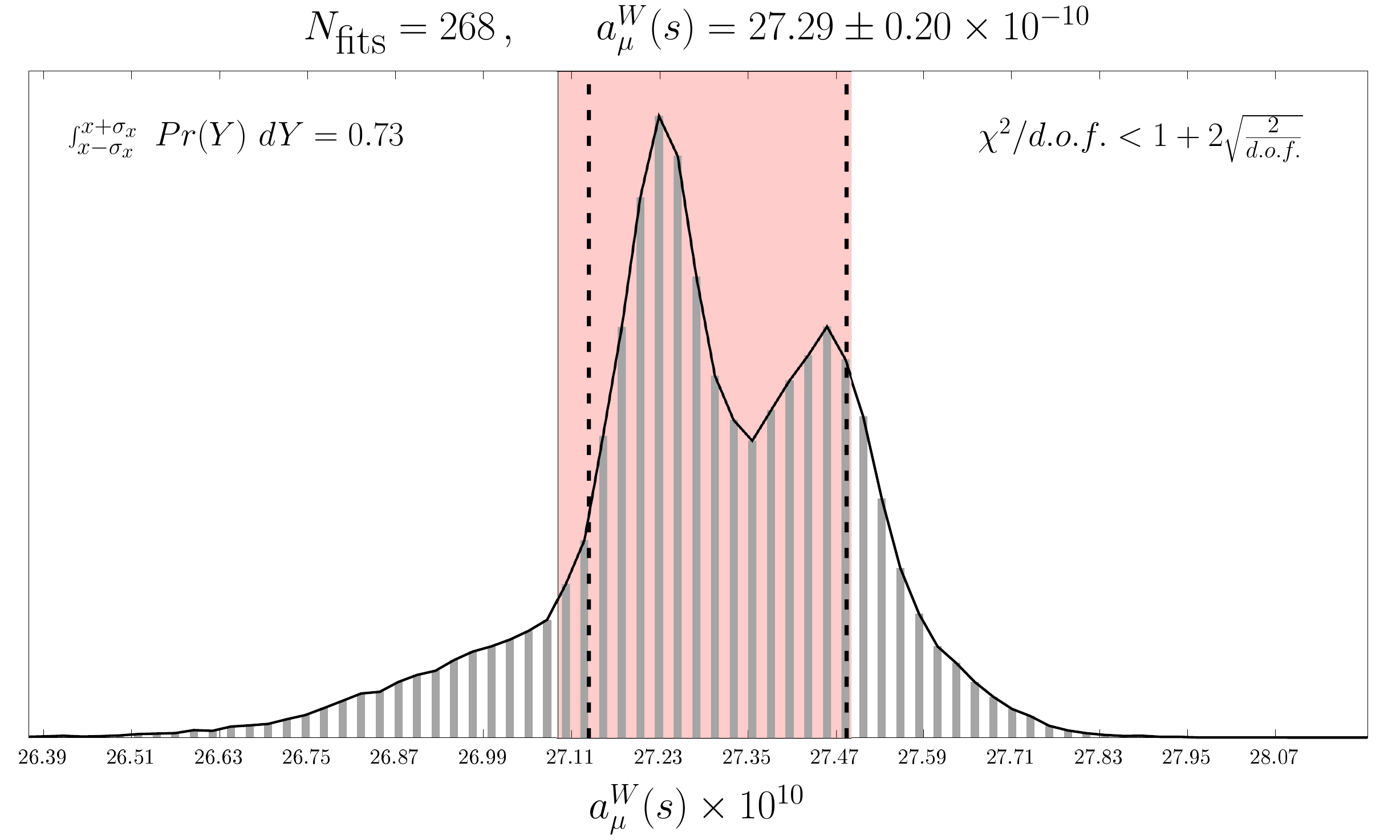}\\
\includegraphics[scale=0.32]{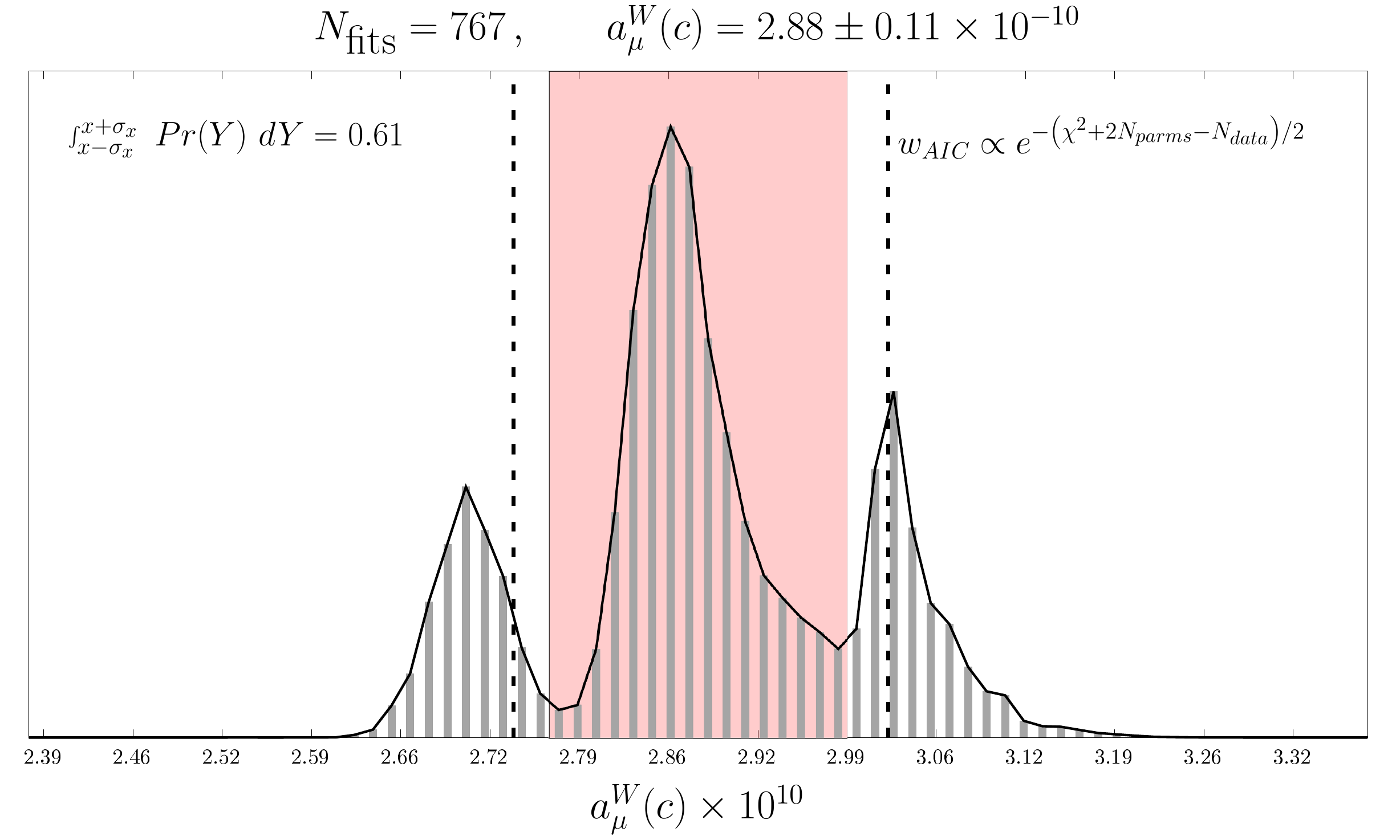} ~ 
\includegraphics[scale=0.32]{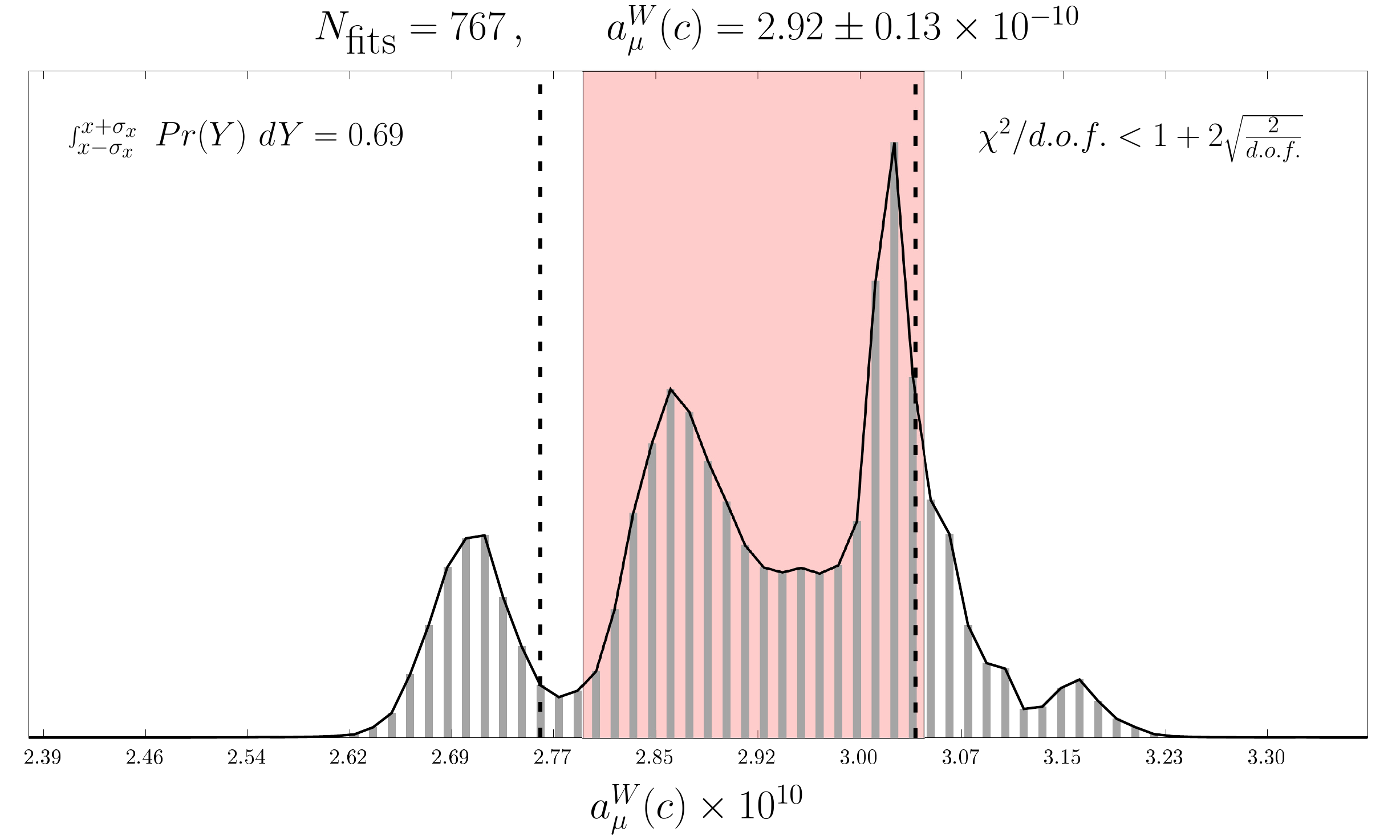}
\vspace{-0.5cm}
\caption{\it \small Histograms of the results at the physical point obtained by our fitting procedures based on Eq.\,(\ref{eq:amuw_fit}) and Eqs.\,(\ref{eq:diff})-(\ref{eq:ratio}) applied to our lattice data of $a_\mu^{\rm W}(\ell, L_{ref})$ (top panels), $a_\mu^{\rm W}(s)$ (middle panels) and $a_\mu^{\rm W}(c)$ (bottom panels) adopting for the weights $\omega_k$ either the AIC (left panels) or the step function (right panels), described respectively by Eqs.\,(\ref{eq:AIC}) and (\ref{eq:stepF}). The red bands correspond to our final results\,(\ref{eq:amuW_ell_Lref})-(\ref{eq:amuW_charm_final}).  In each panel we show the number of fits, the average ($x$) and the error ($\sigma_x$) evaluated according to Eq.\,(\ref{eq:averaging}) and the cumulative probability corresponding to the interval $[x - \sigma_x, x + \sigma_x]$. The vertical short-dashed lines correspond to the 16-th and 84-th percentiles of the p.d.f.~$Pr(Y)$.}.
\label{fig:amuW}
\end{center}
\end{figure}

Using Eq.\,(\ref{eq:amuw_fit}) and Eqs.\,(\ref{eq:diff})-(\ref{eq:ratio}) we have carried out hundreds of combined fits of our lattice data for the two regularizations ``tm" and ``OS" by minimizing a correlated $\chi^2$-variable. In this case, the entries of the correlation matrix are in the range $0.5-0.7$ for the light-quark contribution, and typically larger (reaching up to $0.99$) for the heavier flavours.
Also here we have excluded from the averaging procedure given by Eq.\,(\ref{eq:averaging}) all fits having $d.o.f. = 1$ in order to avoid overfitting.
In Fig.\,\ref{fig:amuW} we show the histograms of the results obtained at the physical point for the two choices\,(\ref{eq:AIC}) and (\ref{eq:stepF}) of the weights $\omega_k$ appearing in Eq.\,(\ref{eq:averaging}).
As in the case of the short-distance windows, the distributions of Fig.\,\ref{fig:amuW} exhibit multiple peaks, which implies that the statistical uncertainties are significantly smaller than the systematic ones due to lattice artifacts. 

At the physical point we get
\bea
    \label{eq:amuW_ell_Lref}
    a_\mu^{\rm W}(\ell, L_{ref}) & = & 205.5 ~ (0.7)_{stat} ~ (1.1)_{syst} \cdot 10^{-10} = 205.5 ~ (1.3) \cdot 10^{-10} ~ , ~ \\[2mm]
    \label{eq:amuW_strange_final}
    a_\mu^{\rm W}(s) & = & 27.28 ~ (13)_{stat} ~ (15)_{syst} \cdot 10^{-10} = 27.28 ~ (20) \cdot 10^{-10} ~ . ~ \\[2mm]
    \label{eq:amuW_charm_final}    
    a_\mu^{\rm w}(c) & = & 2.90 ~ (3)_{stat} ~ (12)_{syst} \cdot 10^{-10} = 2.90 ~ (12) \cdot 10^{-10} ~ . ~ 
\eea

To the result of Eq.\,(\ref{eq:amuW_ell_Lref}) we must add the FSE correction $-\Delta a_\mu^{\rm W}(\ell, L_{ref})$ evaluated within the MLLGS model according to Eq.\,(\ref{eq:Delta_amuw}) of Appendix\,\ref{sec:appF} with $w = W$ and $L_{ref} = 5.46$ fm in the continuum limit and at the physical pion mass point. We get
\be
\label{eq:FSE}
   \Delta a_\mu^{\rm W}(\ell, L_{ref}) = - 1.00 ~ (20) \cdot 10^{-10} ~ , ~
\ee
which leads to
\be
    \label{eq:amuW_ell_final}
    a_\mu^{\rm W}(\ell) = 206.5 ~ (1.3) \cdot 10^{-10} ~ . ~
\ee
We expect to get a substantial reduction of the error in Eq.\,(\ref{eq:amuW_ell_final}) using the results from a new ETMC ensemble at the physical pion mass point with a finer lattice
spacing currently under production.

We point out that our result given in Eq.~(\ref{eq:amuW_ell_final}) is consistent at the level of $1.5\,\sigma$ with the previous ETMC estimate $a_\mu^{\rm W}(\ell) = 202.2 (2.6) \cdot 10^{-10}$~\cite{Giusti:2021dvd}, but it improves the precision by a factor $\simeq 2$. This result is mainly related to the improvement of the statistical precision and to the reduction by a factor of $\approx 10$ of the discretization systematics as compared to Ref.\,\cite{Giusti:2021dvd}.


\section{Disconnected contributions}
\label{sec:disconnected}

In this Section we address the calculation of the quark disconnected contributions to the vector correlator $V(t)|_{MA}^{OS}$ (see Eq.\,(\ref{eq:Vrecomb})), which are {\em the sum} of the six relevant quark disconnected (label $(D)$) correlators displayed in Eq.\,(\ref{eq:Vregdetail}) weigthed by the appropriate charge factors, and may globally denoted as $V_{disc.}(t)|_{MA}^{OS}$. The currents involved in the individual correlators are defined in Eq.\,(\ref{eq:MAVcurr}) within the ``OS" regularization. From the vector correlator $V_{disc.}(t)|_{MA}^{OS}$ the values of $a_\mu^{\rm SD}(disc.)$ and $a_\mu^{\rm W}(disc.)$ are straightforwardly evaluated according to Eq.\,(\ref{eq:amu_w}).

The disconnected contributions are computed for the light, strange and charm quark mass using three ensembles close to the physical quark masses, namely cB211.072.64, cC211.060.80 and cD211.054.96. Due to the high cost of the calculation, we do not compute disconnected contributions using the larger volume cB211.072.96 ensemble, since FSEs are expected to be negligible within statistical errors.

The strange and charm quark loops are computed at a quark mass obtained by tuning the $\Omega$ and $\Lambda_c$ baryons, respectively, to their physical value. The values of the bare masses for the strange, $a\mu_s$, and for the charm, $a\mu_c$, quarks are given in Appendix\,\ref{sec:disc}.

Various noise-reduction techniques are employed to improve the signal-to-noise ratio of disconnected loops. These are the one-end-trick~\cite{McNeile:2006bz}, the exact deflation of low-modes~\cite{Gambhir:2016uwp} and hierarchical probing~\cite{Stathopoulos:2013aci}. The one-end-trick is used for all loops; hierarchical probing with distance 8 is used for all loops, except the charm-quark loops for the cB211.072.64 ensemble, where instead distance 4 is used; and deflation of the low-modes is used for the light quark loops for the cB211.072.64 and cC211.060.80 ensembles. The latter method is not  employed for the cD211.054.96 ensemble because of the prohibitively large memory requirements. Instead  multiple stochastic sources are used.
\begin{table}[htb!]
    \centering
    \begin{tabular}{ || c|c|c|c| c| c|c | c|| } 
    \hline
    Ensemble & $\ell \ell$ & $s s$ & $c c$ & $\ell s$ & $\ell c$ & $s c$ & total \\
    \hline
    cB211.072.64 & $-3.37~(13)$ & $-2.090~(59)$ & $-1.18~(14)$ & $+5.29~(15)$ & $-1.52~(24)$  & $+1.67~(13)$ & $-1.20~(23)$ \\
    cC211.060.80 & $-3.36~(16)$ & $-2.090~(73)$ & $-0.78~(11)$ & $+5.53~(17)$ & $-1.48~(20)$ & $+1.37~(15)$ & $-0.80~(18)$ \\
    cD211.054.96 & $-3.54~(16)$ & $-2.084~(75)$ & $-0.71~(14)$ & $+5.60~(18)$ & $-1.51~(21)$  & $+1.27~(18)$ & $-0.96~(20)$ \\
    \hline    
    \end{tabular}
    \caption{\it \small Summary of the various flavour contributions to $a_{\mu}^{\rm SD}($disc.$)$ in units of $10^{-12}$ for the cB211.072.64, cC211.060.80 and cD211.054.96 ensembles. The symbols $\ell \ell$, $s s$ and $c c$ denote respectively the flavour-diagonal light, strange and charm contributions, while $\ell s,$ $\ell c$ and $s c$ denote the off-diagonal light-strange, light-charm and strange-charm contributions,  respectively.}
    \label{tab:amuSD_disco}
\end{table}
\begin{table}[htb!]
    \centering
    \begin{tabular}{ || c|c|c|c| c| c|c |c || } 
    \hline
    Ensemble & $\ell \ell$ & $s s$ & $c c$ & $\ell s$ & $\ell c$ & $s c$ & total  \\
    \hline
    cB211.072.64 & $-1.087~(49)$ & $-0.149~(22)$ & $-0.030~(53)$ & $+0.635~(58)$ & $+0.00~(8)$  & $-0.02~(6)$ & $-0.651~(93)$ \\
    cC211.060.80 & $-1.300~(69)$ & $-0.159~(27)$ & $-0.033~(49)$ & $+0.726~(81)$ & $-0.03~(7)$ & $+0.04~(7)$ & $-0.762~(75)$  \\
    cD211.054.96 & $-1.201~(73)$ & $-0.149~(29)$ & $+0.018~(54)$ & $+0.627~(81)$ & $+0.02~(8)$  & $-0.02~(7)$ & $-0.701~(80)$ \\
    \hline
    \end{tabular}
    \caption{\it \small The same as in Table\,\ref{tab:amuSD_disco}, but for the various flavour contributions to $a_\mu^{\rm W}($disc.$)$ in units of  $10^{-10}$.}
    \label{tab:amuW_disco}
\end{table}

\begin{figure}[htb!]
\begin{center}
\includegraphics[scale=0.60]{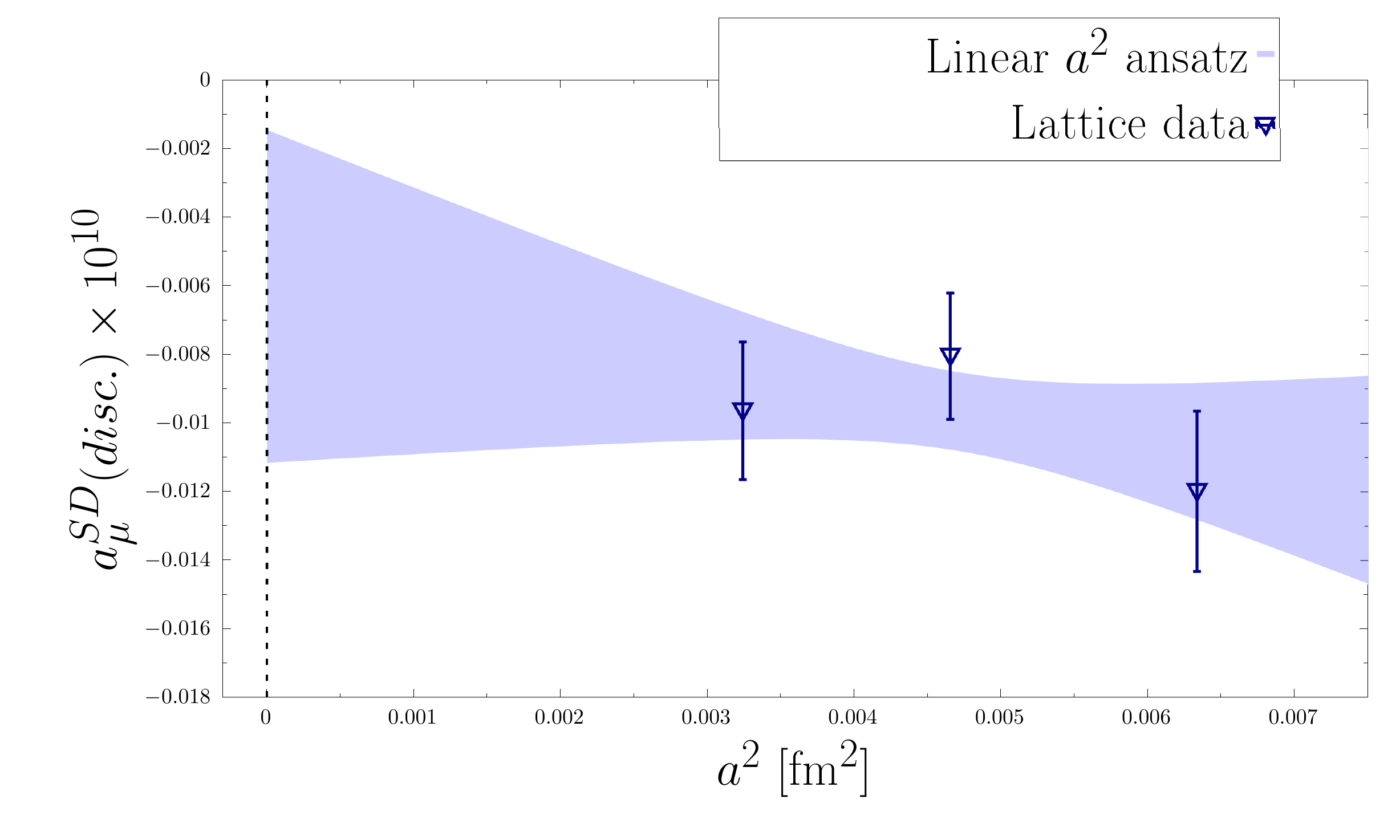}
\includegraphics[scale=0.60]{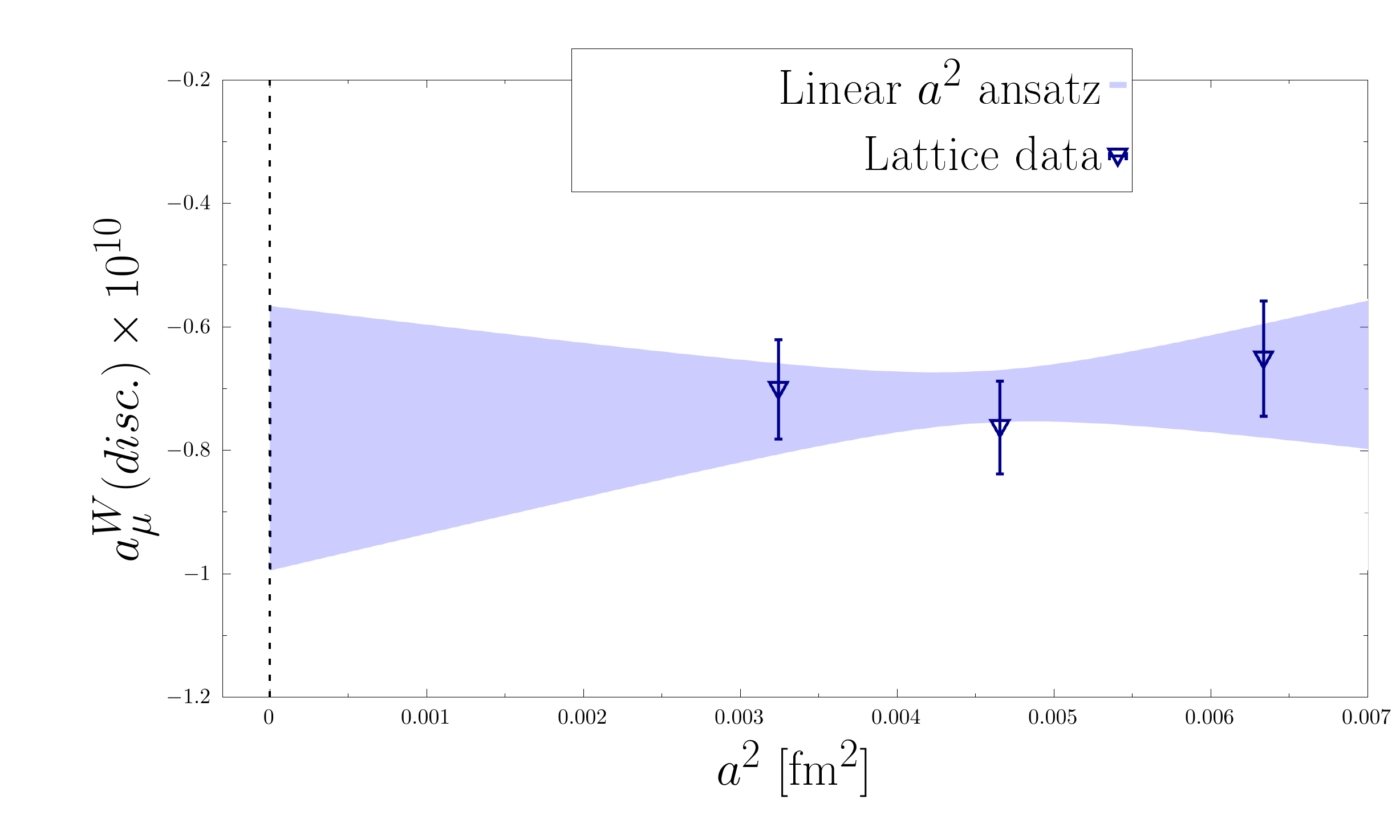}
\caption{\it \small Top panel: The quark-loop disconnected contribution to the short time-distance window, $a_\mu^{\rm SD}$, versus the squared lattice spacing $a^2$ in physical units. Bottom panel: the same as in the top panel, but for the intermediate window $a_\mu^{\rm W}$. The blue band corresponds to the extrapolation performed using a linear fit Ansatz in $a^2$.}
\label{fig:SDW_disco_cont_lim}
\end{center}
\end{figure}

The results for the diagonal and off-diagonal disconnected contributions are summarized in Table~\ref{tab:amuSD_disco} for $a_\mu^{\rm SD}$ and in Table~\ref{tab:amuW_disco} for $a_\mu^{\rm W}$. In Fig.~\ref{fig:SDW_disco_cont_lim}, we show the continuum limit extrapolation for the disconnected contributions to $a_\mu^{\rm SD}$ and $a_\mu^{\rm W}$. 
Qualitatively, for $a_\mu^{\rm W}$ the light-light contribute +150\% of the total disconnected contribution, the strange-light -80\% and the strange-strange +30\%. All other combinations are consistent with zero within the errors. We do not observe sizable cutoff effects at this level of precision.
The disconnected contribution to $a_\mu^{\rm SD}$ is very small, being approximately forty times smaller as compared to our error on the light-connected contribution to $a_\mu^{\rm SD}$. Given the available data, we  perform only a single continuum extrapolation, using a linear fit Ansatz in $a^2$. 

Our results for $a_\mu^{\rm SD}(disc.)$ and $a_\mu^{\rm W}(disc.)$ are
\bea
    \label{eq:amuSD_disc_final}
    a_\mu^{\rm SD}(disc.) & = & - 0.006~(5) \cdot 10^{-10} ~ , ~ \\[2mm]
    \label{eq:amuW_disc_final}
    a_\mu^{\rm W}(disc.) & = & - 0.78~(21) \cdot 10^{-10} ~ . ~
\eea


\section{Comparison  with dispersive $e^+ e^-$ results and other lattice QCD calculations}
\label{sec:comparison}

Our results obtained in the isospin-symmetric limit for the quark connected contributions from the light, strange and charm quarks, as well as, the sum of all quark disconnected flavour diagonal and off-diagonal contributions to the short and intermediate time-distance windows are listed in Eqs.\,(\ref{eq:amuSD_results}) and (\ref{eq:amuW_results}). 
In the case of the intermediate window $a_\mu^{\rm W}$, our findings can be compared with the corresponding ones obtained by the BMW collaboration in Ref.\,\cite{Borsanyi:2020mff}, by the CLS/Mainz group in Ref.\,\cite{Ce:2022kxy}, by Lehner and Meyer in Ref.\,\cite{Lehner:2020crt} and by Aubin et al.~in Ref.\,\cite{Aubin:2022hgm} (which updates their previous result\,\cite{Aubin:2019usy}). We consider also the results obtained by $\chi$QCD collaboration in Ref.\,\cite{Wang:2022lkq}, by ETMC in Ref.\,\cite{Giusti:2021dvd} and by RBC/UKQCD in Ref.\,\cite{RBC:2018dos}, which come from lattice setups that have less than three values of the lattice spacing or do not include ensembles close to the physical pion mass point. All the above results are collected in Table\,\ref{tab:comparison_LQCD}. We observe a remarkable agreement among all lattice QCD results establishing a clear and important success for the computation of this quantity within the framework of lattice QCD. Moreover, very recently the Fermilab lattice, HPQCD, and MILC collaborations published~\cite{FermilabLattice:2022izv} accurate results for one-sided window contributions to $a_{\mu}^{\rm{HVP}}$, quoting in particular a value of $a_\mu^{\rm W} + a_\mu^{\rm SD} = 304.0 (9)(6) $, where the second error accounts for corrections from QED and IB. The above finding is in good agreement with our results (see Eqs.~(\ref{eq:amu_SD_ETM}) and~(\ref{eq:amu_W_ETM}) below).
\begin{table}[htb!]
\begin{center}
    \begin{tabular}{||c||c|c|c|c||}
    \hline
    ~ Ref. ~ & $a_\mu^{\rm W}(\ell)$ & $a_\mu^{\rm W}(s)$ & $a_\mu^{\rm W}(c)$ & $a_\mu^{\rm W}(disc.)$  \\
  \hline \hline
  this work & $~206.5~(1.3)~$ & $~27.28~(0.20)~$ & $~~~2.90~(0.12)~$ & $~-0.78~(0.21)~$ \\
  \hline
 BMW~\mbox{\cite{Borsanyi:2020mff}} & $~207.3~(1.4)~$ & $~27.18~(0.03)~$ & $2.7~~(0.1)$ & $~-0.85~(0.06)~$ \\
  \hline
 ~CLS/Mainz~\mbox{\cite{Ce:2022kxy}} & $~207.0~(1.5)~$ & $~27.68~(0.28)~$ & $~~~2.89~(0.14)~$ & $~-0.81~(0.09)~$ \\
\hline
  ~Lehner and Meyer~\mbox{\cite{Lehner:2020crt}} & $~206.0~(1.2)~$ & $~27.06~(0.22)~$ & -- & --
  \\
 \hline
  ~Aubin et al.~\mbox{\cite{Aubin:2022hgm}} & $~206.8~(2.2)~$ & -- & -- & -- \\
  \hline \hline
  ~$\chi$QCD~\mbox{\cite{Wang:2022lkq}} & $~206.7~(1.5)~$ & $~26.7~~(0.3)~~$ & -- & -- 
  \\
  \hline
  ~ETMC~\mbox{\cite{Giusti:2021dvd}} & $~202.2~(2.6)~$ & $~26.9~~(1.0)~~$ & $~~~2.81~(0.11)~$ & -- 
  \\
  \hline
  ~RBC/UKQCD~\mbox{\cite{RBC:2018dos}} & $~202.9~(1.5)~$ & $~27.0~~(0.2)~~$ & $3.0~~(0.1)$ & -- 
  \\
  \hline \hline
  average & $~206.0~(0.6)~$ & $~27.18~(0.03)~$ & $~~~2.86~(0.06)~$ & $~-0.83~(0.05)~$ \\
  \hline
    \end{tabular}
\end{center}
\vspace{-0.25cm}
\caption{\it \small Contributions to the intermediate time-distance window $a_\mu^{\rm W}$ obtained in this work and in Refs.\,\cite{Borsanyi:2020mff, Ce:2022kxy, Lehner:2020crt, Aubin:2022hgm, Wang:2022lkq, Giusti:2021dvd, RBC:2018dos}, namely the quark connected light ($\ell$), strange (s) and charm (c) diagrams and the sum of the quark disconnected flavour diagonal and off-diagonal diagrams. The last row lists the averages of all the lattice results for each contribution made following the PDG approach. All quantities are in units of $10^{-10}$.}
\label{tab:comparison_LQCD}
\end{table} 

As shown in Sec.\,\ref{sec:windows_E}, the time-window contributions $a_\mu^{\rm SD}$ and $a_\mu^{\rm W}$ can be evaluated using Eq.\,(\ref{eq:amu_w_E}), which involves the energy-modulating functions $\widetilde{\Theta}^{\rm SD}(E)$ and $\widetilde{\Theta}^{\rm W}(E)$, related to the time-modulating functions $\Theta^{\rm SD}(t)$ and $\Theta^{\rm W}(t)$ of Eq.\,(\ref{eq:ME_w}), and the experimental data available for the $e^+ e^-$ ratio $R^{\rm had}(E)$, given in Eq.\,(\ref{eq:Rhad}).

Using the database of Ref.\,\cite{Keshavarzi:2019abf} one gets the quite precise results\,\cite{KNT}
\bea
     \label{eq:amu_SD_KNT}
    a_\mu^{\rm SD}(e^+ e^-) & = & 68.44 (48) \cdot 10^{-10} ~ , ~ \\
    \label{eq:amu_W_KNT}
    a_\mu^{\rm W}(e^+ e^-) & = & 229.51 (87) \cdot 10^{-10} ~ . ~
\eea
More recently, starting from the analyses of Refs.\,\cite{Keshavarzi:2018mgv, Colangelo:2018mtw, Hoferichter:2019mqg, Keshavarzi:2019abf} and adopting the merging procedure of Ref.\,\cite{Aoyama:2020ynm}, which takes into account tensions in the $e^+ e^-$ database in a more conservative way, the authors of Ref.\,\cite{Colangelo:2022vok} quote the values
\bea
    \label{eq:amu_SD_dispersive}
    a_\mu^{\rm SD}(e^+ e^-) & = & 68.4 (5) \cdot 10^{-10} ~ , ~ \\
   \label{eq:amu_W_dispersive}
    a_\mu^{\rm W}(e^+ e^-) & = & 229.4 (1.4) \cdot 10^{-10} ~ . ~
\eea

To compare with the dispersive results, we need to sum up all the quark connected and disconnected contributions evaluated in the previous Sections. The individual contributions are not fully uncorrelated, since they are determined starting from basically the same gauge configurations. However, since the statistical uncertainty of the vector correlator is not dominated by the gauge error (see, e.g., Section\,\ref{sec:amuSD}) and the spatial stochastic sources employed are different for different flavors, we do not expect to have significant correlations among the various contributions to the time windows. We have checked explicitly this point in the case of the light and strange connected contributions and found a negligible correlation. Thus, the uncertainties of the individual quark connected and disconnected contributions are summed in quadrature.

Following the above strategy, the sum of $a_\mu^{\rm SD}(\ell)$, $a_\mu^{\rm SD}(s)$, $a_\mu^{\rm SD}(c)$ and $a_\mu^{\rm SD}(disc.)$, i.e.~the sum of Eqs.\,(\ref{eq:amuSD_ell_final}), (\ref{eq:amuSD_strange_final}), (\ref{eq:amuSD_charm_final}) and (\ref{eq:amuSD_disc_final}), yields the result $68.91(31) \cdot 10^{-10}$.
Adding also the contribution $a_\mu^{\rm SD}(b) = 0.32 \cdot 10^{-10}$ coming from the bottom quark (see also the lattice results of Ref.\,\cite{Hatton:2021dvg}) and a QED correction $a_\mu^{\rm SD}({\rm QED}) = 0.03 \cdot 10^{-10}$, both estimated using the ``rhad" software package\,\cite{Harlander:2002ur}, we get
\be
    \label{eq:amu_SD_ETM}
    a_\mu^{\rm SD}({\rm ETMC}) = 69.27 (34) \cdot 10^{-10} ~ , ~
\ee
which agrees with the dispersive results\,(\ref{eq:amu_SD_KNT}) and\,(\ref{eq:amu_SD_dispersive}) within $\simeq 1.4 \sigma$.

\begin{figure}[htb!]
\begin{center}
\includegraphics[scale=0.85]{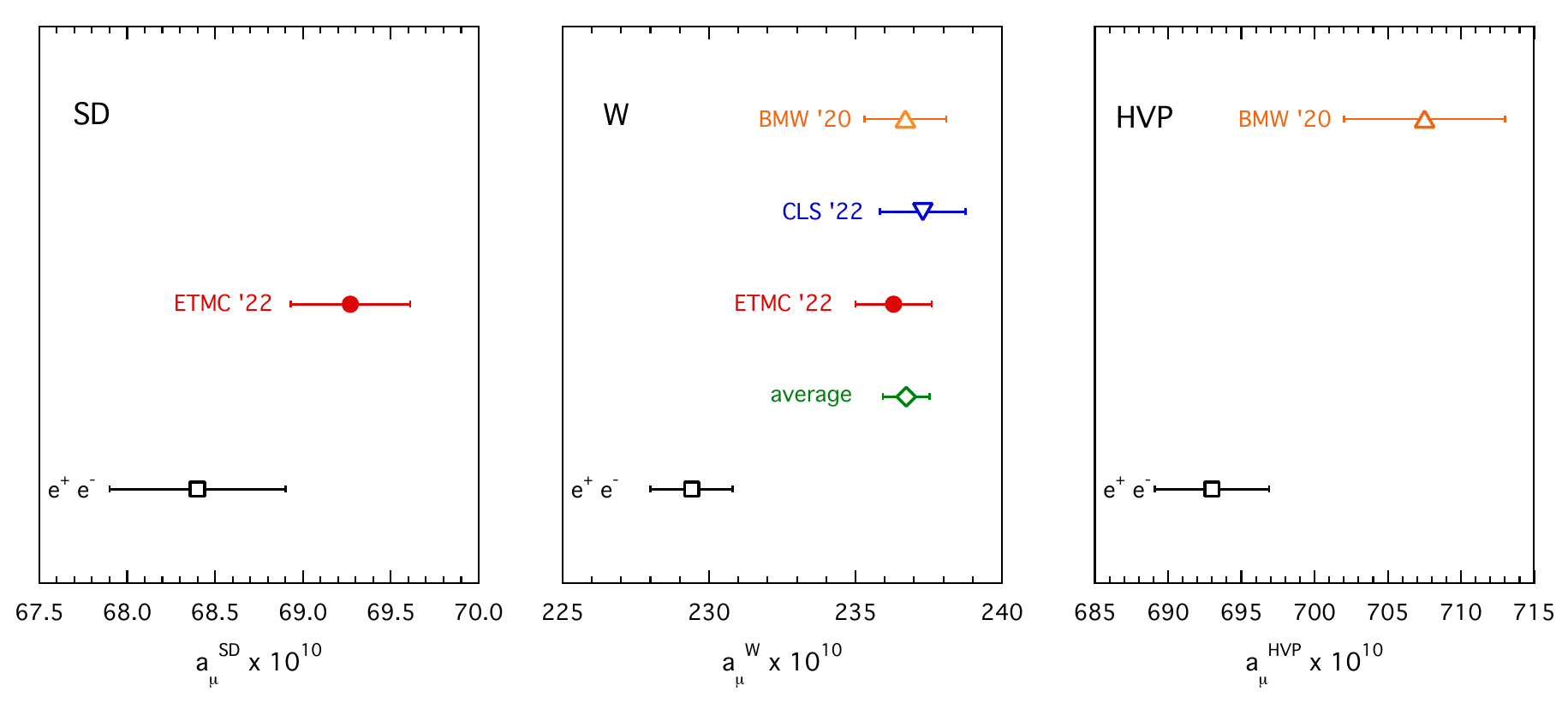}
\vspace{-0.5cm}
\caption{\it \small We show lattice QCD results of the short-distance window $a_\mu^{\rm SD}$ (left panel), intermediate window $a_\mu^{\rm W }$ (central panel), obtained in this work and in Refs.\,\cite{Borsanyi:2020mff,Ce:2022kxy}, and the full HVP term $a_\mu^{\rm HVP}$ (right panel) from Ref.\,\cite{Borsanyi:2020mff}, compared with the corresponding dispersive determinations from Ref.\,\cite{Colangelo:2022vok}, based on experimental $e^+ e^- \to$ hadrons data (see text). In the central panel, the green diamond denotes the average of our result given in Eq.\,(\ref{eq:amu_W_ETM}) with those from Refs.\,\cite{Borsanyi:2020mff,Ce:2022kxy}, namely $a_\mu^{\rm W} = 236.73 (80) \cdot 10^{-10}$.}
\label{fig:comparison}
\end{center}
\end{figure}

In the case of the intermediate window, we have to sum  the results obtained for $a_\mu^{\rm W}(\ell)$, $a_\mu^{\rm W}(s)$, $a_\mu^{\rm W}(c)$ and $a_\mu^{\rm W}(disc.)$, namely the values given in Eqs.\,(\ref{eq:amuW_ell_final}), (\ref{eq:amuW_strange_final}), (\ref{eq:amuW_charm_final}) and (\ref{eq:amuW_disc_final}), obtaining 
$235.9 (1.3) \cdot 10^{-10}$.
Adding the IB contribution $a_\mu^{\rm W}(IB) = 0.43 (4) \cdot 10^{-10}$, estimated using the corresponding BMW results of Ref.\,\cite{Borsanyi:2020mff}, we obtain
\be
    \label{eq:amu_W_ETM}
    a_\mu^{\rm W}({\rm ETMC}) = 236.3 (1.3) \cdot 10^{-10} ~ . ~
\ee
We now compare the above result with other lattice calculations available for the total window contribution satisfying the simple criterion of being based on lattice setups with more than two values of the lattice spacing and at least one ensemble close to the physical pion mass point.
Our value\,(\ref{eq:amu_W_ETM}) is nicely consistent with the result $a_\mu^{\rm W}({\rm BMW}) = 236.7(1.4) \cdot 10^{-10}$\, by the BMW collaboration~\cite{Borsanyi:2020mff} and with the recent one $a_\mu^{\rm W}({\rm CLS}) = 237.30\,(1.46) \cdot 10^{-10}$ \, by the CLS/Mainz group~\cite{Ce:2022kxy} at better than $1\sigma$ level. However, it is in tension with the dispersive result\,(\ref{eq:amu_W_dispersive}) by $3.6 \sigma$.
Averaging our result\,(\ref{eq:amu_W_ETM}) with the one by the BMW collaboration, we obtain $a_\mu^{\rm W} = 236.49 (95) \cdot 10^{-10}$, which disagrees with the dispersive result by $4.2 \sigma$. Taking into account also the very recent result by the CLS/Mainz group\,\cite{Ce:2022kxy}, we get $a_\mu^{\rm W} = 236.73 (80) \cdot 10^{-10}$, which increases the tension with the dispersive result at the level of $\simeq 4.5 \sigma$.
Comparing with the more precise dispersive result\,(\ref{eq:amu_W_KNT}), obtained in Refs.\,\cite{Keshavarzi:2019abf, KNT}, the tension increases further reaching the level of $\simeq 6.1 \sigma$.
The above lattice and dispersive results for the short and intermediate time-distance windows as well as those for the full HVP term are also collected in Fig.\,\ref{fig:comparison}.

The accurate lattice results suggest the possible presence of deviations in the $e^+ e^-$ cross section data with respect to the QCD+QED theory predictions somewhere in the low and/or intermediate energy regions, but not in the high energy region as defined in  Fig.\,\ref{fig:windows_E}.

In Table\,\ref{tab:comparison}, we collect our lattice results for the short and intermediate time-distance windows and the lattice value of the full HVP term taken from Ref.\,\cite{Borsanyi:2020mff}. These lattice results are compared with the corresponding dispersive determination of Ref.\,\cite{Colangelo:2022vok}, based on experimental $e^+ e^- \to$ hadron data. The differences between them denoted by $\Delta a_\mu^w$ for $w = \{ {\rm SD, W, HVP} \}$ are shown in the fourth column.
\begin{table}[htb!]
\begin{center}
    \begin{tabular}{||c||c|c|c||c|c||}
    \hline
    ~ window ($w$) ~ & ~ $a_\mu^w({\rm LQCD})$ ~ & ~ $a_\mu^w(e^+ e^-)$ ~\cite{Colangelo:2022vok} & ~~ $\Delta a_\mu^w$ ~~ & ~ $a_\mu^w(2\pi)$ ~\cite{Colangelo:2022vok} & ~ $\Delta a_\mu^w / a_\mu^w(2\pi)$ ~ \\
  \hline \hline
  SD & $~69.3~(0.3)$~[*] & $~~68.4~(0.5)$ & $~0.9~(0.6)$ & $~13.7~(0.1)$ & $0.066~(43)$  \\
  \hline
 W & $236.3~(13)$~[*] & $229.4~(1.4)$ & $~6.9~(1.9)$ & $138.3~(1.2)$ & $0.050~(14)$  \\
  \hline
 HVP & $707.5~(5.5)$~\cite{Borsanyi:2020mff} & $693.0~(3.9)$ & ~$14.5~(6.7)$~ & $494.3~(3.6)$ & $0.029~(14)$  \\
  \hline
    \end{tabular}
    \flushleft ~~~ [*] = this work
\end{center}
\vspace{-0.5cm}
\caption{\it \small Values of $a_\mu^w$ obtained in this work for the short and intermediate time-distance windows, $w = \{ \rm{SD, W} \}$, and from Ref.\,\cite{Borsanyi:2020mff} for the full HVP term, $w = {\rm HVP}$, compared with the corresponding dispersive determinations of Ref.\,\cite{Colangelo:2022vok}, based on experimental $e^+ e^- \to$ hadrons data (third column). The difference between the second and third columns, $\Delta a_\mu^w$, is given in the fourth column, while the contributions of the $2\pi$ channels $a_\mu^w(2\pi)$ (below a center-of-mass energy of 1 GeV), obtained in Ref.\,\cite{Colangelo:2022vok}, are shown in the fifth column. All quantities are in units of $10^{-10}$ except for the last column, where we list the values of the ratio between $\Delta a_\mu^w$ and the $2 \pi$ contribution $a_\mu^w(2\pi)$.}
\label{tab:comparison}
\end{table} 
The contribution of the $2\pi$ channels (below a center-of-mass energy of 1 GeV) to the various windows, $a_\mu^w(2\pi)$, as determined in Ref.\,\cite{Colangelo:2022vok}, are compared with the differences $\Delta a_\mu^w$. We find that the ratio of $\Delta a_\mu^w / a_\mu^w(2\pi)$ is at the level of $\simeq 3 - 5 \%$ for the three windows albeit with large uncertainties.
This suggests, qualitatively, that the accurate lattice results for the time windows and for the full HVP term could be compatible with an overall few-percent enhancement of the $e^+ e^-$ cross section data in the $2 \pi$ channels at center-of-mass energies below 1 GeV.


\section{Conclusions}
\label{sec:conclusions}

We have presented a lattice determination of the leading-order HVP contribution to the muon anomalous magnetic moment, $a_{\mu}^{\rm HVP}$, in the so-called short- and intermediate-distance windows, $a_{\mu}^{\rm SD}$ and $a_{\mu}^{\rm W}$, defined by the RBC/UKQCD collaboration\,\cite{RBC:2018dos}.  

For this determination we have employed a set of  gauge ensembles produced by ETMC with $N_f = 2 + 1 + 1$ flavours of Wilson-clover twisted-mass sea quarks with masses tuned very close to their physical values\,\cite{Alexandrou:2018egz, ExtendedTwistedMass:2020tvp, ExtendedTwistedMass:2021qui, Finkenrath:2022eon}. 
The gauge ensembles used are simulated at three different values of the lattice spacing, namely $a \simeq 0.057, 0.068, 0.080$ fm, and with spatial lattice sizes up to $L \simeq 7.6$~fm. 

We worked in the isospin-symmetric limit. The quark connected contributions from the light ($u/d$), strange and charm quarks, as well as the sum of all quark disconnected flavour diagonal and off-diagonal contributions are computed. These are then used to evaluate the contribution to the short and intermediate time-distance windows, obtaining the results listed in Eqs.\,(\ref{eq:amuSD_results}) and (\ref{eq:amuW_results}). 
In the case of the intermediate window $a_\mu^{\rm W}$, our findings are in nice agreement with several results obtained by other lattice QCD collaborations, as shown in Table~\ref{tab:comparison_LQCD}. Such a remarkable agreement within small uncertainties represents the robustness of the evaluation of this quantity within the framework of lattice QCD.

Adding the bottom-quark and the QED contribution to the short-distance window, $a_\mu^{\rm SD}(b) + a_\mu^{\rm SD}({\rm QED}) = 0.35 \cdot 10^{-10}$, evaluated in perturbative QCD using the ``rhad" software package\,\cite{Harlander:2002ur}, and the IB contribution to the intermediate window, $a_\mu^{\rm W}({\rm IB}) = 0.43(4) \cdot 10^{-10}$ taken from Ref.\,\cite{Borsanyi:2020mff}, we get 
\bea
    \label{eq:amuSD_ETM_final}
    a_\mu^{\rm SD}({\rm ETMC}) & = & 69.27(34) \cdot 10^{-10} ~ , ~ \\[2mm]
    \label{eq:amuW_ETM_final}
    a_\mu^{\rm W}({\rm ETMC}) & = & 236.3(1.3) \cdot 10^{-10} ~ . ~
\eea

Our result for the short-distance contribution given in Eq.\,(\ref{eq:amuSD_ETM_final}) is consistent with the recent dispersive value $a_\mu^{\rm SD}(e^+ e^-) = 68.4(5) \cdot 10^{-10}$\,\cite{Colangelo:2022vok} within $\simeq 1.4 \sigma$.  In the case of the intermediate window, our value given in Eq.\,(\ref{eq:amuW_ETM_final}) is larger than the dispersive result $a_{\mu}^{\rm W}(e^+ e^-) = 229.4(1.4) \cdot 10^{-10}$\,\cite{Colangelo:2022vok} by $\simeq 3.6 \sigma$. Our value is nicely consistent with the BMW result $a_{\mu}^{\rm W}({\rm BMW}) = 236.7(1.4) \cdot 10^{-10}$\,\cite{Borsanyi:2020mff} and with the recent CLS/Mainz one $a_\mu^{\rm W}({\rm CLS}) = 237.30\,(1.46) \cdot 10^{-10}$ \,\cite{Ce:2022kxy} at better than $1\sigma$ level.
The tension between our value and the dispersive result increases from $\simeq 3.6 \sigma$ to $\simeq 4.2 \sigma$ if we average our result\,(\ref{eq:amuW_ETM_final}) with the one obtained by the BMW collaboration, leading to $a_\mu^{\rm W} = 236.49 (95) \cdot 10^{-10}$. Including in the average also the recent CLS/Mainz result we get $a_\mu^{\rm W} = 236.73 (80) \cdot 10^{-10}$, which is in disagreement with the dispersive result by $\simeq 4.5 \sigma$.

In conclusion, the impact of our lattice computations is twofold. Concerning the intermediate-distance window we confirm the two currently most accurate lattice QCD results, namely those from the BMW collaboration and the CLS/Mainz group, increasing the discrepancy with the corresponding prediction based on $e^+ e^-$ cross section data to the significant level of $\simeq 4.5$ standard deviations. Moreover, we have computed accurately for the first time the short-distance window, finding that there is no significant tension with the corresponding dispersive result. This is a clear indication that any deviation between QCD+QED theory predictions, the framework employed in SM-based lattice calculations, and the $e^+ e^-$ cross section experiments is unlikely to occur at high energy. Instead, it may occur somewhere in the low and/or intermediate energy regions.


\section*{Acknowledgments}

We thank all members of ETMC for the most enjoyable collaboration. We are very grateful to Guido Martinelli and Giancarlo Rossi for many discussions on the lattice setup and the methods employed in this work. We thank Nazario Tantalo for valuable discussions about the physical information that can be obtained by comparing the experimental data on $e^+e^- \to$~hadrons with the lattice predictions for observables related to the HVP term.
We thank the developers of the QUDA~\cite{Clark:2009wm, Babich:2011np, Clark:2016rdz} library for their continued support, without which the calculations for this project would not have been possible.

S.B.~and J.F.~are supported by the H2020 project PRACE 6-IP (grant agreement No.~82376) and the EuroCC project (grant agreement No.~951740). We acknowledge support by the European Joint Doctorate program STIMULATE grant agreement No.~765048. P.D. acknowledges support from the European Unions Horizon 2020 research and innovation programme under the Marie Sk\l{}odowska-Curie grant agreement No.~813942 (EuroPLEx) and also support from INFN under the research project INFN-QCDLAT.
K.H.~is supported by the Cyprus Research and Innovation  Foundation under contract number POST-DOC/0718/0100, under contract number CULTURE-AWARD-YR/0220/0012 and by the EuroCC project (grant agreement No.~951740). 
R.F.~acknowledges partial support from the University of Tor Vergata program “Beyond Borders/ Strong Interactions: from Lattice QCD to Strings, Branes and Holography".
F.S., G.G.~and S.S.~are supported by the Italian Ministry of University and Research (MIUR) under grant PRIN20172LNEEZ. 
F.S.~and G.G.~are supported by INFN under GRANT73/CALAT.
This work is supported by the Deutsche Forschungsgemeinschaft (DFG, German Research Foundation) and the NSFC through the funds provided to the Sino-German Collaborative Research Center CRC 110 “Symmetries and the Emergence of Structure in QCD” (DFG Project-ID 196253076 - TRR 110, NSFC Grant No.~12070131001).
The authors gratefully acknowledge the Gauss Centre for Supercomputing e.V.~(www.gauss-centre.eu) for funding the project pr74yo by providing computing time on the GCS Supercomputer SuperMUC at Leibniz Supercomputing Centre (www.lrz.de), as well as computing time projects 
on the GCS supercomputers JUWELS Cluster and JUWELS Booster~\cite{JUWELS} at the J\"ulich Supercomputing Centre (JSC) and time granted by the John von Neumann Institute for Computing (NIC) on the supercomputers JURECA and JURECA Booster~\cite{Jureca}, also at JSC. Part of the results were created within the EA program of JUWELS Booster also with the help of the JUWELS Booster Project Team (JSC, Atos, ParTec, NVIDIA). We further acknowledge computing time granted on Piz Daint at Centro Svizzero di Calcolo Scientifico (CSCS) via the project with id s702. The authors acknowledge the Texas Advanced Computing Center (TACC) at The University of Texas at Austin for providing HPC resources that have contributed to the research results. The authors gratefully acknowledge PRACE for awarding access to HAWK at HLRS within the project with Id Acid 4886.

\appendix

\section{Lattice setup and simulation details}
\label{sec:simulations}

In this work, we analyze the gauge ensembles produced recently by ETMC in isospin-symmetric QCD (isoQCD) with $N_f = 2 + 1 + 1$ flavors of Wilson-clover twisted-mass quarks and described in Refs.\,\cite{Alexandrou:2018egz, ExtendedTwistedMass:2020tvp, ExtendedTwistedMass:2021qui, Finkenrath:2022eon}.
Our renormalizable lattice theory is specified by the following action
\begin{align}
\label{eq:lattMA}
S = S_{YM}[U] + S_{\rm q,\; sea}[\Psi_{\ell},\Psi_h,U] + S_{\rm q, \; val}[\{q_f, q'_f\}, U] + 
S_{\rm ghost}[\{\phi_f, \phi'_f\}, U]\,,
\end{align}
which corresponds to a mixed action lattice setup employing twisted mass~\cite{Frezzotti:2000nk, Frezzotti:2003ni} and Osterwalder-Seiler fermions\,\cite{Osterwalder:1977pc}. This setup allows to avoid any undesired strange-charm quark mixing through cutoff effects and to preserve the automatic ${\cal{O}}(a)$-improvement of all physical observables\,\cite{Frezzotti:2004wz}. Moreover, it offers the possibility of considering two different regularizations of the current-current correlators relevant for the present study. 

The gluon action $S_{YM}[U]$ is the (mean-field) improved Iwasaki one\,\cite{Iwasaki:1985we}. It contains the bare gauge coupling $\beta=6/g_0^2$ that controls the lattice spacing $a$, as QCD asymptotic freedom implies $a \Lambda_{QCD} \sim \exp (-1/(2b_0 g_0^2))$ with $b_0 >0$.

Concerning the fermionic sector of the action, it can be written as the sum of the sea and valence quark actions. The sea quark action is written in terms of a light $\Psi_\ell = (u_{\rm sea}, d_{\rm sea})$ and a heavy $\Psi_h = (c_{\rm sea}, s_{\rm sea})$ quark doublet, namely
\bea
S_{\rm q,\; sea} & = & a^4\sum_x \left\{ \bar\Psi_\ell(x) [\gamma \cdot \tilde\nabla + \mu_\ell - i\gamma_5 \tau^3 W_{\rm cr}^{\rm cl}] \Psi_\ell(x) 
\right. ~ \nonumber \\
& + & \left. \bar\Psi_h(x) [\gamma \cdot \tilde\nabla + \mu_\sigma 
+ \tau^3 \mu_\delta - i\gamma_5 \tau^1 W_{\rm cr}^{\rm cl}] \Psi_h(x) \right\} \label{eq:Ssea}.
\eea
For the valence quark action,  it is convenient to allow for several replica, labelled by $\eta=1,2,3,...$, of each quark flavour $f$ with different values of the Wilson parameter $r_{f,\eta}$, which in practice we restrict to be $r_{f,\eta}=(-1)^{\eta-1}$. Thus, we have
\begin{align}
S_{\rm  q,\; val} = a^4\sum_x \sum_{f,\eta}  \bar q_{f,\eta}(x) \, [ \, \gamma \cdot \tilde\nabla + m_f - {\rm sign}(r_{f,\eta}) i\gamma_5  W_{\rm cr}^{\rm cl}|_{r=1} \, ] \, q_{f,\eta}(x) \, , \label{eq:Sval}
\end{align}
where $q_{f,\eta}$ is a single flavour field and $f$ runs over the four lightest quark flavours $u,d,s,c$. By $\tilde{\nabla}_{\mu}$ we denote the symmetric gauge covariant lattice derivative, while $\nabla_\mu$ and $\nabla_\mu^*$ stand for the analogous forward and backward lattice derivatives, and (${\rm H}(4)$ covariant) ``spacetime'' indices are omitted when contracted with each other.
In the expressions above, the critical Wilson-clover operator is defined as
\begin{align}
    W_{\rm cr}^{\rm cl}|_{r} = -a\frac{r}{2} \nabla^* \!\cdot\! \nabla + m_{\rm cr}(r) + 
    a\frac{c_{SW}(r)}{32} \gamma_\mu\gamma_\nu a^{-1} [Q_{\mu\nu} - Q_{\nu\mu}] \,,
\end{align}
i.e.\ it includes the critical mass $m_{\rm cr}$ term and a clover term $\propto \gamma_\mu\gamma_\nu a^{-1} [Q_{\mu\nu} - Q_{\nu\mu}] $ (i.e.\ a lattice discretization of the Pauli term $\propto i \sigma_{\mu\nu} F_{\mu\nu})$)\,\cite{Sheikholeslami:1985ij} with a coefficient, $c_{SW}$, that is identical for all sea and valence flavours and is fixed to the value obtained in one-loop tadpole boosted perturbation theory~\cite{Aoki:1998qd}. 
In writing the valence quark action $S_{\rm  q,\; val}$, we have also exploited the known property\,\cite{Frezzotti:2004wz} $W_{\rm cr}^{\rm cl}|_{-r}= - W_{\rm cr}^{\rm cl}|_{r}$. In Eq.~(\ref{eq:Ssea}) for $S_{\rm q,\; sea}$ the operators $W_{\rm cr}^{\rm cl}$ are implicitly defined for $r=1$ and their two-flavour structure is displayed by the Pauli matrices $\tau^3$ and $\tau^1$ acting in flavour space.

In the sea and valence quark action sectors the critical Wilson-clover term, which includes the critical mass counterterm $\propto m_{\rm cr} \sim 1/a$, is taken at maximal twist with respect to the soft quark mass terms in order to guarantee automatic ${\cal{O}}(a)$-improvement of the physical observables\,\cite{Frezzotti:2003ni,Frezzotti:2005gi} and $m_{\rm cr}$ is set to a unique value for all flavours\,\cite{Frezzotti:2004wz}. The inclusion of the clover term turns out to be very beneficial for further reduction of the residual cutoff effects, in particular those on the neutral pion mass, thereby making the Monte Carlo simulations close to the physical pion point numerically stable \,\cite{Alexandrou:2018egz} (see also Ref.\,\cite{ETM:2015ned}). 

The valence ghost action term reflects the form and follows the notation of the valence quark action, viz.\
\begin{equation}
S_{\rm ghost} \! = a^4\sum_x \sum_{f,\eta}  \phi_{f,\eta}^\dagger(x) \, [ \, \gamma \cdot \tilde\nabla + m_f - {\rm sign}(r_{f,\eta}) i\gamma_5  W_{\rm cr}^{\rm cl}|_{r=1} \, ] \, \phi_{f,\eta}(x) \, ,
\end{equation}
with each $\phi_{f,\eta}$ being a complex boson field of spin 1/2 (i.e.\ a ghost), and is included in order to obtain  formally vanishing contributions from all the valence fields to the effective gluonic action. Of course no ghost fields ever occur in our actual computations.

For the light quark doublet  the {\em sea and valence} bare mass, $\mu_\ell$, is unique and takes  values such as to obtain $M_\pi$ close to $M_\pi^{phys} = M_\pi^{isoQCD} = 135.0\,(2)$ MeV\,\cite{ExtendedTwistedMass:2021qui}. Larger values of $M_\pi$ are used to compute observables relevant for scale setting, where the analysis includes a chiral extrapolation to $M_\pi^{phys}$.

The masses of the strange and charm {\em sea} quarks are set within $\sim 5\%$ accuracy to their physical values for each ensemble by carefully tuning the parameters $\mu_\sigma$ and $\mu_\delta$ in $S_{\rm q , \; sea}$ (in the preliminary stage of our simulations) in order to reproduce the renormalization group invariant (RGI) values $M_{D_s}/f_{D_s} = 7.9(0.1)$ and $m_c/m_s = 11.8(0.2)$ adopted in Refs.\,\cite{Alexandrou:2018egz, ExtendedTwistedMass:2020tvp, ExtendedTwistedMass:2021qui}. The above values are consistent with the more precise, recent determinations $M_{D_s}/f_{D_s} = 7.88(0.02)$ and $m_c/m_s = 11.77(0.03)$ from Ref.\,\cite{FlavourLatticeAveragingGroupFLAG:2021npn} and they are sufficiently accurate for the purposes of the present work\footnote{From our study of the light-quark sea mass corrections (see below Table\,\ref{tab:mass_corr}) we observe that a change of $\simeq 5 \%$ in the light-quark sea mass affects $a_\mu^{\rm W}(\ell)$ by less than $\simeq 0.05 \%$. Sea quark effects are in general suppressed as the quark mass is increased and, in the limit where the sea quark mass, $m_{sea}$, of a given flavour is large as compared to $\Lambda_{QCD}$, the relative change of an observable is expected to scale as the relative sea quark mass change times ${\cal{O}}[(\Lambda_{QCD}^2 / m_{sea}^2 )(\alpha_s^2(m_{sea})/\pi^2)]$. Thus, a $\simeq 5 \%$ relative mistuning of strange- and charm-quark sea masses cannot have a significant impact on window observables that are evaluated with a few permille relative uncertainty.}.

The masses of the strange and charm {\em valence} quarks are very accurately fixed by two physical inputs, which can conveniently be chosen as the kaon and $D$-meson masses. In this way, using the ETMC ensembles of Refs.\,\cite{Alexandrou:2018egz, ExtendedTwistedMass:2020tvp, ExtendedTwistedMass:2021qui, Finkenrath:2022eon}, the values of the charm and strange as well as light $u/d$ renormalized quark masses were determined in Ref.\,\cite{ExtendedTwistedMass:2021gbo}. Here, we redetermine the strange and charm valence quark masses by the physically equivalent requirements of reproducing the energy of the $\phi$ resonance (or the mass of the fictitious pseudoscalar meson $\eta_s$ determined accurately in isoQCD at the physical point in Ref.\,\cite{Borsanyi:2020mff}) and the energy of the $J/\psi$ resonance (or the mass of the pseudoscalar meson $\eta_c$). As discussed in Appendix\,\ref{sec:masses}, we find results in nice agreement, up to lattice artifacts in the charm sector, with those obtained from the $M_K$ and $M_D$ inputs in Ref.\,\cite{ExtendedTwistedMass:2021gbo}. 

Following this procedure, we are able to determine the valence quark mass parameters $m_s$ and $m_c$ using high statistics observables that are computed on the same gauge configurations and using the same stochastic sources as the vector current-current correlator $V(t)$ of Eq.~(\ref{eq:VV}). This method is very convenient for minimizing the statistical error on the time windows $a_\mu^w$. In practice, to interpolate our results to the physical strange and charm valence quark masses, we evaluate the contributions to $V(t)$ from vector correlators in the $s$ and $c$ valence sector for a few values of the bare valence quark masses $a \mu_s^{val}$ and $a \mu_c^{val}$, which will be specified later in Tables\,\ref{tab:simulated_ms} and \ref{tab:simulated_mc}.

Essential information on the ETMC ensembles relevant for this work are collected in Table\,\ref{tab:simudetails}. With respect to Refs.~\cite{ExtendedTwistedMass:2020tvp, ExtendedTwistedMass:2021qui, ExtendedTwistedMass:2021gbo} two other dedicated gauge ensembles, cB211.072.96 and cD211.054.96, have been produced for the investigation of FSEs and cutoff effects\,\cite{Finkenrath:2022eon}. The cB211.074.96 ensemble, which has a spatial lattice size $L \approx 7.6$ fm,  is used to estimate FSEs by comparing to  the smaller cB211.074.64 ensemble, while the cD211.054.96 ensemble corresponds to our finest lattice spacing $a \simeq 0.057$~fm. Note that for all ensembles  used in this work the pion mass is simulated quite close to the isoQCD reference value $M_\pi^{phys} = M_\pi^{isoQCD} = 135.0(2)$~MeV, which was also adopted in Refs.\,\cite{ExtendedTwistedMass:2021qui, ExtendedTwistedMass:2021gbo}. For the evaluation of the light-quark connected contribution, the inversions of the Dirac operator have been performed using $N_{hits} = 10^3$ spatial stochastic sources per gauge configuration. The techniques adopted for the calculation of the disconnected diagrams are briefly outlined in Section~\ref{sec:disconnected}.  
\begin{table}[htb!]
\begin{center}
    \begin{tabular}{||c||c|c|c|c|c||c|c||}
    \hline
    ~~~ ensemble ~~~ & ~~~ $\beta$ ~~~ & ~~~ $V/a^{4}$ ~~~ & ~~~ $a$ (fm) ~~~ & ~~~ $a\mu_{\ell}$ ~~~ & ~ $M_{\pi}$ (MeV) ~ & ~ $L$ (fm) ~ & $M_{\pi}L$ ~ \\
  \hline
  cB211.072.64 & $1.778$ & $64^{3}\cdot 128$ & $0.07957~(13)$ & $0.00072$ & $140.2~(0.2)$ & $5.09$ & $3.62$ \\
  
  cB211.072.96 & $1.778$ & $96^{3}\cdot 192$ & $0.07957~(13)$ & $0.00072$ & $140.1~(0.2)$ & $7.64$ & $5.43$ \\
  
  cC211.060.80 & $1.836$ & $80^{3}\cdot 160$ & $0.06821~(13)$ & $0.00060$ & $136.7~(0.2)$ & $5.46$ & $3.78$ \\
  
  cD211.054.96 & $1.900$ & $96^{3}\cdot 192$ & $0.05692~(12)$ & $0.00054$ & $140.8~(0.2)$ & $5.46$ & $3.90$ \\
  \hline
    \end{tabular}
\end{center}
\caption{\it \small Parameters of the ETMC ensembles used in this work. We give the light-quark bare mass, $a \mu_\ell = a \mu_u = a \mu_d$,  the pion mass $M_\pi$, of the lattice size $L$ and the product $M_\pi L$. The values of the lattice spacing are determined as explained in Appendix\,\ref{sec:spacing} using the 2016 PDG value $f_\pi^{phys} = f_\pi^{isoQCD} = 130.4(2)$ MeV\,\cite{ParticleDataGroup:2016lqr} of the pion decay constant for setting the scale.}
\label{tab:simudetails}
\end{table} 

Pion-mass mistuning effects, which are at most of order $5 - 6$ MeV for the ensembles listed in Table~\ref{tab:simudetails}, are relevant for the light-quark contribution to the intermediate window ($a_{\mu}^{\rm W}(\ell)$), and completely negligible within the accuracy for all the other contributions considered in this work. Indeed, $a_{\mu}^{\rm W}(\ell)$ is dominated by $\pi \pi$ and $\pi \pi \pi$ contributions, and hence particularly sensitive to variations of the light-quark mass. In order to minimize the systematic errors related to the (small) difference $M_{\pi} - M_{\pi}^{isoQCD}$, we performed additional simulations enabling to correct our lattice data for the mistuning of $M_{\pi}$. In practice we evaluated the corrections to $a_\mu^{\rm W}(\ell)$ due to the appropriate small change of $\mu_{\ell}$ in the valence and in the sea sector of the lattice action.

The former correction has been determined by performing additional inversions of the light-quark Dirac operator employing a slightly smaller value of the light bare quark mass $a\mu'_{\ell} < a\mu_{\ell}$, keeping the sea quark mass fixed to $a\mu_{\ell}$. The values of the valence light quark mass $a \mu'_{\ell}$ have been chosen, for each ensemble, according to the following relation
\be
    a\mu'_{\ell} \approx a\mu_{\ell}\left(\frac{M_{\pi}^{isoQCD}}{M_{\pi}}\right)^{2} ~ , ~
\ee
where $M_{\pi}$ is the measured value of the pion mass on any given ensemble. Since such corrections are expected to be of the order of few permille, only a limited number of stochastic sources ($\mathcal{O}(100)$) have been used for this calculation. The valence correction $\delta V_{\ell}^{val}(t)$ to the vector correlator has been then determined for both tm and OS regularizations as
\be
    \label{eq:valence_correction}
    \delta V_{\ell}^{val}(t) = V_{\ell}(t, a\mu'_{\ell}, a\mu_{\ell}) - V_{\ell}(t, a\mu_{\ell}, a\mu_{\ell}) ~ , ~
\ee 
where $V_{\ell}(t, a\mu_{\ell}, a\mu_{\ell})$ is the unitary vector correlator, while $V_{\ell}(t, a\mu'_{\ell}, a\mu_{\ell})$ is the one obtained from simulations at the valence quark mass $a \mu'_{\ell}$ and sea quark mass $a\mu_{\ell}$. In order to reduce the statistical noise, Eq.\,(\ref{eq:valence_correction}) has been evaluated using a common set of stochastic sources for both valence masses, $a\mu_{\ell}$ and $a\mu'_{\ell}$.

As for the evaluation of the corrections to $a_\mu^{\rm W}(\ell)$ coming from the sea sector, we rely on the so-called expansion method. At leading order in $\delta(a\mu_{\ell}) \equiv a\mu'_{\ell} - a\mu_{\ell}$ the correction $\delta V^{sea}_{\ell}(t)$ to the vector correlator, can be determined as
\bea
\label{eq:sea_correction}
\delta V^{sea}_{\ell}(t) & = & \frac{ \int [d\Phi] e^{ -S[\Phi, a\mu_{\ell}] - \delta(a\mu_{\ell}) \int \bar{\Psi}_{\ell}\Psi_{\ell}(x)} \mathcal{O}(t)}{ \int [d\Phi] e^{ -S[\Phi, a\mu_{\ell}] -\delta(a\mu_{\ell})\int \bar{\Psi}_{\ell}{\Psi}_{\ell}(x)}} - \frac{ \int [d\Phi] e^{ -S[\Phi, a\mu_{\ell}] } \mathcal{O}(t)}{ \int [d\Phi] e^{ -S[\Phi, a\mu_{\ell}] }} \nonumber \\[2mm]
& = & -\delta(a\mu_{\ell}) \sum_{x} \left[ \langle \bar{\Psi}_{\ell}\Psi_{\ell}(x)~\mathcal{O}(t)     \rangle - \langle \mathcal{O}(t) \rangle \langle \bar{\Psi}_{\ell}\Psi_{\ell}(x) \rangle        \right] + \mathcal{O}\left( \delta^2(a\mu_{\ell}) \right) \nonumber \\[2mm]
\mathcal{O}(t) & \equiv & \frac{a^{3}}{3}\sum_{\vec{x}}\sum_{j=1,2,3} J^{\ell\ell'}_{j, reg}(x) [ J^{\ell\ell'}_{j, reg} ]^{\dagger}(0)
\eea
where, to keep the notation simple, we collectively denote with $\Phi$ all fermionic and gluonic fields, while $S[\Phi, a\mu_{\ell}]$ corresponds to the Wilson-clover twisted mass action in Eq.\,(\ref{eq:lattMA}) with bare light-quark mass $a\mu_{\ell}$. The composite field $\mathcal{O}(t)$ is a product of two light valence quark currents (see Eqs.\,(\ref{eq:Vregdetail})-(\ref{eq:MAVcurr}) for the notation) but no sea quark fields $\Psi_\ell$, $\bar\Psi_\ell$. The expansion method has the clear advantage that no new gauge configurations have to be generated, and allows to compute $\delta V^{sea}_{\ell}(t)$ from the insertion of the (light-quark) scalar density inside the current-current correlator. All VEVs in Eq.~(\ref{eq:sea_correction}) are evaluated in the gauge background generated by $S[\Phi, a\mu_{\ell}]$. Finally, the total correction $\delta a_{\mu}^{\rm W}(\ell)$ to the light-quark contribution to the intermediate window is given by
\be
\delta a_{\mu}^{\rm W}(\ell) = 2\alpha_{em}^{2} \int_{0}^{\infty}~dt~t^{2} K(m_{\mu}t)\Theta^{\rm W}(t) \left[ \delta V_{\ell}^{val}(t) + \delta V_{\ell}^{sea}(t) \right] ~ . ~
\ee

In Table~\ref{tab:mass_corr} we show the values of the correction $\delta a_{\mu}^{\rm W}(\ell)$ (for both ``tm" and ``OS" regularizations) on the four ensembles of Table~\ref{tab:simudetails}, along with the original values of $a_{\mu}^{\rm W}(\ell)$, the simulated values of $\delta(a\mu_{\ell})$, and the values of the pion masses obtained after performing such corrections. It turns out that: ~ i) the correction $\delta a_{\mu}^{\rm W}(\ell)$ shifts upwards the intermediate window by approximately $2\sigma$ (or less) for the cB211.072.64, cB211.072.96 and cD211.054.96 ensembles, while it is completely negligible within the uncertainty for the ensemble cC211.06.80, and ~ ii) the sea quark mass correction is found to be significantly smaller than the one from the valence quark mass.  

\begin{table}[htb!]
\begin{center}
    \begin{tabular}{||c||c|c||c|c||c|c||}
    \hline
    ~~~ ensemble ~~~ & ~~~ $\delta (a\mu_{\ell})$ ~~~ & ~ $M_{\pi}$ (MeV) ~ & ~ $a_{\mu}^{\rm W}(\ell,tm)$ ~ & ~ $\delta a_{\mu}^{\rm W}(\ell, tm)$ ~ & ~ $a_{\mu}^{\rm W}(\ell, OS)$ ~ & ~ $\delta a_{\mu}^{\rm W}(\ell, OS)$ ~\\
  \hline
  cB211.072.64 & $-5.25$ & $135.2~(2)$ & $204.61~(36)$ & $0.13~~(10)$ & $203.36~(33)$ & $0.38~~(7)$ \\
  
  cB211.072.96 & $-5.25$ & $135.2~(2)$ & $205.98~(21)$ & $0.26~(11)$ & $203.18~(20)$ & $0.51~~(8)$ \\
  
  cC211.060.80 & $-1.50$ & $134.9~(3)$ & $204.88~(44)$ & $0.05~~(5)$ & $203.02~(37)$ & $0.02~~(5)$ \\
  
  cD211.054.96 & $-4.36$ & $135.1~(3)$ & $205.67~(42)$ & $0.36~~(9)$ & $203.96~(39)$ & $0.31~(10)$ \\
  \hline 
    \end{tabular}
\end{center} 
\caption{\it \small The values of the light-quark bare mass difference $\delta (a \mu_\ell )$ adopted for each ETMC gauge ensemble, given in units of $10^{-5}$. The third column contains the values of $M_\pi$ in physical units obtained after the mass correction, while the other columns show the original values for $a_{\mu}^{\rm W}(\ell)$ and the resulting $\delta a_{\mu}^{\rm W}(\ell)$, given in units of $10^{-10}$, in the two regularizations ``tm" and ``OS".}
\label{tab:mass_corr}
\end{table}

\subsection{The correlator $V(t)$ in the mixed action setup}
\label{sec:VcorrMA}

In our formulation, we find it convenient to evaluate the vector correlator $V(t)$ (see Eq.\,(\ref{eq:VV})) by employing two different regularizations for the quark connected contributions to the correlator. Moreover, as discussed in the following, using a mixed action (MA) setup, we can define renormalized correlators for each individual quark flavour ($\ell$, $s$, $c$) connected term and for the various quark flavour diagonal and off-diagonal disconnected contributions to $a_\mu^w$ (here $w={\rm SD}, {\rm W}$). This flexibility turns out to be advantageous for  extrapolating independently contributions to $a_\mu^w$  that can have a different magnitude and relative accuracy to the continuum limit.  After taking the continuum limit they are combined to yield the desired results of the unitary isoQCD theory.

The vector correlator $V(t)$ for $N_f=2+1+1$ QCD can be reconstructed, 
in the  continuum limit, by combining a number of renormalized correlators in our MA setup, 
with coefficients dictated by the em-charge of the various quark flavours. Namely
\be
    V(t)|_{MA}^{reg} = \frac{1}{3} a^3\sum_{\bf x} \sum_{i=1,3} V_{ii}(x) |_{MA}^{reg}\; , 
    \qquad \qquad x=({\bf x},t) \; , 
    \label{eq:Vrecomb}
\ee
with
\bea
    V_{ii}(x)|_{MA}^{reg} & = &
    \frac{4+1}{9} \left\langle J_{i, reg}^{\ell\ell'}(x) [J_{i, reg}^{\ell\ell'}]^\dagger(0) \right\rangle^{\!(C)} 
     \!\!\! + \frac{1}{9} \left\langle J_{i, reg}^{ss'}(x) [J_{i, reg}^{ss'}]^\dagger(0) \right\rangle^{\!(C)} \nonumber \\[2mm]
     & + & \frac{4}{9} \left\langle J_{i,reg}^{cc'}(x) [J_{i, reg}^{cc'}]^\dagger(0) \right\rangle^{\!(C)} 
    \!\!\! + \frac{4+1-2-2}{9} 
     \left\langle  J_{i, OS}^{\ell\ell}(x) [J_{i, OS}^{\ell'\ell'}]^\dagger(0) \right \rangle^{\!(D)} \nonumber \\[2mm]
    & + & \frac{1}{9} \left \langle J_{i, OS}^{ss}(x) [J_{i, OS}^{s's'}]^\dagger(0) \right \rangle^{\!(D)} 
   \!\!\! + \frac{4}{9}  \left \langle J_{i, OS}^{cc}(x) [J_{i, OS}^{c'c'}]^\dagger(0) \right \rangle^{\!(D)} \nonumber \\[2mm]
   & - & \frac{1}{9} \left \langle J_{i, OS}^{\ell \ell}(x) [J_{i, OS}^{ss}]^\dagger(0) + {\rm hc} \right \rangle^{\!(D)} 
   \!\!\! + \frac{2}{9} \left \langle  J_{i, OS}^{\ell \ell}(x) [J_{i, OS}^{cc}]^\dagger(0) + {\rm hc} \right \rangle^{\!(D)} \nonumber \\[2mm]
   & - & \frac{2}{9} \left \langle J_{i, OS}^{ss}(x) [J_{i, OS}^{cc}]^\dagger(0) + {\rm hc} \right \rangle^{\!(D)}  \, ,
\label{eq:Vregdetail}     
\eea
where $reg \in \{ {\rm tm,OS} \}$ labels the two regularizations of the single-flavour renormalized vector currents, up to variants that are equivalent in the large statistics limit. Adopting the lighter notation $q_{f,\eta}(x) \equiv f_\eta(x)$ for the single-flavour valence quark fields, the relevant renormalized vector currents read
\small
\bea
    \label{eq:MAVcurr}
    J_{\mu,tm}^{\ell \ell'} & = & Z_A \bar \ell_{1} \gamma_\mu  \ell_{2} \; , \quad 
    J_{\mu,OS}^{\ell \ell'} = Z_V \bar \ell_{1} \gamma_\mu  \ell_{3} \; , \quad 
    J_{\mu,OS}^{\ell \ell} = Z_V \bar \ell_{1} \gamma_\mu  \ell_{1} \; , 
    \quad
    J_{\mu,OS}^{\ell' \ell'} = Z_V \bar \ell_{3} \gamma_\mu  \ell_{3} \; , ~ 
    \nonumber \\[2mm] 
    J_{\mu,tm}^{ss'} & = & Z_A \bar s_{1} \gamma_\mu  s_{2} \; , \quad 
    J_{\mu,OS}^{ss'} = Z_V \bar s_{1} \gamma_\mu  s_{3} \; , \quad 
    J_{\mu,OS}^{ss} = Z_V \bar s_{1} \gamma_\mu  s_{1} \; , \quad
    J_{\mu,OS}^{s's'} = Z_V \bar s_{3} \gamma_\mu  s_{3} \; , \nonumber \\[2mm] 
    J_{\mu,tm}^{cc'} & = & Z_A \bar c_{1} \gamma_\mu  c_{2} \; , \quad 
    J_{\mu,OS}^{cc'} = Z_V \bar c_{1} \gamma_\mu  c_{3} \; , \quad 
    J_{\mu,OS}^{cc} = Z_V \bar c_{1} \gamma_\mu  c_{1} \; , \quad
    J_{\mu,OS}^{c'c'} = Z_V \bar c_{3} \gamma_\mu  c_{3} \; ~ 
\eea
\normalsize
where $Z_A$ or $Z_V$ are the appropriate ultraviolet (UV) finite RCs for the bare local vector currents in Eq.~(\ref{eq:MAVcurr}). We recall that $\ell =u=d$, as in our lattice QCD setup $u$ and $d$ quarks are mass degenerate. For all valence quark flavours $f=\ell,s,c$, we have $r_{f,1}=-r_{f,2}=r_{f,3}=...=1$. The suffixes on the currents (e.g.\ $s s'$,
$ss$ or $s's'$) just remind whether the quark and antiquark field entering in each current belong to different (e.g. $s s'$) or equal (e.g. $ss$ or $s's'$)
valence fermion replica, independently of the Wilson's $r$-values which are specified by the index $\eta=1,2,3.$
The suffix $(C)$ or $(D)$ attached to the correlators contributing to $V_{ii}(x)|_{MA}^{reg}$ in Eq.~(\ref{eq:Vrecomb}) indicates whether these correlators give rise to quark connected or disconnected Wick contractions.

As customary when working with twisted mass lattice fermions, we say that a vector current, e.g.\ $J_{\mu,reg}^{ff'}$, is written in the $tm$ or $OS$ regularization if the two valence quarks, $f$ and $f'$, entering the current appear in the valence quark action $S_{\rm q, val}$ of  Eq.~(\ref{eq:Sval}) with $r_{f'} = -r_f$ or $r_{f'} = r_f$, respectively. As one can see from the examples in Eq.~(\ref{eq:MAVcurr}), this implies that the OS regularization is the unique possible choice for the currents entering in the fermion disconnected correlators. Numerically, they give much smaller contributions to $a_\mu^w$ than the quark connected correlators, for which instead the two lattice discretizations are available. 

We outline here the main steps of the proof showing that one can extract  physical information on the correlator $V(t)$ in QCD from the correlators $V(t)|_{MA}^{tm,OS} $, following the logic that was adopted in Ref.~\cite{Frezzotti:2004wz} 
for correlators relevant to other physical observables.
Let us start from the UV finite RCs of the currents appearing in Eq.~(\ref{eq:MAVcurr}). 

For the currents involving two {\em different} valence-replica quark fields (i.e.\ $J_{\mu,reg}^{\ell \ell'}$, $J_{\mu,reg}^{s s'}$, $J_{\mu,reg}^{c c'}$), which enter in the quark connected correlators, one easily checks that the appropriate RC is $Z_A$ or $Z_V$ for $reg=tm$ or $reg=OS$, respectively. Indeed, since the RC are named following the standard notation for untwisted Wilson lattice fermions, this result is easily obtained by rewriting for the two considered regularizations ($reg$) the current operator in the valence quark basis where the Wilson term appears untwisted, v.i.z.\
\begin{align}
    S_{\rm  q,\; val} = a^4\sum_x \sum_{f,\eta}  \bar \chi_{f,\eta}(x) \, [ \, \gamma \cdot \tilde\nabla + {\rm sign}(r_{f,\eta}) i\gamma_5 m_f +  W_{\rm cr}^{\rm cl}|_{r=1} \, ] \, \chi_{f,\eta}(x) \, . \label{eq:Sval_Wbasis}
\end{align}
Comparing with the form of $S_{\rm  q,\; val}$ in Eq.~(\ref{eq:Sval}), one sees that the relation between the two valence quark field bases reads
\begin{align}
f_\eta \equiv q_{f,\eta} = e^{ i \frac{\pi}{4} \gamma_5 {\rm sign}(r_{f,\eta}) } \chi_{f,\eta} \; , \qquad \bar f_\eta \equiv  \bar q_{f,\eta} = 
\bar\chi_{f,\eta} e^{i \frac{\pi}{4} \gamma_5 {\rm sign}(r_{f,\eta})}  \; .
\end{align}
Taking as an example the currents $J_{\mu,tm}^{s s'}$ and $J_{\mu,OS}^{s s'}$, it follows that since
\begin{align}
     \bar s_{1} \gamma_\mu  s_{2} = \bar \chi_{s,1} \gamma_5 \gamma_\mu \chi_{s,2} \; ,
     \qquad
     \bar s_{1} \gamma_\mu  s_{3} = \bar \chi_{s,1} \gamma_\mu \chi_{s,3} \; ,
\label{eq:barcurex} \end{align}
the first and second bare currents in Eq.(\ref{eq:barcurex}) are renormalized with $Z_A$ 
and $Z_V$, respectively. The same argument  holds for all the other currents having two different valence-replica quark fields in Eq.~(\ref{eq:MAVcurr}), with the relative sign of the Wilson $r$-parameters of the valence quark and antiquark determining whether the proper RC is $Z_V$ or $Z_A$\footnote{For completeness we recall that such scale independent RCs are needed in lattice regularization that break chiral symmetries in order to have the valence quark currents normalized consistently with the chiral Ward identities (WI) of QCD
~\cite{Bochicchio:1985xa, Bhattacharya:2005rb}.}.
As for the vector currents involving two {\em equal} valence-replica quark fields, i.e.\ $J_{\mu,OS}^{\ell \ell}$, $J_{\mu,OS}^{s s}$, $J_{\mu,OS}^{c c}$, $J_{\mu,OS}^{\ell' \ell'}$, $J_{\mu,OS}^{s' s'}$, $J_{\mu,OS}^{c' c'}$, which enter in the quark disconnected correlators, it is easy to check that their form is unchanged upon rewriting them in the quark basis where the valence fermion action takes the form given in Eq.~(\ref{eq:Sval_Wbasis}). 
Thus, the problem is reduced to determining the renormalization pattern for a single flavour vector current in untwisted Wilson lattice QCD.  
Taking, for instance, the case of the charm quark flavour (the argument is unchanged for $s$, $d$ and $u$), what we are after is the relation between the chiral covariantly renormalized current, say $[\bar q_c \gamma_\mu q_c]_R$, and the bare current $\bar q_c \gamma_\mu q_c$. In Appendix\,\ref{sec:appD} we show that for standard Wilson fermions the flavour singlet and non-singlet vector current RCs actually coincide, i.e.\ $Z_{V^0} = Z_V$, from which it follows that the current $\bar q_c \gamma_\mu q_c $ is only {\em multiplicatively renormalized} and $[\bar q_c \gamma_\mu q_c]_R = Z_V \bar q_c \gamma_\mu q_c  $. As a consequence, all the vector currents that involve two equal valence-replica quark fields in our mixed action setup are also multiplicatively renormalized through $Z_V$.

An important result is obtained exploiting renormalizability of our mixed action setup~(\ref{eq:lattMA}) and {\em universality}, provided~\cite{Frezzotti:2004wz}
\begin{itemize}
\item[i)] a suitable renormalization condition (e.g.\ the value of $f_\pi$ in isoQCD) is imposed as $g_0^2 \to 0$, 
\item[ii)] the bare soft mass parameters are matched so as to work with equal {\em sea} and {\em valence} renormalized quark masses for each flavour (see below for more details), and
\item[iii)] all the current-current correlators appearing in Eq.~(\ref{eq:Vrecomb}) are normalized consistently with the chiral WI of $N_f=2+1+1$ QCD.
\end{itemize}
Namely, the correlator $V(t)|^{tm,OS}_{MA}$ in Eq.(\ref{eq:Vrecomb}) admits a continuum limit that coincides with the one of the formally identical correlator, $V(t)|^{GW}_{MA}$, evaluated at equal renormalized quark masses in a chiral-symmetric lattice regularization, e.g. defined using Ginsparg-Wilson {\em sea and valence} quarks, whence the label $GW$, of the same mixed $N_f=2+1+1$ QCD action. 
Moreover, at the given renormalized quark masses, the continuum limit of $V(t)|^{GW}_{MA}$ in the chiral-symmetric lattice regularization  is identical to the continuum limit of $V(t)|^{GW}$, i.e.\ the correlator of Eq.~(\ref{eq:VV}) in the unitary
$N_f=2+1+1$ QCD setup.\,\footnote{This identity is easily checked by noting that the correlators  $V(t)|^{GW}_{MA}$ and  $V(t)|^{GW}$, being defined in the same UV-regularization for all types of quark fields and evaluated at equal renormalized masses, give rise to identical Wick contractions at finite lattice spacing.} The latter is precisely the quantity of interest for extracting $a_\mu^{\rm SD, W}$, as discussed in Section~\ref{sec:definitions}.
This concludes our proof. 

A few remarks are in order about the way of tuning  the  bare soft mass parameters in order to work with equal {\em sea} and {\em valence} renormalized quark masses for each physical flavour in the mixed action setup~(\ref{eq:lattMA}). Based on the results of Ref.~\cite{Frezzotti:2004wz} a simple way of doing so consists in matching the sea and valence bare mass parameters according to
\begin{align}
    m_\ell ~=~ \mu_u = \mu_d   \, , 
    \qquad  m_s ~=~ \mu_\sigma - \frac{Z_S}{Z_P} \mu_\delta \, 
    \qquad  m_c ~=~ \mu_\sigma + \frac{Z_S}{Z_P} \mu_\delta \; , \label{eq:massmatch}
\end{align}
where $Z_P$ ($Z_S$) is the RC of the pseudoscalar (scalar) flavour non-singlet quark bilinear density, and fixing the values of $\mu_u = \mu_d$, $\mu_s$ and $\mu_c$ in order to reproduce the ``physical" values of three observables (sensitive to the light, strange and charm quark masses) in isoQCD.
Of course the definition of such ``physical" values in isoQCD is conventional, since in the physical world SU(2) isospin symmetry is only approximate, but any arbitrariness induced by the conventional definition of an isoQCD world can be removed by evaluating the corresponding QED and strong IB corrections\footnote{As mentioned in Sections~\ref{sec:introduction} and \ref{sec:comparison}, the IB correction to $a_\mu^{\rm SD}$ is negligible, while the one to $a_\mu^{\rm W}$ is estimated using the BMW results of Ref.\,\cite{Borsanyi:2020mff}.} (for a review see, e.g., Ref.~\cite{FlavourLatticeAveragingGroupFLAG:2021npn}).

In practice we tune $\mu_\sigma$ and $\mu_\delta$ by matching them to their valence counterparts $m_s$ and $m_c$ as discussed in Ref.\,\cite{Alexandrou:2018egz}. The values of the valence quark masses $m_{s}$ and $m_{c}$ are in turn fixed in order to reproduce the phenomenologically well known values of $M_{D_s}/f_{D_s}$ and $m_c/m_s$, as discussed above in introducing the mixed action lattice setup~(\ref{eq:lattMA}). Such a tuning step could be performed with an accuracy of few percents for all lattice resolutions in the early stages of the simulation effort without affecting significantly the uncertainty of the final results owing to the very mild dependence of $a_\mu^{\rm SD}$ and $a_\mu^{\rm W}$ on the strange and charm {\em sea} quark masses.

%

Aiming at a few permille determination of the window contributions, it is crucial to have a high precision determination of the RCs $Z_{A}$ and $Z_{V}$ as well as of the values of the lattice spacing. While the accurate evaluation of $Z_{A}$ and $Z_{V}$ will be discussed in Sec.\,\ref{sec:renormalization}, we address now a significant improvement of the determination of the lattice spacing (the result of which is given in Table\,\ref{tab:simudetails}) with respect to the results obtained in Ref.\,\cite{ExtendedTwistedMass:2021qui}.

\subsection{Improved determination of the lattice spacing}
\label{sec:spacing}

In order to reduce the uncertainties on the lattice spacing as compared to the results obtained in Ref.\,\cite{ExtendedTwistedMass:2021qui}, we take advantage of the following improvements: ~ i) pseudoscalar observables are now available with substantial higher accuracy thanks to a huge number of stochastic sources ($N_{source} \simeq 10^3$) per gauge configuration; ~ ii) two new ensembles at the physical point, namely cB211.072.96 and cD211.054.96 are included in the analysis; ~ iii) a significant increase in the number of independent gauge configurations analyzed for the two physical mass point ensembles the cB211.072.64 and cC211.060.80 as compared to what we used in Ref.\,\cite{ExtendedTwistedMass:2021qui}; ~ iv) the inclusion of four ensembles at a coarser lattice spacing of $a \simeq 0.091~fm$, namely cA211.53.24, cA211.40.24, cA211.30.32 and cA211.12.48, taken from Ref.~\cite{ExtendedTwistedMass:2021qui}; they will be referred to as A-type ensembles. The simulated pion masses are in the range $170 - 350$ MeV\,\cite{ExtendedTwistedMass:2021qui} and, therefore, they are not close to the physical pion point. Consequently, in the analyses of the light- and strange-quark contributions to the window observables they would require a significant extrapolation in the pion mass. Nevertheless, they can be useful for the analysis of the charm contribution to the window observables, since the latter ones have a very slight dependence upon the pion mass; ~ v) the inclusion of four B-type ensembles at $a \simeq 0.080$~fm, namely cB211.25.24, cB211.25.32, cB211.25.48 and cB211.14.64 from Ref.~\cite{ExtendedTwistedMass:2021qui}, for getting control over FSEs and pion-mass dependence.

Following Ref.\,\cite{ExtendedTwistedMass:2021qui} the analysis is performed using as input the dimensionless variable 
\be
    \label{eq:xipi}
    \xi_\pi = \frac{M_\pi^2}{16\pi^2 f_\pi^2}
\ee
corrected for finite size effects (FSEs) using the resummed formulae from Ref.\,\cite{Colangelo:2005gd}. Then, we fit the pion decay constant in lattice units, $a f_\pi$, determined on the A- and B-type (second largest lattice spacing) ensembles, using the following Ansatz inspired by chiral perturbation theory (ChPT)
\bea
    \label{eq:afpi_fit}
    af_\pi^j(\xi_\pi, L ) = af_\pi^j(\xi_\pi^{phys}, \infty) & \cdot & 
        \left\{ 1 - 2\xi_\pi \mbox{log}(\xi_\pi / \xi_\pi^{phys}) + \left[ P + P_{disc} (af_\pi^j)^2 \right] \cdot (\xi_\pi - \xi_\pi^{phys}) \right\} \nonumber \\[2mm] 
    & \cdot & \left\{ 1 + P_{FSE} \xi_\pi \frac{e^{-M_\pi L}}{(M_\pi L)^{3/2}}              \right\} ~ , ~ 
\eea
where $j = A, B$ and $af_\pi^A$, $af_\pi^B, P, P_{disc}, P_{FSE}$ are free fitting parameters. We use a total of 10 ensembles\footnote{For the ensemble cA211.12.48 we correct the value of $af_{\pi}$ accounting for violation from maximal twist condition following Ref.\,\cite{ExtendedTwistedMass:2021qui}.}. 

Using the values obtained for $P, P_{disc}$ and $P_{FSE}$ we can correct the lattice data of $a f_\pi$ for the mistuning in $\xi_\pi$ and for FSEs on all the ETMC ensembles, i.e.~also on the $C$ and $D$ ensembles. After applying such corrections, which are small on the physical point $B$, $C$ and $D$ ensembles, the lattice spacing is determined from
\be
    a^X = af_\pi^X(\xi_\pi^{phys}, \infty)/ f_\pi^{phys} ~ , ~ 
          \qquad X = A,B,C,D ~ , ~  
\ee
where $f_\pi^{phys} = f_\pi^{isoQCD} = 130.4~(2)$ MeV\,\cite{ParticleDataGroup:2016lqr} (as in Ref.\,\cite{ExtendedTwistedMass:2021qui}). As no use is made here of the gradient flow quantity $w_0/a$, the above scale setting procedure is equivalent to that of Ref.\,\cite{ExtendedTwistedMass:2021qui} only up to relative O($a^2$) effects on the lattice spacing.

\begin{figure}[htb!]
    \centering
    \includegraphics[scale=0.6]{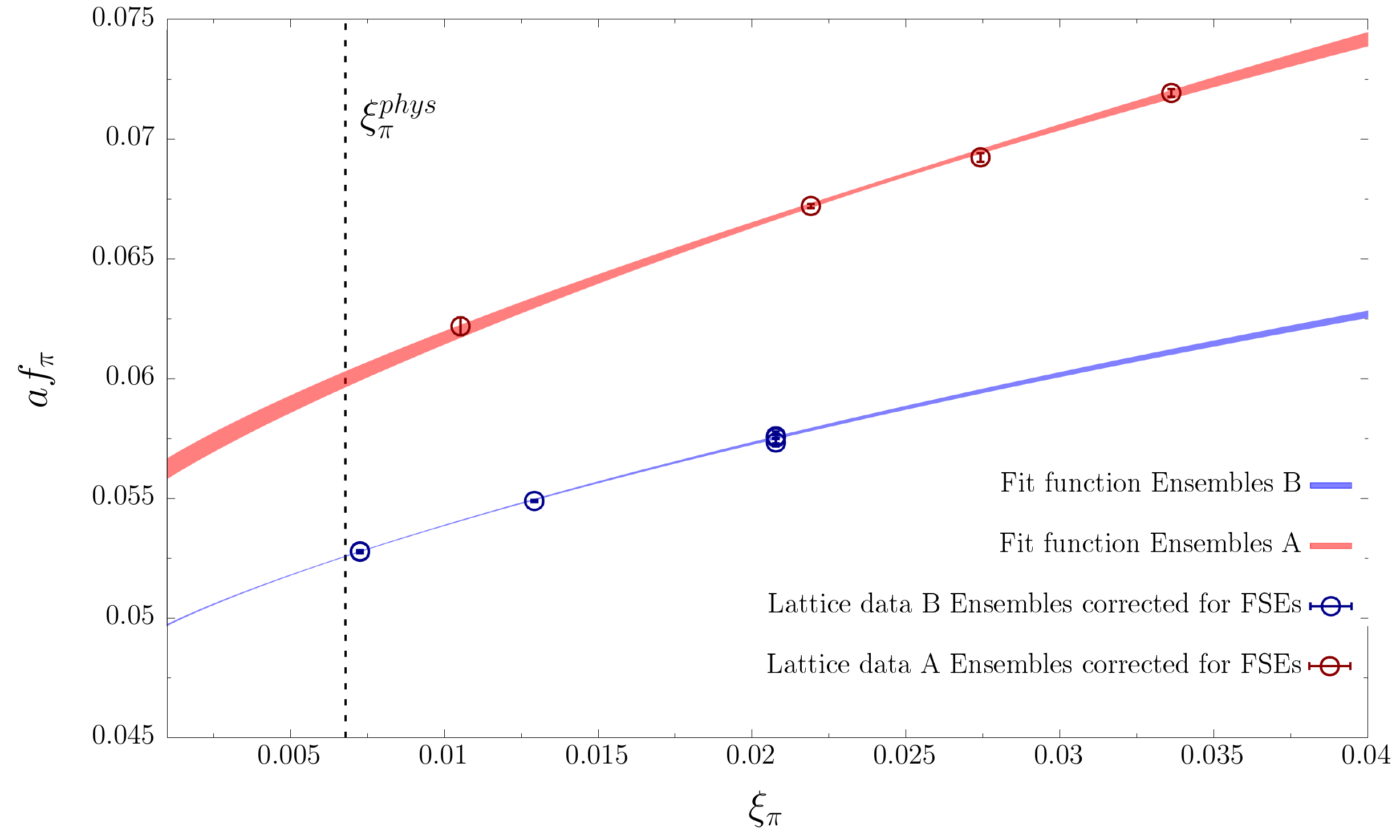}
    \caption{\it \small The pion decay constant in lattice units, $a f_\pi$, as determined on the A and B ensembles, versus the dimensionless variable $\xi_\pi$ given in Eq.\,(\ref{eq:xipi}). The smaller markers represent the results obtained using the combined fit\,(\ref{eq:afpi_fit}).
    Both data points and fit curves are shown after removing FSE using the result for $P_{FSE}$. The vertical dotted line corresponds to the physical value $\xi_\pi = \xi_\pi^{phys} = \xi_\pi^{isoQCD} \simeq 0.0068$.}
    \label{fig:afpi}
\end{figure}

The reduced $\chi^2$ of the fit based on the Ansatz~(\ref{eq:afpi_fit}) is $\chi^2 / {\rm d.o.f.} \simeq 1.6$, with 10 measurements and 5 parameters. Fit stability is checked by including or excluding the cA211.12.48 ensemble, and by including or excluding all the $A$-type ensembles. In the latter case we set $P_{disc} = 0$. 
The quality of the fitting procedure is illustrated in Fig.\,\ref{fig:afpi}. The values of the lattice spacing for the various ETMC ensembles are collected in Table\,\ref{tab:spacing} and compared with the ones from Ref.\,\cite{ExtendedTwistedMass:2021qui}.
\begin{table}[htb!]
\centering
\begin{tabular}{|c||c|c|}
\hline
~ Ensembles ~ & ~ a (fm) ~ [this work] ~  & ~ a (fm) ~ [from  Ref.\,\cite{ExtendedTwistedMass:2021qui}] ~ \\ \hline
A & $0.09076~(54)$   & $0.09471~(39)$ \\ \hline
B & $0.07957~(13)$   & $0.08161~(30)$ \\ \hline
C & $0.06821~(13)$   & $0.06942~(26)$ \\ \hline
D & $0.05692~(12)$   & $0.05770~(20)$ \\ \hline

\end{tabular}
\caption{\it \small In the first column we give the  ensemble type according to its lattice spacing, in the second column we give the updated values of the lattice spacing $a$ and the third column our previous determination using a smaller set of ensembles and statistics~\cite{ExtendedTwistedMass:2021qui}.}
\label{tab:spacing}
\end{table}
 
\begin{figure}[htb!]
\centering
\includegraphics[scale=0.6]{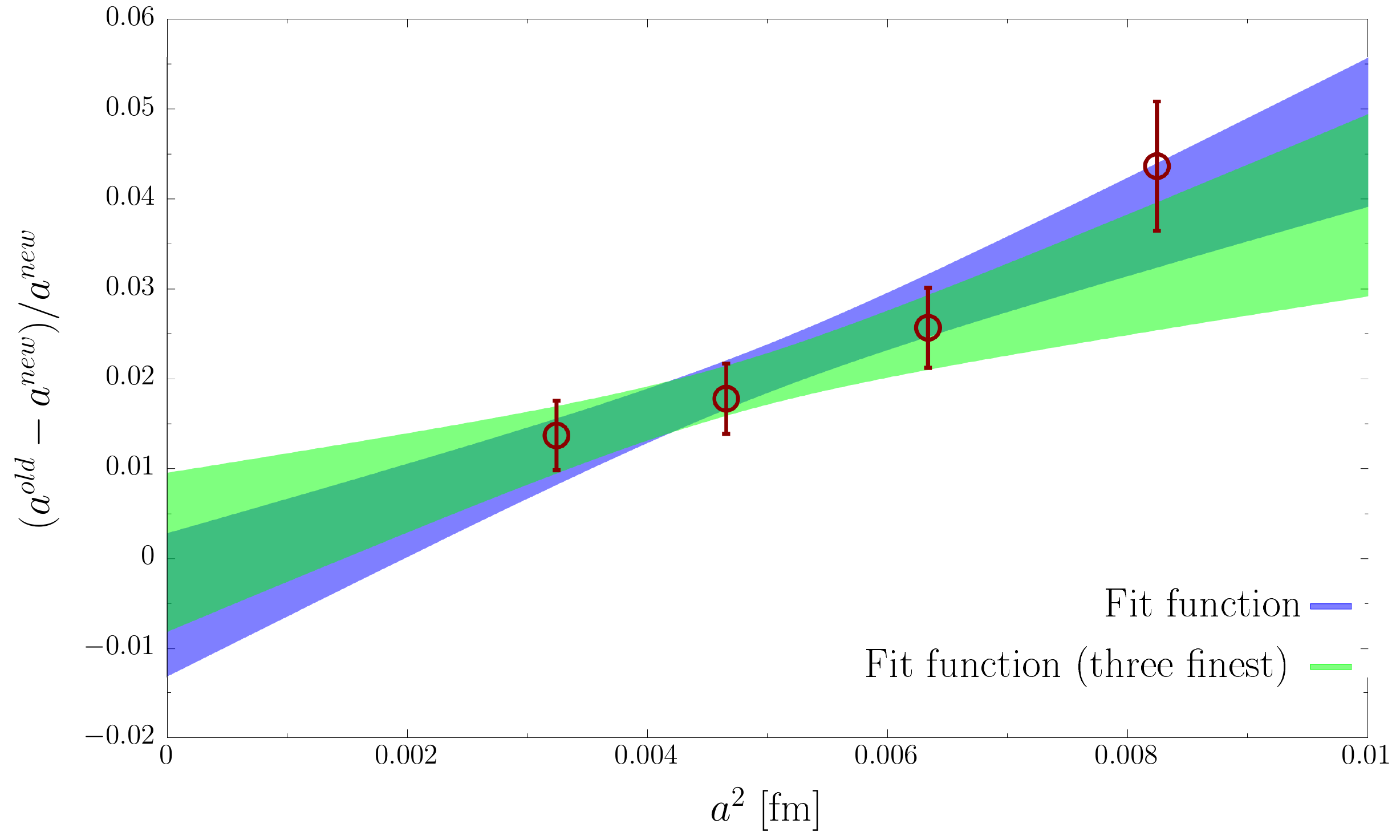}
\caption{\it \small $a^2$-scaling behaviour of the relative difference between the two determinations of the lattice spacing given in Table\,\ref{tab:spacing}. The violet and green bands correspond to a linear fit in $a^2$ applied respectively to all the four data points and to the three finest ones only. The width of the bands represents one standard deviation.}
\label{fig:a2scaling}
\end{figure}

It can be seen that, except for the A-type ensembles, the updated values of the lattice spacing are more precise than those obtained in Ref.\,\cite{ExtendedTwistedMass:2021qui} by a factor of $\simeq 2$. We reach a precision better than $\simeq 0.2 \%$ for the ensembles B, C and D, while for the A ensembles the relative uncertainty of $a$ is equal to $\simeq 0.6 \%$. The reason why the accuracy of the lattice determination for the A-type ensembles is not at the same level is that, unlike for the rest of the ensembles, we do not have simulations very close to the physical pion point. This means that we have to extrapolate from larger values of $\xi_\pi$ to reach the physical value increasing the error, as demonstrated in Fig.~\ref{fig:afpi}.
In Fig.\,\ref{fig:a2scaling}, we plot the relative difference between the two determinations of the lattice spacing that exhibits a nice $a^2$-scaling behaviour.

In this work we use the values of the lattice spacing given in the central column of Table\,\ref{tab:spacing} (see also Table\,\ref{tab:simudetails}).

Finally, a relevant question which we want to address, is the sensitivity of the short and intermediate windows to the uncertainty of the lattice spacing, focusing on the (most relevant) case of the light-quark contribution. While the windows are dimensionless quantities, the lattice spacing enters their calculation explicitly through the fact that the two dimensionful quantities entering the leptonic kernel, the muon mass $m_{\mu}$ and the window parameter $\Delta$, must be converted in lattice units. The impact that the relative uncertainty $\Delta a/a$ on the lattice spacing produces in $a_{\mu}^{w}(\ell)$ (w=SD, W), is given by
\be
\frac{ \Delta a_{\mu}^{w}(\ell)}{a_{\mu}^{w}(\ell)} = \bigg| \frac{ \partial \log{ (a_{\mu}^{w}(\ell))   }}{\partial\log{(a)}}\bigg|\cdot \frac{ \Delta a}{a} \equiv C^{w}\cdot  \frac{ \Delta a}{a} ~ . ~
\ee
The coefficient $C^{w}$ can be computed numerically from the knowledge of the lattice vector correlator, and from the derivative of the integration kernel w.r.t.~the lattice spacing $a$. In the case of the full HVP this coefficient turns out to be around $1.8$, as already pointed out in Ref.~\cite{DellaMorte:2017dyu}. For the short and intermediate windows, we find instead $C^{\rm SD} \sim 0.1$ and $C^{\rm W} \sim 0.4$. The short-distance window is therefore largely insensitive to the uncertainty on the scale setting, while for the intermediate window the impact is a factor of four smaller than that for the full HVP. 


\section{Hadronic determination of $Z_{V}$ and $Z_{A}$}
\label{sec:renormalization}

In order to reach a high precision determination of the two scale-invariant RCs $Z_{V}$ and $Z_{A}$ we employ a hadronic method based on the WI combined with a high statistics determination of the relevant suitable correlators. This allows us to obtain an accuracy of $\simeq 0.03 - 0.10 \%$ for $Z_A$ and of $\simeq 0.001 \%$ for $Z_V$, thus reaching the desired accuracy. We collect the  values of $Z_{A}$ and $Z_{V}$ used in this work for each of the ETMC ensembles of Table\,\ref{tab:simudetails}  in Table\,\ref{tab:RCs}.

\begin{table}[htb!]
    \centering
    \begin{tabular}{||c||c|c||}
    \hline
    ensemble & $Z_{V}$ & $Z_{A}$ \\
    \hline
    ~ cB211.072.64 ~ & ~ $0.706379~(24)$ ~ & ~ $0.74294~(24)$ ~ \\
    \hline
    ~ cB211.072.96 ~ & ~ $0.706405~(17)$ ~ & ~ $0.74267~(17)$ ~ \\
    \hline
    ~ cC211.060.80 ~ & ~ $0.725404~(19)$ ~ & ~ $0.75830~(16)$ ~  \\
    \hline
    ~ cD211.054.96 ~ & ~ $0.744108~(12)$ ~ & ~ $0.77395~(12)$ ~ \\
    \hline
    \end{tabular}
    \caption{\it \small The values of $Z_{V}$ and $Z_{A}$ used in this work for each of the ETMC ensembles of Table\,\ref{tab:simudetails}, determined by employing the WI-based hadronic method described in the next subsections.}
    \label{tab:RCs} 
\end{table}


We proceed to illustrate in detail the method, based on WI and universality, that enables us to obtain the two RCs $Z_{V}$ and $Z_{A}$ with very high precision. The derivation relies on two main ingredients, namely an exact conserved current relation holding in the Wilson twisted-mass regularization and the fact that the critical Wilson term is a truly dimension five irrelevant operator.

\subsection{ Case of $Z_{V}$}

In order to discuss the evaluation of the RC of the flavour non-singlet vector and
axial currents, in the context of the mixed action setup for $N_f=2+1+1$ LQCD
described in Sect.~III it is convenient (and enough) to focus on the Lagrangian
for just two different {\em valence} quark flavours, which can be taken with a 
common soft mass parameter, say $\mu_F$, and are denoted by $F$ and $F^\prime$.

Let us start by considering the case where the two distinct valence quark flavors
are taken at maximal twist with opposite $r$-Wilson parameters, i.e.\ $r_{F} = -
r_{F^\prime} =1$. In order to lighten notation and ease a number of algebraic steps in the 
following let us collect the two valence quark field in a two-flavour valence field, $\psi_{val,-}$,  i.e.\
\begin{align}
    \psi_{val,-} \equiv (q_F, q_{F^\prime}) \; , \qquad 
    \bar \psi_{val,-} \equiv (\bar q_F , \bar q_{F^\prime}) \; ,
\end{align}
with the suffix ``$-$'' reminding of $r_F r_{F^\prime} <0$. The $FF^\prime$ sector of
the valence quark action~(\ref{eq:Sval}) reads 
\begin{align}
S_{\rm q, val} \; \supset \; S_{\rm q, val}^{F ^\prime} = a^4\sum_x  \bar \psi_{val,-} (x) \, [ \, \gamma \cdot \tilde\nabla + \mu_F - i\gamma_5 \tau^3  W_{\rm cr}^{\rm cl}|_{r=1} \, ] \,
\psi_{val,-} (x) \, , \label{eq:SvalFF'} \; ,
\end{align}
with the $\tau^3$ matrix acting in the $FF^\prime$ flavor space.
%
%
Since RC are named after the quark basis where the Wilson term is untwisted, it is useful
to write also in this basis,
\begin{align}
    \chi_{val,-} \equiv e^{-i\pi\gamma_{5}\tau^{3}/4}\psi_{val,-} , \qquad
    \bar{\chi}_{val,-} \equiv \bar{\psi}_{val,-}  e^{-i\pi\gamma_{5}\tau^{3}/4}
\end{align}
the $FF^\prime$ sector of the valence quark action (see Eq.~(\ref{eq:SvalFF'})), namely
\begin{align}
    S_{\rm q, val}^{F F^\prime} = a^4\sum_x  \bar \chi_{val,-} (x) \, [ \, \gamma \cdot \tilde\nabla + i \gamma_5 \tau^3 \mu_F + W_{\rm cr}^{\rm cl}|_{r=1} \, ] \, \chi_{val,-} (x) \, , . \label{eq:SvalFF'chi} \; 
\end{align}
The expression of axial and vector currents, as well as scalar and pseudoscalar densities,
is different in the two different bases, while the physical meaning of these operators is manifest in the $\psi$-basis where the quark mass term ($\propto \mu_F$) takes its canonical form. In that basis, also referred to as the ``physical'' one, we use here for the operators symbols in calligraphic style and write
\begin{align}
  \mathcal{A}_\mu^{a} &\equiv \bar{\psi}_{val,-}\gamma_\mu\gamma_5\frac{\tau^a}{2}\psi_{val,-}
  =\;
  \begin{cases}
    \epsilon^{3ab}V_\mu^b = 
    \epsilon^{3ab} \bar\chi_{val,-} \gamma_\mu \frac{\tau^b}{2} \chi_{val,-}
     & \text{$(a=1,2)$},\\
    A_\mu^3 = \bar\chi_{val,-} \gamma_\mu\gamma_5 \frac{\tau^3}{2} \chi_{val,-} 
    & \text{$(a=3)$},
  \end{cases} \label{eq:axial_current_rot}\\
  \mathcal{V}_\mu^a  &\equiv \bar{\psi}_{val, -} \gamma_\mu\frac{\tau^a}{2}\psi_{val, -} \,\,\,\,\,=\;
  \begin{cases}
    \epsilon^{3ab}A_\mu^b = 
    \epsilon^{3ab} \bar\chi_{val,-} \gamma_\mu\gamma_5 \frac{\tau^b}{2} \chi_{val,-}
     & \text{$(a=1,2)$},\\
    V_\mu^3 =   \bar\chi_{val,-} \gamma_\mu \frac{\tau^3}{2} \chi_{val,-}
    & \text{$(a=3)$},
  \end{cases} \label{eq:vector_current_rot} \\
  \mathcal{P}^a &\equiv \bar{\psi}_{val, -} \gamma_5\frac{\tau^a}{2}\psi_{val, -} \,\,\,\,\,=\;
  \begin{cases}
     P^a = \bar{\chi}_{val, -} \gamma_5\frac{\tau^a}{2}\chi_{val, -}
     & \text{$(a=1,2)$},\label{eq:axial_density_rot}\\
    i\frac12 S^0 = i\frac12 \bar{\chi}_{val, -} \chi_{val, -} & \text{$(a=3)$},
  \end{cases} \\
   \mathcal{S}^0 &\equiv \bar{\psi}_{val,-} \psi_{val,-} = 2iP^3 =
   2i\bar{\chi}_{val, -} \gamma_5\frac{\tau^3}{2}\chi_{val, -}  ,
   \label{eq:scalar_density_rot}    
\end{align}

\begin{figure}[htb!]
\begin{center}
\includegraphics[scale=0.85]{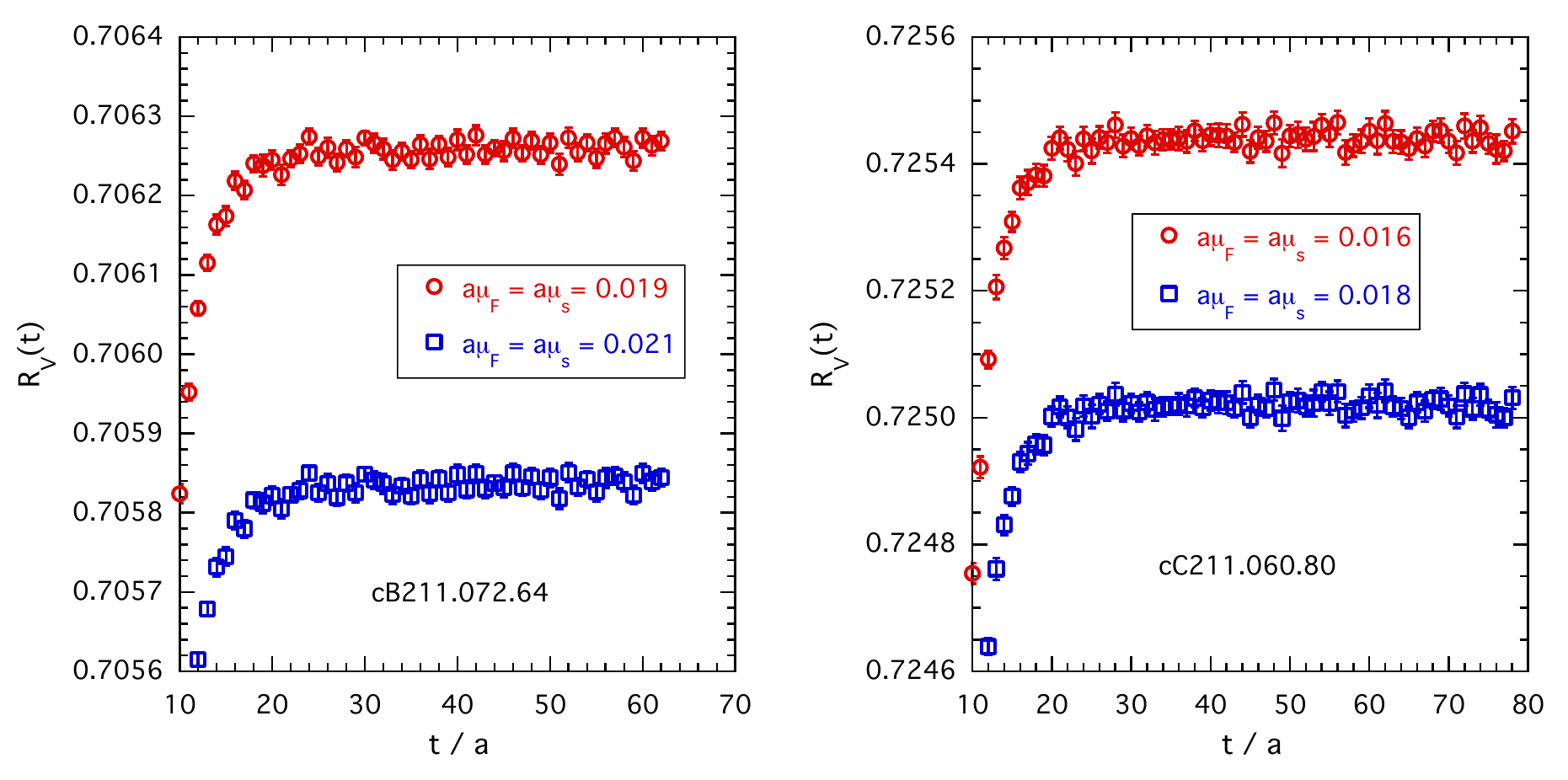}
\vspace{-0.5cm}
\caption{\it \small Time behavior of the estimator $R_V(t)$, given by Eq.\,(\ref{eq:est_RV}), for the cB211.072.64 (left panel) and cC211.060.80 (right panel) ensembles. The red circles and the blue squares correspond to two simulated values of $\mu_F$ shown in the inset.}
\label{fig:RV}
\end{center}
\end{figure}

Owing to the exact flavour symmetry of Wilson fermions for massless quarks ($\mu_F = 0$), the WI
\begin{align}
\partial_\mu^* \langle \widetilde{V}^a_\mu(x) O(0) \rangle= -2 \mu_F \epsilon^{3ab}
\langle P^b(x) O(0)\rangle~,
\label{eq:PCVC}
\end{align}
where $\partial^{*}_{\mu}$ is the lattice backward derivative,
holds true exactly at finite lattice spacing. In Eq.~(\ref{eq:PCVC})     $\widetilde{V}_{\mu}$ is the exactly conserved point-split lattice current,
\bea
    \widetilde{V}^a_\mu(x) & = &  
 \frac12 \Big\{\bar{\chi}_{val,-}(x)(\gamma_\mu\!-\!1)\frac{\tau^a}{2}
 U_\mu(x)\chi_{val,-}(x+a\hat\mu) \nonumber \\
 & + & \bar{\chi}_{val,-}(x+a\hat\mu)(\gamma_\mu\!+\!1)\frac{\tau^a}{2} 
 U_\mu^{-1}(x)\chi_{val,-}(x)\Big\}~.   
\eea
The lattice WI of Eq.~(\ref{eq:PCVC}) can be used to determine the finite RC $Z_{V}$ of the point-like current $V_{\mu}^{a},~a=1,2$, i.e. of the axial current $\mathcal{A}_{\mu}^{a},~ a=1,2$ in the ``physical'' $\psi$-basis. Making use of Eq.~(\ref{eq:PCVC}) and of the transformation law of Eqs.~(\ref{eq:axial_current_rot})-(\ref{eq:axial_density_rot}), it is easy to show that $Z_{V}$ can be extracted using 
\begin{align}
Z_{V} = \lim_{\mu_F\to 0} 2\mu_F\frac{ \sum_{x}\langle \mathcal{P}^{1}(x)\mathcal{P}^{1}(0)\rangle}{\sum_{x}\tilde{\partial}_{\mu}\langle\mathcal{A}_{\mu}^{1}(x)\mathcal{P}^{1}(0)\rangle}~, 
\end{align}
where $\tilde{\partial}_{\mu}$ is the lattice symmetric derivative and operators are written in the physical basis, see Eqs.~(\ref{eq:axial_current_rot}) and~(\ref{eq:axial_density_rot}). Moreover, the limit $\mu_F \to 0$, it is strictly speaking unnecessary, since the difference $Z_{V}(\mu_F)- Z_{V}(0)$ amounts only to lattice artifacts of order $\mathcal{O}( a^2\mu_F^2)$ or $\mathcal{O}( a^2\mu_F\Lambda_{QCD})$ . Any choice of $\mu_F$, which of course must be set to the same value in physical units for all ensembles, is legitimate, and we can use this freedom to evaluate $Z_{V}$ at a convenient value of $\mu_F$.  
According to the discussion above, for each ensemble we extract $Z_{V}$ from the large time behavior, $t/a \gg 1$, of the following estimator\footnote{The value of the estimator at small times are affected by relatively larger lattice artifacts, while at large times no significant deterioration of the signal-to-noise ratio is expected as the one-pion state dominates the correlators.}
\begin{align}
\label{eq:est_RV}
R_{V}(t)  \equiv 2\mu_F \frac{ C_{PP}^{\rm{tm}}(t)}{\tilde{\partial}_{t} C_{VP}^{\rm{tm}}(t)}~, \end{align}
where (correlators are named here after the unphysical $\chi$-basis)
\begin{align}
C_{PP}^{\rm tm}(t) &= \frac{1}{L^{3}}{\displaystyle \sum_{x,z}} \langle 0 | \mathcal{P}^{1}(x)~\mathcal{P}^{1}(z) | 0 \rangle~\delta_{t, (t_x - t_z )} \\
C_{VP}^{\rm tm}(t) &= \frac{1}{L^{3}}{\displaystyle \sum_{x,z}} \langle 0 | \mathcal{A}_{0}^{1}(x)~\mathcal{P}^{1 \dag}(z) | 0 \rangle~\delta_{t, (t_x - t_z )} \; . 
\end{align}
We recall that $\mathcal{P}^1(x) = \bar{\psi}_{val,-}(x)\gamma_{5}\frac12 \tau^{1}\psi_{val,-}(x)$ and $\mathcal{A}_{0}^{1}(x) = \bar{\psi}_{val,-}(x)\gamma_{0}\gamma_{5}\frac12 \tau^{1}\psi_{val,-}(x)$ are the pseudoscalar and axial point-like bare (non-singlet) currents in the ``physical'' basis, while 
in Eq.~(\ref{eq:est_RV}) the suffix $\rm {tm}$ reminds that all the quark bilinear operators appearing in the correlators $C_{PP}^{\rm{tm}}$ and $C_{VP}^{\rm{tm}}$ involve a quark of flavour $F$ and an antiquark of flavour $F^\prime$, or viceversa, with $r_F=-r_{F^\prime}$. Because of the WI in Eq.~(\ref{eq:PCVC}) one must have
\be
    R_V(t) \longrightarrow Z_{V} + \mathcal{O}(a^2\mu_F^2, a^2\mu_F\Lambda_{QCD}) ~ . ~
\ee
In Fig.\,\ref{fig:RV} we show the time behavior of the estimator $R_V(t)$ for the cB211.072.64 and cC211.060.80 ensembles, and for two values of $a\mu_F$. It can be seen that the WI method allows to determine the RC $Z_{V}$ from the plateau of $R_V(t)$ with remarkably high precision. Regarding the value of $\mu_F$, we fix it by the requirement that the ground state mass of the correlator $C_{PP}^{\rm tm}(t)$ matches the mass of the fictitious $\eta_{s}$ meson (made out of a strange quark and a strange antiquark), i.e. $M_{\eta_{s}} = 689.89~(50)~{\rm MeV}$\,\cite{Borsanyi:2020mff}. 
The values of $Z_V$ corresponding to this choice are collected in Table\,\ref{tab:RCs}.

\subsection{Case of $Z_{A}$}

Let us consider now the case where the two distinct valence quark flavors
are taken at maximal twist with equal $r$-Wilson parameters, i.e.\ $r_{F} = r_{F^{\prime\prime}} =1$. In order to lighten notation and ease some algebra we collect the two valence quark field in a two-flavour valence field, $\psi_{val,+}$,  i.e.\
\begin{align}
    \psi_{val,+} \equiv (q_F, q_{F^{\prime\prime}}) \; , \qquad 
    \bar \psi_{val,+} \equiv (\bar q_F , \bar q_{F^{\prime\prime}}) \; ,
\end{align}
with the suffix ``$+$'' reminding of $r_F r_{F^{\prime\prime}} >0$. The $FF^{\prime\prime}$ sector of the valence quark action~(\ref{eq:Sval}) reads 
\begin{align}
S_{\rm q, val} \; \supset \; S_{\rm q, val}^{F F^{\prime\prime}} = a^4\sum_x  \bar \psi_{val,+} (x) \, [ \, \gamma \cdot \tilde\nabla + \mu_F - i\gamma_5 \mathbbm{1} W_{\rm cr}^{\rm cl}|_{r=1} \, ] \,
\psi_{val,+} (x) \, , \label{eq:SvalFF"} \; ,
\end{align}
with the  $2 \times 2$ identity matrix $\mathbbm{1}$ acting in the $FF^{\prime\prime}$ flavor space.
As RCs are named after the quark basis where the Wilson term is untwisted, we introduce in this basis the notations
\begin{align}
    \chi_{val,+} \equiv e^{-i\pi\gamma_{5}/4}\psi_{val,+} , \qquad
    \bar{\chi}_{val,+} \equiv \bar{\psi}_{val,+}  e^{-i\pi\gamma_{5}/4}
\end{align}
for the $FF^{\prime\prime}$ sector of the valence quark action (see Eq.~(\ref{eq:SvalFF"})), namely
\begin{align}
    S_{\rm q, val}^{F F^{\prime\prime}} = a^4\sum_x  \bar \chi_{val,+} (x) \, [ \, \gamma \cdot \tilde\nabla + i \gamma_5 \mathbbm{1} \mu_F + W_{\rm cr}^{\rm cl}|_{r=1} \, ] \, \chi_{val,+} (x) \, , . \label{eq:SvalFF"chi} \; 
\end{align}
It follows that in the $FF^{\prime\prime}$ valence sector the non-singlet axial currents and the pseudoscalar densities in the ``physical'' $\psi$-basis (for which we use calligraphic style symbols) are given by
\begin{align}
 \mathcal{A}_\mu^{a} &\equiv \bar{\psi}_{val,+}\gamma_\mu\gamma_5\frac{\tau^a}{2}\psi_{val,+}
  = A_\mu^{a,OS} = \bar{\chi}_{val,+}\gamma_\mu\gamma_5\frac{\tau^a}{2}\chi_{val,+} \; ,
  \qquad (a=1,2,3) \label{eq:OSaxcurr}
\end{align}
and 
\begin{align}
 \mathcal{P}_\mu^{a} &\equiv \bar{\psi}_{val,+}\gamma_5\frac{\tau^a}{2}\psi_{val,+}
  = iS_\mu^{a,OS} = i \bar{\chi}_{val,+}\frac{\tau^a}{2}\chi_{val,+} \; ,
  \qquad (a=1,2,3) \; , \label{eq:OSpsdens}
\end{align}
where the suffix ``OS'' on the $\chi$-basis expression of the operators reminds that they are made out of a valence quark and antiquark having equal $r$ parameters in the action, see Eqs.~(\ref{eq:SvalFF"}) and~(\ref{eq:SvalFF"chi}).

In order to determine the RC $Z_{A}$, let us start by defining the following ratio
\begin{align}
R_{A}(t)  \equiv 2\mu_{q} \frac{ C_{SS}^{\rm{OS}}(t)}{\tilde{\partial}_{t} C_{AS}^{\rm{OS}}(t)}~,    \end{align}
where (correlators are named here after the unphysical $\chi$-basis)
\begin{align}
C_{SS}^{\rm OS}(t) &= \frac{1}{L^{3}}{\displaystyle \sum_{x,z}} \langle 0 | \mathcal{P}^{1}(x)~\mathcal{P}^{1}(z) | 0 \rangle~\delta_{t, (t_x - t_z )} \\
C_{AS}^{\rm OS}(t) &= \frac{1}{L^{3}}{\displaystyle \sum_{x,z}} \langle 0 | \mathcal{A}_{0}^{1}(x)~\mathcal{P}^{1}(z) | 0 \rangle~\delta_{t, (t_x - t_z )}\, ,    
\end{align}
with $\mathcal{A}_{0}^{1}$ and $\mathcal{P}^{1}$ given by Eqs.~(\ref{eq:OSaxcurr}) 
and~(\ref{eq:OSpsdens}). The suffix ``OS'' in the correlators reminds that $C_{SS}^{\rm{OS}}$ and $C_{AS}^{\rm{OS}}$ involve a quark of flavour $F$ and an antiquark of flavour $F^\prime$, or viceversa, with $r_F=r_{F^\prime}$.

At large time distances $t/a \gg 1$ one has the following asymptotic behavior
\begin{align}
C_{SS}^{\rm OS}(t) &\to |G_{\pi}^{\rm OS}|^{2} \frac{ e^{-M_{\pi}^{\rm OS}t} + e^{-M_{\pi}^{\rm{OS}}(T-t)}}{2 M_{\pi}^{\rm{OS}}} \\
a\tilde{\partial}_{t}C_{AS}^{\rm OS}(t) &\to \frac{f_{\pi}^{\rm OS}}{Z_{A}} M_{\pi}^{\rm OS}\sinh{( aM_{\pi}^{\rm OS})}( G_{\pi}^{\rm{OS}})^{*}\frac{ e^{-M_{\pi}^{\rm OS}t} + e^{-M_{\pi}^{\rm{OS}}(T-t)}}{2 M_{\pi}^{\rm{OS}}},
\end{align}
where $M_{\pi}^{\rm OS}$ is the mass of the valence OS pion $\pi$, i.e.\ the ground state mass extracted from $C_{SS}^{\rm OS}(t)$ at the given quark mass $\mu_F$, while $f_{\pi}^{\rm OS}$ is related to the pion decay constant, $f_\pi$, through:
\begin{align}
f_{\pi}^{\rm OS} = f_{\pi} + \mathcal{O}(a^{2})~.    
\end{align}
The previous equations imply the following asymptotic large time behavior for $R_{A}(t)$
\begin{align}
R_{A}(t) \to 2a\mu_F \frac{ Z_{A}}{f_{\pi}^{\rm OS}}\frac{ G_{\pi}^{\rm OS}}{M_{\pi}^{\rm OS}\sinh{ (a M_{\pi}^{\rm OS} )}}~.    
\end{align}

In order to determine $Z_{A}$ from the estimator $R_{A}(t)$, it is necessary to have an independent way to extract $f_{\pi}^{\rm OS}$, since both $G_{\pi}^{\rm OS}$ and $M_{\pi}^{\rm OS}$ can be determined from $C_{SS}^{\rm OS}(t)$ alone. This can be achieved exploiting the fact that, as a consequence of the WI given in Eq.~(\ref{eq:PCVC}), the meson decay constant $f_{\pi}$ can be extracted in the ``tm'' regularization without the knowledge of any RCs, from the large time behaviour of $C_{PP}^{\rm tm}(t)$, namely using:
\begin{align}
C_{PP}^{\rm tm}(t) \to |G_{\pi}^{\rm tm}|^{2} \frac{ e^{-M_{\pi}^{\rm tm}t} + e^{-M_{\pi}^{\rm{tm}}(T-t)}}{2 M_{\pi}^{\rm{tm}}},\quad f_{\pi}^{\rm tm} = 2a\mu_F \frac{ G_{\pi}^{\rm tm}}{M_{\pi}^{\rm tm}\sinh{ (a M_{\pi}^{tm})}}~,
\end{align}
where again $f_{\pi}^{\rm tm} = f_{\pi} + \mathcal{O}(a^{2})$. By imposing $f_{\pi}^{\rm OS}= f_{\pi}^{\rm tm}$, which is true up to lattice artifacts, we can determine $Z_{A}$ using
\begin{align}
\label{eq:est_RA}
\bar{R}_{A}(t) \equiv R_{A}(t)\frac{ M_{\pi}^{\rm OS}\sinh{( aM_{\pi}^{\rm OS} )}}{M_{\pi}^{\rm tm}\sinh{( aM_{\pi}^{\rm tm} )}}\frac{Z_{S}}{Z_{P}} \to Z_{A}~,
\end{align}
where 
\begin{align}
\label{eq:ZPZS}
\frac{Z_{P}}{Z_{S}} = \frac{G_{\pi}^{\rm OS}}{G_{\pi}^{\rm tm}}~,    
\end{align}
is the ratio between the flavor non-singlet pseudoscalar ($Z_{P}$) and scalar ($Z_{S}$) RCs.

\begin{figure}[htb!]
\begin{center}
\includegraphics[scale=0.85]{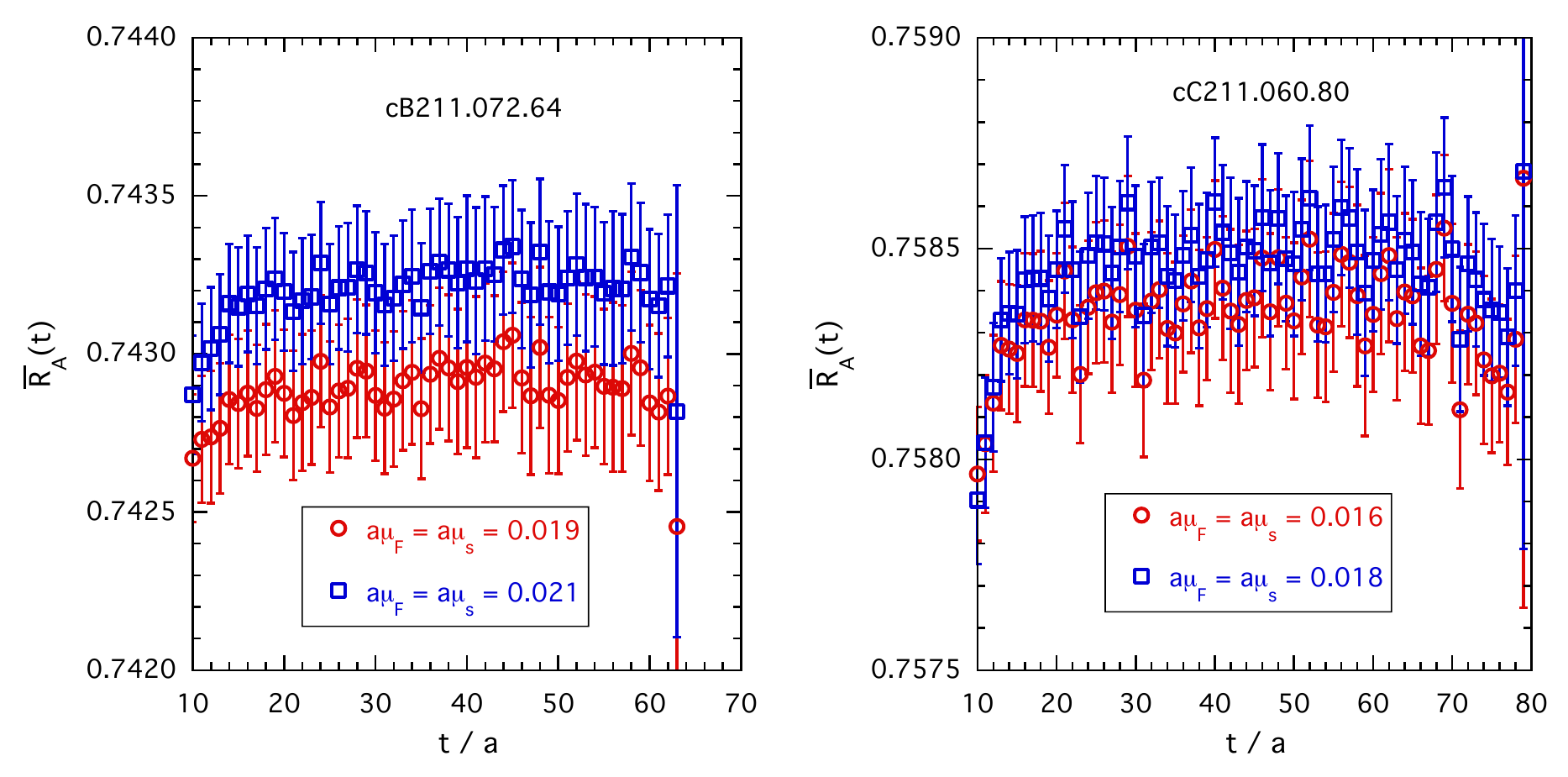}
\vspace{-0.5cm}
\caption{\it \small The same as in Fig.\,\ref{fig:RV}, but for the estimator $\overline{R}_A(t)$ given by Eq.\,(\ref{eq:est_RA}).}
\label{fig:RA}
\end{center}
\end{figure}

As in the case of $Z_{V}$, there is freedom in choosing the valence quark mass $\mu_F$ at which the correlators are evaluated, provided this value is kept fixed in physical units for all ensembles. Indeed, the difference $Z_{A}(\mu_F) - Z_{A}(0)$ represents a mere $\mathcal{O}(a^2\mu_F^2, a^2\mu_F\Lambda_{QCD})$ cut-off effect. In Fig.\,\ref{fig:RA} we show our determination of the estimator $\bar{R}_{A}(t)$ for the cB211.072.64 and cC211.06.80 ensembles, and for two values of $a\mu_F$. We adopt for $Z_{A}$ the same choice made for $Z_{V}$, and fix $\mu_F$ to the strange mass by the requirement that the ground state mass of the $C_{PP}^{\rm tm}(t)$ correlator, matches the one of the $\eta_{s}$ meson, i.e. $M_{\eta_{s}} = 689.89~(50)~{\rm MeV}$. 
The values of $Z_A$ corresponding to this choice, and obtained from the plateaux of the estimator $\overline{R}_A(t)$, are collected in Table\,\ref{tab:RCs}.

For sake of completeness we collect in Table\,\ref{tab:ratios_RCs} the values of the ratios of RCs $Z_A / Z_V$ corresponding to the results of Table\,\ref{tab:RCs}, and $Z_P / Z_S$ obtained from Eq.\,(\ref{eq:ZPZS}).
\begin{table}[htb!]
    \centering
    \begin{tabular}{||c||c|c||}
    \hline
    ensemble & $Z_A / Z_V $ & $Z_P / Z_S$ \\
    \hline
    ~ cB211.072.64 ~ & ~ $1.05176~(35)$ ~ & ~ $0.79018~(35)$ ~ \\
    \hline
    ~ cB211.072.96 ~ & ~ $1.05134~(24)$ ~ & ~ $0.79066~(23)$ ~ \\
    \hline
    ~ cC211.060.80 ~ & ~ $1.04535~(22)$ ~ & ~ $0.82308~(23)$ ~  \\
    \hline
    ~ cD211.054.96 ~ & ~ $1.04011~(16)$ ~ & ~ $0.85095~(18)$ ~ \\
    \hline

    \end{tabular}
    \caption{\it \small The values of $Z_A / Z_V$ and $Z_P / Z_S$ obtained from the hadronic method discussed in the text for each of the ETMC ensembles of Table\,\ref{tab:simudetails}.}
    \label{tab:ratios_RCs} 
\end{table}



\section{The physical strange- and charm-quark masses}
\label{sec:masses}

In this Appendix we describe our strategy to reach the physical values of the valence strange- and charm-quark masses, $m_s^{phys}$ and $m_c^{phys}$, using various hadronic inputs. Our results are well consistent with those obtained in Ref.~\cite{ExtendedTwistedMass:2021gbo} using the the kaon mass to determine $m_s^{phys}$ and the $D_s$-meson mass to determine $m_c^{phys}$.

In Sections\,\ref{sec:strange_mass} and \ref{sec:charm_mass} we list respectively the values of the valence bare strange- and charm-quark masses, $a\mu_s$ and $a\mu_c$, used for each gauge ensemble to interpolate 
the simulations of Section\,\ref{sec:connected}
to the physical strange- and charm-quark masses.
In Section\,\ref{sec:disc} we list the values of $a\mu_s$ and $a\mu_c$ used to evaluate the strange and charm quark loops of Section\,\ref{sec:disconnected}.

\subsection{The physical strange-quark mass}
\label{sec:strange_mass}

\begin{figure}[htb!]
  \includegraphics[width=0.90\linewidth]{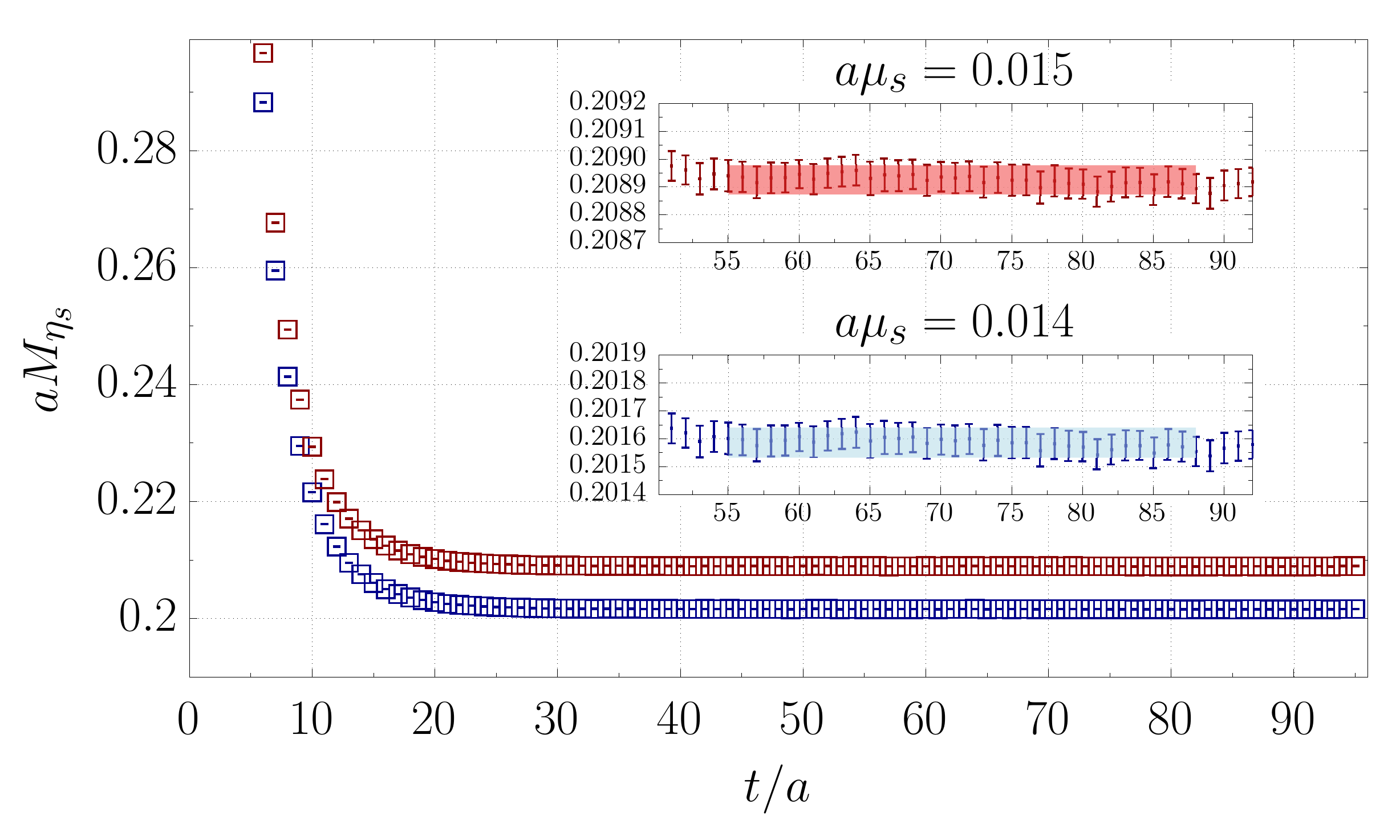}  
  \includegraphics[width=0.90\linewidth]{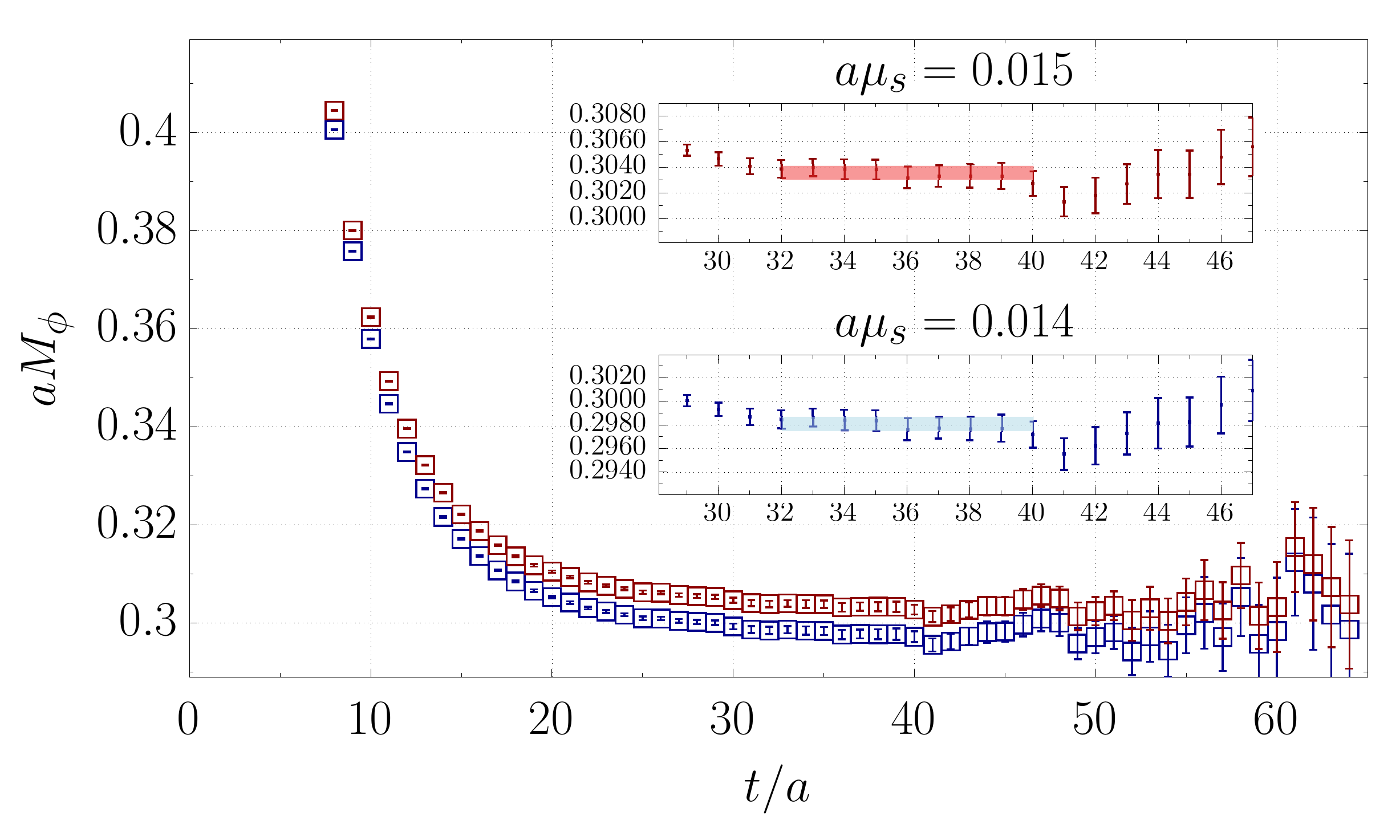}  
\caption{\it \small Effective masses $aM_{\eta_s}$ (top) and $aM_\phi$ (bottom) obtained, respectively, from the strange pseudoscalar and vector correlators evaluated in the ``tm" regularization in the case of the cD211.054.96 ensemble. The horizontal bands indicate the results of a constant fit in the plateaux regions, where the ground-state dominates.}
\label{fig:eff_mass_phi_etas}
\end{figure}

In order to reach the physical strange-quark mass $m_s^{phys}$, we made use of two different hadronic inputs, namely the mass of a fictitious $\eta_s$ meson, made of two mass-degenerate strange-like quarks of different flavors, and the mass of the $\phi$ vector meson. The physical value of the fictitious $\eta_s$-meson mass, $M_{\eta_s}$, was determined with sub-permille precision in Ref.\,\cite{Borsanyi:2020mff}, so that throughout this work we make use of the value
\be
    \label{eq:Metas_phys}
    M_{\eta_s}^{phys} = 689.89~(49)~{\rm MeV} ~ , ~      
\ee
while for the mass of the $\phi$ meson we rely on the PDG\,\cite{ParticleDataGroup:2020ssz} value
\be
    \label{eq:Mphi_phys}
    M_\phi^{phys} = 1019.461~(16)~{\rm MeV} ~ . ~
\ee
Within the lattice QCD formulation, we extract $aM_{\eta_s}$ and $aM_\phi$ from the connected part of the strange pseudoscalar and vector correlators, respectively, evaluated in the ``tm" regularization, which guarantees that discretization effects are of order ${\cal{O}}(a^2 \mu_s)$. Thus, in the case of the $\phi$ meson we neglect the contribution from quark disconnected diagrams, which are expected to yield a tiny correction of order ${\cal{O}}(\alpha_s^3)$. In Fig.\,\ref{fig:eff_mass_phi_etas} we show the quality of our determination of  $aM_{\eta_s}$ and $aM_\phi$ on our finest ensemble cD211.054.96.

\begin{figure}[htb!]
    \centering
    \includegraphics[scale=0.55]{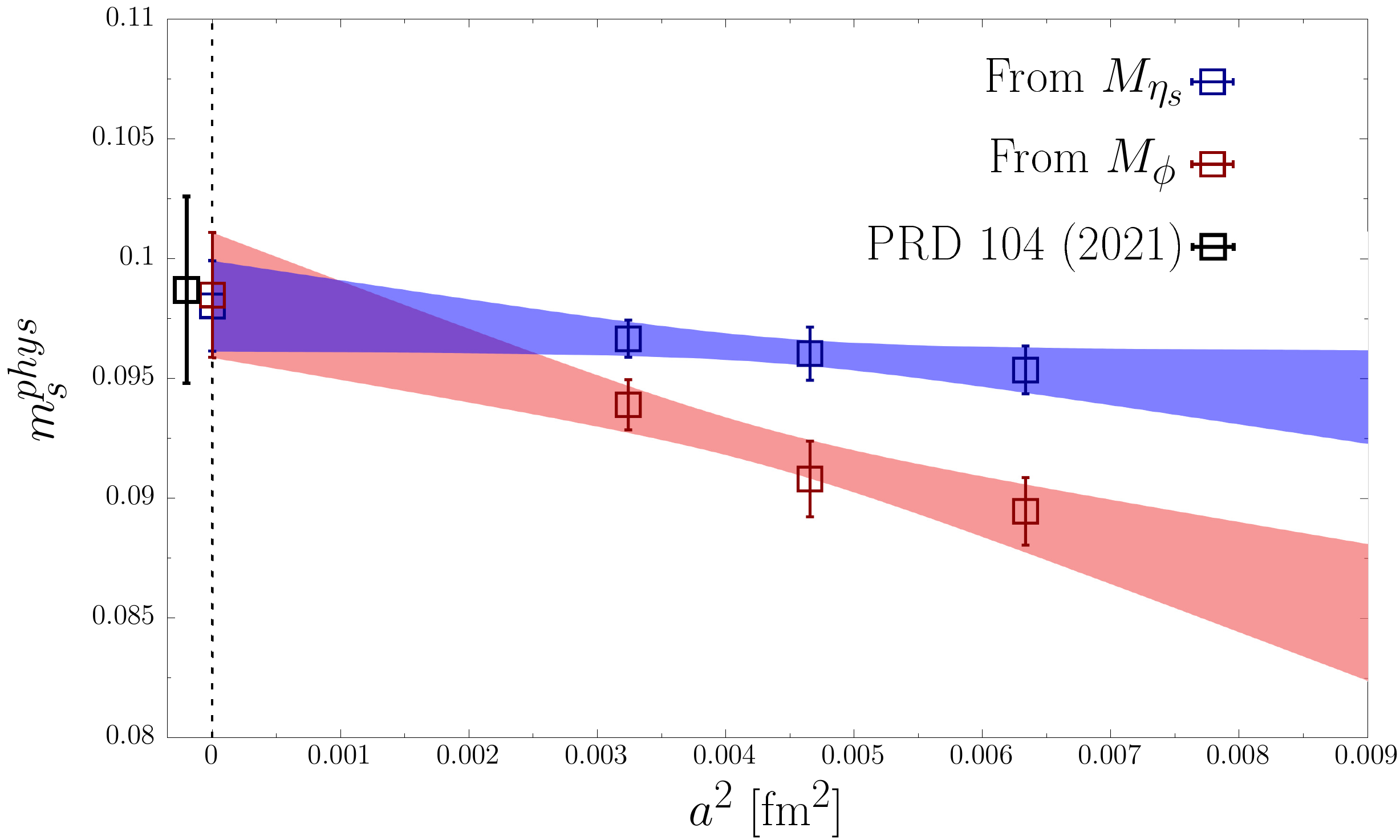}
    \caption{\it \small Continuum limit extrapolation of $m_s^{phys}$ in the $\overline{MS}(2\,{\rm GeV})$ scheme, determined using the $\eta_s$ meson (blue squares) and the $\phi$-meson (red squares) masses as hadronic input. The two determinations agree in the continuum limit within one standard deviation. The black square at $a^2 = 0$ corresponds to the result obtained in Ref.~\cite{ExtendedTwistedMass:2021gbo} using the kaon mass to determine $m_s^{phys}$.}
    \label{fig:cont_lim_ms}
\end{figure}

In order to determine the physical strange-quark mass $am_s^{phys}$ in lattice units, we interpolate/extrapolate our lattice data for $aM_{\eta_s}$ and $aM_{\phi}$ using the following linear Ansatz
\be
    \label{eq:aMP_fit}
    aM_P = aM_P^{phys} + \kappa \cdot \left( \frac{a\mu_s}{Z_P}  - am_s^{phys} \right) ~ , ~ \qquad P = \{ \eta_s, \phi \} ~ , ~     
\ee
with $\kappa$ and $a m_s^{phys}$ being fitting parameters for each ensemble. $aM_P^{phys}$ is obtained from Eqs.\,(\ref{eq:Metas_phys})-(\ref{eq:Mphi_phys}) using the improved determination of the lattice spacings the values of which are listed Table\,\ref{tab:spacing}. 
The results obtained for $am_s^{phys}$ in the $\overline{MS}(2\,{\rm GeV})$ scheme are collected in Table\,\ref{tab:ams_phys} and shown in Fig.~\ref{fig:cont_lim_ms} versus the squared lattice spacing. No significant FSEs are visible and the data exhibit a nice $a^2$-scaling behavior. 
The continuum limit extrapolations for $m_s^{phys}$ corresponding to the use of the $\eta_s$- and $\phi$-meson masses as hadronic inputs agree very well within one standard deviation and, moreover, they are consistent with the result obtained in Ref.~\cite{ExtendedTwistedMass:2021gbo} using the kaon mass to determine $m_s^{phys}$.
\begin{table}[htb!]
    \centering
    \begin{tabular}{||c||c|c||}
    \hline
        ~~ Ensemble ~~ & ~~ $am_s^{phys}(\eta_s)$ ~~ & ~~ $am_s^{phys}(\phi)$ ~~ \\
        \hline
        cB211.072.64 &0.03846(41) &0.03608(58) \\
        cB211.072.96 &0.03845(41) &0.03533(51) \\
        cC211.060.80 &0.03320(40) &0.03139(58) \\
        cD211.054.96 &0.02788(25) &0.02709(32) \\
        \hline
    \end{tabular}
    \caption{\it \small Values of $a m_s^{phys}$ in the $\overline{MS}(2\,{\rm GeV})$ scheme\,\cite{ExtendedTwistedMass:2021gbo,ExtendedTwistedMass_RCs} in lattice units determined using in Eq.\,(\ref{eq:aMP_fit}) either the $\eta_s$-meson mass\,(\ref{eq:Metas_phys}) or the $\phi$-meson mass\,(\ref{eq:Mphi_phys}) as the physical hadronic input.}
    \label{tab:ams_phys}
\end{table}

For each gauge ensemble we perform simulations at two values of the valence bare strange-quark mass, $a\mu_s$, in order to interpolate the results for $a_\mu^{\rm SD}(s)$ and $a_\mu^{\rm W}(s)$ to the physical strange-quark mass $m_s^{phys}$. The simulated values of $a\mu_s$ are collected in Table\,\ref{tab:simulated_ms} together with the values of the RC $Z_P$ of the pseudoscalar density obtained in the RI-MOM scheme and converted in the $\overline{MS}(2\,{\rm GeV})$ one in Refs.\,\cite{ExtendedTwistedMass:2021gbo,ExtendedTwistedMass_RCs}. 
\begin{table}[htb!]
    \centering
    \begin{tabular}{|| c || c | c || c ||}
    \hline
        ~~ Ensemble ~~ & ~~ $a\mu_s^L$ ~~ & ~~ $a\mu_s^H$ ~~ & ~~ $Z_P[\overline{MS}(2\,{\rm GeV})]$ ~~ \\
        \hline
        cB211.072.64 & $0.019$ & $0.021$ & $0.4788~(54)$\\
        cB211.072.96 & $0.019$ & $0.021$ & $0.4788~(54)$\\
        \hline
        cC211.060.80 & $0.016$ & $0.018$ & $0.4871~(49)$\\
        \hline
        cD211.054.96 & $0.014$ & $0.015$ & $0.4894~(44)$\\
        \hline
    \end{tabular}
    \caption{\it \small Values of the bare strange-quark mass $a \mu_s$ and of the RC $Z_P$ (evaluated in the RI-MOM scheme and converted in the $\overline{MS}(2\,{\rm GeV})$ one\,\cite{ExtendedTwistedMass:2021gbo, ExtendedTwistedMass_RCs}) for each of the four ensembles of Table\,\ref{tab:simudetails}. We indicate with $a\mu_s^L$ ($a\mu_s^H$) the lightest (heaviest) bare strange-quark mass used for each ensemble.}
    \label{tab:simulated_ms}
\end{table}

\subsection{The physical charm-quark mass}
\label{sec:charm_mass}

\begin{figure}[htb!]
  \includegraphics[width=0.90\linewidth]{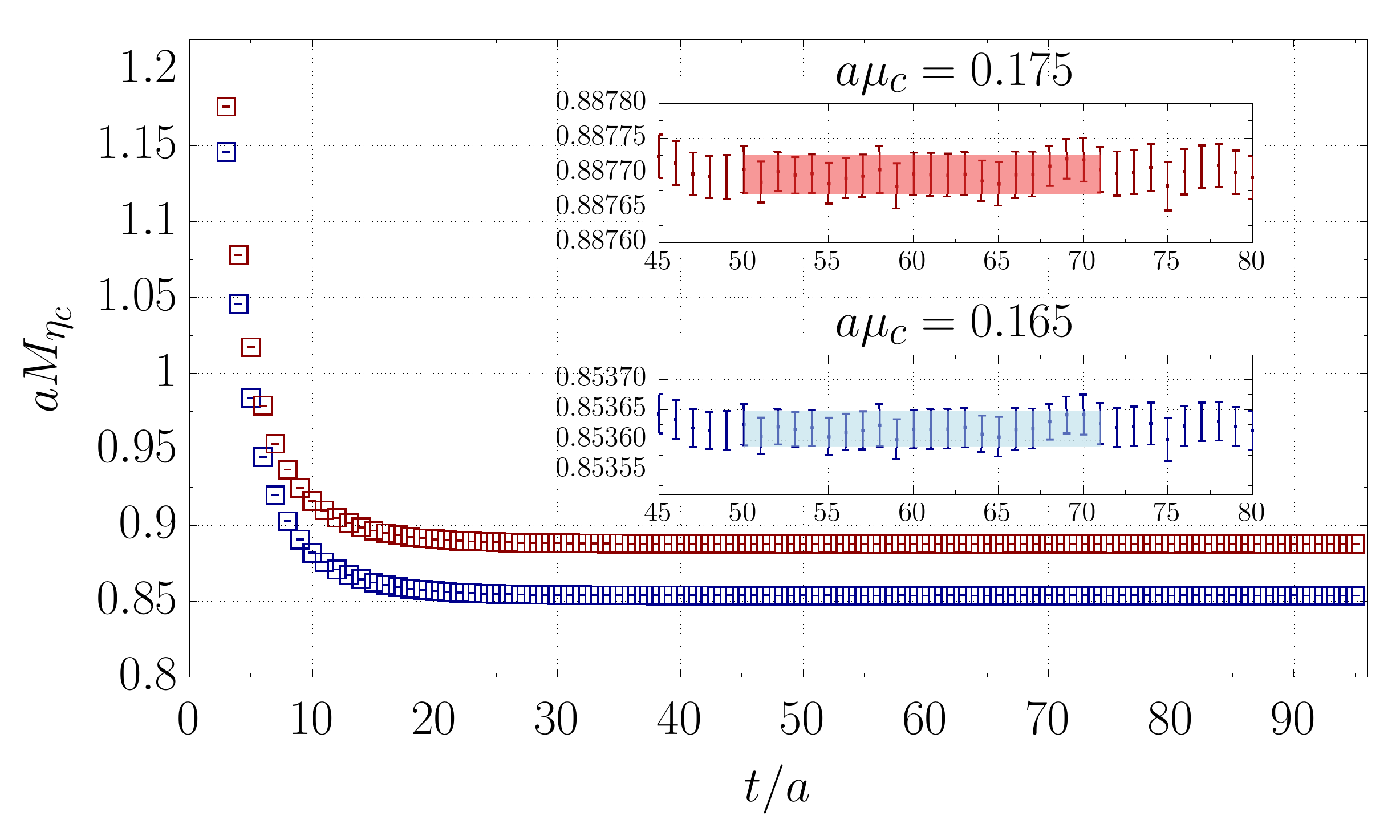}  
  \includegraphics[width=0.90\linewidth]{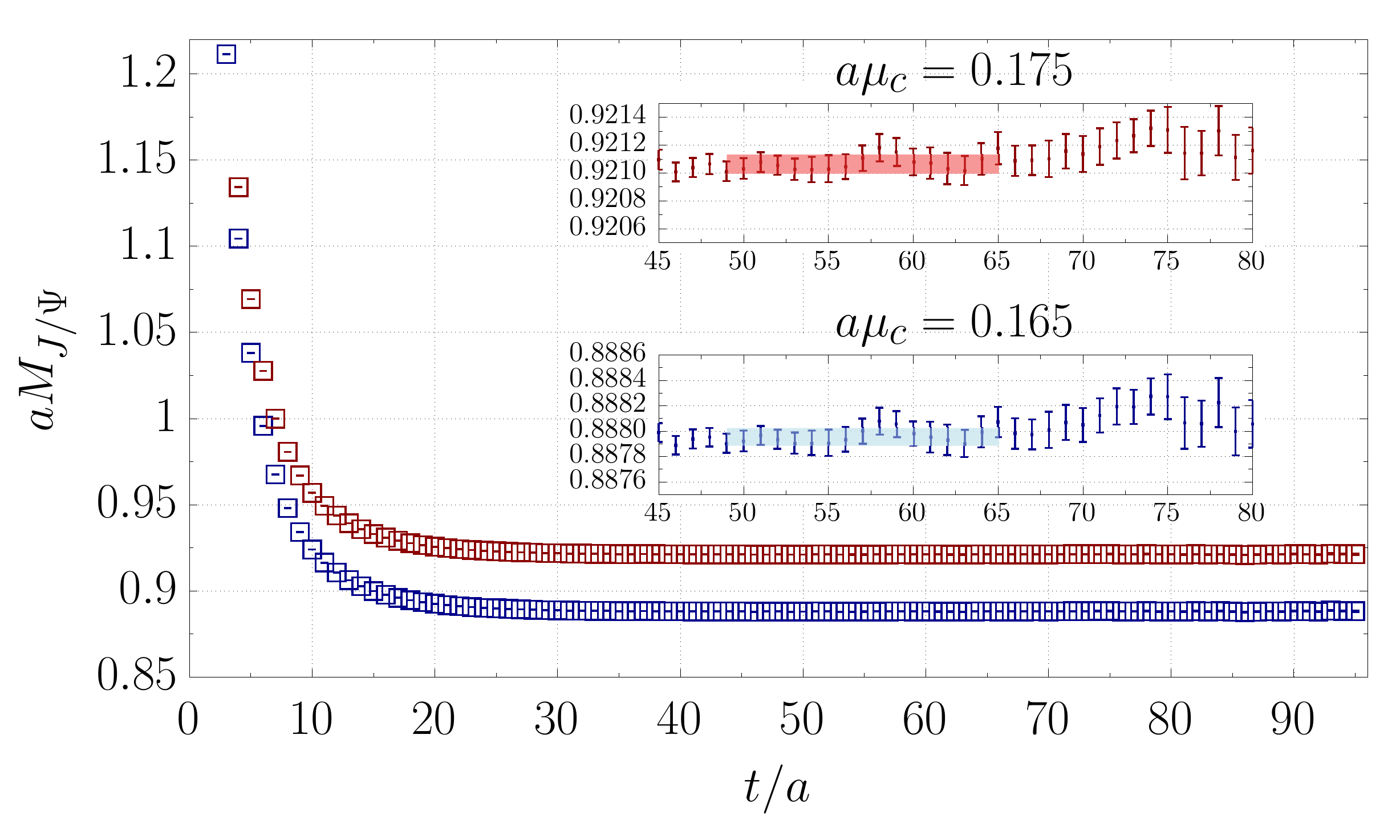}
\caption{\it \small Effective masses $aM_{\eta_c}$ (top) and $aM_{J/\Psi}$ (bottom) obtained, respectively, from the charm pseudoscalar and vector correlators evaluated in the ``tm" regularization in the case of the cD211.054.96 ensemble. The horizontal bands indicate the results of a constant fit in the plateaux regions, where the ground-state dominates.}
\label{fig:eff_mass_Jpsi_etac}
\end{figure}

In order to reach the physical charm-quark mass $m_c^{phys}$, we use two different hadronic inputs, namely the masses of the pseudoscalar $\eta_c$ and vector $J/\Psi$ mesons. In this work, we adopt the PDG values\,\cite{ParticleDataGroup:2020ssz}
\bea
    \label{eq:Metac_phys}
    M_{\eta_c}^{phys} & = & 2.984~(4)~{\rm GeV} ~ , ~ \\[2mm]      
    \label{eq:MJpsi_phys}
    M_{J/\Psi}^{phys} & = & 3.097~(1)~{\rm GeV} ~ , ~
\eea
where the errors include the estimate of the quark disconnected contributions made in Refs.\,\cite{Hatton:2020qhk, Zhang:2021xrs}.
We extract $aM_{\eta_c}$ and $aM_{J/\Psi}$ from the connected part of the charm pseudoscalar and vector correlators, respectively. In Fig.\,\ref{fig:eff_mass_Jpsi_etac}, we show the quality of our determination of $aM_{\eta_c}$ and $aM_{J/\Psi}$ using our finest ensemble cD211.054.96.

\begin{figure}[htb!]
    \centering
    \includegraphics[scale=0.55]{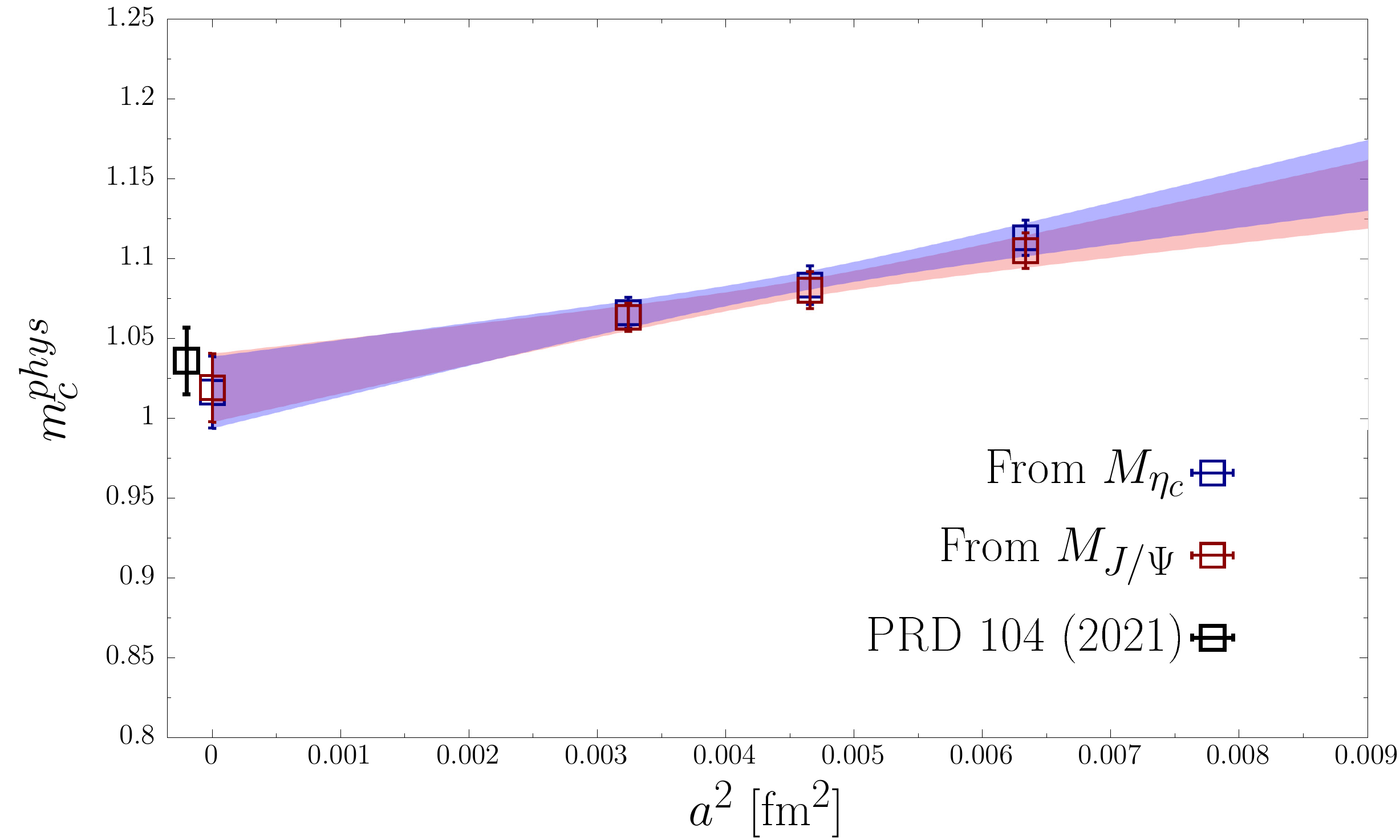}
    \caption{\it \small Continuum limit extrapolation of $m_c^{phys}$ in the $\overline{MS}(3\,{\rm GeV})$ scheme, determined using either the $\eta_c$-meson (blue squares) or the $J/\Psi$-meson (red squares) masses as hadronic input. The two determinations  agree in the continuum limit. The black square at $a^2 = 0$  corresponds to the result obtained in Ref.~\cite{ExtendedTwistedMass:2021gbo} using the mass of the $D_s$ meson to determine $m_c^{phys}$.}
    \label{fig:cont_lim_mc}
\end{figure}

In order to determine the physical charm-quark mass $am_c^{phys}$ in lattice units, we interpolate/extrapolate our lattice data for $aM_{\eta_c}$ and $aM_{J/\Psi}$ using the following linear Ansatz
\be
    \label{eq:aMP_fit_charm}
    aM_P = aM_P^{phys} + \overline{\kappa} \cdot \left( \frac{a\mu_c}{Z_P}  - am_c^{phys} \right) ~ , ~ \qquad P = \{ \eta_c, J/\Psi \} ~ , ~     
\ee
with $\overline{\kappa}$ and $a m_c^{phys}$ being fitting parameters for each ensemble. $aM_P^{phys}$ is obtained from Eqs.\,(\ref{eq:Metac_phys})-(\ref{eq:MJpsi_phys}) using the improved determination of the lattice spacing.
The results obtained for $am_c^{phys}$ in the $\overline{MS}(3\,{\rm GeV})$ scheme are collected in Table\,\ref{tab:amc_phys} and shown in Fig.~\ref{fig:cont_lim_mc} versus the squared lattice spacing. No significant FSEs are visible and the data exhibit a nice $a^2$-scaling behavior. 
The continuum limit extrapolations for $m_c^{phys}$ corresponding to the use of the $\eta_c$- and $J/\Psi$-meson masses as hadronic inputs agrees very well within one standard deviation and, moreover, they are consistent with the result obtained in Ref.~\cite{ExtendedTwistedMass:2021gbo} using the $D_s$-meson mass to determine $m_c^{phys}$.
\begin{table}[htb!]
    \centering
    \begin{tabular}{||c||c|c||}
    \hline
        Ensemble & $am_c^{phys}(\eta_c)$ & $am_c^{phys}(J/\Psi)$ \\
        \hline
        cA211.53.24  &0.5210(81) &0.5128(83) \\
        cA211.40.24  &0.5213(82) &0.5133(83) \\
        cA211.30.32  &0.5218(81) &0.5145(83) \\
        cB211.072.64 &0.4489(45) &0.4457(46) \\
        cC211.060.80 &0.3746(43) &0.3735(42) \\
        cD211.054.96 &0.3076(29) &0.3068(27) \\
        \hline
    \end{tabular}
    \caption{\it \small Values of $a m_c^{phys}$ in the $\overline{MS}(3\,{\rm GeV})$ scheme\,\cite{ExtendedTwistedMass:2021gbo,ExtendedTwistedMass_RCs} in lattice units determined using in Eq.\,(\ref{eq:aMP_fit_charm}) either the $\eta_c$-meson mass\, from Eq.~(\ref{eq:Metac_phys}) or the $J/\Psi$-meson mass\, from Eq.~(\ref{eq:MJpsi_phys}) as the physical hadronic input.}
    \label{tab:amc_phys}
\end{table}

For each ensemble, we perform simulations at three values of the valence bare charm-quark mass, $a\mu_c$, in order to interpolate the results for $a_\mu^{\rm SD}(c)$ and $a_\mu^{\rm W}(c)$ to the physical charm-quark mass $m_c^{phys}$. The values of $a\mu_c$ used are collected in Table\,\ref{tab:simulated_mc} together with the values of the RC $Z_P$ of the pseudoscalar density obtained in the RI-MOM scheme and converted in the $\overline{MS}(3\,{\rm GeV})$ one in Refs.\,\cite{ExtendedTwistedMass:2021gbo, ExtendedTwistedMass_RCs}. 
\begin{table}[htb!]
    \centering
    \begin{tabular}{|| c || c | c | c || c ||}
    \hline
        ~~ Ensemble ~~ & ~~ $a\mu_c^L$ ~~ & ~~ $a\mu_c^M$ ~~ & ~~ $a\mu_c^H$ & ~~ $Z_P[\overline{MS}(3\,{\rm GeV})]$ \\
        \hline
        cA211.53.24 & $0.265$ & $0.290$ & $0.300$ & $0.5267~(54)$\\
        cA211.40.24 & $0.265$ & $0.290$ & $0.300$ & $0.5267~(54)$\\
        cA211.30.32 & $0.265$ & $0.290$ & $0.300$ & $0.5267~(54)$\\
        \hline
        cB211.072.64 & $0.210$ & $0.230$ & $0.250$ & $0.5314~(59)$\\
        \hline
        cC211.060.80 & $0.175$ & $0.195$ & $0.215$ & $0.5406~(54)$\\
        \hline
        cD211.054.96 & $0.165$ & $0.175$ & -- & $0.5431~(48)$\\
        \hline
    \end{tabular}
    \caption{\it \small Values of the bare charm-quark mass $a\mu_c$ in lattice units and of the RC $Z_P$  (evaluated in the RI-MOM scheme and converted in the $\overline{MS}(3\,{\rm GeV})$ one\,\cite{ExtendedTwistedMass:2021gbo, ExtendedTwistedMass_RCs}) for each of the ETMC ensembles employed in the charm sector. We indicate with $a\mu_c^L$, $a\mu_c^M$ and $a\mu_c^H$, respectively,  the lightest, the intermediate, and the heaviest bare charm-quark masses used for each ensemble.}
    \label{tab:simulated_mc}
\end{table}

\subsection{The strange- and charm-quark masses in disconnected contributions}
\label{sec:disc}

\begin{figure}[htb!]
    \centering
    \includegraphics[width=0.475\linewidth]{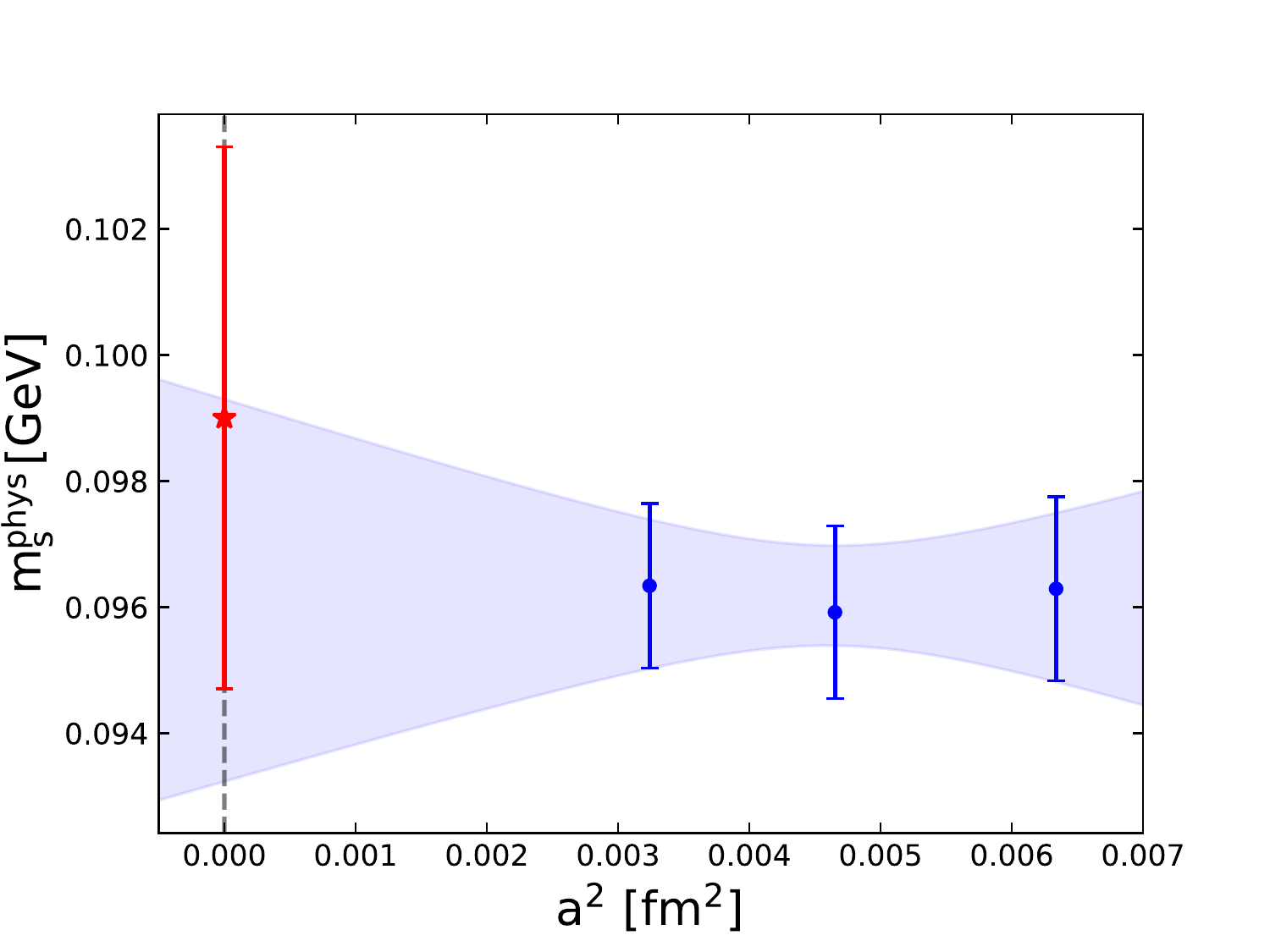}
    \includegraphics[width=0.475\linewidth]{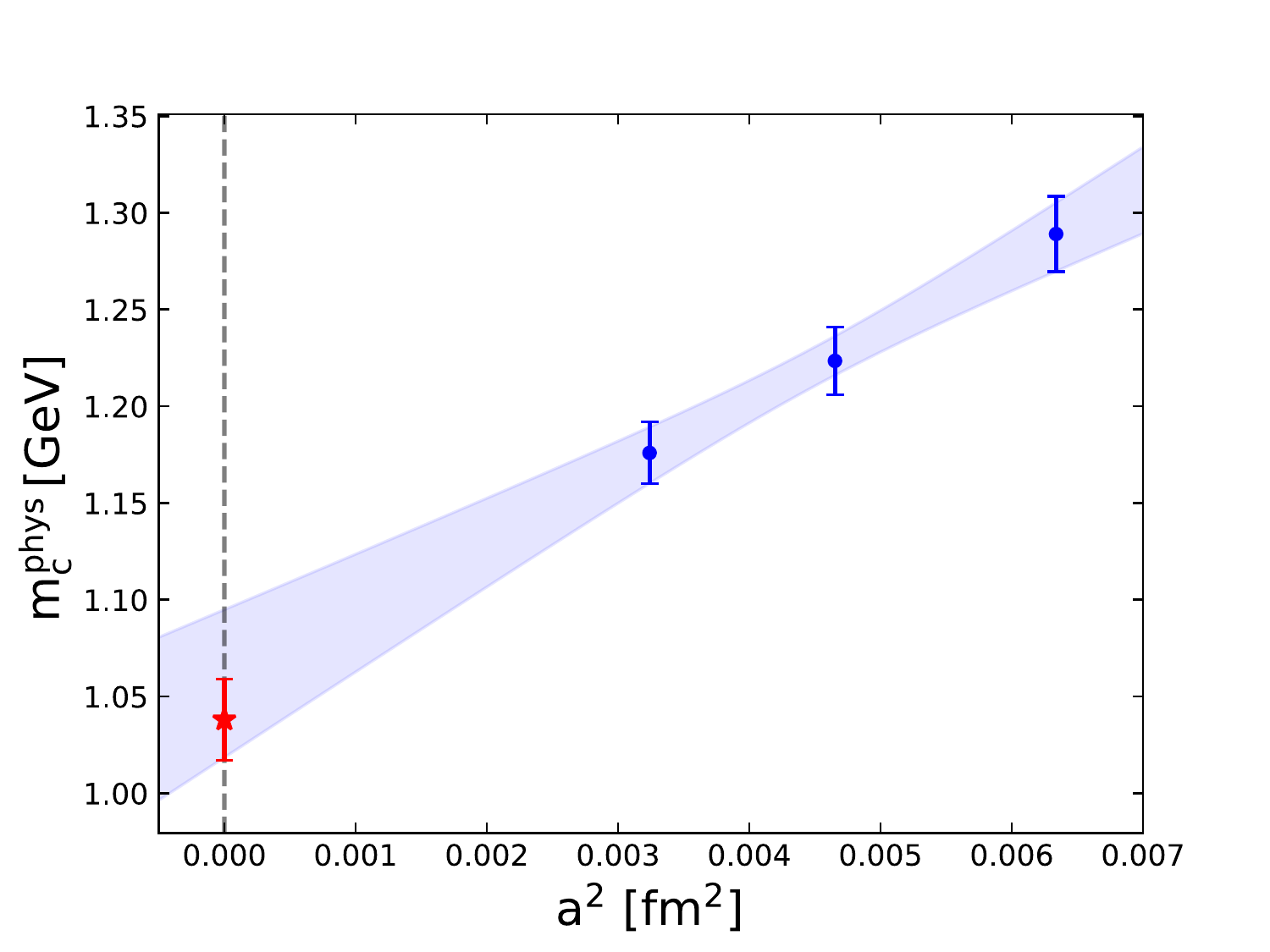}
    \caption{\it \small Renormalized strange (left) and charm (right) quark mass, given respectively in the $\overline{MS}(2\,{\rm GeV})$ and $\overline{MS}(3\,{\rm GeV})$ scheme\,\cite{ExtendedTwistedMass:2021gbo,ExtendedTwistedMass_RCs}, obtained in this work using the RCs $Z_P$ from Tables\,\ref{tab:simulated_ms} and~\ref{tab:simulated_mc} and tuning the $\Omega$ and $\Lambda_c$ baryon masses, versus the squared lattice spacing. The red stars are the results of the continuum limit extrapolation carried out in Ref.\,\cite{ExtendedTwistedMass:2021gbo}.}
    \label{fig:mu_disconnected}
\end{figure}

In Section\,\ref{sec:disconnected} the strange and charm quark loops are computed at a quark mass obtained by tuning the $\Omega$ and $\Lambda_c$ baryons, respectively, to their physical value. The values of the bare masses for the strange, $a\mu_s$, and for the charm, $a\mu_c$, quarks are listed in Table~\ref{tab:mu_disconnected}.
\begin{table}[htb!]
    \centering
    \begin{tabular}{||c||c|c||}
    \hline
        Ensemble & $a \mu_s$ & $a \mu_c$ \\
        \hline
        cB211.072.64 & 0.01860 & 0.249 \\
        cC211.060.80 & 0.01615 & 0.206 \\
        cD211.054.96 & 0.01360 & 0.166 \\
        \hline
    \end{tabular}
    \caption{\it \small Values of the bare quark masses $a\mu_s$ and $a\mu_c$ used for the calculation of strange and charm disconnected contributions.}
    \label{tab:mu_disconnected}
\end{table}
 In Fig.~\ref{fig:mu_disconnected}, we show the continuum limit of the renormalized strange and charm quark masses in the $(\overline{MS}(2\,{\rm GeV})$ and $\overline{MS}(3\,{\rm GeV})$ scheme\,\cite{ExtendedTwistedMass:2021gbo, ExtendedTwistedMass_RCs}, respectively. 
 We compare them against the results computed in the continuum limit in Ref.\,\cite{ExtendedTwistedMass:2021gbo}.
 Note that the values of the renormalized strange-quark mass $m_s^{phys} = Z_{P}^{-1}(\overline{MS}, 2\,{\rm GeV})\mu_{s} $ do not show sizable cut-off effects, while $m_{c}^{phys}= Z_{P}^{-1}(\overline{MS}, 3\,{\rm GeV})\mu_{c}$ does.


\section{Flavor-singlet renormalization constants}
\label{sec:appD}

In this Appendix we show that in LQCD with Wilson quarks the (UV finite) RCs of the singlet ($Z_{V^0}$) and non-singlet ($Z_V$) {\em point-like} vector currents coincide. 
This property follows from the exact invariance of the massless theory under flavor singlet and non-singlet vector transformations and it holds generally with any number $N_f$ of Wilson fermions of arbitrary mass, for all possible values of the twist angle and of the clover improvement coefficient.

For definiteness let us consider LQCD with $N_f=4$ flavours of Wilson quarks, say $u$, $d$, $s$ and $c$, with renormalized masses $\hat m_f = Z_m m_f \equiv Z_m (m_0 - m_{\rm cr})$, $f = u, d, s, c$, with or without a flavour singlet clover term. 
Defining $D^W_{\rm cr}(U) = \gamma \cdot \tilde\nabla - (a/2) \nabla^* \nabla + m_{\rm cr} + (i/4) c_{sw} \sigma \cdot F$ the critical Dirac--Wilson operator, with $\nabla_\mu$ ($\nabla_\mu^*$) the forward (backward) gauge covariant lattice derivative, which implicitly depends on the gauge--links field $U$, and $\tilde\nabla_\mu = \frac{1}{2} [ \nabla_\mu + \nabla_\mu^*]$, the lattice action reads
\be
S_{LQCD} = S_{YM}[U] + a^4 \sum_x \sum_{f=u,d,s,c} \bar q_f(x) [D^W_{\rm cr}(U)+ m_f] q_f(x) \; ,
\label{Waction} 
\ee
where $S_{YM}[U]$ denotes the pure gauge action term~\footnote{Both here and in Appendix\,\ref{sec:simulations} the fields $q_f$ and $\bar q_f$ refer to the quark of flavour $f$ in the ``physical'' basis where its soft mass term takes the canonical form $m_f \bar q_f q_f$. However the lattice regularizations are different, though related in the 
chiral limit by an axial rotation, since the critical Wilson term here is taken aligned to the soft mass term in 
the chiral internal space, while it is maximally twisted in the lattice setup of Appendix\,\ref{sec:simulations}, implying that the RCs of an operator with a given physical meaning in the two lattice regularizations are in general related through an axial rotation (depending of course on the details of $r_f$, $r_f'$ for the valence fields).}.
In the limit of degenerate quark masses ($m_u=m_d=m_s=m_c$) the lattice action is manifestly invariant under both flavour singlet $U(1)$ and flavor non-singlet $SU(4)$ global vector transformations of the quark fields $q_f$, $f = u, d, s, c$, with exactly conserved {\em one-point split} Noether currents given by
\be
\hat V_\mu^{(0)}(x) = \frac{1}{2} \sum_{f=u,d,s,c} [ \bar q_f(x+a\hat\mu) (1+\gamma_\mu) U_\mu^\dagger(x) q_f(x)
- \bar q_f(x) (1-\gamma_\mu) U_\mu(x) q_f(x+a\hat\mu) ]
\label{exSVcurr}
\ee
and
\be
\hat V_\mu^{(b)}(x) = \frac{1}{2} \sum_{f,h} [ \bar q_h(x+a\hat\mu) \lambda^b_{hf} (1+\gamma_\mu) U_\mu^\dagger(x) q_f(x) - \bar q_h(x) \lambda^b_{hf} (1-\gamma_\mu) U_\mu(x) q_f(x+a\hat\mu) ] \; ,
\label{exNSVcurr} 
\ee
with $\lambda^b$ ($b=1,2,..., 15$) being the generators of $SU(4)$. The corresponding conserved charges are
\be
Q_V^{(0)} = a^3 \sum_{\vec{x}} \hat V_0^{(0)}(x) = 3 Q_{\rm bar}  \; , \qquad
Q_V^{(b)} = a^3 \sum_{\vec{x}} \hat V_0^{(b)}(x)  \; , b=1,2,..., 15 \; , \label{exVch}
\ee
implying that the one-point split lattice currents~(\ref{exSVcurr}) and~(\ref{exNSVcurr}) are {\em not renormalized}. 
The baryon number charge is $Q_{\rm bar} = \frac{1}{3} Q_V^{(0)}$.
Note also that single flavor {\em one-point split} conserved vector currents $\hat V_\mu^{(f)}$  exist, for $f=u,d,s,c$, that are obtained by the appropriate combinations of the identity and the diagonal $SU(4)$ generator matrices. From these currents the corresponding single flavor conserved charges, $Q_V^{f}$, can be defined, in analogy with Eq.~(\ref{exVch}). For instance, taking $\lambda^{(15)} = {\rm diag}(-1,-1,-1,+3)$, one has $Q_V^c = \frac{1}{4} [Q_V^{0} + Q_V^{15}]$.

For non-degenerate quark masses, e.g. in $N_f=2+1+1$ LQCD, the U(1) vector symmetry remains exact while the $SU(4)$ flavor symmetry is only softly broken and the charges $Q_V^{(b)} $ ($b=1,2,...,15$), though being in general time-dependent, satisfy the same SU(4) charge algebra as in the mass-degenerate case. This implies that all the one-point split lattice currents\,(\ref{exSVcurr}) and\,(\ref{exNSVcurr}) still admit unit RC.

Here our focus is on the renormalization properties of the {\em point-like} bare vector currents
\be
V_\mu^{(0)}(x) = \sum_f \bar q_f(x) \gamma_\mu q_f(x) \; ,\qquad
V_\mu^{(b)}(x) = \sum_f \bar q_f(x) \lambda^b \gamma_\mu q_f(x) \; , 
\quad (b=1,2,..., 15) \, .
\label{pointVcurr} 
\ee
Owing to the presence of the Wilson term in the lattice action the point-like currents~(\ref{pointVcurr}) are not conserved, while, based on the exact vector $U(1)$ and $SU(4)$ symmetries, one expects that their properly renormalized flavor singlet and non-singlet counterparts read
\be
[V_\mu^{(0)}]_R(x) = Z_{V^0} V_\mu^{(0)}(x)  \; , \qquad
[V_\mu^{(b)}]_R(x) = Z_{V} V_\mu^{(b)}(x)  \; , \quad (b=1,2,..., 15) \, ,
\ee
with $Z_{V^0}$ and $Z_V$ non--trivial dimensionless, UV finite functions of the bare gauge coupling $g_0^2$. 

As far as the renormalization  of vector singlet and non-singlet currents is concerned, the values of the individual quark masses play no role, because they can at most affect $Z_V$ and $Z_{V^0}$ through immaterial O($am$) lattice artifacts~\footnote{It is well known that by adding soft mass terms, i.e.~$m_f \bar q_f q_f$, to the action density of massless QCD no new UV divergencies, apart from those that are reabsorbed in the usual quark mass renormalization ($\hat m_f = Z_m m_f$), appear in the theory~\cite{Weinberg:1973xwm}.}. 
The latter will actually be O($a^2 m^2$) and O($a^2 m\Lambda_{QCD}$) if the correlators from which the RCs are determined are O($a$) improved. Hence with no loss of generality in the following we can assume fully degenerate quark masses and set 
\begin{align}
    m_{u,d,s,c} \equiv m \; .   \label{degqmass}
\end{align}
The proof that in LQCD with Wilson quarks $Z_{V^0} = Z_V$ will proceed in two steps. 
\begin{itemize}
\item[1)] We observe that $Z_{V^0}= Z_V$ if and only if the insertion of the {\em point-like} bare vector charge $a^3 \sum_{\vec{x}} \bar q_h(x) \gamma_0 q_h(x)$ of a certain fixed flavour $h$ in the correlation functions of multilocal operators with no $h$-flavour valence quarks (or zero total $h$-flavour quantum number) vanishes.
\item[2)] We  prove that the aforementioned operator insertion, which of course gives rise only to quark disconnected diagrams, is actually vanishing.

\end{itemize}

For the second step we find it convenient to employ, as a proof-technical tool, a {\em mixed action LQCD} (MALQCD) setup, which has no direct relation to the twisted mass mixed action framework adopted in the paper and in no way restricts the validity of the result. 
As far as we know, the result $Z_{V^0} = Z_V$ for LQCD with Wilson quarks is currently established in perturbation theory only up to the two-loop level
(included)~\cite{Constantinou:2014rka}.

\vspace{0.5cm}

{\bf Step 1)} \\
For definiteness let us identify the flavour $h$ with the charm, i.e.\ $h=c$.
To lighten notation we write $q_f \equiv f$.
Assuming no special relation between $Z_{V^0}$ and $Z_V$ in LQCD we have (see e.g. Ref.~\cite{Bhattacharya:2005rb})
\bea
&& [\bar c \gamma_\mu c]_R(x) = \frac{1}{4} ([V_\mu^{(0)}]_R(x) + [V_\mu^{(15)}]_R(x) ) = \nonumber \\
&& = \frac{1}{4} (Z_{V^0} + 3 Z_{V}) [\bar c \gamma_\mu c](x) +
\frac{1}{4} (Z_{V^0}-Z_{V}) [\bar u \gamma_\mu u +\bar d \gamma_\mu d +\bar s \gamma_\mu s](x)
\; . \label{charmVloc} 
\eea 
Inserting the corresponding charm vector charge in correlation functions of local fields, $\Phi_\alpha$, interpolating states with zero charm number and non-zero baryon number, say e.g.\ proton states, i.e. $\Phi_\alpha(y) = [(\bar u^C \gamma_5 d) u]_\alpha(y)$, the conservation of the charm number charge $Q_V^c$ evidently implies
\bea
&& \!\!\!\!\!\!\!\! 0 = 
\langle \Phi_\beta(z) a^3\sum_{\vec x}[\bar c \gamma_0 c]_R(x) \Phi_\alpha^\dagger(y) \rangle
= \frac{1}{4} (Z_{V^0} + 3 Z_{V}) \langle \Phi_\beta(z) a^3\sum_{\vec x} [\bar c \gamma_0 c](x)  \Phi_\alpha^\dagger(y) \rangle +  \nonumber \\
&& 
 \!\!\!\!\!\!\!\! + \frac{1}{4} (Z_{V^0}-Z_{V}) \langle \Phi_\beta(z) a^3\sum_{\vec x} [\bar u \gamma_0 u + \bar d \gamma_0 d + \bar s \gamma_0 s]_R(x)   \Phi_\alpha^\dagger(y) \rangle \; .
\label{step1rel} 
\eea 
Since the correlator with coefficient $(Z_{V^0} - Z_{V})$ is non-vanishing (already at the classical level), we see that the insertion of the bare charm charge  $a^3\sum_{\vec x} [\bar c \gamma_0 c](x)$ between operators containing no $c$ (valence) quark vanishes if and only if $Z_{V^0}-Z_{V} = 0$. As  there is no loss of generality in taking $h=c$ the statement of Step 1) is hence proved within LQCD.

\vspace{0.5cm}

{\bf Step 2)} \\
We now want to prove that in LQCD with $N_f=4 $ mass degenerate quarks one has
\be
\langle \Phi_\beta(z) a^3\sum_{\vec x} [\bar c \gamma_0 c](x)  \Phi_\alpha^\dagger(y) 
\rangle_{LQCD} = 0 
\label{step2rel} \ee
with $\langle \dots \rangle_{LQCD}$ reminding that the lattice path integral is evaluated with the action~(\ref{Waction}).

A key point is that the plain LQCD correlator in the l.h.s.\ of Eq.~(\ref{step2rel}) satisfies the identity
\be
\langle \Phi_\beta(z) a^3\sum_{\vec x} [\bar c \gamma_0 c](x)  \Phi_\alpha^\dagger(y)
\rangle_{LQCD} \; =  \;
\langle \Phi_\beta^\prime(z) a^3\sum_{\vec x} [\bar c \gamma_0 c](x)  \Phi_\alpha^{\prime \, \dagger}(y)
\rangle_{MALQCD}
\label{correlID1} 
\ee
where the correlator on the r.h.s.\ is instead evaluated in a mixed action LQCD setup, as reminded by the notation $\langle \dots \rangle_{MALQCD}$, with lattice action (recall Eq.~(\ref{degqmass}))
\bea
\!\!\!\!\!\!&& S_{MALQCD} = S_{YM}[U] + a^4 \sum_x \sum_{f=u,d,s,c} \bar f(x) [D^W_{\rm cr}(U)+ m\,] f(x) + \nonumber \\
\!\!\!\!\!\!&& + a^4 \sum_x \sum_{f' = u', d', ...} \left\{ \bar f'(x) [D^W_{\rm cr}(U)+ m\,] f'(x) + 
\bar \chi_{f'}(x) [D^W_{\rm cr}(U)+ m\, ] \chi_{f'}(x) \right\} \; ,
\label{SMAWLQCD} 
\eea 
where $u'$, $d'$, ...~are mere valence quark fields and $\chi_{u'}$, $\chi_{u'}$, ...~are the corresponding valence ghost spin-1/2 fields (obeying Bose statistics so as to cancel all virtual sea contributions from the ``primed'' fields), while the proton interpolating valence operator is $\Phi_\alpha^\prime(y) = [(\bar u^{\prime \, C} \gamma_5 d^\prime) u^\prime]_\alpha(y)$. 
It is known~\cite{Frezzotti:2004wz} that the critical mass parameter $m_{\rm cr}$ for Wilson lattice valence quark and ghost fields coincides with the one for plain LQCD Wilson quarks, so that $D^W_{\rm cr}(U)$ in Eq.~(\ref{SMAWLQCD}) is the same lattice Dirac operator as in Eq.~(\ref{Waction}).

The identity~(\ref{correlID1}) follows from the fact the plain LQCD and the mixed action LQCD correlators, in view of the specific flavour content of the fields involved, give rise to identical Wick contractions. Indeed by construction the plain LQCD and the mixed action LQCD formulations lead to identical vertices and identical fermion propagators, evaluated on identical gauge configurations, because in both lattice setups the gauge effective action is given by $S_L^{eff}[U] = S_{YM}[U] - \sum_{f=u,d,s,c}\log \det [(D^W_{\rm cr}(U)+ m \, ]$. 

Moreover the identity ~(\ref{correlID1}) evidently implies an analogous identity where the charge insertion $a^3 \sum_{\vec{x}} [\bar c \gamma_0 c](x)$ is replaced by $a^3 \sum_{\vec{x}} \frac{1}{4} \sum_{f=u,d,s,c} [\bar f \gamma_0 f](x)$:
\be
\langle \Phi_\beta(z) a^3\sum_{\vec x} [\bar c \gamma_0 c](x)  \Phi_\alpha^\dagger(y)
\rangle_{LQCD} \; =  \;
\langle \Phi_\beta^\prime(z) a^3\sum_{\vec x} \frac{1}{4} \sum_{f=u,d,s,c} [\bar f \gamma_0 f](x) \Phi_\alpha^{\prime \, \dagger}(y) \rangle_{MALQCD} \; .
\label{correlID1bis} 
\ee

On the other hand, in the MALQCD setup, owing to the {\em exact invariance} of the action under $U(1)$ vector transformations acting only on the fields $u, \, d, \, s, \, c, \bar u, \,\bar d,  \,\bar s, \,\bar c$, there exists a {\em conserved} baryon charge, given by either the one-point split current $\hat V_0^{(0)}$ (see Eq.\,(\ref{exSVcurr})) or the local (multiplicatively renormalized through $Z_{V^0}$) vector current $V_0^{(0)} = \sum_{f=u,d,s,c} \left[ \bar f(x) \gamma_0 f(x) \right]$:\footnote{On the contrary in LQCD one cannot say that the point-like current $\bar c \gamma_0 c$ is only multiplicatively renormalized, unless $Z_{V^0} = Z_V$ holds (see Eq.\,(\ref{charmVloc})).} 
\be 
 Q_{\rm bar}^{u,d,s,c} = a^3 \sum_{\vec{x}} \hat V_0^{0} (x) \frac{1}{3} 
= a^3 \sum_{\vec{x}} \sum_{f=u,d,s,c} [\bar f \gamma_0 f](x) Z_{V^0}  \frac{1}{3} \, . 
\label{QbarNf4MA} 
\ee 
The identity~(\ref{correlID1bis}) can hence be cast in a form where the occurrence of {\em conserved} charge $Q_{\rm bar}^{u,d,s,c}$ is explicit, i.e.\
\be
\langle \Phi_\beta(z) a^3\sum_{\vec x} [\bar c \gamma_0 c](x)  \Phi_\alpha^\dagger(y)
\rangle_{LQCD} \; =  \;
\frac{3}{4 Z_{V^0}}
\langle \Phi_\beta^\prime(z) \, Q_{\rm bar}^{u,d,s,c} \, \Phi_\alpha^{\prime \, \dagger}(y)
\rangle_{MALQCD} \; = \; 0 \; ,
\label{correlID2} 
\ee
and the last (key) equality follows from the fact that the operators $\Phi_\alpha^{\prime \, \dagger}$ and $\Phi_\beta^\prime$ involving only valence quark fields $u'$ and $d'$ 
commute with $Q_{\rm bar}^{u,d,s,c}$. 
Indeed, inserting intermediate states in the MALQCD correlator of Eq.~(\ref{correlID2}) it
is clear that all the states created by the action of $\Phi_\alpha^{\prime \, \dagger}$ are
inert under the action of $Q_{\rm bar}^{u,d,s,c}$-charge~\footnote{This property holds due to exact conservation of $Q_{\rm bar}^{u,d,s,c}$ in the MALQCD renormalizable (but non-unitary) theory even if the underlying Hilbert-Fock space globally has (owing to states with $\chi_{f'}$ ghosts) an indefinite metric.}. The statement of Step 2), i.e.~Eq.\,(\ref{step2rel}), is thus proven.

Combining Step~2) with Step~1) one concludes that $Z_{V^0} = Z_{V}$ in LQCD with Wilson quarks.

An alternative proof of $Z_{V^0} = Z_{V}$ might be given by relying on large $N_c$ arguments in Wilson lattice $SU(N_c)$-QCD. Working at arbitrary values of $N_c$ and at fixed values of the renormalized coupling $u = g_R^2 N_C$, one can infer from the existence of exactly conserved $U(1)$ and $SU(4)$ vector charges the vanishing of the {\em quark disconnected} Wick contractions that would otherwise lead to $Z_{V^0} \neq Z_{V}$. We omit here the details of such a proof, which, although being technically different, appears conceptually equivalent to the one given above.


A comment is in order about why similar relations, of the form $Z_{\Gamma^0} = Z_{\Gamma}$, are not expected to hold in general for $\Gamma = A,\, S,\,P, \,T$, i.e.\ for bilinear operator other than vector ones, at least in LQCD with Wilson quarks or in other lattice formulations breaking chiral symmetries. This situation is at variance with respect to what happens in UV regularizations respecting all the non-anomalous chiral symmetries (such as lattice QCD with overlap quarks), where it is known that $Z_A = Z_V = Z_{V^0}=1$,
while $Z_P = Z_{S^0}$, $Z_S= Z_{P^0}$ and, owing to identical multiplicative renormalization of all quark masses, $Z_S = Z_{S^0}$. 

For the case of $\Gamma=V$, that we discussed above, our proof of the relations~(\ref{step1rel}) and~(\ref{step2rel}) relies on the fact that even in a lattice formulation breaking chiral symmetries the flavour singlet and non-singlet vector transformations are exact invariances of the lattice action, enabling to define single-flavour conserved charges for each flavour $f=u,\, d\, ,s\, ,c$.  
The existence of such conserved charges was in fact exploited to prove the vanishing of the quantities in Eqs.\,(\ref{step1rel}) and\,(\ref{step2rel}).

But similar symmetry properties hold true neither for the axial currents (case $\Gamma=A$), owing to the $U_A(1)$ anomaly in the flavour singlet sector, nor for the scalar ($\Gamma=S$), pseudoscalar ($\Gamma = P$) and tensor ($\Gamma=T$) densities, which are not related to any conserved currents, too. 




\section{Free-theory calculation of the leading lattice artifacts at short distance}
\label{sec:appE}

In this appendix we will show some of the details of the calculation of the $a^{2}/t^{2}$ lattice artifacts appearing in the vector correlator at short distance. The calculation is performed in lattice perturbation theory at order $\alpha_{s}^{0}$ with $N_{f}=2$ massless fermions. The approach which we use is similar to the one adopted in Ref.~\cite{Ce:2021xgd}, where the free-theory isovector correlator was computed using one local and one conserved current. Here we analyze the case in which $V_{ud}(t)$ is computed using both the twisted-mass and the Osterwalder-Seiler local currents of Eq.~(\ref{eq:MAVcurr}).   
For non-interacting massless twisted-mass fermions, the up/down quark propagator is given by:
\begin{align}
\label{eq:prop_mom}
\langle \psi_{\ell}(p)\bar{\psi}_{\ell}(-p)\rangle = \frac{-i\gamma_{\mu}\tilde{p}_{\mu} -ir\gamma_{5}\frac{a}{2}\sum_{\mu}\hat{p}_{\mu}^{2}}{ \sum_{\mu} \tilde{p}_{\mu}^{2} + \frac{a^{2}}{4}\left( \sum_{\mu}\hat{p}_{\mu}^{2}\right)^{2} }  ~,   
\end{align}
where $r=1$ if $\ell=u$ and $r=-1$ if $\ell = d$, and
\begin{align}
\tilde{p}_{\mu} = \frac{1}{a}\sin{(ap_{\mu})},\qquad \hat{p}_{\mu} = \frac{2}{a}\sin{(\frac{ap_{\mu}}{2})}~.   
\end{align}
The coordinate-space quark propagator is then given by
\begin{align}
\label{eq:prop_coord}
\langle \psi_{\ell}(x)\bar{\psi}_{\ell}(y)\rangle = \int_{-\frac{\pi}{a}}^{\frac{\pi}{a}} \frac{dp_{0}}{2\pi} \int_{-\frac{\pi}{a}}^{\frac{\pi}{a}} \frac{d^{3}\bm{p}}{(2\pi)^{3}}e^{ip_{0}t} e^{i\bm{p}\cdot(\bm{x-y})}\cdot\frac{-i\gamma_{\mu}\tilde{p}_{\mu} -ir\gamma_{5}\frac{a}{2}\sum_{\mu}\hat{p}_{\mu}^{2}}{ \sum_{\mu} \tilde{p}_{\mu}^{2} + \frac{a^{2}}{4}\left( \sum_{\mu}\hat{p}_{\mu}^{2}\right)^{2} }~,
\end{align}
where $t = x_{0}-y_{0}$. The integral over $p_{0}$ can be computed exactly using the {bf residue} theorem. The denominator in Eq.~(\ref{eq:prop_mom}) can be written as:
\begin{align}
\sum_{\mu} \left[ \tilde{p}_{\mu}^{2} + \frac{a^{2}}{4}\left( \sum_{\mu}\hat{p}_{\mu}^{2}\right)^{2}\right] = -\frac{2}{a^{2}}A(\bm{p})\left( \cosh{( iap_{0})} - \cosh{(aE_{\bm{p}})}\right)~,  
\end{align}
where $A(\bm{p})$ and $E_{\bm{p}}$ are defined as
\bea
A(\bm{p}) & = & 1 + \frac{1}{2}a^{2}\hat{\bm{p}}^{2} ~ , ~ \nonumber \\ 
\cosh{(aE_{\bm{p}})}& = & \sqrt{ \frac{a^{2}B(\bm{p})(4A(\bm{p}) + a^{2}B(\bm{p}))}{4A^{2}(\bm{p})} + 1 } ~ , ~ \\
B(\bm{p}) & = & \hat{\bm{p}}^{2} + \frac{a^{2}}{2}\sum_{i<j} \hat{p}_{i}^{2}\hat{p}_{j}^{2} ~ . ~ \nonumber  
\eea
The momentum-space lattice quark propagator has two poles in the complex plane at $ip_{0} = \pm E_{\bm{p}}$, and the corresponding residue can be computed using
\begin{align}
D^{-1}(\bm{p}) \equiv \lim_{ip_{0} \to  E_{\bm{p}}} a^{2}\frac{(ip_{0} - E_{\bm{p}})}{ 2A(\bm{p})\left( \cosh{( iap_{0})} - \cosh{(aE_{\bm{p}})}\right)  } =  \frac{1}{\sqrt{B(\bm{p})\cdot\left( 4A(\bm{p}) + a^{2}B(\bm{p})\right)}}~.   
\end{align}
Using the previous results,  Eq.~(\ref{eq:prop_coord}) can be written as
\bea
\label{eq:corr_res}
\langle \psi_{\ell}(x)\bar{\psi}_{\ell}(y)\rangle & = & \int_{-\frac{\pi}{a}}^{\frac{\pi}{a}} \frac{d^{3}\bm{p}}{(2\pi)^{3}}e^{-E_{\bm{p}}|t|}~\frac{e^{i\bm{p}\cdot(\bm{x-y})}}{D(\bm{p})} \nonumber \\
& \cdot & \left[ \textrm{sgn}(t)\frac{\gamma_{0}}{a}\sinh{(a E_{\bm{p}})} -i\bm{\gamma}\cdot\tilde{\bm{p}} - ir\gamma_{5}\frac{a}{2}\left( \hat{\bm{p}}^{2} - \frac{ B(\bm{p})}{A(\bm{p})}\right)\right] ~ ,  ~  
\eea
which is valid for $t\ne 0$. The light-connected vector correlator $V_{ud}(t)$ can be readily computed using Eq.~(\ref{eq:corr_res}). The result is
\begin{align}
\label{eq:V_free}
V_{ud}(t) = 4N_{c}(q_{em, u}^{2}+q_{em,d}^{2})\int_{-\frac{\pi}{a}}^{\frac{\pi}{a}} \frac{d^{3}\bm{p}}{(2\pi)^{3}} \frac{e^{-2E_{\bm{p}}|t|}}{D^{2}(\bm{p})}\left[ \frac{1}{a^{2}}\sinh^{2}{(aE_{\bm{p}})} +\frac{1}{3}\tilde{\bm{p}}^{2} \pm \frac{a^{2}}{4}\left(\hat{\bm{p}}^{2} - \frac{ B(\bm{p})}{A(\bm{p})}    \right)^{2}      \right]~,    \end{align}
where the plus sign corresponds to the result obtained using the current $J_{\mu}^{\rm OS}$, while the minus sign to the one obtained using $J_{\mu}^{\rm tm}$. The dangerous $\mathcal{O}(a^{2}\log({a}))$ artifacts in $a_{\mu}^{\rm SD}$ (see the discussion in Sec.~\ref{sec:amuSD}), which are generated upon integration in the short-distance window, stem from the $\mathcal{O}(a^{2}/t^{2})$ artifacts of the vector correlator. Expanding the integrand of Eq.~(\ref{eq:V_free}) in powers of the lattice spacing up to order $\mathcal{O}(a^{2})$, we get
\begin{align}
\label{eq:corr_expanded}
V_{ud}(t) = 4N_{c}(q_{em,u}^{2}+q_{em,d}^{2})\int_{-\infty}^{\infty}\frac{d^{3}\bm{p}}{(2\pi)^{3}} e^{-2|\bm{p}|\cdot|t|}\left[~\frac{1}{3}  -\frac{1}{3}a^{2}|\bm{p}|^{2} + a^{2}|\bm{p}|^{3}|t|G(\bm{p})  + \mathcal{O}(a^{4})         ~\right]~,   
\end{align}
where $G(\bm{p})$ is a dimensionless function given by
\begin{align}
G(\bm{p}) = \frac{2}{9}\left( 1 - \frac{ \sum_{i < j} p^{2}_{i}p^{2}_{j}}{|\bm{p}|^{4}}\right)~.  
\end{align}
The result of Eq.~(\ref{eq:corr_expanded}) does not depend upon the chosen values of $r$, and therefore the current $J_{\mu}^{\rm OS}$ and $J_{\mu}^{\rm tm}$ produce the same $a^{2}\log{(a)}$ discretization effects in $a_{\mu}^{\rm SD}$. The integrals appearing in Eq.~(\ref{eq:corr_expanded}) can be computed analytically. We obtain:
\begin{align}
V_{ud}(t) = (q_{em,u}^{2} + q_{em,d}^{2})\cdot \frac{ 4N_{c}}{24\pi^{2}}\cdot\frac{1}{t^{3}}\cdot\left( 1+ \frac{a^{2}}{t^{2}} + \mathcal{O}(a^{4})  \right)~.    
\end{align}


\section{Parameterization of FSEs in the windows}
\label{sec:appF}

Following Ref.~\cite{Giusti:2018mdh} the isovector part of the correlator $V_{ud}(t)$ can be analytically represented as the sum of two terms, $V_{dual}(t) + V_{\pi \pi}(t) $, where $V_{\pi \pi}(t)$ represents the two-pion contribution in a finite box, while $V_{dual}(t)$ is the ``dual'' representation of the tower of the contributions coming from the excited states above the two-pion ones.
Therefore, $V_{\pi \pi}(t)$ is expected to dominate at large and intermediate time distances, let's say $t \gtrsim 1$ fm, while the contribution of $V_{dual}(t)$ is important at short time distances, as firstly observed in Ref.~\cite{Giusti:2017jof}.
The FSEs on the correlator $V_{ud}(t)$ were analyzed in Ref.~\cite{Giusti:2018mdh} using the above representation and it was found that the main contribution comes from the two-pion states. Thus, we make use of these findings to construct our parameterization of FSEs for the windows.

As it is well known after Refs.~\cite{Luscher:1985dn, Luscher:1986pf, Luscher:1990ux, Luscher:1991cf}, the energy levels $\omega_n$ of two pions in a finite box of volume $L^3$ are given by
 \be
     \omega_n = 2 \sqrt{M_\pi^2 + k_n^2} ~ ,
     \label{eq:omegan}
 \ee
where the discretized values $k_n$ should satisfy the L\"uscher condition, which for the case at hand (two pions in a $P$-wave with total isospin $1$) reads as 
 \be
     \delta_{11}(k_n) + \phi\left( \frac{k_nL}{2\pi} \right) = n \pi ~ ,
     \label{eq:kn}
 \ee
with $\delta_{11}$ being the (infinite volume) scattering phase shift and $\phi(z)$ a known kinematical function given by
\be
    \mbox{tan}\phi(z) = - \frac{2 \pi^2 z}{\sum_{\vec{m} \in \mathcal{Z}^3} \left( |\vec{m}|^2 - z^2 \right)^{-1}} ~ .
    \label{eq:phi}
\ee

The two-pion contribution $V_{\pi \pi}(t)$ can be written as~\cite{Lellouch:2000pv, Meyer:2011um, Francis:2013fzp}
 \be
     V_{\pi \pi}(t) = \sum_n \nu_n |A_n|^2 e^{-\omega_n t} ~ ,
     \label{eq:V2pi}
 \ee
where $\nu_n$ is the number of vectors $\vec{z} \in \mathcal{Z}^3$ with norm $|\vec{z}|^2 = n$ and the squared amplitudes $|A_n|^2$ are related to the timelike pion form factor $F_\pi(\omega_{n}) = |F_\pi(\omega_n)| e^{i\delta_{11}(k_n)}$ by
 \be
     \nu_n |A_n|^2 = \frac{2 k_n^5}{3 \pi \omega_n^2} |F_\pi(\omega_n)|^2\left[ k_n \delta_{11}^\prime(k_n) + 
                               \frac{k_nL}{2\pi} \phi^\prime\left( \frac{k_nL}{2\pi} \right) \right]^{-1} ~ .
     \label{eq:An}
 \ee
Following Ref.~\cite{Giusti:2018mdh} we adopt the Gounaris-Sakurai (GS) parameterization~\cite{Gounaris:1968mw} of the timelike pion form factor $F_\pi(\omega_{n}) = |F_\pi(\omega_n)| e^{i\delta_{11}(k_{n})}$, where the form factor phase coincides with the scattering phase shift according to the Watson theorem. The GS Ansatz is based on the dominance of the $\rho$ resonance in the amplitude of the pion-pion P-wave elastic scattering (with total isospin $1$), namely
\be
     F_\pi^{(GS)}(\omega) = \frac{M_\rho^2 - A_{\pi \pi}(0)}{M_\rho^2 - \omega^2 - A_{\pi \pi}(\omega)} ~ ,
     \label{eq:Fpi_GS}
\ee
where the (twice-subtracted~\cite{Gounaris:1968mw}) pion-pion amplitude $A_{\pi \pi}(\omega)$ is given by
\be
    A_{\pi \pi}(\omega) = h(M_\rho) + (\omega^2 - M_\rho^2) \frac{h^\prime(M_\rho)}{2 M_\rho} - h(\omega) + i \omega \Gamma_{\rho \pi \pi}(\omega)
    \label{eq:Apipi}
\ee
with
\bea
    \label{eq:Gamma_rhopipi}
    \Gamma_{\rho \pi \pi}(\omega) & = & \frac{g_{\rho \pi \pi}^2}{6 \pi} \frac{k^3}{\omega^2} ~ , \\[2mm]
    \label{eq:homega}
    h(\omega) & = & \frac{g_{\rho \pi \pi}^2}{6 \pi} \frac{k^3}{\omega} \frac{2}{\pi}\mbox{log}\left( \frac{\omega + 2k}{2M_\pi} \right) ~ , \\[2mm]
    \label{eq:hpomega}
    h^\prime(\omega) & = & \frac{g_{\rho \pi \pi}^2}{6 \pi} \frac{k^2}{\pi \omega} \left\{ 1 + \left(1 + \frac{2 M_\pi^2}{\omega^2} \right) 
                                           \frac{\omega}{k} \mbox{log}\left(\frac{\omega + 2k}{2 M_\pi} \right) \right\} ~ , \\[2mm]
    \label{eq:Apipi0}
     A_{\pi \pi}(0) & = & h(M_\rho) - \frac{M_\rho}{2} h^\prime(M_\rho) + \frac{g_{\rho \pi \pi}^2}{6 \pi} \frac{M_\pi^2}{\pi}                                   
\eea
and $k \equiv \sqrt{\omega^2 / 4 - M_\pi^2}$.
By analytic continuation the GS form factor at $\omega = 0$ is normalized to unity, i.e.~$F_\pi^{(GS)}(\omega = 0) = 1$.
The scattering phase shift $\delta_{11}(k)$, i.e.~the phase of the pion form factor according to the Watson theorem, is given by
\be
    \mbox{cot}\delta_{11}(k) = \frac{M_\rho^2 - \omega^2 - h(M_\rho) - (\omega^2 - M_\rho^2) h^\prime(M_\rho) / (2 M_\rho) +
                                              h(\omega)}{\omega \Gamma_{\rho \pi \pi}(\omega)} ~ .
    \label{eq:delta11}
\ee

The GS form factor (\ref{eq:Fpi_GS}) contains two parameters: the resonance mass $M_\rho$ and its strong coupling with two pions $g_{\rho \pi \pi}$. Since the ETM ensembles of Table\,\ref{tab:simudetails} are quite close to the physical pion point, we fix the $\rho$ mass and the strong coupling $g_{\rho \pi \pi}$ at their physical values, namely $M_\rho = 0.775$ GeV and $g_{\rho \pi \pi} = \sqrt{48 M_\rho^2 \Gamma_\rho / (M_\rho^2 - 4 M_\pi^2)^{3/2}} = 5.95$\,\cite{ParticleDataGroup:2020ssz}. 

As well known, the infinite volume limit of Eq.\,(\ref{eq:V2pi}) is given by
\bea
     V_{\pi \pi}^\infty(t) = \frac{1}{48 \pi^2} \int_{2M_{\pi}}^\infty d\omega ~ \omega^2 
     \left[ 1 - \frac{4 M_{\pi}^2}{\omega^2} \right]^{3/2} |F_{\pi}(\omega)|^2 e^{-\omega t} ~ 
     \label{eq:V2pi_vol}
\eea
and, therefore, the FSEs on the window contribution $a_\mu^w(\ell)$ for $w = \{\rm SD, W, LD \}$ can be written as 
\be
    \label{eq:Delta_amuw}
    \Delta a_\mu^w(L) \equiv a_\mu^w(\ell) \Bigl|_L - a_\mu^w(\ell) \Bigl|_{L = \infty} = 2 \alpha_{em}^2 \frac{10}{9} \int_0^\infty dt \, t^2 \, K(m_\mu t) \, \Theta^w(t) \left[ V_{\pi \pi}(t) - V_{\pi \pi}^\infty(t)\right] ~ , ~
\ee
where the charge factor $10/9$ takes into account the proportionality between the light-quark connected and the isovector correlators in isosymmetric QCD, while the correlators $V_{\pi \pi}(t)$ and $V_{\pi \pi}^\infty(t)$ are given by Eqs.\,(\ref{eq:V2pi}) and (\ref{eq:V2pi_vol}), respectively.

As a check of our parameterization\,(\ref{eq:Delta_amuw}) we consider the estimate of the (continuum) FSEs in the isovector channel made by the BMW Collaboration in the intermediate window\,\cite{Borsanyi:2020mff}, viz.
\be
    \label{eq:Delta_amuW_BMW}
    \Delta a_\mu^{{\rm W}; I=1}(L_{ref}^{\rm BMW}) = a_\mu^{{\rm W}; I=1}(\ell) \Bigl|_{L_{ref}^{\rm BMW}} - a_\mu^{{\rm W};I=1}(\ell) \Bigl|_{L = \infty} = - 0.49(2)(4) \cdot 10^{-10} ~
\ee
with $L_{ref}^{\rm BMW}= 6.272$ fm. 
Using $50$ two-pion states in Eq.\,(\ref{eq:V2pi_vol}) at $L = L_{ref}^{\rm BMW}$ we obtain
\begin{align}
\Delta a_{\mu}^{{\rm W};I=1}(L_{ref}^{\rm BMW}) = a_{\mu}^{{\rm W};I=1}(\ell) \Bigl|_{L_{ref}^{\rm BMW}} - a_\mu^{{\rm W};I=1}(\ell) \Bigl|_{L = \infty} = - 0.37 \cdot 10^{-10}~,     
\end{align}
which roughly corresponds to $75~(7)\%$ of the BMW result of Eq.~(\ref{eq:Delta_amuW_BMW}). We devise to extrapolate to the infinite volume limit, employing the GS parameterization. However to take into account the deviation from the BMW result, we enhance the GS correlator by a factor $1.25$ and associate to this correction a relative error of $20\%$.

For completeness we report our determination of $\Delta a_\mu^{\rm W}(L_{ref})$ evaluated using Eq.\,(\ref{eq:Delta_amuw}) for the intermediate window at the physical pion mass point and at the reference lattice size $L_{ref} = 5.46$ fm, namely
\be
   \Delta a_\mu^{\rm W}(L_{ref}) = -1.00 ~ (20) \cdot 10^{-10} ~ , ~
\ee
which is used in Section\,\ref{sec:amuW} to correct FSEs on our data for $a_\mu^W(\ell, L_{ref})$.


%
%
%


\bibliography{biblio}

\providecommand{\href}[2]{#2}\begingroup\raggedright\begin{thebibliography}{10}

\bibitem{RBC:2018dos}
{\scshape RBC, UKQCD} collaboration, \emph{{Calculation of the hadronic vacuum
  polarization contribution to the muon anomalous magnetic moment}},
  \href{https://doi.org/10.1103/PhysRevLett.121.022003}{\emph{Phys. Rev. Lett.}
  {\bfseries 121} (2018) 022003}
  [\href{https://arxiv.org/abs/1801.07224}{{\ttfamily 1801.07224}}].

\bibitem{Colangelo:2022vok}
G.~Colangelo, A.~El-Khadra, M.~Hoferichter, A.~Keshavarzi, C.~Lehner,
  P.~Stoffer et~al., \emph{Data-driven evaluations of euclidean windows to
  scrutinize hadronic vacuum polarization},
  \href{https://doi.org/https://doi.org/10.1016/j.physletb.2022.137313}{\emph{Physics
  Letters B} {\bfseries 833} (2022) 137313}.

\bibitem{Borsanyi:2020mff}
S.~Borsanyi et~al., \emph{{Leading hadronic contribution to the muon magnetic
  moment from lattice QCD}},
  \href{https://doi.org/10.1038/s41586-021-03418-1}{\emph{Nature} {\bfseries
  593} (2021) 51} [\href{https://arxiv.org/abs/2002.12347}{{\ttfamily
  2002.12347}}].

\bibitem{Ce:2022kxy}
M.~C\`e et~al., \emph{{Window observable for the hadronic vacuum polarization
  contribution to the muon $g-2$ from lattice QCD}},
  \href{https://arxiv.org/abs/2206.06582}{{\ttfamily 2206.06582}}.

\bibitem{Muong-2:2021ojo}
{\scshape Muon g-2} collaboration, \emph{{Measurement of the Positive Muon
  Anomalous Magnetic Moment to 0.46 ppm}},
  \href{https://doi.org/10.1103/PhysRevLett.126.141801}{\emph{Phys. Rev. Lett.}
  {\bfseries 126} (2021) 141801}
  [\href{https://arxiv.org/abs/2104.03281}{{\ttfamily 2104.03281}}].

\bibitem{Muong-2:2021ovs}
{\scshape Muon g-2} collaboration, \emph{{Magnetic-field measurement and
  analysis for the Muon $g-2$ Experiment at Fermilab}},
  \href{https://doi.org/10.1103/PhysRevA.103.042208}{\emph{Phys. Rev. A}
  {\bfseries 103} (2021) 042208}
  [\href{https://arxiv.org/abs/2104.03201}{{\ttfamily 2104.03201}}].

\bibitem{Muong-2:2021xzz}
{\scshape Muon g-2} collaboration, \emph{{Beam dynamics corrections to the
  Run-1 measurement of the muon anomalous magnetic moment at Fermilab}},
  \href{https://doi.org/10.1103/PhysRevAccelBeams.24.044002}{\emph{Phys. Rev.
  Accel. Beams} {\bfseries 24} (2021) 044002}
  [\href{https://arxiv.org/abs/2104.03240}{{\ttfamily 2104.03240}}].

\bibitem{Muong-2:2021vma}
{\scshape Muon g-2} collaboration, \emph{{Measurement of the anomalous
  precession frequency of the muon in the Fermilab Muon $g-2$ Experiment}},
  \href{https://doi.org/10.1103/PhysRevD.103.072002}{\emph{Phys. Rev. D}
  {\bfseries 103} (2021) 072002}
  [\href{https://arxiv.org/abs/2104.03247}{{\ttfamily 2104.03247}}].

\bibitem{Muong-2:2006rrc}
{\scshape Muon g-2} collaboration, \emph{{Final Report of the Muon E821
  Anomalous Magnetic Moment Measurement at BNL}},
  \href{https://doi.org/10.1103/PhysRevD.73.072003}{\emph{Phys. Rev. D}
  {\bfseries 73} (2006) 072003}
  [\href{https://arxiv.org/abs/hep-ex/0602035}{{\ttfamily hep-ex/0602035}}].

\bibitem{Abe:2019thb}
M.~Abe et~al., \emph{{A New Approach for Measuring the Muon Anomalous Magnetic
  Moment and Electric Dipole Moment}},
  \href{https://doi.org/10.1093/ptep/ptz030}{\emph{PTEP} {\bfseries 2019}
  (2019) 053C02} [\href{https://arxiv.org/abs/1901.03047}{{\ttfamily
  1901.03047}}].

\bibitem{Davier:2017zfy}
M.~Davier, A.~Hoecker, B.~Malaescu and Z.~Zhang, \emph{{Reevaluation of the
  hadronic vacuum polarisation contributions to the Standard Model predictions
  of the muon $g-2$ and ${\alpha (m_Z^2)}$ using newest hadronic cross-section
  data}}, \href{https://doi.org/10.1140/epjc/s10052-017-5161-6}{\emph{Eur.
  Phys. J. C} {\bfseries 77} (2017) 827}
  [\href{https://arxiv.org/abs/1706.09436}{{\ttfamily 1706.09436}}].

\bibitem{Keshavarzi:2018mgv}
A.~Keshavarzi, D.~Nomura and T.~Teubner, \emph{{Muon $g-2$ and $\alpha(M_Z^2)$:
  a new data-based analysis}},
  \href{https://doi.org/10.1103/PhysRevD.97.114025}{\emph{Phys. Rev. D}
  {\bfseries 97} (2018) 114025}
  [\href{https://arxiv.org/abs/1802.02995}{{\ttfamily 1802.02995}}].

\bibitem{Colangelo:2018mtw}
G.~Colangelo, M.~Hoferichter and P.~Stoffer, \emph{{Two-pion contribution to
  hadronic vacuum polarization}},
  \href{https://doi.org/10.1007/JHEP02(2019)006}{\emph{JHEP} {\bfseries 02}
  (2019) 006} [\href{https://arxiv.org/abs/1810.00007}{{\ttfamily
  1810.00007}}].

\bibitem{Hoferichter:2019mqg}
M.~Hoferichter, B.-L.~Hoid and B.~Kubis, \emph{{Three-pion contribution to
  hadronic vacuum polarization}},
  \href{https://doi.org/10.1007/JHEP08(2019)137}{\emph{JHEP} {\bfseries 08}
  (2019) 137} [\href{https://arxiv.org/abs/1907.01556}{{\ttfamily
  1907.01556}}].

\bibitem{Keshavarzi:2019abf}
A.~Keshavarzi, D.~Nomura and T.~Teubner, \emph{{$g-2$ of charged leptons,
  $\alpha (M^2_Z)$ , and the hyperfine splitting of muonium}},
  \href{https://doi.org/10.1103/PhysRevD.101.014029}{\emph{Phys. Rev. D}
  {\bfseries 101} (2020) 014029}
  [\href{https://arxiv.org/abs/1911.00367}{{\ttfamily 1911.00367}}].

\bibitem{Davier:2019can}
M.~Davier, A.~Hoecker, B.~Malaescu and Z.~Zhang, \emph{{A new evaluation of the
  hadronic vacuum polarisation contributions to the muon anomalous magnetic
  moment and to $\mathbf{\boldsymbol\alpha(m_Z^2)}$}},
  \href{https://doi.org/10.1140/epjc/s10052-020-7792-2}{\emph{Eur. Phys. J. C}
  {\bfseries 80} (2020) 241}
  [\href{https://arxiv.org/abs/1908.00921}{{\ttfamily 1908.00921}}].

\bibitem{Aoyama:2020ynm}
T.~Aoyama et~al., \emph{{The anomalous magnetic moment of the muon in the
  Standard Model}},
  \href{https://doi.org/10.1016/j.physrep.2020.07.006}{\emph{Phys. Rept.}
  {\bfseries 887} (2020) 1} [\href{https://arxiv.org/abs/2006.04822}{{\ttfamily
  2006.04822}}].

\bibitem{Alexandrou:2018egz}
C.~Alexandrou et~al., \emph{{Simulating twisted mass fermions at physical
  light, strange and charm quark masses}},
  \href{https://doi.org/10.1103/PhysRevD.98.054518}{\emph{Phys. Rev. D}
  {\bfseries 98} (2018) 054518}
  [\href{https://arxiv.org/abs/1807.00495}{{\ttfamily 1807.00495}}].

\bibitem{ExtendedTwistedMass:2020tvp}
{\scshape Extended Twisted Mass} collaboration, \emph{{Quark masses and decay
  constants in $N_f=2+1+1$ isoQCD with Wilson clover twisted mass fermions}},
  \href{https://doi.org/10.22323/1.363.0181}{\emph{PoS} {\bfseries LATTICE2019}
  (2020) 181} [\href{https://arxiv.org/abs/2001.09116}{{\ttfamily
  2001.09116}}].

\bibitem{ExtendedTwistedMass:2021qui}
{\scshape Extended Twisted Mass} collaboration, \emph{{Ratio of kaon and pion
  leptonic decay constants with Nf=2+1+1 Wilson-clover twisted-mass fermions}},
  \href{https://doi.org/10.1103/PhysRevD.104.074520}{\emph{Phys. Rev. D}
  {\bfseries 104} (2021) 074520}
  [\href{https://arxiv.org/abs/2104.06747}{{\ttfamily 2104.06747}}].

\bibitem{Finkenrath:2022eon}
J.~Finkenrath et~al., \emph{{Twisted mass gauge ensembles at physical values of
  the light, strange and charm quark masses}},
  \href{https://doi.org/10.22323/1.396.0284}{\emph{PoS} {\bfseries LATTICE2021}
  (2022) 284} [\href{https://arxiv.org/abs/2201.02551}{{\ttfamily
  2201.02551}}].

\bibitem{Hatton:2021dvg}
D.~Hatton, C.T.H.~Davies, J.~Koponen, G.P.~Lepage and A.T.~Lytle,
  \emph{{Bottomonium precision tests from full lattice QCD: Hyperfine
  splitting, $\Upsilon$ leptonic width, and b quark contribution to $e^+e^-
  \to$ hadrons}},
  \href{https://doi.org/10.1103/PhysRevD.103.054512}{\emph{Phys. Rev. D}
  {\bfseries 103} (2021) 054512}
  [\href{https://arxiv.org/abs/2101.08103}{{\ttfamily 2101.08103}}].

\bibitem{Harlander:2002ur}
R.V.~Harlander and M.~Steinhauser, \emph{{rhad: A Program for the evaluation of
  the hadronic R ratio in the perturbative regime of QCD}},
  \href{https://doi.org/10.1016/S0010-4655(03)00204-2}{\emph{Comput. Phys.
  Commun.} {\bfseries 153} (2003) 244}
  [\href{https://arxiv.org/abs/hep-ph/0212294}{{\ttfamily hep-ph/0212294}}].

\bibitem{KNT}
A.~Keshavarzi, D.~Nomura and T.~Teubner, \emph{private communication},  2022.

\bibitem{Keshavarzi:2020bfy}
A.~Keshavarzi, W.J.~Marciano, M.~Passera and A.~Sirlin, \emph{{Muon $g-2$ and
  $\Delta \alpha$ connection}},
  \href{https://doi.org/10.1103/PhysRevD.102.033002}{\emph{Phys. Rev. D}
  {\bfseries 102} (2020) 033002}
  [\href{https://arxiv.org/abs/2006.12666}{{\ttfamily 2006.12666}}].

\bibitem{Crivellin:2020zul}
A.~Crivellin, M.~Hoferichter, C.A.~Manzari and M.~Montull, \emph{{Hadronic
  Vacuum Polarization: $(g-2)_\mu$ versus Global Electroweak Fits}},
  \href{https://doi.org/10.1103/PhysRevLett.125.091801}{\emph{Phys. Rev. Lett.}
  {\bfseries 125} (2020) 091801}
  [\href{https://arxiv.org/abs/2003.04886}{{\ttfamily 2003.04886}}].

\bibitem{Malaescu:2020zuc}
B.~Malaescu and M.~Schott, \emph{{Impact of correlations between $a_{\mu }$ and
  $\alpha _\text {QED}$ on the EW fit}},
  \href{https://doi.org/10.1140/epjc/s10052-021-08848-9}{\emph{Eur. Phys. J. C}
  {\bfseries 81} (2021) 46} [\href{https://arxiv.org/abs/2008.08107}{{\ttfamily
  2008.08107}}].

\bibitem{DiLuzio:2021uty}
L.~Di~Luzio, A.~Masiero, P.~Paradisi and M.~Passera, \emph{{New physics behind
  the new muon g-2 puzzle?}},
  \href{https://doi.org/10.1016/j.physletb.2022.137037}{\emph{Phys. Lett. B}
  {\bfseries 829} (2022) 137037}
  [\href{https://arxiv.org/abs/2112.08312}{{\ttfamily 2112.08312}}].

\bibitem{Ce:2022eix}
M.~C\`e, A.~G\'erardin, G.~von Hippel, H.B.~Meyer, K.~Miura, K.~Ottnad et~al.,
  \emph{{The hadronic running of the electromagnetic coupling and the
  electroweak mixing angle from lattice QCD}},
  \href{https://doi.org/10.1007/JHEP08(2022)220}{\emph{JHEP} {\bfseries 08}
  (2022) 220} [\href{https://arxiv.org/abs/2203.08676}{{\ttfamily
  2203.08676}}].

\bibitem{Giusti:2017jof}
D.~Giusti, V.~Lubicz, G.~Martinelli, F.~Sanfilippo and S.~Simula,
  \emph{{Strange and charm HVP contributions to the muon ($g - 2)$ including
  QED corrections with twisted-mass fermions}},
  \href{https://doi.org/10.1007/JHEP10(2017)157}{\emph{JHEP} {\bfseries 10}
  (2017) 157} [\href{https://arxiv.org/abs/1707.03019}{{\ttfamily
  1707.03019}}].

\bibitem{Giusti:2018mdh}
D.~Giusti, F.~Sanfilippo and S.~Simula, \emph{{Light-quark contribution to the
  leading hadronic vacuum polarization term of the muon $g-2$ from twisted-mass
  fermions}}, \href{https://doi.org/10.1103/PhysRevD.98.114504}{\emph{Phys.
  Rev. D} {\bfseries 98} (2018) 114504}
  [\href{https://arxiv.org/abs/1808.00887}{{\ttfamily 1808.00887}}].

\bibitem{Giusti:2019xct}
D.~Giusti, V.~Lubicz, G.~Martinelli, F.~Sanfilippo and S.~Simula,
  \emph{{Electromagnetic and strong isospin-breaking corrections to the muon $g
  - 2$ from Lattice QCD+QED}},
  \href{https://doi.org/10.1103/PhysRevD.99.114502}{\emph{Phys. Rev. D}
  {\bfseries 99} (2019) 114502}
  [\href{https://arxiv.org/abs/1901.10462}{{\ttfamily 1901.10462}}].

\bibitem{Bernecker:2011gh}
D.~Bernecker and H.B.~Meyer, \emph{{Vector Correlators in Lattice QCD: Methods
  and applications}},
  \href{https://doi.org/10.1140/epja/i2011-11148-6}{\emph{Eur. Phys. J. A}
  {\bfseries 47} (2011) 148} [\href{https://arxiv.org/abs/1107.4388}{{\ttfamily
  1107.4388}}].

\bibitem{ExtendedTwistedMass:2021gbo}
{\scshape Extended Twisted Mass} collaboration, \emph{{Quark masses using
  twisted-mass fermion gauge ensembles}},
  \href{https://doi.org/10.1103/PhysRevD.104.074515}{\emph{Phys. Rev. D}
  {\bfseries 104} (2021) 074515}
  [\href{https://arxiv.org/abs/2104.13408}{{\ttfamily 2104.13408}}].

\bibitem{deDivitiis:2011eh}
G.M.~de~Divitiis et~al., \emph{{Isospin breaking effects due to the up-down
  mass difference in Lattice QCD}},
  \href{https://doi.org/10.1007/JHEP04(2012)124}{\emph{JHEP} {\bfseries 04}
  (2012) 124} [\href{https://arxiv.org/abs/1110.6294}{{\ttfamily 1110.6294}}].

\bibitem{deDivitiis:2013xla}
{\scshape RM123} collaboration, \emph{{Leading isospin breaking effects on the
  lattice}}, \href{https://doi.org/10.1103/PhysRevD.87.114505}{\emph{Phys. Rev.
  D} {\bfseries 87} (2013) 114505}
  [\href{https://arxiv.org/abs/1303.4896}{{\ttfamily 1303.4896}}].

\bibitem{Luscher:1985dn}
M.~Luscher, \emph{{Volume Dependence of the Energy Spectrum in Massive Quantum
  Field Theories. 1. Stable Particle States}},
  \href{https://doi.org/10.1007/BF01211589}{\emph{Commun. Math. Phys.}
  {\bfseries 104} (1986) 177}.

\bibitem{Luscher:1986pf}
M.~Luscher, \emph{{Volume Dependence of the Energy Spectrum in Massive Quantum
  Field Theories. 2. Scattering States}},
  \href{https://doi.org/10.1007/BF01211097}{\emph{Commun. Math. Phys.}
  {\bfseries 105} (1986) 153}.

\bibitem{Luscher:1990ux}
M.~Luscher, \emph{{Two particle states on a torus and their relation to the
  scattering matrix}},
  \href{https://doi.org/10.1016/0550-3213(91)90366-6}{\emph{Nucl. Phys. B}
  {\bfseries 354} (1991) 531}.

\bibitem{Luscher:1991cf}
M.~Luscher, \emph{{Signatures of unstable particles in finite volume}},
  \href{https://doi.org/10.1016/0550-3213(91)90584-K}{\emph{Nucl. Phys. B}
  {\bfseries 364} (1991) 237}.

\bibitem{Lellouch:2000pv}
L.~Lellouch and M.~Luscher, \emph{{Weak transition matrix elements from finite
  volume correlation functions}},
  \href{https://doi.org/10.1007/s002200100410}{\emph{Commun. Math. Phys.}
  {\bfseries 219} (2001) 31}
  [\href{https://arxiv.org/abs/hep-lat/0003023}{{\ttfamily hep-lat/0003023}}].

\bibitem{Meyer:2011um}
H.B.~Meyer, \emph{{Lattice QCD and the Timelike Pion Form Factor}},
  \href{https://doi.org/10.1103/PhysRevLett.107.072002}{\emph{Phys. Rev. Lett.}
  {\bfseries 107} (2011) 072002}
  [\href{https://arxiv.org/abs/1105.1892}{{\ttfamily 1105.1892}}].

\bibitem{Francis:2013fzp}
A.~Francis, B.~Jaeger, H.B.~Meyer and H.~Wittig, \emph{{A new representation of
  the Adler function for lattice QCD}},
  \href{https://doi.org/10.1103/PhysRevD.88.054502}{\emph{Phys. Rev. D}
  {\bfseries 88} (2013) 054502}
  [\href{https://arxiv.org/abs/1306.2532}{{\ttfamily 1306.2532}}].

\bibitem{Gounaris:1968mw}
G.J.~Gounaris and J.J.~Sakurai, \emph{{Finite width corrections to the vector
  meson dominance prediction for $\rho \to e^+ e^-$}},
  \href{https://doi.org/10.1103/PhysRevLett.21.244}{\emph{Phys. Rev. Lett.}
  {\bfseries 21} (1968) 244}.

\bibitem{DellaMorte:2008xb}
M.~Della~Morte, R.~Sommer and S.~Takeda, \emph{{On cutoff effects in lattice
  QCD from short to long distances}},
  \href{https://doi.org/10.1016/j.physletb.2009.01.059}{\emph{Phys. Lett. B}
  {\bfseries 672} (2009) 407}
  [\href{https://arxiv.org/abs/0807.1120}{{\ttfamily 0807.1120}}].

\bibitem{Ce:2021xgd}
M.~C\`e, T.~Harris, H.B.~Meyer, A.~Toniato and C.~T\"or\"ok, \emph{{Vacuum
  correlators at short distances from lattice QCD}},
  \href{https://doi.org/10.1007/JHEP12(2021)215}{\emph{JHEP} {\bfseries 12}
  (2021) 215} [\href{https://arxiv.org/abs/2106.15293}{{\ttfamily
  2106.15293}}].

\bibitem{Husung:2019ytz}
N.~Husung, P.~Marquard and R.~Sommer, \emph{{Asymptotic behavior of cutoff
  effects in Yang\textendash{}Mills theory and in Wilson\textquoteright{}s
  lattice QCD}},
  \href{https://doi.org/10.1140/epjc/s10052-020-7685-4}{\emph{Eur. Phys. J. C}
  {\bfseries 80} (2020) 200}
  [\href{https://arxiv.org/abs/1912.08498}{{\ttfamily 1912.08498}}].

\bibitem{EuropeanTwistedMass:2014osg}
{\scshape European Twisted Mass} collaboration, \emph{{Up, down, strange and
  charm quark masses with N$_f$ = 2+1+1 twisted mass lattice QCD}},
  \href{https://doi.org/10.1016/j.nuclphysb.2014.07.025}{\emph{Nucl. Phys. B}
  {\bfseries 887} (2014) 19} [\href{https://arxiv.org/abs/1403.4504}{{\ttfamily
  1403.4504}}].

\bibitem{Akaike}
H.~Akaike, \emph{{A new look at the statistical model identification}},
  {\emph{IEEE Transactions on Automatic Control} {\bfseries 19} (1974) 716}.

\bibitem{Neil:2022joj}
E.T.~Neil and J.W.~Sitison, \emph{{Improved information criteria for Bayesian
  model averaging in lattice field theory}},
  \href{https://arxiv.org/abs/2208.14983}{{\ttfamily 2208.14983}}.

\bibitem{Giusti:2021dvd}
D.~Giusti and S.~Simula, \emph{{Window contributions to the muon hadronic
  vacuum polarization with twisted-mass fermions}},
  \href{https://doi.org/10.22323/1.396.0189}{\emph{PoS} {\bfseries LATTICE2021}
  (2022) 189} [\href{https://arxiv.org/abs/2111.15329}{{\ttfamily
  2111.15329}}].

\bibitem{McNeile:2006bz}
{\scshape UKQCD} collaboration, \emph{{Decay width of light quark hybrid meson
  from the lattice}},
  \href{https://doi.org/10.1103/PhysRevD.73.074506}{\emph{Phys. Rev. D}
  {\bfseries 73} (2006) 074506}
  [\href{https://arxiv.org/abs/hep-lat/0603007}{{\ttfamily hep-lat/0603007}}].

\bibitem{Gambhir:2016uwp}
A.S.~Gambhir, A.~Stathopoulos and K.~Orginos, \emph{{Deflation as a Method of
  Variance Reduction for Estimating the Trace of a Matrix Inverse}},
  \href{https://doi.org/10.1137/16M1066361}{\emph{SIAM J. Sci. Comput.}
  {\bfseries 39} (2017) A532}
  [\href{https://arxiv.org/abs/1603.05988}{{\ttfamily 1603.05988}}].

\bibitem{Stathopoulos:2013aci}
A.~Stathopoulos, J.~Laeuchli and K.~Orginos, \emph{{Hierarchical probing for
  estimating the trace of the matrix inverse on toroidal lattices}},
  \href{https://arxiv.org/abs/1302.4018}{{\ttfamily 1302.4018}}.

\bibitem{Lehner:2020crt}
C.~Lehner and A.S.~Meyer, \emph{{Consistency of hadronic vacuum polarization
  between lattice QCD and the R-ratio}},
  \href{https://doi.org/10.1103/PhysRevD.101.074515}{\emph{Phys. Rev. D}
  {\bfseries 101} (2020) 074515}
  [\href{https://arxiv.org/abs/2003.04177}{{\ttfamily 2003.04177}}].

\bibitem{Aubin:2022hgm}
C.~Aubin, T.~Blum, M.~Golterman and S.~Peris, \emph{{Muon anomalous magnetic
  moment with staggered fermions: Is the lattice spacing small enough?}},
  \href{https://doi.org/10.1103/PhysRevD.106.054503}{\emph{Phys. Rev. D}
  {\bfseries 106} (2022) 054503}
  [\href{https://arxiv.org/abs/2204.12256}{{\ttfamily 2204.12256}}].

\bibitem{Aubin:2019usy}
C.~Aubin, T.~Blum, C.~Tu, M.~Golterman, C.~Jung and S.~Peris, \emph{{Light
  quark vacuum polarization at the physical point and contribution to the muon
  $g-2$}}, \href{https://doi.org/10.1103/PhysRevD.101.014503}{\emph{Phys. Rev.
  D} {\bfseries 101} (2020) 014503}
  [\href{https://arxiv.org/abs/1905.09307}{{\ttfamily 1905.09307}}].

\bibitem{Wang:2022lkq}
{\scshape chiQCD} collaboration, \emph{{Muon g-2 with overlap valence
  fermion}},  \href{https://arxiv.org/abs/2204.01280}{{\ttfamily 2204.01280}}.

\bibitem{FermilabLattice:2022izv}
{\scshape Fermilab Lattice, MILC, HPQCD} collaboration, \emph{{Windows on the
  hadronic vacuum polarization contribution to the muon anomalous magnetic
  moment}}, \href{https://doi.org/10.1103/PhysRevD.106.074509}{\emph{Phys. Rev.
  D} {\bfseries 106} (2022) 074509}
  [\href{https://arxiv.org/abs/2207.04765}{{\ttfamily 2207.04765}}].

\bibitem{Clark:2009wm}
M.A.~Clark, R.~Babich, K.~Barros, R.C.~Brower and C.~Rebbi, \emph{{Solving
  Lattice QCD systems of equations using mixed precision solvers on GPUs}},
  \href{https://doi.org/10.1016/j.cpc.2010.05.002}{\emph{Comput. Phys. Commun.}
  {\bfseries 181} (2010) 1517}
  [\href{https://arxiv.org/abs/0911.3191}{{\ttfamily 0911.3191}}].

\bibitem{Babich:2011np}
R.~Babich, M.A.~Clark, B.~Joo, G.~Shi, R.C.~Brower and S.~Gottlieb,
  \emph{{Scaling Lattice QCD beyond 100 GPUs}},  in \emph{{SC11 International
  Conference for High Performance Computing, Networking, Storage and Analysis
  Seattle, Washington, November 12-18, 2011}}, 2011,
  \href{https://doi.org/10.1145/2063384.2063478}{DOI}
  [\href{https://arxiv.org/abs/1109.2935}{{\ttfamily 1109.2935}}].

\bibitem{Clark:2016rdz}
M.A.~Clark, B.~Jo{\'o}, A.~Strelchenko, M.~Cheng, A.~Gambhir and R.C.~Brower,
  \emph{{Accelerating Lattice QCD Multigrid on GPUs Using Fine-Grained
  Parallelization}},  in \emph{SC '16: Proceedings of the International
  Conference for High Performance Computing, Networking, Storage and Analysis},
  pp.~795--806, 2016, \href{https://doi.org/10.1109/SC.2016.67}{DOI}
  [\href{https://arxiv.org/abs/1612.07873}{{\ttfamily 1612.07873}}].

\bibitem{JUWELS}
{J\"{u}lich Supercomputing Centre}, \emph{{JUWELS: Modular Tier-0/1
  Supercomputer at the J\"{u}lich Supercomputing Centre}},
  \href{https://doi.org/10.17815/jlsrf-5-171}{\emph{Journal of large-scale
  research facilities} {\bfseries 5} (2019) }.

\bibitem{Jureca}
{J\"{u}lich Supercomputing Centre}, \emph{{JURECA: Modular supercomputer at
  J\"{u}lich Supercomputing Centre}},
  \href{https://doi.org/10.17815/jlsrf-4-121-1}{\emph{Journal of large-scale
  research facilities} {\bfseries 4} (2018) }.

\bibitem{Frezzotti:2000nk}
{\scshape Alpha} collaboration, \emph{{Lattice QCD with a chirally twisted mass
  term}}, \href{https://doi.org/10.1088/1126-6708/2001/08/058}{\emph{JHEP}
  {\bfseries 08} (2001) 058}
  [\href{https://arxiv.org/abs/hep-lat/0101001}{{\ttfamily hep-lat/0101001}}].

\bibitem{Frezzotti:2003ni}
R.~Frezzotti and G.C.~Rossi, \emph{{Chirally improving Wilson fermions. 1. O(a)
  improvement}},
  \href{https://doi.org/10.1088/1126-6708/2004/08/007}{\emph{JHEP} {\bfseries
  08} (2004) 007} [\href{https://arxiv.org/abs/hep-lat/0306014}{{\ttfamily
  hep-lat/0306014}}].

\bibitem{Osterwalder:1977pc}
K.~Osterwalder and E.~Seiler, \emph{{Gauge Field Theories on the Lattice}},
  \href{https://doi.org/10.1016/0003-4916(78)90039-8}{\emph{Annals Phys.}
  {\bfseries 110} (1978) 440}.

\bibitem{Frezzotti:2004wz}
R.~Frezzotti and G.C.~Rossi, \emph{{Chirally improving Wilson fermions. II.
  Four-quark operators}},
  \href{https://doi.org/10.1088/1126-6708/2004/10/070}{\emph{JHEP} {\bfseries
  10} (2004) 070} [\href{https://arxiv.org/abs/hep-lat/0407002}{{\ttfamily
  hep-lat/0407002}}].

\bibitem{Iwasaki:1985we}
Y.~Iwasaki, \emph{{Renormalization Group Analysis of Lattice Theories and
  Improved Lattice Action: Two-Dimensional Nonlinear O(N) Sigma Model}},
  \href{https://doi.org/10.1016/0550-3213(85)90606-6}{\emph{Nucl. Phys. B}
  {\bfseries 258} (1985) 141}.

\bibitem{Sheikholeslami:1985ij}
B.~Sheikholeslami and R.~Wohlert, \emph{{Improved Continuum Limit Lattice
  Action for QCD with Wilson Fermions}},
  \href{https://doi.org/10.1016/0550-3213(85)90002-1}{\emph{Nucl. Phys. B}
  {\bfseries 259} (1985) 572}.

\bibitem{Aoki:1998qd}
S.~Aoki, R.~Frezzotti and P.~Weisz, \emph{{Computation of the improvement
  coefficient c(SW) to one loop with improved gluon actions}},
  \href{https://doi.org/10.1016/S0550-3213(98)00742-1}{\emph{Nucl. Phys. B}
  {\bfseries 540} (1999) 501}
  [\href{https://arxiv.org/abs/hep-lat/9808007}{{\ttfamily hep-lat/9808007}}].

\bibitem{Frezzotti:2005gi}
R.~Frezzotti, G.~Martinelli, M.~Papinutto and G.C.~Rossi, \emph{{Reducing
  cutoff effects in maximally twisted lattice QCD close to the chiral limit}},
  \href{https://doi.org/10.1088/1126-6708/2006/04/038}{\emph{JHEP} {\bfseries
  04} (2006) 038} [\href{https://arxiv.org/abs/hep-lat/0503034}{{\ttfamily
  hep-lat/0503034}}].

\bibitem{ETM:2015ned}
{\scshape ETM} collaboration, \emph{{First physics results at the physical pion
  mass from $N_f=2$ Wilson twisted mass fermions at maximal twist}},
  \href{https://doi.org/10.1103/PhysRevD.95.094515}{\emph{Phys. Rev. D}
  {\bfseries 95} (2017) 094515}
  [\href{https://arxiv.org/abs/1507.05068}{{\ttfamily 1507.05068}}].

\bibitem{FlavourLatticeAveragingGroupFLAG:2021npn}
{\scshape Flavour Lattice Averaging Group (FLAG)} collaboration, \emph{{FLAG
  Review 2021}},
  \href{https://doi.org/10.1140/epjc/s10052-022-10536-1}{\emph{Eur. Phys. J. C}
  {\bfseries 82} (2022) 869}
  [\href{https://arxiv.org/abs/2111.09849}{{\ttfamily 2111.09849}}].

\bibitem{ParticleDataGroup:2016lqr}
{\scshape Particle Data Group} collaboration, \emph{{Review of Particle
  Physics}}, \href{https://doi.org/10.1088/1674-1137/40/10/100001}{\emph{Chin.
  Phys. C} {\bfseries 40} (2016) 100001}.

\bibitem{Bochicchio:1985xa}
M.~Bochicchio, L.~Maiani, G.~Martinelli, G.C.~Rossi and M.~Testa, \emph{{Chiral
  Symmetry on the Lattice with Wilson Fermions}},
  \href{https://doi.org/10.1016/0550-3213(85)90290-1}{\emph{Nucl. Phys. B}
  {\bfseries 262} (1985) 331}.

\bibitem{Bhattacharya:2005rb}
T.~Bhattacharya, R.~Gupta, W.~Lee, S.R.~Sharpe and J.M.S.~Wu, \emph{{Improved
  bilinears in lattice QCD with non-degenerate quarks}},
  \href{https://doi.org/10.1103/PhysRevD.73.034504}{\emph{Phys. Rev. D}
  {\bfseries 73} (2006) 034504}
  [\href{https://arxiv.org/abs/hep-lat/0511014}{{\ttfamily hep-lat/0511014}}].

\bibitem{Colangelo:2005gd}
G.~Colangelo, S.~Durr and C.~Haefeli, \emph{{Finite volume effects for meson
  masses and decay constants}},
  \href{https://doi.org/10.1016/j.nuclphysb.2005.05.015}{\emph{Nucl. Phys. B}
  {\bfseries 721} (2005) 136}
  [\href{https://arxiv.org/abs/hep-lat/0503014}{{\ttfamily hep-lat/0503014}}].

\bibitem{DellaMorte:2017dyu}
M.~Della~Morte, A.~Francis, V.~G\"ulpers, G.~Herdo\'\i{}za, G.~von Hippel,
  H.~Horch et~al., \emph{{The hadronic vacuum polarization contribution to the
  muon $g-2$ from lattice QCD}},
  \href{https://doi.org/10.1007/JHEP10(2017)020}{\emph{JHEP} {\bfseries 10}
  (2017) 020} [\href{https://arxiv.org/abs/1705.01775}{{\ttfamily
  1705.01775}}].

\bibitem{ParticleDataGroup:2020ssz}
{\scshape Particle Data Group} collaboration, \emph{{Review of Particle
  Physics}}, \href{https://doi.org/10.1093/ptep/ptaa104}{\emph{PTEP} {\bfseries
  2020} (2020) 083C01}.

\bibitem{ExtendedTwistedMass_RCs}
{\scshape ETM} collaboration, \emph{Non-perturbative renormalisation of quark
  bilinear operators with $n_f = 2+1+1$ wilson-clover twisted-mass fermions}, .

\bibitem{Hatton:2020qhk}
{\scshape HPQCD} collaboration, \emph{{Charmonium properties from lattice
  $QCD$+QED : Hyperfine splitting, $J/\psi$ leptonic width, charm quark mass,
  and $a^c_\mu$}},
  \href{https://doi.org/10.1103/PhysRevD.102.054511}{\emph{Phys. Rev. D}
  {\bfseries 102} (2020) 054511}
  [\href{https://arxiv.org/abs/2005.01845}{{\ttfamily 2005.01845}}].

\bibitem{Zhang:2021xrs}
R.~Zhang, W.~Sun, F.~Chen, Y.~Chen, M.~Gong, X.~Jiang et~al.,
  \emph{{Annihilation diagram contribution to charmonium masses}},
  \href{https://doi.org/10.1088/1674-1137/ac3d8c}{\emph{Chin. Phys. C}
  {\bfseries 46} (2022) 043102}
  [\href{https://arxiv.org/abs/2110.01755}{{\ttfamily 2110.01755}}].

\bibitem{Weinberg:1973xwm}
S.~Weinberg, \emph{{New approach to the renormalization group}},
  \href{https://doi.org/10.1103/PhysRevD.8.3497}{\emph{Phys. Rev. D} {\bfseries
  8} (1973) 3497}.

\bibitem{Constantinou:2014rka}
M.~Constantinou, M.~Hadjiantonis and H.~Panagopoulos, \emph{{Renormalization of
  Flavor Singlet and Nonsinglet Fermion Bilinear Operators}},
  \href{https://doi.org/10.22323/1.214.0298}{\emph{PoS} {\bfseries LATTICE2014}
  (2014) 298} [\href{https://arxiv.org/abs/1411.6990}{{\ttfamily 1411.6990}}].

\end{thebibliography}\endgroup
\bibliographystyle{JHEP}

\end{document}